\documentclass[12pt]{report}
\usepackage{newcuthesis}
\usepackage{graphicx}
\usepackage{horowitzThesis}
\usepackage{caption} 
\usepackage{slashed}
\usepackage{setspace} 
\usepackage{fancyhdr} 
\usepackage{amsmath}
\usepackage{mathrsfs}
\usepackage{amssymb}
\usepackage{revsymb}
\usepackage[numbers,square,sort&compress]{natbib} 
\setcitestyle{citesep={,}}
\usepackage{notoccite} 
\usepackage{color}
\usepackage[pagebackref=true]{hyperref} 

\newcommand{\thesisauthor}{William A.\ Horowitz}
     
\newcommand{\thesisyear}{2008}
\newcommand{\thesistitle}{Probing the Frontiers in QCD}

\definecolor{darkgreen}{rgb}{0,.7,0}

\hypersetup{
    unicode=true,          
    pdftitle={\thesistitle},    
    pdfauthor={\thesisauthor},     
    pdfnewwindow=true,      
    bookmarksnumbered=true,
    colorlinks=true,       
    linkcolor=red,          
    citecolor=darkgreen,        
    filecolor=magenta,      
    urlcolor=cyan,           
    menucolor=blue,
    baseurl=http://www.arxiv.org/abs/,
    linktocpage=true
}

\doublespace
\newcommand{\FMslash}[1]{\slashed{#1}}

\begin{document}
\pagenumbering{alph} 
\pdfbookmark[0]{Preamble}{Preamble}
\thesistitlepage 

\thesiscopyrightpage  


\begin{thesisabstract}
With the energy scales opened up by \rhic and \lhc the age of \highpt physics is upon us.  This has created new opportunities and novel mysteries, both of which will be explored in this thesis.  The possibility now exists experimentally to exploit these high momentum particles to uniquely probe the unprecedented state of matter produced in heavy ion collisions.  At the same time na\"ive theoretical expectations have been dashed by data.

The first puzzle we confront is that of the enormous \intermediatept azimuthal anisotropy, or \vtwocomma, of jets observed at \rhiccomma.  Typical lines of reasoning lead to an anticorrelation between \vtwo and the overall jet suppression, \raacomma; the larger the \vtwo the smaller the \raacomma.  By simultaneously plotting the two this relationship becomes manifest and it is clear that usual energy loss mechanisms cannot reproduce the observed pattern---while jets are suppressed, the \vtwo is anomalously large compared to the quenching.  
We argue that the data can be reproduced by a focusing of the partonic jets caused by processes associated with a deconfined quark-gluon plasma.

The second puzzle is the surprisingly similar suppression of light mesons and nonphotonic electrons, which precludes perturbative predictions predicated on gluon bremsstrahlung radiation as the dominant energy loss channel.  Near qualitative agreement results from including collisional energy loss and integrating over the fluctuating jet pathlengths.  

Another conjecture for heavy quark energy loss comes via explicit construction using the AdS/CFT correspondence; the momentum loss of a hanging dragging string moving through the deconfined plasma leads to qualitative agreement with heavy quark decay data.  We propose a robust test to experimentally differentiate these two competing ideas: the ratio of charm to bottom suppression rapidly approaches 1 for pQCD but is independent of momentum and well below 1 for AdS/CFT.

Finally as a warmup problem to calculating the photon bremsstrahlung associated with jet energy loss we quantify improvements to the perturbative estimates of the Ter-Mikayelian effect.  By not neglecting the interference from away-side jets we find agreement with our results and the classical limit, regulate divergences at low momenta, and note the importance of terms neglected in previous in-medium radiative energy loss derivations.

\end{thesisabstract}

\pagenumbering{roman}
\setcounter{page}{1}
\pagestyle{plain}
\pdfbookmark[1]{Table of Contents}{TableofContents}
\tableofcontents
\newpage
\phantomsection
\addcontentsline{toc}{section}{List of Figures}
\listoffigures

\begin{thesisacknowledgments}
\thispagestyle{plain}

Just the other day a new neighbor moved into my apartment building.  He too grew up in Atlanta, and will enter as a graduate student in the Math Ph.\ D.\ program.  He looked so young!  Somehow more than writing my thesis and preparing to move to Ohio, seeing him made me reflect back on my time at Columbia.  I've changed so much!  Beyond even my development as a scientist I realized I am so much more self confident and mature than when I came here, five years ago.  And I know I owe much of these changes to Miklos.  It's not been an easy road.  I always excelled in the formal structure of the classroom; I'm amazed by how Miklos helped to guide my difficult transition to independent researcher.  I'm grateful to take away some of his love for esoteria, hard work, and simple analytic estimation.  And I don't know of any other advisor who so actively supports his students' nascent careers.  In the recent \TECHQM phonecalls I haven't even needed to announce my name when talking; everyone in the field knows me by voice---although there is the possibility that this is more of a statement about me!

I've also felt and appreciated the support for me as an apprentice scholar throughout the field, first and foremost from the Columbia group: I've enjoyed and benefited from the discussions with Al Mueller, Bill Zajc, and Brian Cole.  I wouldn't have had the opportunity to stay in Frankfurt without the help of Walter Greiner, Carsten Greiner, Dirk Rischke, and especially Horst St\"ocker, all of whom I learned from.  I want to thank Ivan Vitev for arranging for me to stay, work, and study with him at Los Alamos.  And to all the people who've given me a chance to visit and be enlightened outside Columbia in my graduate career: John Harris and Helen Caines at Yale; everyone at BNL, but especially Dima Kharzeev, Larry McLerran, Rob Pisarski, and Raju Venugopalan; Ralf Rapp at Texas A\&M; Volcker Koch, Nu Xu, and Xin-Nian Wang at LBNL; Urs Wiedemann and Karel Safarik in \textsc{Cern}; Peter Levai in Budapest; Charles Gale, Sangyong Jeon, and Guy Moore at McGill; and of course Ulrich Heinz and Yuri Kovchegov at OSU for giving me a job!  Then there's pretty much everyone else in a field filled with generally kind and perspicacious people.  
Of especial note are Magdalena and Denes, Joerg Aichelin, Nestor Armesto, Steffan Bass, Jorge Casalderrey-Solana, Hendrik van Hees, Abhijit Majumder, Berndt Mueller, Jamie Nagle, Andre Peshier, Carlos Salgado, and Derek Teaney.    

Azfar Adil, Simon Wicks, Kurt Hinterbichler, Tatia Englemore, Ali Hanks, and Eric Vazquez, my Columbia colleagues, I am so grateful for your support and illuminating conversations.  I've valued your friendship over these years, and I'll miss all of you.  

And for always being there, my friends Ian, Phil, and Young from back home.  

Finally, to Mr.\ Owens for inspiring me to a career in physics with girls swinging pianos in circles around their heads, Porsches racing off starting lines, and cones tipping over ever-steepening ledges: thank you.
\end{thesisacknowledgments}

\thesisdedicationpage

\pagenumbering{arabic}
\setcounter{page}{1}

\pagestyle{fancy}
\renewcommand{\chaptermark}[1]{\markboth{#1}{}}
\rhead{\thepage}
\lhead{Chapter \arabic{chapter}: \leftmark}
\cfoot{}
\renewcommand{\footrulewidth}{0.4pt}

\fancypagestyle{plain}{
\renewcommand{\headrulewidth}{0pt}
\renewcommand{\footrulewidth}{0pt}
\lhead{}
\chead{}
\rhead{\thepage}
\lfoot{}
\rfoot{}
\cfoot{}
\renewcommand{\footrulewidth}{0.4pt}
}

\mychapter{Introduction}{chapter:intro}
\section{Philosophy of Physics}
It's hard to overemphasize the importance of the rational-reductionist tradition begun when Thales 
claimed that ``All things are made of water'' \cite{aristotle}.  However the ancient Greek philosophers did not believe in the usefulness of experimentation; their investigations into Nature came from reasoning alone.  Of course this emphasis on logic over the physical caused certain difficulties, culminating most famously with Zeno and his arguments against the possibility of motion \cite{aristotle:zeno}.  

Modern philosophy of science places primacy in experiment: the usefulness of a scientific theory is measured by its falsifiability \cite{popper}.  And while many in the string theory community have reverted to measuring progress based on aesthetics \cite{newyorker}, I am a devout Popperite.

This thesis is broadly organized as follows.  The experimental and theoretical advances leading to QCD as the unique theory of the strong nuclear force are reviewed.  The existence of a transition from normal nuclear matter to a novel state is motivated, and some of the traditional theoretical tools and signatures of such are described.  This places the work of the first half of the thesis in context.  Then the recent application of conjectured strongly-coupled methods of AdS/CFT to heavy ion physics is detailed, which the latter half of this thesis proposes to test experimentally.

\section{Experimental Measurements Leading \texorpdfstring{\\}{} to QCD}
Any history of science is necessarily revisionist.  The path to our current understanding was not straight and certainly not chronological; in fact, there were plenty of dead ends.  This is of course not what we as physicists would like to think, and in our attempt to make sense 
of the past the work of many is elided while a lucky few  
are picked 
out as the representatives for the discovery of now-obvious results.  As this is not a thesis on the history of 
physics I will, with some regret, continue in this tradition; additionally I will focus on the developments leading to QCD as the fundamental theory of the strong force.  For a more detailed narrative of physics from the late 19th century and beyond see, e.g., \cite{pais,neeman,Griffiths:1987tjB} and references therein.

Reaching all the way back to the Greeks again, Leucippus and his pupil Democritus were the first to postulate atomic theory: matter (as opposed to the void) is made up of indestructible, individual particles \cite{aristotle:democritus}.  
It took until the early 19$^\textrm{th}$ century for Dalton, inspired by his own experiments and the experimentally derived laws of conservation of mass \cite{lavoisier} and definite proportions \cite{proust} in chemical reactions, to propose the precursor to the modern, scientific atomic theory \cite{dalton1,dalton2}.

About a hundred years later, atoms began falling apart.  In 1897 J.\ J.\ Thomson definitively demonstrated that cathode rays are made up of negatively charged particles, electrons \cite{Thomson:1897cm}.  
Soon afterward, Rutherford's \cite{Rutherford:1911zz} observation of large angle scattering from gold foil showed that the majority of atomic mass is found in a minuscule, positively charged nucleus.  
It became clear from the transmutation experiments begun by Rutherford that the hydrogen nucleus was one of the fundamental building blocks of all other nuclei; 
as such he named it the proton \cite{rutherford:proton}
.  Chadwick's discovery of the neutron, simultaneously explaining the charge-mass asymmetry and providing a means of keeping the positive charge within the nucleus, completed the discovery of the constituents of atoms \cite{Chadwick:1932ma}.

Yukawa \cite{Yukawa:1935xg} developed a theory for the force binding nuclei together, positing the existence of an as-yet unmeasured particle whose mass in natural units is of the order of the nuclear radius, 1 fm$^{-1}$.  
After a period of confusion, Powell conducted the definitive experiments \cite{Lattes:1947mw,Lattes:1947mx} that disentangled the pion from the muon.  Then things got weird.  In December of that same year a kaon was first seen in a cosmic ray cloud chamber photograph \cite{Rochester:1947mi}.  Strange particles proliferated: the $\eta$, $\phi$, and $\omega$ mesons were found as was the $\Lambda$ baryon.  

During the time that the muon-pion puzzle was being sorted out, St\"uck\-el\-berg proposed the conservation of baryon number to explain the stability of protons.  Experimentally, strange particles are produced on short timescales but decay relatively slowly; Pais suggested \cite{Pais:1952} that their production and decay mechanisms were different.  Gell-Mann \cite{Gell-Mann:1953,Gell-Mann:1956} and Nishijima \cite{Nakano:1953} expanded on this and St\"uckelberg's idea by positing a new conserved quantity: strangeness.    

Gell-Mann, and independently Ne'eman (see \cite{Gell-Mann:1964} and references therein), 
began to understand the proliferation of hadrons by organizing them in multiplets that are representations of the group SU(3), famously predicting the existence of the $\Omega^-$ baryon (discovered in 1964 \cite{Barnes:1964pd}).  Even before this particle was observed, several authors (see \cite{Lichtenberg:1980,Greenberg:1980} and references therein) improved upon the phenomenological Eightfold Way by boldly postulating the existence of subnuclear structure.  Gell-Mann named these smaller, fundamental building blocks of hadrons quarks.

The quark model was initially enormously successful.  By taking these constituents to have spin-1/2 and fractional charge 2/3 for the $u$ and -1/3 for the $d$ and $s$, the spin, charge, and strangeness of all known hadrons was understood.  Moreover the multiplets and their mass-ordering were well described.  Using current algebra techniques scaling laws were proposed \cite{Bjorken:1968dy,Callan:1969uq} and found \cite{Bloom:1969kc,Breidenbach:1969kd}.  These two SLAC papers also found large-angle scattering from protons, which, like Rutherford's earlier experiments with atoms \cite{Rutherford:1911zz}, showed decisively that nucleons have substructure.

\begin{figure}[!htb]
\centering
$\begin{array}{cc}
\includegraphics[width=2.55in]{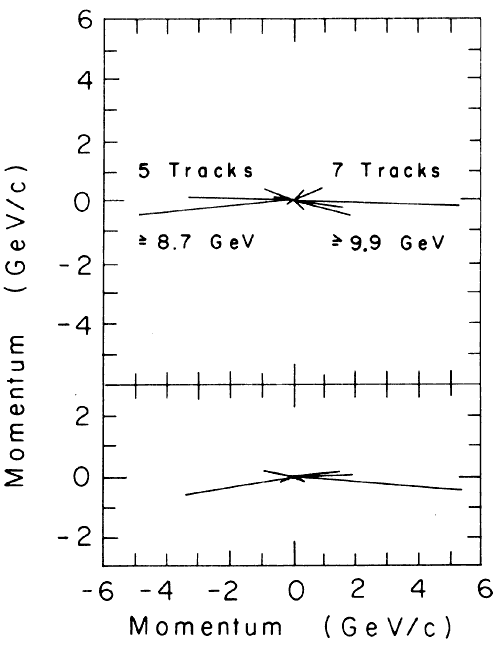} &
\includegraphics[width=2.5in]{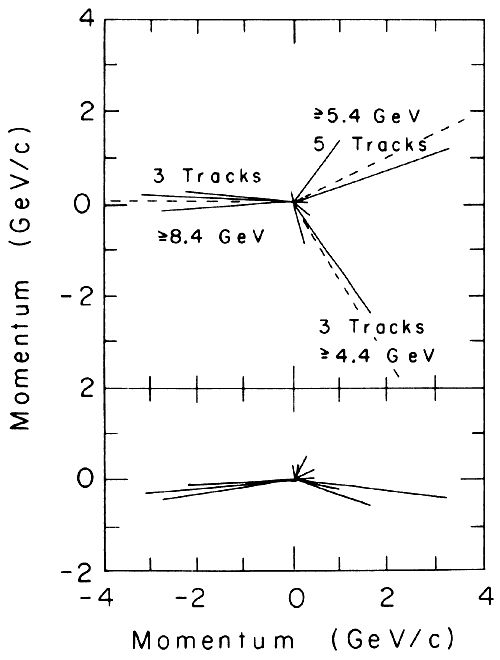} \\[-.05in]
\textrm{{\figsize(a)}} & \textrm{{\figsize(b)}}
\end{array}$
\caption{\label{intro:twoandthreejetevent}Momentum representation of two (a) two-jet events and (b) three-jet events.  Figure adapted from \cite{Brandelik:1979bd}.}
\end{figure}

There were two problems with the quark model at this time: quarks had not been observed and some baryons apparently violated the Pauli principle.  Although the terminology came later in \cite{Lichtenberg:1970}, Nambu and Han \cite{Han:1965pf} proposed a solution to the latter problem: quarks come not only in different flavors but also different colors.  Phenomenologically one could posit that nature requires color neutrality to explain the former.  In the same paper Nambu and Han introduced gauge vector fields associated with the quark color charge.  Feynman rederived one of the previously mentioned scaling relations after developing the parton theory of nucleons: at high $q^2$ nucleons look like they are made up of point-like objects \cite{Feynman:1969ej,Yang:1969,Feynman:1973xcB,Kogut:1972di}.  To agree with data nucleons have to be made up of not just the three valence quarks, but also sea quarks \cite{Bjorken:1969ja} and gluons \cite{Kuti:1971ph}.  
While no experiment has yet to directly observe a bare quark or gluon (nor do we expect them to) there is strong indirect evidence for quarks from two jet events \cite{Hanson:1975fe} and gluons from three \cite{Brandelik:1979bd} and four \cite{Abreu:1990ce,Decamp:1992ip} jet events; see \fig{intro:twoandthreejetevent}.

To describe the strong force binding quarks in nuclei, then, one wants a quantum field theory that becomes weaker at short distances.  The idea of a non-Abelian gauge theory was introduced by Yang and Mills \cite{Yang:1954ek}; quantization was achieved by Faddeev and Popov \cite{Faddeev:1967fc}, and 't Hooft proved their renormalizability \cite{'tHooft:1971fh}.  After asymptotic freedom was first shown for non-Abelian theories \cite{Gross:1973id,Politzer:1973fx}, it was proved that in four dimensions this property is unique to non-Abelian theories \cite{Coleman:1973sx}.  The infrared divergences of non-Abelian theories led to a natural explanation of confinement \cite{Weinberg:1973un,Gross:1973ju}, the inability to directly observe quarks or gluons.

Experimental evidence for QCD as the theory of strong interactions is legion.  At high momenta, for which perturbative methods are applicable, log violations of Bjorken scaling in deep inelastic scattering were predicted \cite{Gross:1973ju,Gross:1974cs,Georgi:1951sr} and observed \cite{Watanabe:1975su,Chang:1975sv}.  Next-to-leading order (NLO) calculations reproduce the world data of prompt photon production \cite{Aurenche:2006vj}.  Heavy quark jet production rates are calculable and agree with experiment (see \cite{Cacciari:2007hn} and references therein).  For a review comparing QCD theory and data at $e^+e^-$ colliders see \cite{Kluth:2006bw}.

\begin{figure}[htb!]
\centering
$\begin{array}{cc}
\includegraphics[width=3.3 in]{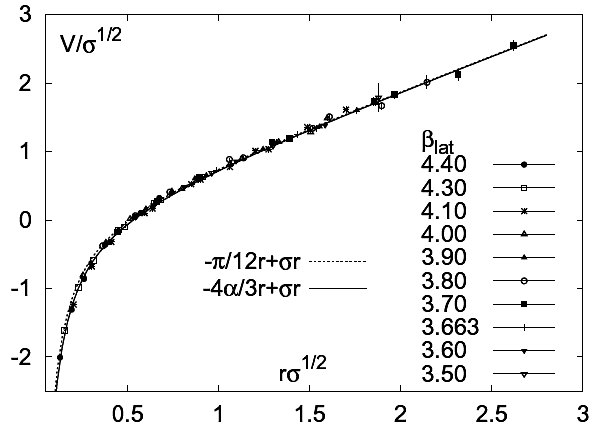} &
\includegraphics[width=1.83 in]{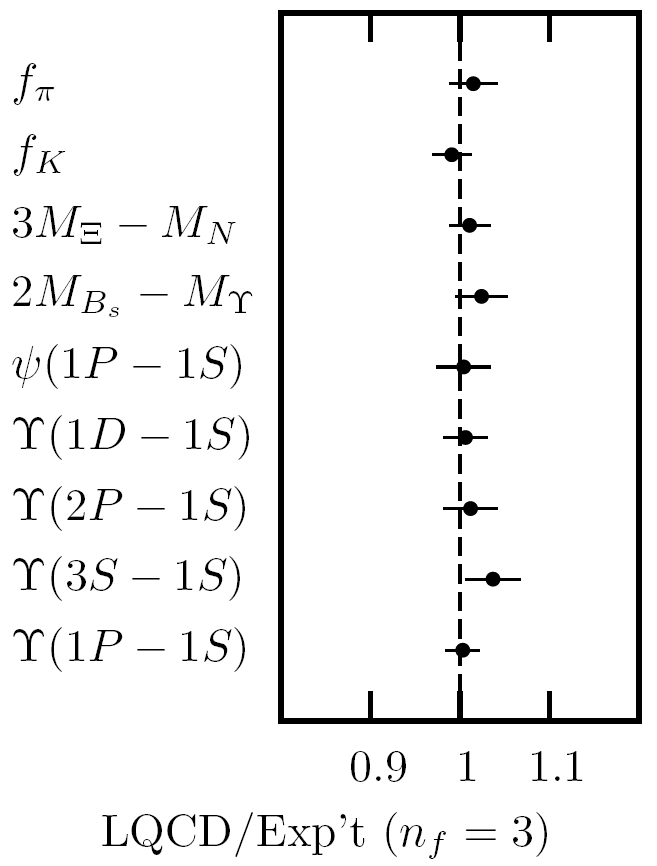} \\[-.05in]
\textrm{{\figsize(a)}} & \textrm{{\figsize(b)}}
\end{array}$
\caption{
\label{intro:lattice}
(a) Heavy quark potential at $T=0$ from LQCD calculations in \cite{Karsch:2000kv}.  The lattice results are well approximated by the Cornell potential, $V(r)=-\alpha/r+r$.  Figure adapted from \cite{Kaczmarek:2005ui}.  (b) Lattice QCD results divided by experimental data for a range of nonperturbative quantities; the values agree to within statistical and systematic errors of 3\% or less.  Figure adapted from \cite{Davies:2003ik}.}
\end{figure}

At low momenta lattice calculations \cite{Wilson:1974sk} show additional evidence for QCD.  $q\bar{q}$ pairs experience a linear potential, numerically suggesting a mechanism for confinement; see \fig{intro:lattice} (a).  Hadron masses found from lattice QCD are converging to those seen in the lab.  \fig{intro:lattice} (b) compares experimental measurements of nonperturbative quantities computed on the lattice and finds agreement to data to within statistical and systematic errors of 3\% or less.  

\begin{figure}[htb!]
\centering
\includegraphics[width=\textwidth]{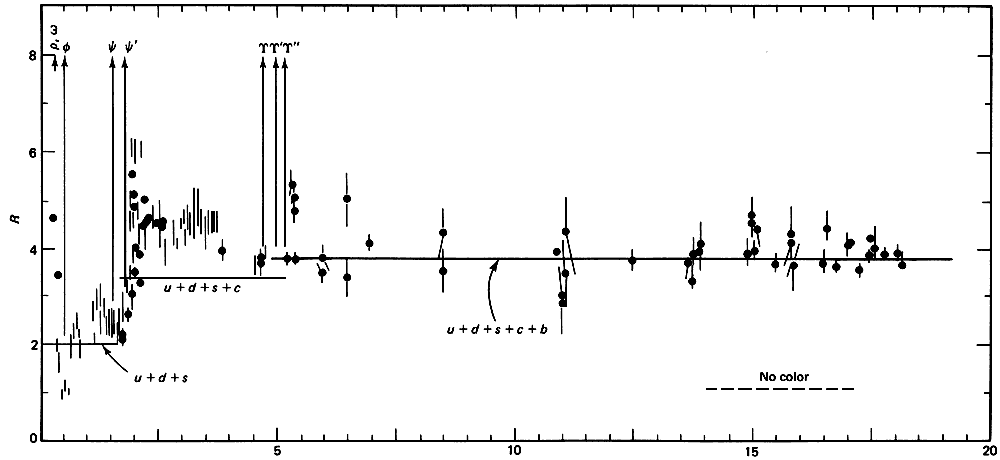}
\caption{
\label{intro:R}
$R=\sigma(e^+e^-\rightarrow\textrm{hadrons})/\sigma(e^+e^-\rightarrow\mu^+\mu^-)$ plotted against electron energy in GeV.  Above 5 GeV, $R$ is well approximated by $N_c\sum Q_f^2$ with $N_c=3$.  Figure adapted from \cite{Griffiths:1987tjB}. 
}
\end{figure}

\begin{figure}[htb!]
\centering
\includegraphics[width=4.25 in]{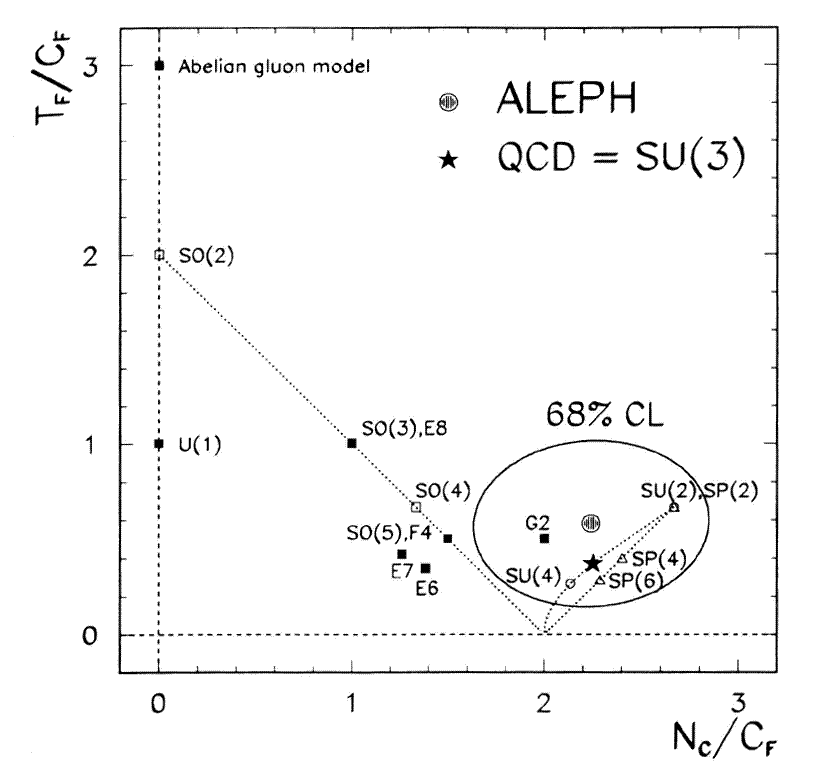}
\caption{\label{intro:su3}Experimental verification of SU(3) as the gauge group of QCD from four jet events measured at LEP.  $N_c$, $C_F$, and $T_F$ are properties of the group structure functions $f$ and generators $T$ defined by $f^{abc}f^{*abd}=\delta^{cd}N_c$, $(T^a T^{\dag a})_{ij}=\delta_{ij} C_F$, and $Tr(T^a T^{\dag b}) = \delta^{ab} T_F$, where repeated indices are summed over.  Groups with $N_c\ne3$ are excluded by decay ratios while S0(3) and U(1)$_3$ do not have the correct ratios of group theoretic factors.  Figure adapted from \cite{Decamp:1992ip}.}
\end{figure}

To resolve the Pauli problem mentioned earlier one needs at least three colors; several measurements demonstrate that Nature in fact uses no more than three (for a review see \cite{Narison:2004} and \cite{Halzen:1984}).  The ratio of the cross sections,
$\sigma(e^+e^-\rightarrow\textrm{hadrons})/\sigma(e^+e^-\rightarrow\mu^+\mu^-)=N_c\sum Q_f^2$,
tests both the number of colors ($N_c$) and active flavors ($N_f$) as a function of center of mass energy, shows that at energies above the bottom mass but below the top that $N_c=3$ and $N_f=5$; see \fig{intro:R}.  The decay rate of the $Z^0$ to hadrons compared to $e^+e^-$ depends on $N_c$; again, experimentally $N_c=3$.  Similarly, the inclusive semi-hadronic decay rate of the $\tau$ lepton to its semi-leptonic one is a direct measure of $N_c$ with $N_c=3$.  And the decay of neutral pions to two photons is a direct measure of the square of the number of colors; the theoretical prediction 
\be
\Gamma(\pi^0\rightarrow\gamma\gamma)\;=\;[N_c(Q_u^2-Q_d^2)]^2\left(\frac{\alpha^2}{64\pi^2}\right)\frac{m_\pi^3}{f_\pi^2} \;=\; 7.7\;\;\mathrm{eV},
\ee
where $f_\pi=92.4$ MeV is the pion decay constant controlling the $\pi^-\rightarrow\mu\nu$, is in remarkable agreement with data ($7.7\pm0.6$ eV).  Four jet measurements identified directly the triple gluon vertex and found that the gauge group is SU(3) instead of SO(3) or perhaps U(1)$_3$ \cite{Abreu:1990ce,Decamp:1992ip}; see \fig{intro:su3}.  Similarly the running of the coupling agrees well for $N_c=3$ \cite{Tung:2001cv}.

\section{QCD Phase Diagram}
\begin{figure}[!htb]
\centering
$\begin{array}{cc}
\includegraphics[width=2.5in]{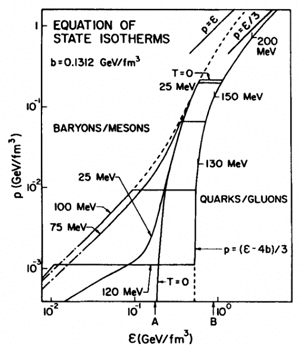} &
\includegraphics[width=2.65in]{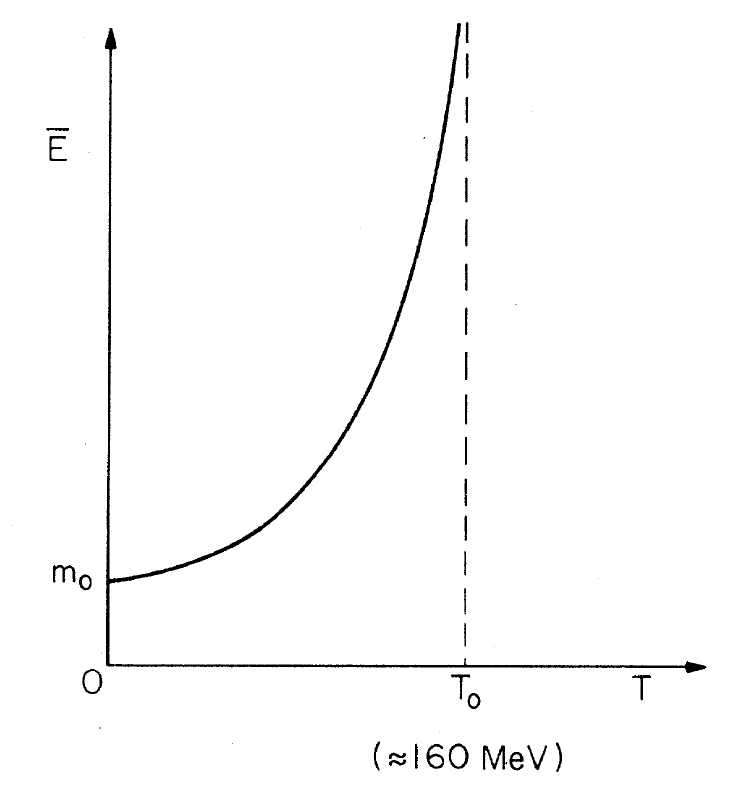} \\[-.05in]
\textrm{{\figsize(a)}} & \textrm{{\figsize(b)}}
\end{array}$
\caption{\label{intro:hagedorn}(a) Phase transition from confined baryonic/mesonic matter to quark/gluon matter from the mean field quantum hadrodynamics (QHD) approach of Walecka \cite{Walecka:1974qa}; figure adapted from \cite{Walecka:2004}.  (b) A sharp rise in energy reminiscent of a phase transition is seen at $T\approx160$ MeV in Hagedorn's statistical model \cite{Hagedorn:1965st}; figure adapted from \cite{Frautschi:1971ij}.}
\end{figure}

The possibility of a new state of nuclear matter accessible through the collisions of heavy nuclei has a long history (see \cite{Baym:2001in} and references therein).  From the low energy side, Lee and Wick showed that scalar field theories could support an `abnormal' nuclear state with properties far different from those in the usual vacuum \cite{Lee:1974ma}.  In more modern language there is a nonzero vacuum expectation value for a quark condensate which one expects to melt at higher temperatures, thus restoring chiral symmetry.  Both the statistical model of Hagedorn \cite{Hagedorn:1965st,Frautschi:1971ij} and the hadronic mean field approach of Walecka \cite{Walecka:1974qa,Chin:1974sa,Baym:1975mf} predict a phase transition; see \fig{intro:hagedorn}.  

From the QCD side, and before the advent of asymptotic freedom, Itoh was the first to suggest the possibility of deconfined quark matter \cite{Itoh:1970uw}, followed by Carruthers (as cited in \cite{Baym:2001in}).  Collins and Perry \cite{Collins:1974ky} were the first to recognize the importance of asymptotic freedom, leading at large energies to a `quark soup.'  Shuryak coined the name `quark-gluon plasma' (QGP) \cite{Shuryak:1980tp} as noted in \cite{Adcox:2004mh}. 

\begin{figure}[!htb]
\centering
$\begin{array}{cc}
\includegraphics[width=2.95 in]{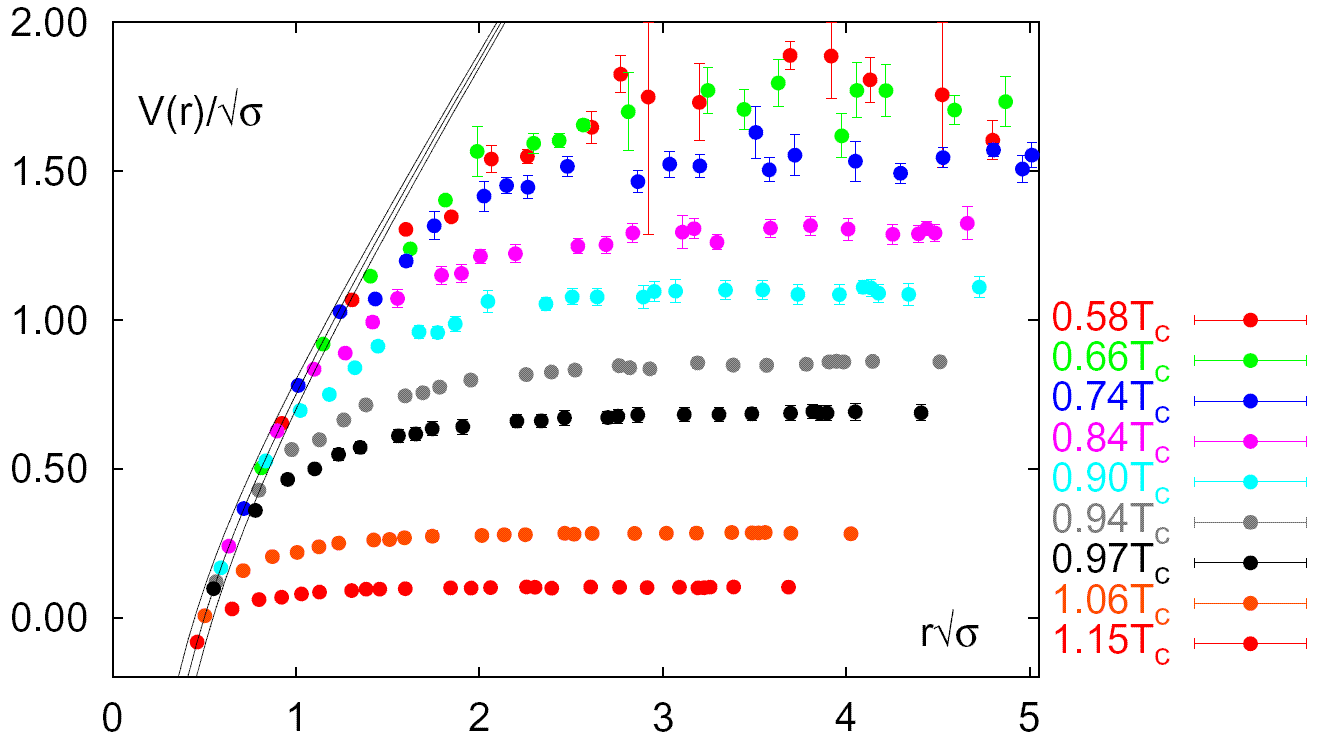} &
\includegraphics[width=2.25in]{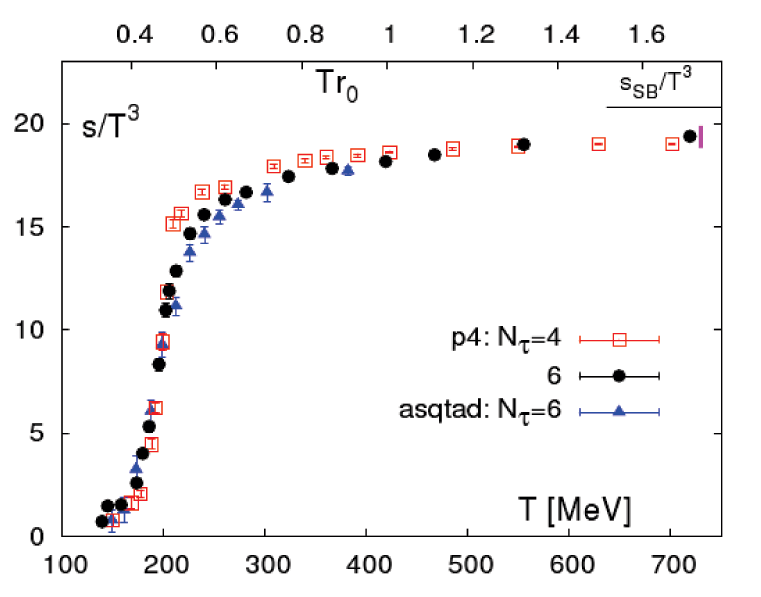} \\[-.05in]
\textrm{{\figsize(a)}} & \textrm{{\figsize(b)}}
\end{array}$
\caption{\label{intro:latticePT}(a) The $q\bar{q}$ potential from LQCD for heavy quarks shows evidence of plasma formation at high temperatures.  For $T>T_c$ the potential becomes independent of distance for separations longer than a temperature-dependent characteristic length; figure adapted from \cite{Kaczmarek:1999mm,Karsch:2001vs,Doring:2005mt,Adil:2007}.  (b) Lattice calculations show a sharp rise in entropy density as a function of temperature, indicating a significant change in the number of degrees of freedom at $T_c\sim180$ MeV; figure adapted from \cite{Cheng:2007jq}.}
\end{figure}

The characteristic length for the polarization of the medium caused by color charges is the inverse of the Debye mass, $\mu_D\sim gT$, which is related to the temperature of the plasma.  LQCD calculations of the static $q\bar{q}$ potential as a function of distance and temperature show that for $T>T_c$ the potential is screened; see \fig{intro:latticePT} (a).  Lattice calculations show a sharp rise of entropy density as a function of temperature, with the density afterward given by an ideal gas with quark and gluon degrees of freedom to within 20\%; recent results using almost physical quark masses \cite{Cheng:2007jq} are shown in \fig{intro:latticePT} (b).  There is disagreement within the lattice community as to whether the chiral transition and the deconfinement transition occur simultaneously.  Nevertheless all these very different lines of reasoning point to something very interesting occuring in nuclear matter at high temperatures and densities.

Interestingly these disparate descriptions of nuclear matter T high energy density all obtain similar values for the transition temperature $T_c$.  Simple dimensional analysis leads one to expect the temperature of such a QGP to be around $\Lambda_{QCD}\approx200$ MeV, as this is a necessary momentum scale to resolve distances of order the nucleon and at the same time is the scale of normal nuclear energy densities of 1 GeV/fm$^3$.  Walecka yields $\sim 150$ MeV \cite{Walecka:2004}.  
Frautschi found that Hagedorn's bootstrap gives $T_c\approx160$ MeV \cite{Frautschi:1971ij}.  
Current data for $T_c$ from the lattice are in qualitative agreement with the previous results, although they are inconsistent quantitatively.  The Wuppertal group found \cite{Aoki:2006br} chiral restoration at $151(3)(3)$ MeV and a crossover phase transition at $176(3)(4)$ MeV while the BNL/Bielefeld group found \cite{Cheng:2007jq} $T_c=192(7)(4)$ with the chiral limit smaller by 3\%.  

In order to explore the new physics of nuclear matter \emph{in extremis} an ambitious program of heavy ion collisions began with Bevalac and continued through \agscomma, \spscomma, \rhiccomma, and soon will commence at \lhccomma.

\clearpage
\mychapter{Motivation}{chapter:signs}
Since there are so many theoretical indications for a phase transition in QCD at low baryon chemical potential and at $T>T_c\simeq160$ MeV, and since that unexplored phase would be a truly novel and interesting state of matter in which the ordinarily confined quarks and gluons---and not protons, neutrons, pions, etc.---are the pertinent degrees of freedom, one naturally asks how one might observe the creation of such conditions.  Due to the techniques used to calculate them, the experimentally measured quantities associated with heavy ion collisions naturally separate themselves into low momentum, or bulk, and high momentum jet observables.

Although heavy ion experiments have a long pedigree this discussion will focus mainly on \rhic data (see the white papers from the four experimental collaborations at \rhic for a review \cite{Adcox:2004mh,Arsene:2004fa,Adams:2005dq,Back:2004je}); this thesis focuses on jet observables, for which cross sections were simply too small at previous experiments, and the data from \rhiccomma, as discussed below, is qualitatively different from these previous experiments in exciting new ways.

\section{Bulk Observables}
The most basic bulk observable is the energy deposited by all particles.  Surprisingly, this elementary quantity provides a qualitative estimate of the energy density created in heavy ion collisions \cite{Adcox:2004mh}.  More differential measurements have the potential to provide much more information on the medium.  Simply taking ratios of particle species gives insight into the thermal properties of their creation \cite{Rafelski:1982pu,Koch:1986ud,Becattini:1995xt,Becattini:1997rv,Cleymans:1992hy,Kaneta:2004zr,Letessier:2005qe,Castorina:2007eb}.  Single particle spectra and their distribution over the reaction plane, through the use of hydrodynamics, hold the hope of determining the bulk evolution and its equation of state.  Two particle correlations, often quoted as Hanbury Brown-Twiss (HBT) radii \cite{HanburyBrown:1956pf,HanburyBrown:1954wr}, measure the physical dimensions of the fireball at freezeout and provide a consistency check for hydrodynamics, a test which it has consistently failed (for a review see \cite{Lisa:2005dd}).
\subsection{Bulk Evolution}\label{bulk}
\begin{figure}[!htb]
\centering
\includegraphics[width=4.5 in]{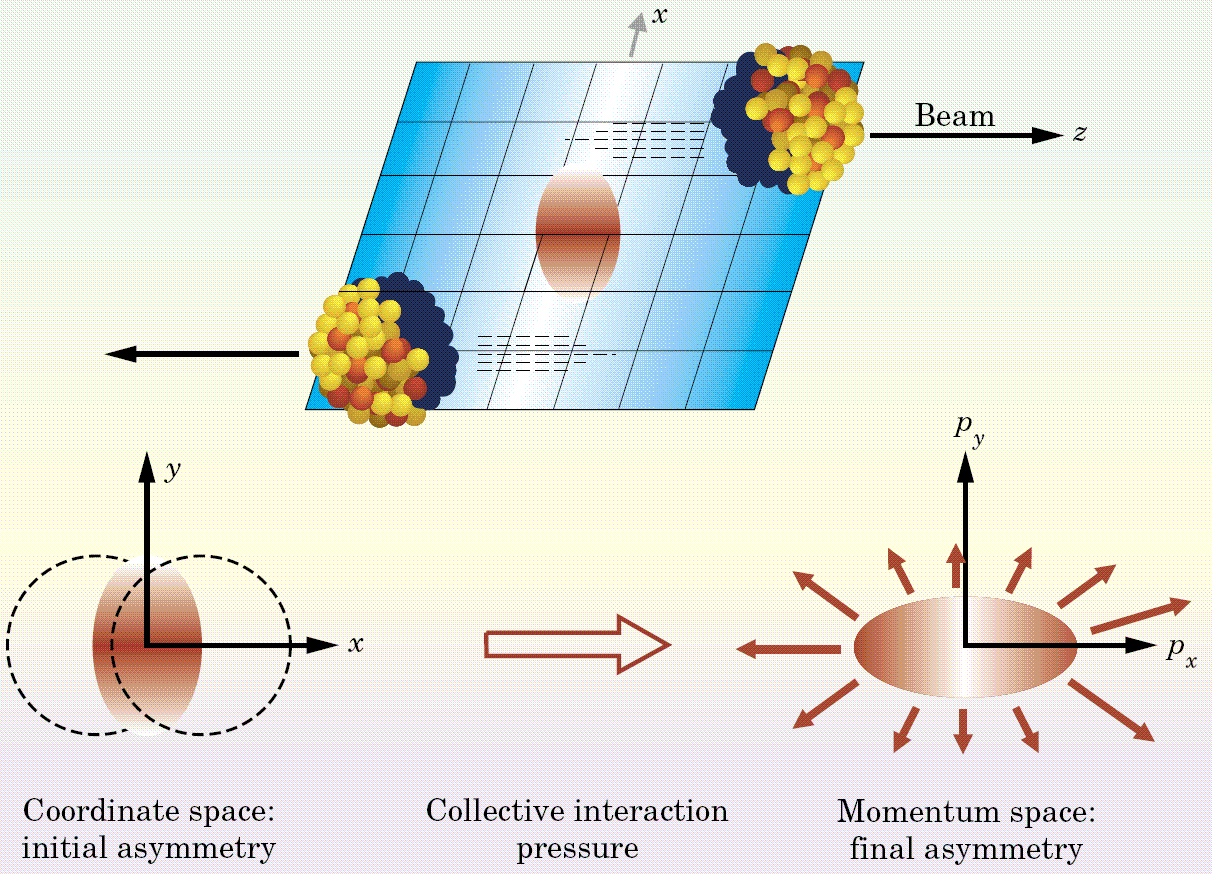}
\caption{\label{intro:hydropic}
In non-central heavy ion collisions there is initial spatial anisotropy that the resultant collective pressure gradients evolve into anisotropy in momentum space.  Hydrodynamics aims to quantitatively model this process to gain information on the medium and its properties.  Figure adapted from \cite{Ludlam:2003rh}.}
\end{figure}

A thermalized medium with a small mean free path compared to the system size can be understood using (relativistic) hydrodynamics (see \cite{Rischke:1998fq,Kolb:2003dz,Huovinen:2006jp,Hirano:2008hy} for reviews).  One then might hope to learn about the physics of a heavy ion collision by measuring bulk observables and comparing them to results from hydrodynamic evolution.  Schematically, hydrodynamics takes a given set of initial conditions and evolves them according to the known conservation laws of the system and a set of externally specified equations, for instance the equation of state for ideal hydrodynamics and the stress fields for second order viscous Israel-Stewart hydrodynamics; see \fig{intro:hydropic}.  Strong evidence for a deconfined QGP could come from a robust result whereby hydrodynamics with a QGP equation of state (EOS) reproduces experimental data while hydro with a hadronic EOS does not.  
It is interesting to note that the original conception of the QGP was of a weakly interacting plasma of deconfined quarks and gluons.  For hydrodynamics to be useful, though, thermalization after the violent nuclear collisions must be both rapid and sustained.  This of course requires a strongly coupled quark gluon plasma, or sQGP, as named in \cite{Gyulassy:2004vg,Gyulassy:2004zy}; here `strong' refers not to the strong force, but rather to the ratio of potential to kinetic energy, $\Gamma$, being greater than 1.  However, this is not a universally accepted naming convention.  \cite{Adams:2005dq} takes these properties as assumed and refers only to a QGP; for a more detailed discussion see \cite{Adcox:2004mh}.

\begin{figure}[!htb]
\centering
$\begin{array}{cc}
\includegraphics[width=2.85in]{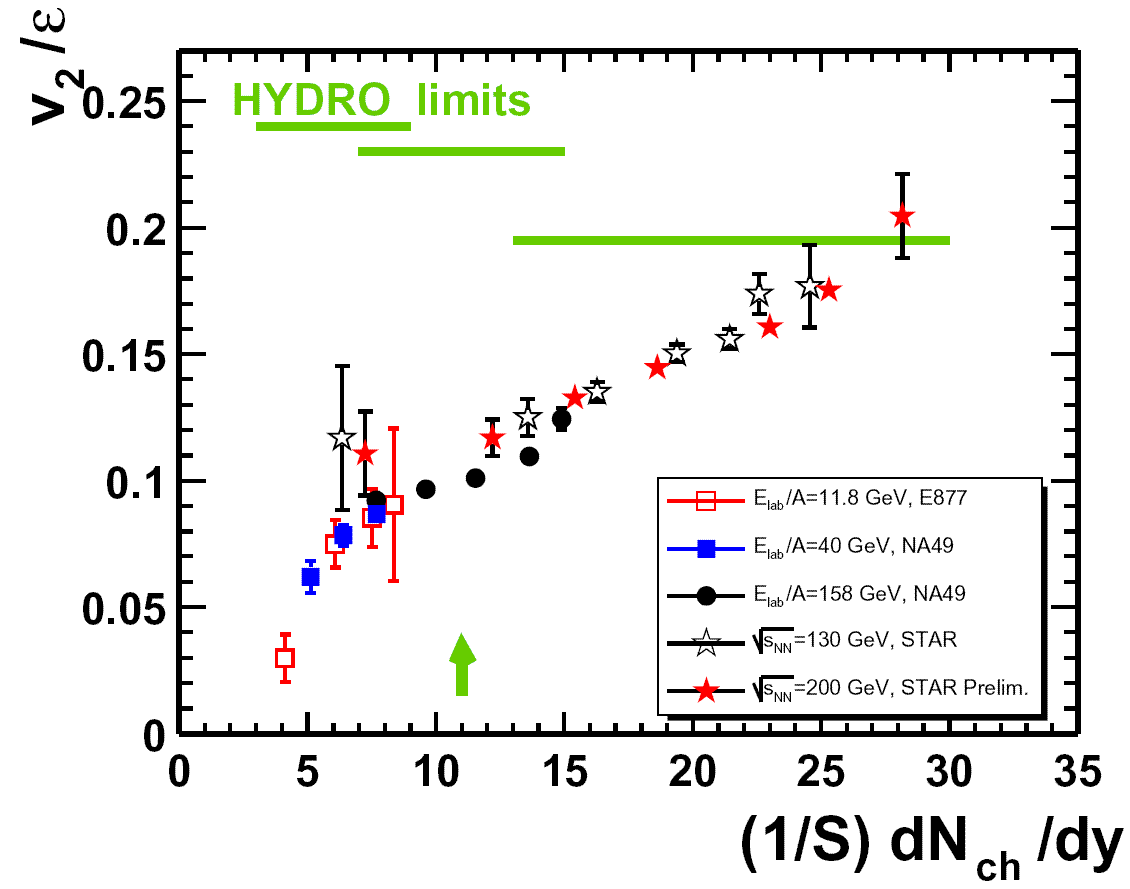} &
\includegraphics[width=2.2in]{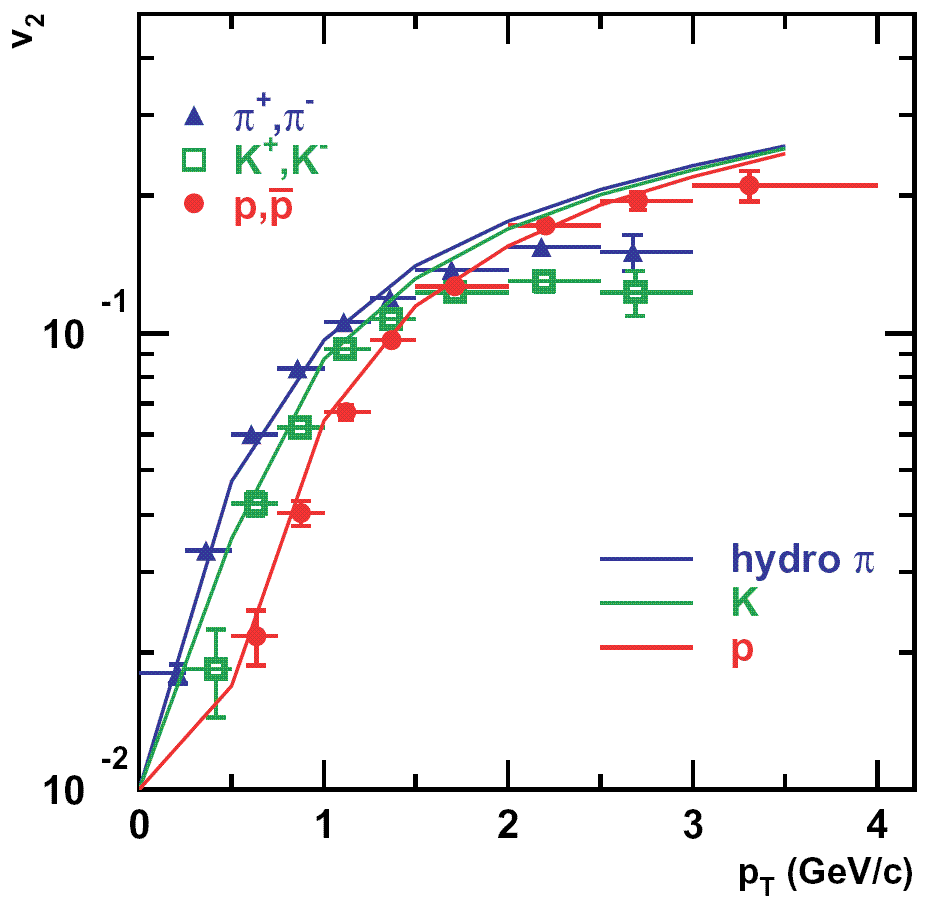} \\[-.05in]
\textrm{{\figsize(a)}} & \textrm{{\figsize(b)}}
\end{array}$
\caption{\label{intro:hydro}(a) \vtwo scaled to the initial spacial eccentricity, $\epsilon$, as a function of the charge particle density per unit transverse area.  \lowpt flow at previous colliders fell far short of the hydrodnamic limit, which is reached for the first time at \rhiccomma.  Figure adapted from \cite{Alt:2003ab}. (b) Early azimuthal asymmetry, \vtwoptcomma, predictions from \cite{Huovinen:2001cy} compared to \rhic data \cite{Adler:2003kt} as adapted from \cite{Adcox:2004mh}.}
\end{figure}

\begin{figure}[!htb]
\centering
\includegraphics[width=5. in]{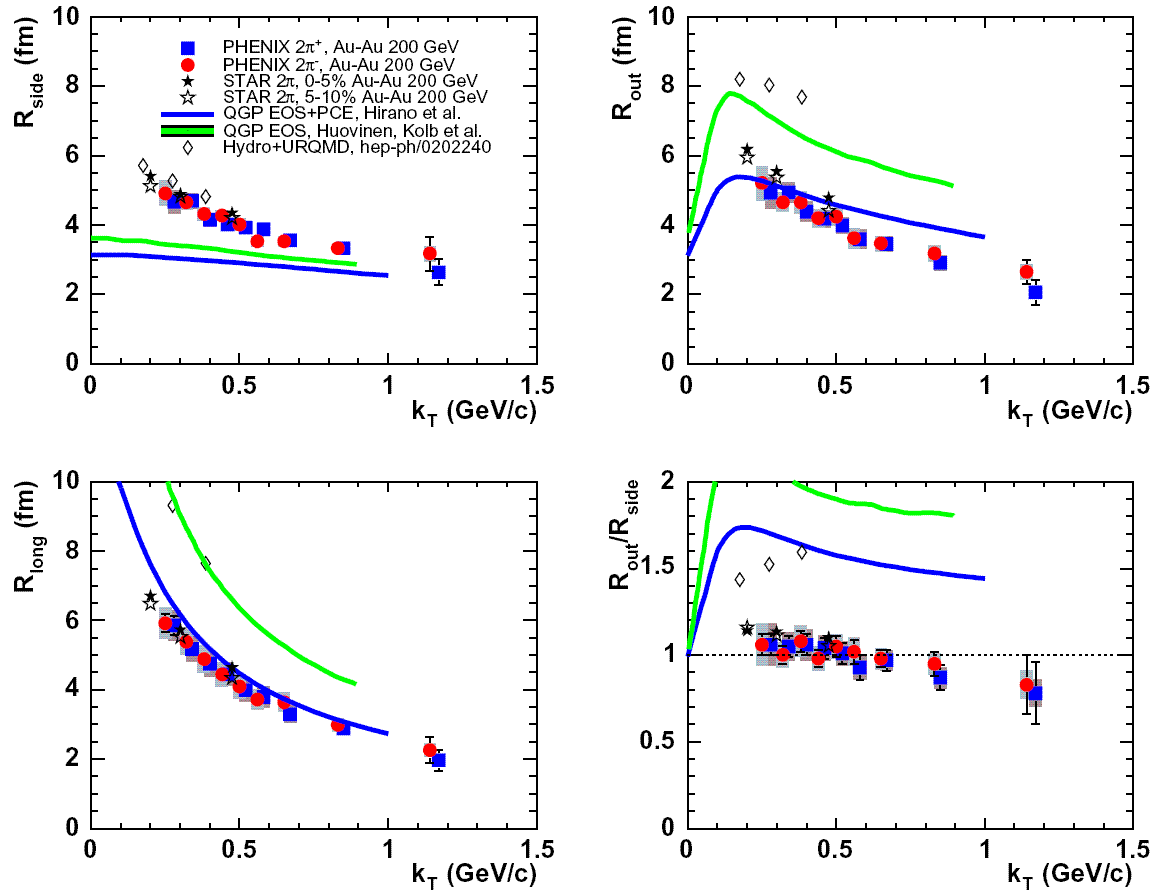}
\caption{\label{intro:hydrohbt}
HBT parameters \cite{Adler:2004rq,Adams:2003ra} compared to various hydrodynamics calculations \cite{Hirano:2002ds,Heinz:2002un,Soff:2002qw}; none describe the data well.  Figure adapted from \cite{Adcox:2004mh}.}
\end{figure}

Since hydrodynamics predicts the evolution of the entirety of the bulk there are a number of observables that can be compared to data.  The simplest is single the particle spectra and their angular distribution with respect to the reaction plane.  
It is useful to Fourier expand the detected distribution,
\be
dN(\eqnpt,\phi) = dN(\eqnpt)\left( 1 + 2 \sum_{n} v_n(\eqnpt) \cos(n\phi) \right),
\ee
where the normalization $dN(\eqnpt)$ and \vtwocomma, or azimuthal anisotropy, are the two most important moments for heavy ion collisions.  Early results from \rhic and ideal hydrodynamics were quite promising: while the \vtwo of particle yields generated by hydro was too large when compared to previous heavy ion experiments, those at \rhic matched quite well \cite{Huovinen:2001cy}; see \fig{intro:hydro}.  Unfortunately the HBT radii were not so well reproduced; see \fig{intro:hydrohbt}. 

\begin{figure}[!htb]
\centering
\includegraphics[width=4 in]{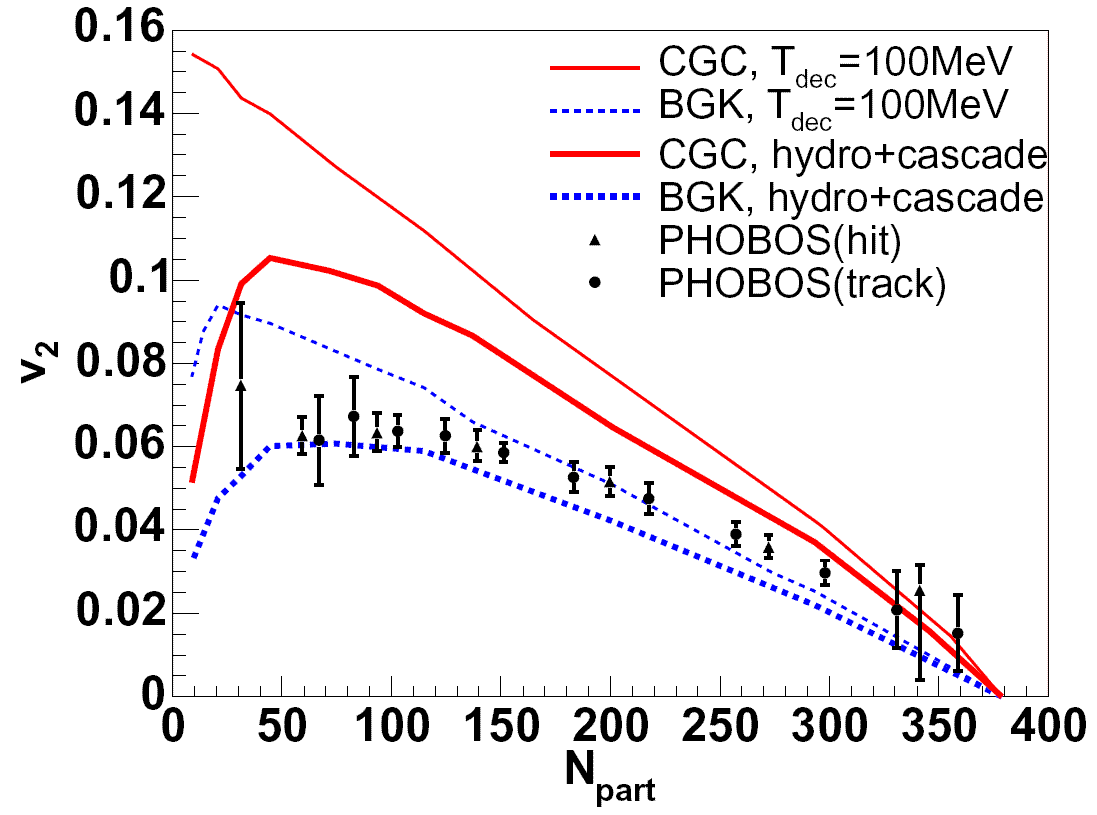}
\caption{\label{intro:hirano}Recent calculations including hadronic rescattering \cite{Hirano:2005xf} that show the influence of the initial conditions (IC) on the output.  For diffuse, Glauber-like IC ideal hydrodynamics slightly underpredicts data whereas sharper CGC-like IC require viscous effects to follow the experimental trend.}
\end{figure}

However the picture can never be this crystal clear.  Hydrodynamics is a set of evolution equations; the initial conditions are input, and, not surprisingly, what comes out of hydro is highly dependent on what goes in; see \fig{intro:hirano}.  There have been some suggestions for testing the initial state \cite{Adil:2005qn,Adil:2005bb}; without good theoretical or, better, experimental control over the initial conditions statements drawn from hydrodynamical modeling will be inconclusive at best.

There are two additional major complications in interpreting hydrodynamics results.  The first comes from evolution beyond a thermalized medium into a viscous gas of hadronic particles.  As the system expands with time it necessarily becomes too dilute for equilibrium to be maintained and hydrodynamics can no longer be a valid description.  Often taken to occur at the same time is the breakup of the quark gluon plasma into ordinary hadronic matter.  But the transition from quark and gluon degrees of freedom to baryons and mesons is not well understood.  Unlike \highpt fragmentation---discussed later---this hadronization is in the \lowpt sector, and there is no reason to believe that experimental measurements in, say, $p+p$ collisions can provide independent information on the process.  
Similarly the low momenta involved makes it nonperturbative; there is little theoretical insight, either, although recombination \cite{Das:1977cp,Hwa:1979pn} and quark coalescence \cite{Biro:1994mp,Csizmadia:1998vp} are two proposed mechanisms for hadronization.  More sophisticated $3+1D$ hydrodynamics treatments that include hadronic rescattering (a so-called hadronic `afterburner') have had good success in reproducing both single particle spectra and azimuthal anisotropy \cite{Hirano:2005xf}.  Unfortunately even these highly computationally intensive calculations that follow evolution all the way until free streaming still do not describe the observed HBT radii.  

The second open issue is the role of viscous corrections.  Past $3+1D$ relativistic hydrodynamic treatments of \rhic were all ideal; the viscous terms were explicitly set to zero.  There have been some attempts to quantify the magnitude of these corrections \cite{Teaney:2003kp}.  
However the rigorous treatment of viscosity in relativistic hydrodynamics is still an open problem.  The inclusion of only the first order correction terms leads to unstable solutions with acausal modes \cite{Hiscock:1983zz,PhysRevD.31.725,PhysRevD.35.3723}.  The second order Israel-Stewart formalism was shown to be stable and causal \cite{Muller:1967,Israel:1976tn,Stewart:1977,Israel:1979wp}, but the need to specify 10 independent stress field component initial conditions remains a formidable problem.  Much work is underway on modeling, with $2+1D$ viscous treatments recently published \cite{Romatschke:2007mq,Song:2008si}.  And there is some hope that with the inclusion of these viscous terms hydrodynamics will finally reproduce the HBT correlation radii as well as the single particle distributions \cite{Teaney:2003kp}.  However fully $3+1D$ relativistic viscous hydrodynamics models are still unavailable, and will be so for some time.  

Nonetheless a truly quantitative understanding of the size of viscous effects allowed by data in heavy ion collisions is crucial for progress in the field.  Teaney \cite{Teaney:2003kp} estimated that in \rhic collisions the viscosity to entropy density ratio, $\eta/s$, often shortened to viscosity to entropy ratio, could be no more than about 0.1.  Utilizing elastic channels with perturbative cross sections only, weakly coupled pQCD gives $\eta/s\sim1$ \cite{Danielewicz:1984ww,Baym:1990uj,Arnold:2000dr,Arnold:2003zc}.  The most famous and promising result of AdS/CFT is the calculation of $\eta/s = 1/4\pi$ in a strongly coupled plasma; for more on the AdS/CFT conjecture see Section \ref{intro:adscft}.  

Currently the best bridge between $3+1D$ ideal and viscous hydrodynamics comes from parton transport theory \cite{Zhang:1999rs,Molnar:2000jh,Xu:2004mz}.  For parton cascades including only $2\rightarrow2$ processes, the ideal hydrodynamics limit is only approached when using very large elastic cross sections \cite{Molnar:2001ux}.  Nevertheless first comparisons between viscous hydrodynamics calculations and those from the cascade have been made \cite{Molnar:2008xj}.  Recent work by Xu and Greiner \cite{Xu:2004mz} has incorporated $2\leftrightarrow3$ processes; utilizing this promising new code they have found, using perturbative ideas alone, $\eta/s\sim.1$ \cite{Xu:2007ns} and also hope to include both bulk and jet dynamics within a single theoretical framework \cite{Fochler:2008ts}.  There are plenty of difficulties, though, as including gluon splitting and merging makes numerical acausal artifacts more pronounced (their work has yet to be confirmed by other groups) and the gluon radiation from jets does not model coherence effects well.  
As noted previously it is hard to overemphasize the importance of initial conditions (IC) when considering hydrodynamics output.  On the crucial issue of viscosity the two are necessarily coupled: consistency with data with more diffuse IC requires a more ideal fluid; sharper edges a more viscous one.  Hopefully by independently experimentally testing the IC---see \cite{Adil:2005bb} for a proposed discerning observable---and by theoretically motivating the fluid viscosity we can arrive at a self-consistent picture of the bulk dynamics in heavy ion collisions.  

\section{High-\texorpdfstring{\pt}{pT} Observables}
The exciting first evidence of jets came from \isr in 1972 \cite{Jacob:1972ym,Jacob:1972yc,Busser:1973hs,Alper:1973nv,Banner:1973nu}.  Orders of magnitude more \highpt pions were observed than were expected from a \lowpt extrapolation; the production spectrum had turned over from exponential to power law.  Soon afterward, two jet events were explicitly seen at $e^+e^-$ colliders \cite{Hanson:1975fe,Brandelik:1979bd}.  While Bjorken was the first to suggest using jet suppression to learn about a QCD medium \cite{Bjorken:1982tu}, the precision pQCD predictions of production rates \cite{Aversa:1988vb,Adler:2003pb,Jager:2002xm,deFlorian:2002az} held out the possibility for jet tomography: much like in medical applications such as a \textsc{Pet} scan, a careful measurement of the jet quenching pattern would reveal information on the medium through which the probe traveled.  As high transverse momentum parton jets are produced early in the collision (by Heisenberg's Uncertainty Principle) and preferentially deep within the fireball---more on this later---they are potentially excellent probes of the medium.  

Naturally then before investing the tremendous time and resources necessary to investigate \highpt particles theoretically and experimentally, one should ruminate on 
the epistemology of hard probes.  There were claims in \cite{Adcox:2004mh} that jet suppression can merely give information on the density of scattering centers.  This is an important measurement, as noted previously a high density is a prerequisite for QGP formation, but jets have the potential to in fact reveal much more than simply a mean density.   
The possibility of testing, e.g.\ deconfinement, depends heavily on the energy loss model and the mapping made between its input parameters and the physical medium.  This mapping is a critical, although often overlooked, component of any model attempting tomography.  As an example, the WHDG model \cite{Wicks:2005gt} (see Chapter \ref{chapter:WHDG}) explicitly assumes a connection between the medium density, its temperature, and its Debye mass: light quarks and gluons are taken to have masses proportional to $\mu_D$, and the energy loss results are surprisingly sensitive to its changes; the elastic energy loss is derived from classical considerations that assume plasma polarization.
Additionally Chapter \ref{punch} argues from a phenomenological approach to jet suppression that the large magnitude of the observed anisotropy at \intermediatept is a result of deconfinement physics.

\subsection{Factorization in \texorpdfstring{$p+p$ and $A+A$}{p+p and A+A}}
Factorization provides the theoretical framework within which pQCD calculations are made and states that reactions with large momentum transfer can be factorized separately into long distance and short distance pieces \cite{Libby:1978qf,Ellis:1978sf,Ellis:1978ty,Amati:1978aa,Amati:1978bb,Curci:1980uw,Collins:1983ju,Collins:1992xw}.  There are two crucial components to factorization: (1) the (presently) incalculable nonperturbative low momentum contributions are universal; i.e.\ they are the same for any QCD process, and (2) asymptotic freedom guarantees that the high momentum contribution can be reliably found using pertubative methods in QCD.  A factorization scale, $\mu_F$, is introduced separating the short and long distance contributions; it is of order the hard scale of the problem, but is not specified further from within the theory.  Following \cite{Jager:2002xm} the cross section for producing a hadron of type $h$ in a $p+p$ collision is then
\bea
\label{intro:eq:factorization}
d\sigma^h & = & \sum_{a,b,c}\int dx_a dx_b dz_c f_{a/p}(x_a,\mu_F) f_{b/p}(x_b,\mu_F)D_c^h(z_c,\mu_F') \nonumber\\
& & \times d\hat{\sigma}_{ab}^c(x_a P_A,x_b P_B,P_h/z_c,\mu_R,\mu_F,\mu_F'),
\eea
where the $f$s are parton distribution functions (PDFs), $D_c^h$ is a fragmentation function (FF), and $d\hat{\sigma}_{ab}^c$ is the perturbative partonic cross section.  Using standard notation the partonic variables are in lower case and the hadronic ones are in upper case.  In this way a PDF gives the probability of finding parton $a$ in a proton with momentum fraction between $x_a$ and $x_a+dx$, where $x_a=p_a/P_p$ and something of an abuse of notation occurred by which $p_a$ stands for the momentum of parton $a$ and $P_p$ stands for the momentum of the proton.  The FF gives the probability of parton $c$ hadronizing into $h$ with fractional momentum between $z_c$ and $z_c+dz$, where $P_h=z_c p_c$.  All allowed combinations of $a$, $b$, and $c$ are summed over where $a+b\rightarrow c+X$ such that $c$ may hadronize into $h$.  We note that we have slightly extended the factorization theorem to include the fragmentation process, which introduces the new scale $\mu_F'$.  Additionally a renormalization scale $\mu_R$ associated with the running coupling $\alpha_s$ is included in the perturbative cross section.

Parameterizations of PDFs may be found from, e.g., the \textsc{CTEQ} collaboration \cite{Pumplin:2002vw}, GRV \cite{Gluck:2007ck}, MRST \cite{Martin:2007bv}, or Alekhin \cite{Alekhin:2006zm}.  Hadronic fragmentation functions have similarly been parameterized by, e.g., DSS \cite{deFlorian:2007aj}, AKK \cite{Albino:2008fy}, HKNS \cite{Hirai:2007cx}, and KKP \cite{Kniehl:2000fe}; \cite{Bourhis:1997yu} provides a photon FF.  We note that the posted code for KKP has a bug in its leading order (LO) kaon output that has still not been corrected
; ported \emph{Mathematica} codes for the DSS and KKP fragmentation functions, as well as a corrected \textsc{Fortran} code for KKP can be found online \cite{Horowitz:code}.  Although PDFs and FFs at a specific scale must be found from experiment, their form is constrained by QCD and general sum rule considerations and their evolution is governed by pQCD and DGLAP dynamics \cite{Dokshitzer:1977sg,Gribov:1972ri,Altarelli:1977zs}.

Derivations of perturbative cross sections vary depending on the specific reaction considered.  For those involving gluons and light guarks NLO calculations exist \cite{Jager:2002xm,deFlorian:2002az}; \cite{Binoth:1999qq,Catani:2002ny} give derivations to NLO for prompt photon production; fixed order next to leading log (FONLL) production spectra of heavy quarks can be found in \cite{Cacciari:1998it,Cacciari:2001td}.

\begin{figure}[!htb]
\centering
\includegraphics[width=2.5 in]{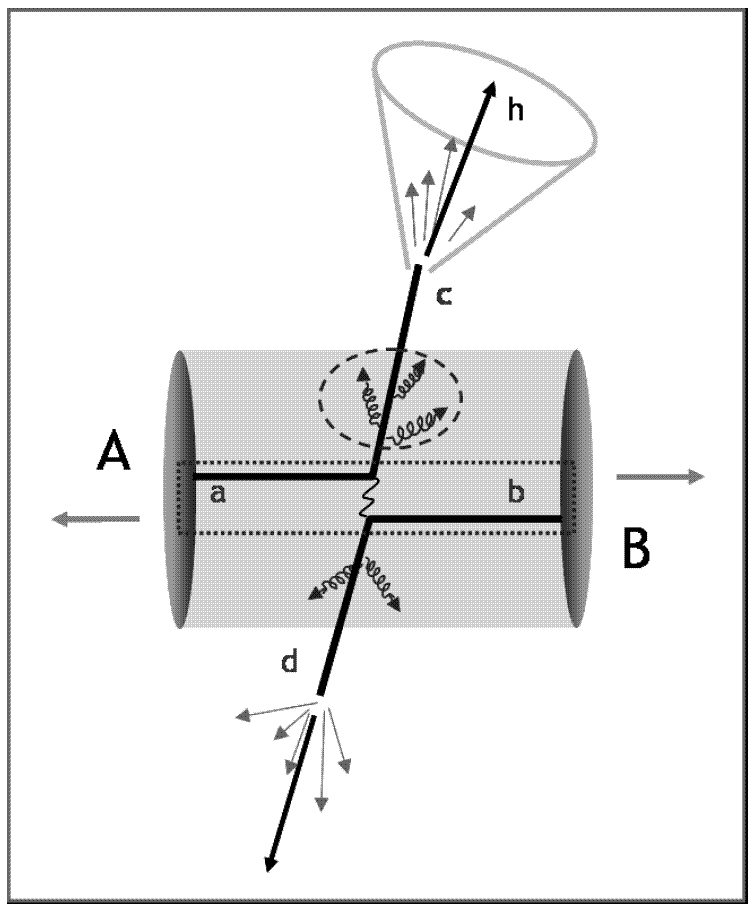}
\caption{\label{intro:mediumloss}
An illustration of $A+B\rightarrow h+X$ in a heavy ion collision.  We assume factorization still holds such that the main alteration to $p+p$ collisions, \protect\eq{intro:eq:factorization}, is the addition of a theoretically calculable $P(\epsilon)$ encapsulating the in-medium energy loss.}
\end{figure}

In heavy ion collisions \eq{intro:eq:factorization} is altered in three significant ways.  First, the geometry of collisions of large nuclei of mass number $A$ is very different from $p+p$; for a review of heavy ion collision geometry, see Appendix \ref{app:geometry}.  Large nuclei suffer a number of hard $p+p$-like reactions as they overlap, 
which is a function of the nuclei involved and the impact parameter $b$.  Second, the PDFs used must be exchanged from nucleon to nuclear.  To lowest order these new distributions should account for the isospin asymmetry of nuclei.  In discussing these in detail beyond this simple correction it is useful to define the quantity
\be
R_a^A(x,Q)\equiv\frac{F_2^{a/A}(x,Q)}{F_2^{a/p}(x,Q)},
\ee
where we have used the structure functions defined as the sum over all flavors of the PDF's weighted by $x$ and the quark charge squared, $F_2(x,Q)\equiv\sum_f xQ_f^2 f_f(x,Q)$.  In the shadowing region $x\lesssim.1$, $R_a^A<1$ \cite{Mueller:1985wy,Eskola:1993mb,Frankfurt:2002kd,Nikolaev:1990yw}.  $.1\lesssim x\lessim.3$ is the anti-shadowing region in which $R_a^A>1$ \cite{Brodsky:1989qz}.  In the EMC $.3\lesssim x\lesssim.7$ region $R_a^A<1$ (\cite{Geesaman:1995yd} and references therein), and in the Fermi motion $x\gtrsim.7$ region $R_a^A>1$ \cite{Bodek:1980ar,Bodek:1981wr,Peilert:1992hv}.  \cite{Eskola:1998df,Eskola:1998iy,Hirai:2001np,deFlorian:2003qf} give some parameterizations of nuclear structure functions.  Finally, and most importantly for this thesis, as the \highpt parton travels through the fireball it interacts with colored scattering centers in the medium.  This is encapsulated in a probability of energy loss $P(\epsilon)$, where $p_T^f=(1-\epsilon)p_T^i$.  Although it is called a probability of energy loss there are processes---e.g.\ $3\rightarrow2$ detailed balance and rare elastic events---in which momentum is gained and $\epsilon<0$ ($\epsilon$ is certainly restricted from above by 1; more on this later).  Given this \eq{intro:eq:factorization} becomes
\bea
\label{intro:eq:hic}
d\sigma^h(b) & = & \sum_{a,b,c}\int dx_a dx_b dz_c f_{a/A}(x_a,\mu_F) f_{b/B}(x_b,\mu_F)D_c^h(z_c,\mu_F') \nonumber\\
& & \times \int d\vec{x}_0 T_{AB}(\vec{x}_0;b) \int \frac{d\phi}{2\pi} d\epsilon P(\epsilon;\vec{x}_0,\phi,\{p\}_i) \nonumber\\
& & \times d\hat{\sigma}_{ab}^c(x_a P_A,x_b P_B,P_h/(1-\epsilon)z_c,\mu_R,\mu_F,\mu_F'),
\eea
where $P(\epsilon)$ in general depends on the position of hard parton production $\vec{x}_0$, its direction of propagation $\phi$, and the properties $\{p\}_i$ of the medium; we will often refer to $P(\epsilon;\vec{x}_0,\phi,\{p\}_i)$ in shorthand as $P(\epsilon)$.  See \fig{intro:mediumloss} for a schematic picture of \eq{intro:eq:hic}.

\subsection{\texorpdfstring{$R_{AA}$}{R\_AA} and Final State Suppression}
A common means of reexpressing the spectrum of hadrons produced in a nuclear collision $A+B\rightarrow h+X$ is to divide out by the trivially binary scaled proton-proton spectrum:
\be
R^h_{AB}(\{x\}_i)\equiv \frac{d\sigma^h_{AB}(\{x\}_i)}{N_\textrm{coll}(\{x\}_i)d\sigma^h_{pp}(\{x\}_i)}.
\ee
For symmetric collisions (such as $Au+Au$ or $Pb+Pb$) we will denote the above quantity as $R_{AA}$.  Often $R_{AA}$ is given as a function of transverse momenta $p_T$ and the angle with respect to the reaction plane $\phi$.  Older, lower statistics data is sometimes reported as $R_{AA}(N_\textrm{part})$, although \cite{Drees:2003zh} demonstrated its limited utility.  

$R_{AA}$ is a nice quantity to work with as it displays the effect of nuclear collisions at a glance: $R_{AA}=1$ means no modification from trivially scaled $p+p$ collisions, $R_{AA}>1$ means enhancement, and $R_{AA}<1$ means suppression.  For this reason, $R_{AA}$ is known as the nuclear modification factor.  When the partonic production spectra are power laws of approximately constant power, which is the case at \rhic and \lhccomma, $R_{AA}$ is approximately given by an especially simple relation; see Appendix \ref{punchappendix}.  

\begin{figure}[!htb]
\centering
\includegraphics[width=3.8 in]{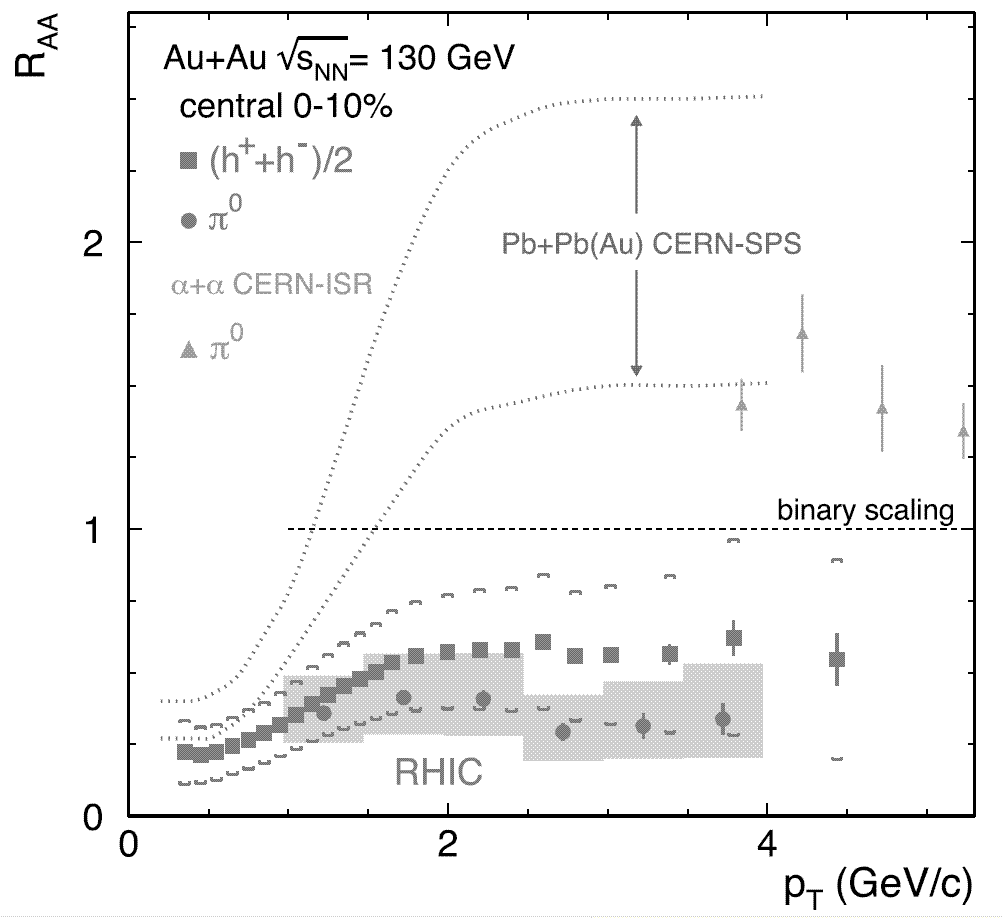}
\caption{\label{intro:suppression}
Suppression of $\pi^0$ and charged hadron jets in central $Au+Au$ collisions compared to binary scaled $p+p$ reactions at $\sqrt{s}=130$ GeV at \rhic from 2001.  Figure adapted from \cite{Adcox:2001jp}.}
\end{figure}

Early data from \rhic showed a stunning suppression of hadronic jets; see \fig{intro:suppression}.  The immediate suspicion was that this was due to final state energy loss.  However the large quenching of jets could result from an alteration in other piece of \eq{intro:eq:hic}: binary scaling, initial state, hard cross section, or fragmentation.  Taking these in reverse order, uncertainty arguments make changes to the fragmentation function unlikely for \highpt pions: in its rest frame Heisenberg gives that a $\sim140$ MeV pion should take $\sim1.4$ fm/$c$ to form; a measured 5 GeV pion has a boost factor of $\gamma\sim36$ which means that in the lab frame it is formed $\sim50$ fm away from where it was originally produced.  This is well outside the medium and at a time much later than any QGP is reasonably expected to exist so it seems the use of vacuum fragmentation functions is quite safe for light mesons.  We note that the same analysis for a 5 GeV detected proton yields a distance of just $\sim1$ fm, most likely well within the medium.  As noted previously, recombination \cite{Das:1977cp,Hwa:1979pn} and quark coalescence \cite{Biro:1994mp,Csizmadia:1998vp} are two proposed models of medium-modified hadronization; there are even indications that heavy quark mesonic states, whose large mass also precludes a necessary vacuum fragmentation, might also survive in the QGP phase \cite{Datta:2003ww,Mocsy:2007jz}.  Changes to the hard cross section are also unlikely as for large momentum transfers asymptotic freedom must make the coupling small.  

\begin{figure}[!htb]
\centering
$\begin{array}{cc}
\includegraphics[width=2.35in]{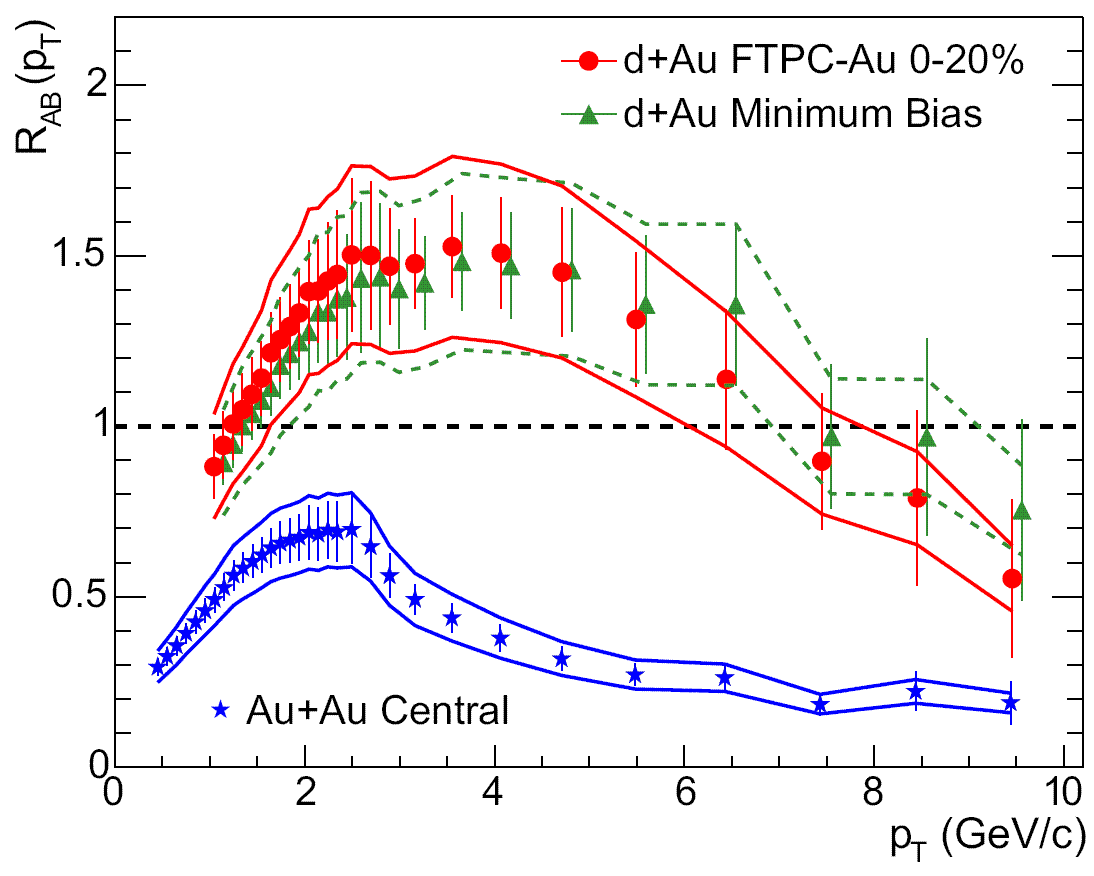} &
\includegraphics[width=2.8in]{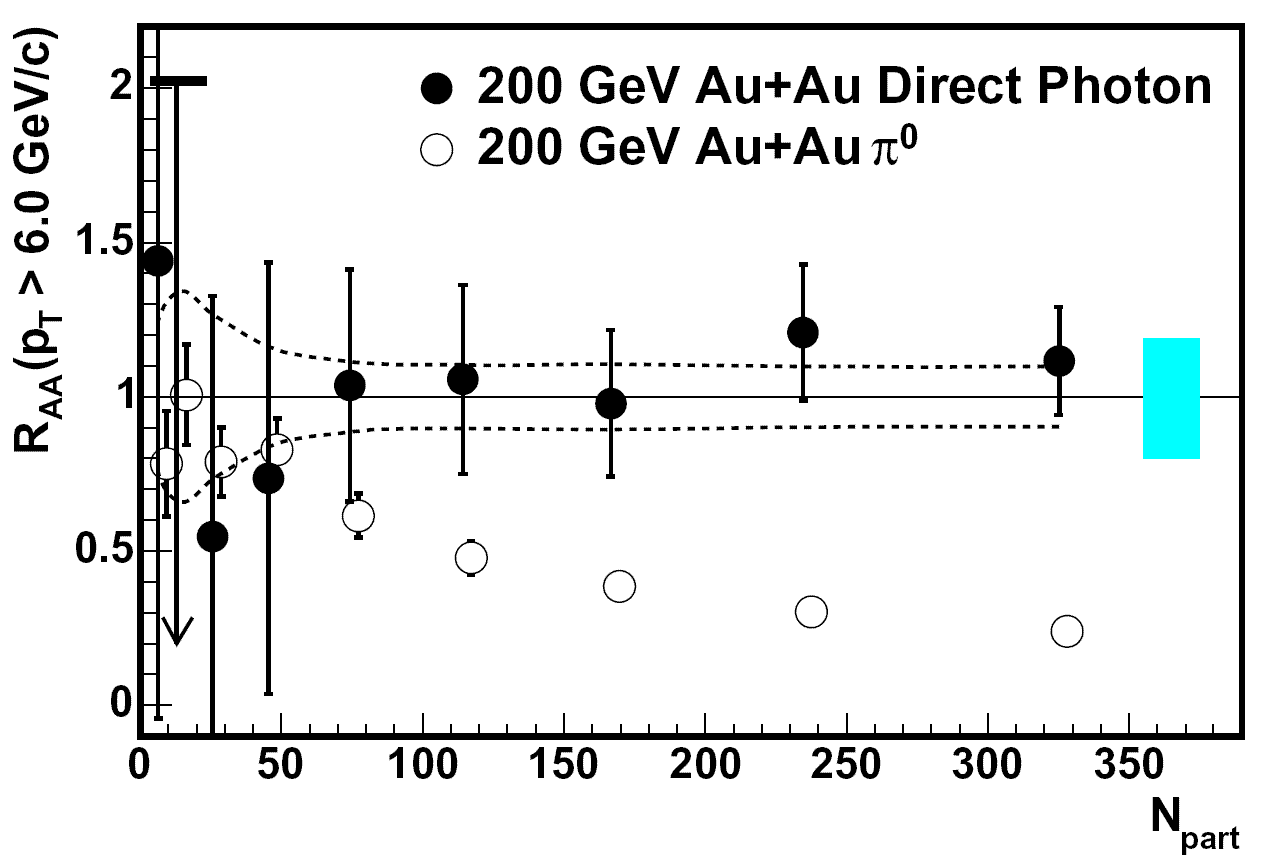} \\[-.05in]
\textrm{{\figsize(a)}} & \textrm{{\figsize(b)}}
\end{array}$
\caption{\label{intro:prod2}(a) $d+A$ results from STAR \cite{Adams:2003im} show that \highpt suppression is not due to initial state effects; figure adapted from \cite{Adams:2005dq}.  (b) $R_{AA}^\gamma(\eqnnpart)\sim1$ extends the results of (a) to show that jet suppression is not due to initial state effects in full $A+A$ collisions.  NLO prompt photon predictions \cite{Aurenche:1983ws,Aurenche:1987fs,Baer:1990ra,Gordon:1993qc} compare well to the PHENIX data \cite{Adler:2005ig}; figure adapted from \cite{Adler:2005ig}.  
}
\end{figure}

While it seemed unlikely that binary scaling did not hold in \rhic collisions many believed the initial state PDFs were significantly altered due to, for instance, the color glass condensate (CGC) \cite{Gribov:1981kg,Gribov:1984tu,Laenen:1994gh,McLerran:1993ni,McLerran:1993ka,McLerran:1994vd,Balitsky:1995ub,Kovchegov:1999ua,Iancu:2000hn,JalilianMarian:1996xn,JalilianMarian:1997jx,JalilianMarian:1997gr,JalilianMarian:1997dw,JalilianMarian:1998cb,Kovner:2000pt,Weigert:2000gi,Iancu:2000hn,Ferreiro:2001qy,Blaizot:2004px,Andersson:2002cf,Szczurek:2003fu,Lonnblad:2004zp,Nikolaev:2004cu}.  The effects of large nuclei on all the aforementioned possibilities were tested by the deuteron-gold, $d+Au$, so-called `control run.'  By smashing these together at the same energy per nucleon as the $A+A$ run changes to binary scaling, hard cross sections, and nuclear distributions were minimized while removing the medium, and thus final state energy loss or any modifications of the fragmentation functions.  \fig{intro:prod2} (a) shows that $R_{dA}\sim1$ for \highpt hadrons (with a rather large Cronin enhancement at lower momenta \cite{Cronin:1974zm}), thus falsifying the claims of CGC-type suppression and showing that the hard production in heavy ion collisions is under theoretical control.  

The \highpt conclusions based on the $d+Au$ control-run were independently verified and extended by the measurement of direct photons \cite{Adler:2005ig}.  As electromagnetic probes are only weakly coupled ($\alpha_\textrm{E\&M}\approx1/137$) to the medium they do not suffer appreciable final state energy loss, even in $A+A$ collisions.  \fig{intro:prod2} (b) compares trivially scaled pQCD predictions for direct photons from $p+p$ reactions \cite{Aurenche:1983ws,Aurenche:1987fs,Baer:1990ra,Gordon:1993qc} to those measured in $Au+Au$.  The striking consistency shows that even in $A+A$ collisions the initial state production is well understood; the variation in nuclear PDFs from $p+p$ to $d+A$ to $A+A$ do not affect \highpt observables much.  

Since the observed jet suppression cannot come from other sources, it must be due to final state energy loss.  Now \emph{\textbf{IF}} energy loss is under theoretical control and \emph{\textbf{IF}} the combined theoretical and experimental system is not fragile, then jet tomography is possible and the observed suppression pattern can be inverted to gain knowledge about the medium.  
\subsection{Elastic Energy Lost History}
Final state suppression calculations began with Bjorken's estimate of the energy lost by a \highpt parton through elastic, $2\rightarrow2$, processes \cite{Bjorken:1982tu}.  He considered the simple $t$-channel diagram associated with a parton traveling through a thermalized, deconfined QGP of temperature $T$.  In this case $|\mathcal{M}|^2\sim1/t^2\sim1/(q^2+M^2)^2$, where $M\sim\mu_D\sim gT$ is the infrared cutoff given by the Debye mass.  With some simplifying assumptions, weighed with the appropriate flux and kinematic factors this gives an austere analytic formula for the energy lost per unit distance,
\be
\frac{dE}{dx} = \pi C_R \alpha_s^2 T^2 \left( 1+\frac{N_f}{6} \right) \log \frac{2 \langle k \rangle T}{M^2},
\ee
where $C_R$ is the jet Casimir, 4/3 for a quark or 3 for a gluon, $N_f$ is the number of active flavors in the medium, and $\langle k \rangle \approx 3T$ is the average momentum of the plasma particles.

This estimate has been improved upon by a number of subsequent papers.  Thoma and Gyulassy \cite{Thoma:1990fm} calculated the linear response to a jet in classical Abelianized QCD, incorporating the hard thermal loop (HTL) work of Braaten and Pisarksi \cite{Braaten:1989kk,Braaten:1989mz,Braaten:1990az} (for a review of HTL, see \cite{Blaizot:2001nr}); the energy loss was found by deriving the work done on the source current by the induced fields in the dielectric medium.  Braaten and Thoma \cite{Braaten:1991jj,Braaten:1991we}, expanding upon the work of Svetitsky \cite{Svetitsky:1987gq}, separately evaluated the high momentum part of the dynamics using the vacuum matrix element and the low momentum piece with linear response and connected the two at a scale $q^*$ which, to leading order, drops out of the problem \cite{Romatschke:2004au}.

These leading log results were compared to asymptotic radiative loss approximations and found to be small \cite{Gyulassy:1990bh,Wang:1994fx}, not surprising as this is a well known result from E\&M \cite{Jackson:1998}.  Further work on elastic mechanisms slowed until \cite{Mustafa:2004dr} showed that with realistic kinematic limits at \rhic energies radiative and elastic losses are of the same order.  This and the experimental evidence of surprisingly strong heavy quark quenching \cite{Adler:2005xv,Akiba:2005bs,Bielcik:2005wu,Dong:2005nm} motivated work to incorporate both radiative and collisional processes in jet quenching models (see Chapter \ref{chapter:WHDG}) and has generated a flurry of interest in improving elastic calculations. 

Some more recent developments in elastic loss include considerations of running coupling, finite creation time effects, and the importance of the higher moments of the distribution.  %
Peshier \cite{Peshier:2006hi} saw a large energy loss enhancement when he allowed the coupling to run; \cite{Braun:2006vd} found a qualitatively similar result.  Wicks \cite{Wicks:2008}, however, showed that for \rhic and \lhc kinematics this is actually a small effect.  All previously mentioned calculations used asymptotic jets created in the infinite past.  Peigne, Gossiaux, and Gousset \cite{Peigne:2005rk} claimed that finite formation time effects were large and persisted far beyond the expected single Debye length.  Consideration of only the elastic pole contributions by Adil \emph{et al.} \cite{Adil:2006ei} reproduced the intuitive result; subsequent work by Gossiaux \emph{et al.} \cite{Gossiaux:2007gd} confirmed this.  Djordjevic tackled the problem from a quantum mechanical standpoint and came to the same conclusion \cite{Djordjevic:2006tw}.  Nevertheless, the effects of finite time and off mass-shell creation are still under investigation \cite{Ayala:2007cq}.

Svetitsky \cite{Svetitsky:1987gq} was the first to include fluctuations about the mean in elastic energy loss; he employed the Fokker-Planck equation.  Moore and Teaney \cite{Moore:2004tg} thoroughly investigated the relativistic Langevin and Fokker-Planck equations in heavy ion collisions.  This work was applied to heavy quark energy loss with additional nonperturbative mesonic resonances by Rapp and van Hees \cite{vanHees:2005wb}.  As shown by Wicks, however, for the pathlengths and densities at \rhic and \lhc the number of $2\rightarrow2$ collisions is of order a few, and the Gaussians resulting from these applications of the Central Limit Theorem are not a good approximation to the elastic energy loss distribution \cite{Wicks:2008}.

\subsection{Background Radiation}
Rigorously deriving radiative energy loss is a tough business.  Even in Abelian QED tremendous effort and theoretical contortions (`reinterpretation') were required to satisfactorily deal with the infrared and ultraviolet divergences (c.f., e.g., \cite{Peskin:1995}).  Undaunted, a number of nuclear theorists have devoted their lives to conquering nonabelian QCD radiative bremsstrahlung.  The formalisms created to tackle this problem can be roughly categorized into four groups: (1) BDMPS-Z-ASW (often simply BDMPS) \cite{Baier:1996vi,Baier:1996kr,Baier:1996sk,Baier:1998yf,Baier:1998kq,Baier:1999ds,Baier:2000mf,Baier:2001qw,Baier:2001yt,Zakharov:1996fv,Zakharov:1996cm,Zakharov:1997uu,Zakharov:2002ik,Kovner:2001vi,Wiedemann:2000ez,Wiedemann:2000za,Wiedemann:2000tf,Salgado:2003gb,Armesto:2003jh}, (2) DGLV (GLV) \cite{Gyulassy:1999ig,Gyulassy:1999zd,Gyulassy:2000fs,Gyulassy:2000er,Gyulassy:2001nm,Gyulassy:2002yv,Vitev:2002pf,Djordjevic:2003zk,Djordjevic:2008iz}, (3) WWOGZ (Higher-Twist) \cite{Wang:2000uj,Guo:2000nz,Wang:2001ifa,Wang:2001cs,Osborne:2002dx,Zhang:2003wk}, and (4) AMY \cite{Arnold:2000dr,Arnold:2003zc,Arnold:2002ja,Turbide:2005fk}.

Gunion and Bertsch \cite{Gunion:1981qs} began the translation of QED into QCD by deriving the strong force field theory diagrams associated with the nuclear analog of incoherent Bethe-Heitler radiation,
\be
\frac{dN_g}{d\eta d^2\wv{q}_\perp} = \frac{C_c^2\alpha_s}{\pi}\frac{\wv{q}_\perp^2}{\wv{q}_\perp^2(\wv{k}_\perp-\wv{q}_\perp)^2},
\ee
where $q=(q^0,\vec{q})=(q^0,q_{||},\wv{q}_\perp)$, and $C_c$ are the color algebra matrix elements.  Multiple coherent scattering over lengths shorter than the radiation formation time leads to interference that the suppresses the emission of radiation; this is the well-known LPM effect in electromagnetism, named after Landau and Pomeranchuk \cite{Landau:1953um} and Migdal \cite{Migdal:1956tc}.  Brodsky and Hoyer \cite{Brodsky:1992nq} began the work of including these effects in QCD calculations, and it was continued by Gyulassy and Wang \cite{Gyulassy:1993hr}.  This paper also introduced the notion of a static color scattering center with a Yukawa-like screened potential and noted the importance of the unique nonabelian extension of the LPM effect in QCD, by which the radiation reinteracts with the medium via the 3-gluon vertex.  Baier, Dokshitzer, Mueller, Peigne, and Schiff \cite{Baier:1996kr,Baier:1996sk,Baier:1998yf} were the first to include this effect in an energy loss calculation.  Unlike the GW paper that examined the thin plasma limit, like Landau and Pomeranchuk they built up the soft gluon radiation from single hard scatterings, BDMPS examined the multiple soft scattering limit, similar to Migdal and Moli\`ere scattering \cite{Moliere:1947,Moliere:1948,Bethe:1953va}.  Contemporaneous to the BDMPS papers, Zakharov developed his own formalism employing path integration \cite{Zakharov:1996fv,Zakharov:1996cm,Zakharov:1997uu,Zakharov:2002ik}, later shown to be equivalent to the BDMPS approach \cite{Baier:1998kq}.

The thin plasma limit of GW was extended in the opacity expansion work of Gyulassy, Levai, and Vitev \cite{Gyulassy:1999ig,Gyulassy:1999zd,Gyulassy:2000fs,Gyulassy:2000er}.  The reaction operator approach they developed allowed the derivation of a closed form solution of the resummed single inclusive gluon radiation distribution $dN_g/dxd^2\wv{k}$ to all orders in opacity, $\chi=L/\lambda$.  At a similar time, Wiedemann examined the dipole path integral, opacity expansion, and the Zakharov and BDMPS limits \cite{Wiedemann:2000ez,Wiedemann:2000za,Wiedemann:2000tf}.  He and Salgado \cite{Salgado:2003gb} numerically investigated the BDMPS and GLV results, publishing a popular public code for calculating BDMPS energy loss `quenching weights.'

In QED the infrared divergences from single scattering bremsstrahlung exponentiate into a Poisson distribution of multiphoton fluctuations about the semiclassical expectation value \cite{Peskin:1995}.  Multigluon fluctuations were incorporated in \cite{Gyulassy:2001nm} by \emph{assuming} this Poisson form: $P(\epsilon|\eqnpt)=\sum_n P_n(\epsilon|\eqnpt)$, $P_1(x|\eqnpt) = \exp(-\langle N_g(\eqnpt) \rangle)dN_g/dx(x;\eqnpt)$, and
\be
\label{intro:eq:poisson}
P_{n+1}(\epsilon|\eqnpt) = \frac{1}{n+1}\int dx P_{n}(x|\eqnpt)\frac{dN_g}{dx}(\epsilon-x|\eqnpt),
\ee
where the final momentum is expressed in terms of the initial momentum as $p_T^f=(1-\epsilon)p_T^i$.  
The importance of multigluon correlation effects is currently unknown and is an important open theoretical problem.  

Also associated with the Poisson convolution is probability leakage, in which $P(\epsilon>1;\eqnpt)$ unphysically has nonzero weight.  For large regions of parameter space at \rhic and \lhc this leakage is quite large due to kinematic constraints neglected in the Poisson approximation.  Two common approaches to dealing with this leakage, $P_\textrm{excess}=\int_1^\infinity P(\epsilon)d\epsilon$, are to reweigh or truncate the distribution.  When reweighing the excess is redistributed evenly by renormalizing the probability: $P_r(\epsilon)=P(\epsilon)/P_\textrm{excess}$.  Truncation takes the excess as total jet absorption: $P_t(\epsilon)=P(\epsilon)\theta(1-\epsilon)+P_\textrm{excess}\delta(\epsilon-1)$.  Since the Salgado-Wiedemann (SW) quenching weights \cite{Salgado:2003gb} use infinite jet energy, calculations for realistic jets with finite energy suffer the same problem.  Dainese, Loizides, and Paic \cite{Dainese:2004te} took two methods for redistributing this excess probability and compared the resulting $R_{AA}$s; these were taken as an error band, which turned out to be quite large.  

Neither of these proposed schemes is particularly satisfying.  While $P(\epsilon)$ with weight for $\epsilon>1$ implies that the estimated number of emitted gluons is high, there is no reason to expect the shape of the true distribution to be the same but with a different normalization.  And obviously the true shape is different from the truncation method.  Relative branching is another Poisson approximation that is an improvement on truncation and renormalization.  In this case
\be
P_{n+1}(\epsilon|\eqnpt) = \frac{1}{n+1}\int \frac{dz}{1-z} P_{n}(z|\eqnpt)\frac{dN_g}{dx}(\frac{\epsilon-z}{1-z}|\eqnpt),
\ee
where $p^n=(1-z)p^i$ and $p^f=(1-\epsilon)/(1-z)p^n=(1-\epsilon)p^i$.  In \cite{Wicks:2008} relative branching results closely resembled those from truncation; redistributing probability evenly gives excess weight to low $\epsilon$ probabilities, which---due to the steeply falling spectrum---disproportionately affects the results.  As can be seen from the above equations the Poisson convolution must have all intermediate energy loss steps evaluated at the same \ptcomma, or else it will not be properly normalized.  For large fractional energy loss, the apparent case at \rhiccomma, this is not a good approximation; it is not clear how large an effect this has on physical observables.

Around the same time Wang began the development of the `Higher-Twist' formalism \cite{Wang:2000uj,Guo:2000nz,Wang:2001ifa,Wang:2001cs,Osborne:2002dx}.  Similarly to GLV, the derivation builds up the energy loss from single hard scatterings but differs by making some alternative assumptions in their evaluation.  Most important, the arbitrary potential in GLV---usually taken as Yukawa---is replaced by an arbitrary gluon distribution function.  This obscures the relation between jet suppression patterns and physical medium quantities such as density and temperature.  For a review comparing the two results see \cite{Gyulassy:2003mc}.  

Motivated by the `dead cone' work of Dokshitzer and Kharzeev \cite{Dokshitzer:2001zm}---just as in QED, in QCD a massive charge also radiates less---and hoping for a consistent theoretical description of gluon, light quark, and heavy flavor suppression these groups' work was extended.  Zhang, Wang, and Wang \cite{Zhang:2003wk} and Armesto, Salgado, and Wiedemann \cite{Armesto:2003jh} included heavy quark mass effects in the WWGO-Z and BDMPS-Z-ASW formalisms.  Djordjevic and Gyulassy included both the effects of a heavy quark jet and a gluon mass term in D-GLV \cite{Djordjevic:2003zk}.

Since then Arnold, Moore, and Yaffe \cite{Arnold:2000dr,Arnold:2003zc,Arnold:2002ja} developed a thermal field theoretic derivation of energy loss in which dominant and subdominant contributions are carefully tracked; \cite{Jeon:2003gi} found the result to be similar to BDMPS.  A large advantage of this formalism is its simultaneous treatment of both gluon and photon bremsstrahlung, providing an added experimental consistency check.  Unfortunately their use of asymptotic states neglects the large interference effects from the initial production radiation, making comparison to data problematic.

Besides in-medium inelastic energy loss two other radiation effects have been studied: transition radiation and Ter-Mikayelian radiation reduction.  Transition radiation occurs in E\&M when a relativistic charged particle propagates through an inhomogeneous medium, in particular the boundary between two spaces with different electrical properties \cite{Baier:1998ej,Schildknecht:2005sc}.  In heavy ion collisions just such a boundary forms between the deconfined QGP medium and the vacuum; \cite{Djordjevic:2005nh} quantified the extra energy loss caused in this transition and detailed its regulation of infrared divergences ordinarily absorbed into DGLAP evolution.  

The Ter-Mikayelian (TM) effect \cite{Ter-Mikayelian:1954,Ter-Mikayelian:1972} is a direct result of radiative quanta gaining mass in a plasma.  As in beta decay, production of high momentum charged particles also has radiation associated with the process.  For QCD this infrared divergent vacuum radiation is absorbed into fragmentation functions, but in-medium the finite gluon mass regulates and suppresses this radiation.  Djordjevic and Gyulassy \cite{Djordjevic:2003be} calculated the QCD analog of the TM effect for single quark pairs.  Chapter \ref{HQPB} extends this work to include back-to-back jet production.
We find the away-side jet, necessary for charge conservation, fills in the dead cone for heavy quarks, further suppresses the production radiation, and naturally regulates the momentum loss as the jet momentum approaches zero.

\subsection{Fragility}\label{intro:fragility}
Previously we argued that by comparing theoretical model results to data jets have the potential to provide information on not just the density of scattering centers but also of the nature of the matter in the medium.  The process would be as follows: models qualitatively inconsistent with experimental observations are falsified; models in quantitative agreement with data set limits on medium properties by their input parameter range allowed by data.  
Clearly the more precise the experimental observable, the more tightly constrained the medium property.  Similarly the more sensitive the theoretical results to changes in medium property the stronger the statement about the medium.  
Of great concern but poorly investigated is the influence of theoretical imprecision---more on this later.

\cite{Eskola:2004cr} was the first to coin the term `fragility,' which we will take to mean the limit on information about the medium that can be learned by inverting the experimental data with theoretical modeling.  The paper raised the issues of surface emission, sensitivity, and fragility, and used these terms rather interchangeably.  In this thesis each term has a separate meaning and will be addressed individually.

\begin{figure}[!htb]
\centering
\includegraphics[width=4.5 in]{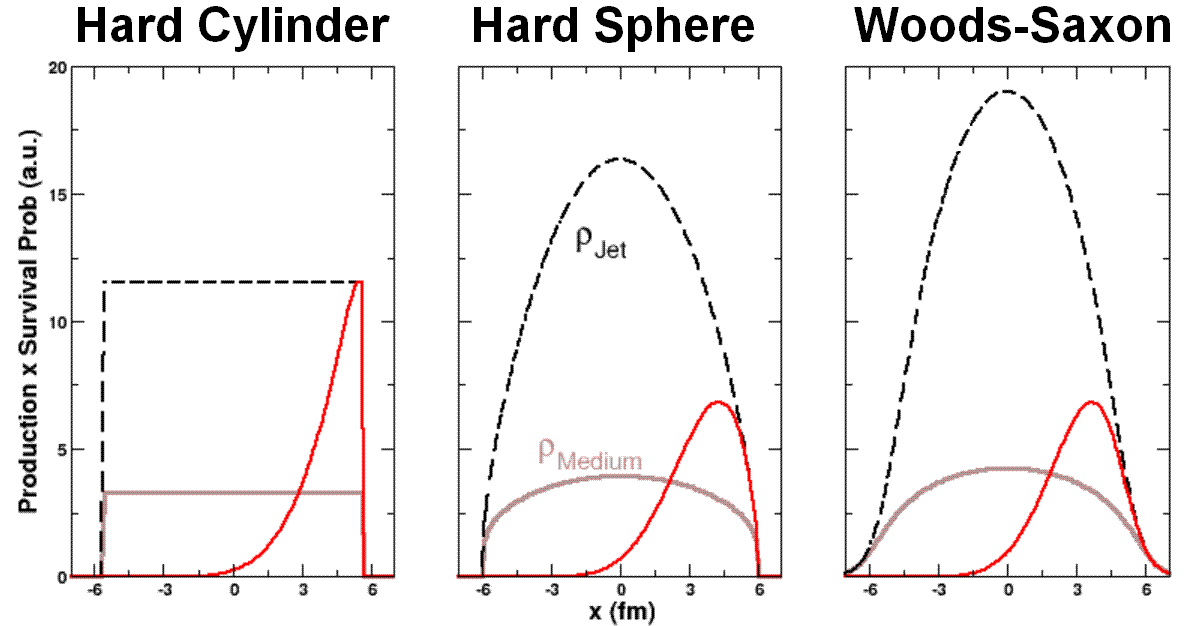}
\caption{\label{intro:surfbias}
Examples of surface bias in different nuclear geometries.  Simplified nuclear geometries such as hard spheres, and especially hard cylinders, create greater surface bias.}
\end{figure}

As noted before the production of \highpt partons is biased toward the center of a heavy ion collision because the process scales like the number of binary collisions.  Due to the large background of \lowpt particles full jet reconstruction algorithms are not yet available for \rhiccomma, although both \star and \phenix are actively researching them \cite{Lai:2008zp,Salur:2008,Putschke:2008}.  Limited to measuring the leading hadron of a jet, experimental cuts combined with the in-medium energy loss and steeply-falling production spectra bias the production point of observed \highpt particles toward the surface of the medium; see \fig{intro:surfbias}.  Na\"ively one expects that there will always be a corona of emission, or surface emission: near the very edge of the fireball the medium is so thin \highpt partons escape with little or no energy loss.  This naturally leads to the idea of a minimum nuclear modification value, $R^\textrm{min}_{AA}$, below which jets cannot be further suppressed.  In fact this is not a correct conclusion for systems with realistic nuclear densities because at the outer edge of the medium no hard scatterings occur: the thinness of the nuclear overlap results in no binary collisions.  In fact Section \ref{pqcdvsadscft} explicitly shows $R_{AA}\rightarrow0$ for large energy loss.  Even worse the phrase `surface emission' leads to a Boolean mindset: Is there surface emission or not?  This is not a useful question for a qualitative, let alone quantitative study of data; a better question would be `How biased are the jets?'  Surface emission as a term will not be used further in this thesis.  

\begin{figure}[!htb]
\centering
$\begin{array}{cc}
\includegraphics[width=2.3in]{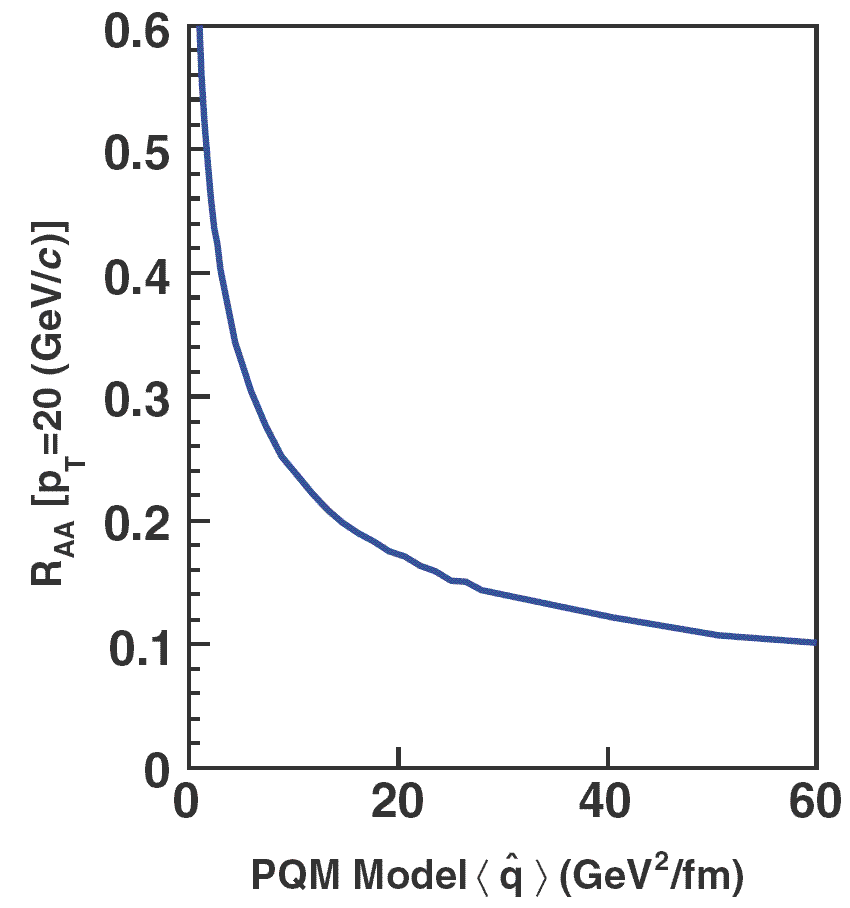} &
\includegraphics[width=2.3in]{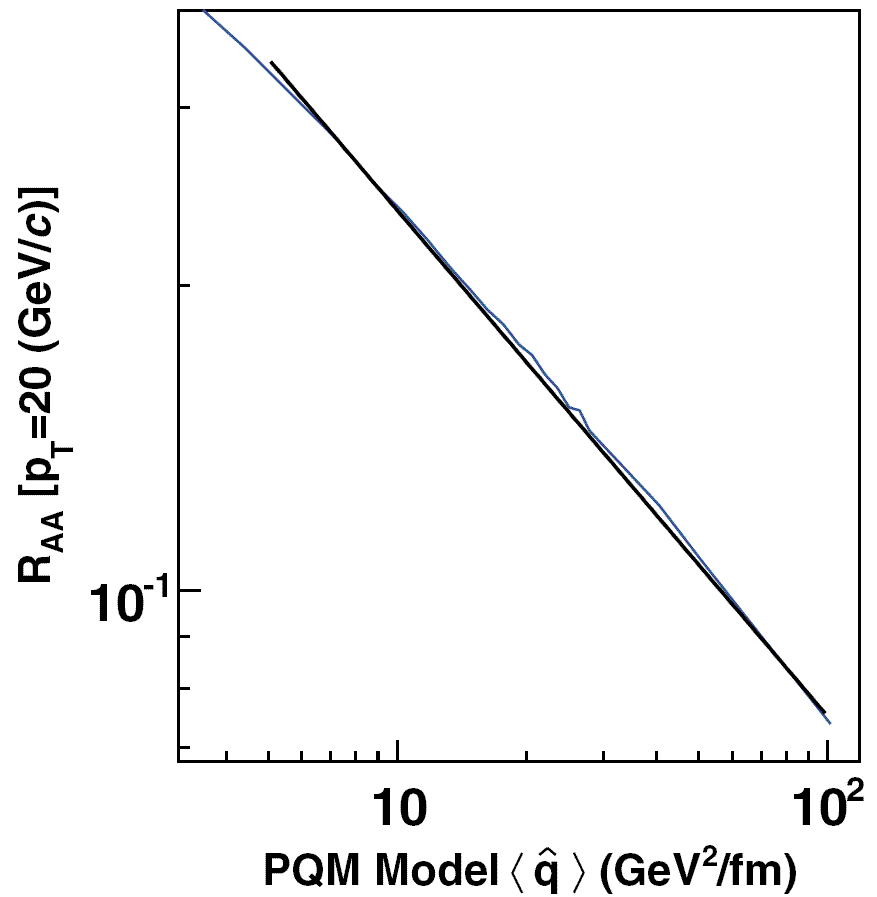} \\[-.05in]
\textrm{{\figsize(a)}} & \textrm{{\figsize(b)}}
\end{array}$
\caption{\label{intro:linlin}
(a) The energy loss model on a linear-linear scale appears to be losing sensitivity; increasing \qhat appears to be less and less effective at decreasing \raacomma. (b) The same plot but on a log-log scale.  It is clear from the fit that $R_{AA}(\eqnqhat)\sim1/\eqnqhat^{1/2}$ and is equally sensitive at all values of $\hat{q}$ explored here.  Figure adapted from \cite{Adare:2008cg}.}
\end{figure}

Theoretical sensitivity quantifies the change in a model-predicted observable given a change in the input parameter(s).  Sensitivity is a slippery slope that may easily slide into the idea of an insensitive model.  Just this type of Boolean thinking led to ironic conclusions from \cite{Eskola:2004cr} who claimed that their calculation of $R_{AA}$ became insensitive to increases in \qhatcomma.  But by quantifying the decrease in \raa as a function of \qhat in a very similar model \cite{Adare:2008cg} found a power law relationship in the `insensitive' results, $\eqnraa\sim\eqnqhat^{-1/2}$.  In this way a fixed fractional change in \qhat leads to a fixed fractional decrease in \raa, which is in agreement with the na\"ive interpretation of a sensitive theory.  The error in interpretation becomes clearer upon looking at plots of \raa vs. \qhatcomma; on a linear-linear scale \raa does appear to saturate (\fig{intro:linlin} (a)), but is clearly a power law on a log-log plot (\fig{intro:linlin} (b)).  

Fragility is another Boolean concept.  By accepting the use of that word one must conclude either one way or another than the generic convolution of theoretical sensitivity$\otimes$theoretical precision$\otimes$experimental precision is fragile or not.  A more quantitative approach would be to determine how much the model$\oplus$data system constrains \emph{physical attributes of the medium}.  The latter is emphasized to underscore the important, but often neglected, physics of mapping theoretical input parameters back to usable information on plasma properties.  Recent work by \phenix has attempted to rigorously statistically quantify the knowledge of the controlling model input parameters to be gained from data \cite{Adare:2008cg,Adare:2008qa}.  For WHDG the result is $dN_g/dy=1400^{+200}_{-375}$ and $^{+600}_{-540}$ at the 68\% and the 95\% confidence levels, respectively.  
While the 1-sigma constraint is within $\sim20\%$, the parameter range consistent with data at the 2-sigma level varies by a factor of 2; this turns out to be roughly true for all the energy loss models.

\begin{figure}[!htb]
\centering
\includegraphics[width=3 in]{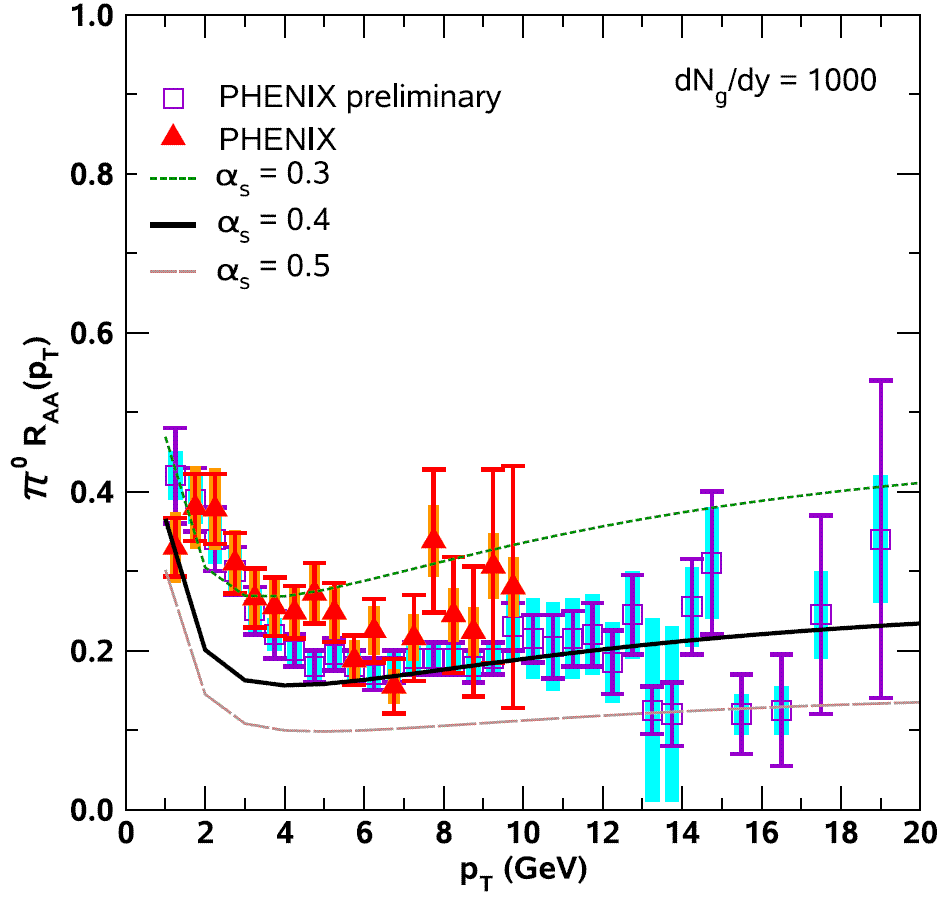}
\caption{\label{intro:pi0runalf}
A simple estimate of the large systematic theoretical error due to the running of the coupling.  Convolved collisional $+$ radiative energy loss calculations \cite{Wicks:2008} of $\pi^0$ \raapt at \rhic for fixed $\eqnalphas=.3$, .4, and .5.  Figure adapted from \cite{Wicks:2008}.}
\end{figure}

Regrettably this calculation was not able to take into account any theoretical error, which has not been extensively researched by any group.  Recent estimates of the error solely due to the running coupling are not promising; see \fig{intro:pi0runalf}.  And there is the large systematic error of vastly different approximations going into the modeling.  This suggests three courses of research for theoretical \highpt physics: (1) search for observables that are more theoretically sensitive, (2) more rigorous estimates of major sources of theoretical error and the possibility of minimizing them, or (3) search for qualitative tests of theoretical formalisms and approximations.  We pursue the latter option in this thesis.

\section{Introduction to AdS/CFT}\label{intro:adscft}
The AdS/CFT conjecture is possibly the most important new theoretical tool developed in the past two decades.  It claims a correspondence, or duality, between $SU(N)$ field theories defined in $d$ dimensions with Type IIB string theory on $d+1$ dimensional anti de-Sitter ($AdS$) space and a $9-d$ dimensional compact manifold \cite{Maldacena:1997re}.  The promise of the conjecture comes from the two different analytically tractable limits of the duals: field theories for small $\lambda=g_{YM}^2 N\ll1$; classical supergravity for $\lambda\gg1$ and $N\gg1$.  In this way strongly coupled string theory problems are approachable via weakly coupled perturbative field theoretic techniques.  More important for heavy ion physics, strongly coupled field theory calculations, previously only analyzable through equilibrium Euclidean numerics on the lattice, are now analytically solvable using classical supergravity.  But what makes the conjecture so tantalizing also makes it so enigmatic: no proof yet exists, and the different regimes of tractable applicability mean one may never be found.

\subsection{History of String Theory}
There's a certain irony that string theory is again purported to describe the strong nuclear force.  Created in the 1960's as a phenomenological model to explain the mass spectrum of hadrons, string theory originally had some successes.  A number of consequences unseen in Nature falsified string theory as a theory of hadrons.  Undiscouraged, the practitioners audaciously proposed string theory as the theory of everything.  Having possibly lost its usefulness as a theory in the landscape, the theory is back as a tool to analytically calculate, appropriately, properties of strongly coupled QCD.  
Surprisingly, string theory has rather humble roots (for a very nice historical overview of string theory developments until 1985 from one of its founders, see Schwarz \cite{Schwarz:2007yc} and references therein).  In the early 1960's there was a successful quantum field theory of electrodynamics, but Geoffrey Chew argued that a weak-coupling field theory approach was inappropriate for the strong force.  Even worse, no hadron appeared more fundamental than another: there was a paralyzing nuclear democracy (while Gell-Mann had already published his Eightfold Way quarks were considered merely a mathematical construct; the pivotal Rutherfordian \slac measurements had not yet been made).

The order of the day was to focus on physical quantities, especially the S Matrix of on-shell asymptotic particles, as opposed to off-shell physics.  General physical arguments of causality and nonnegative probabilities implied unitarity and maximal analyticity of the S matrix.  Chew and Frautschi argued for the further requirement of maximal analyticity in the angular momentum. Then the partial wave amplitudes $a_l(s)$, become analytic functions $a(l,s)$ of angular momentum and center of mass energy with isolated poles, Regge poles.  Regge trajectories, $l=\alpha(s)$, then give the position of these poles.  

\begin{figure}[!htb]
\centering
\includegraphics[width=4.75 in]{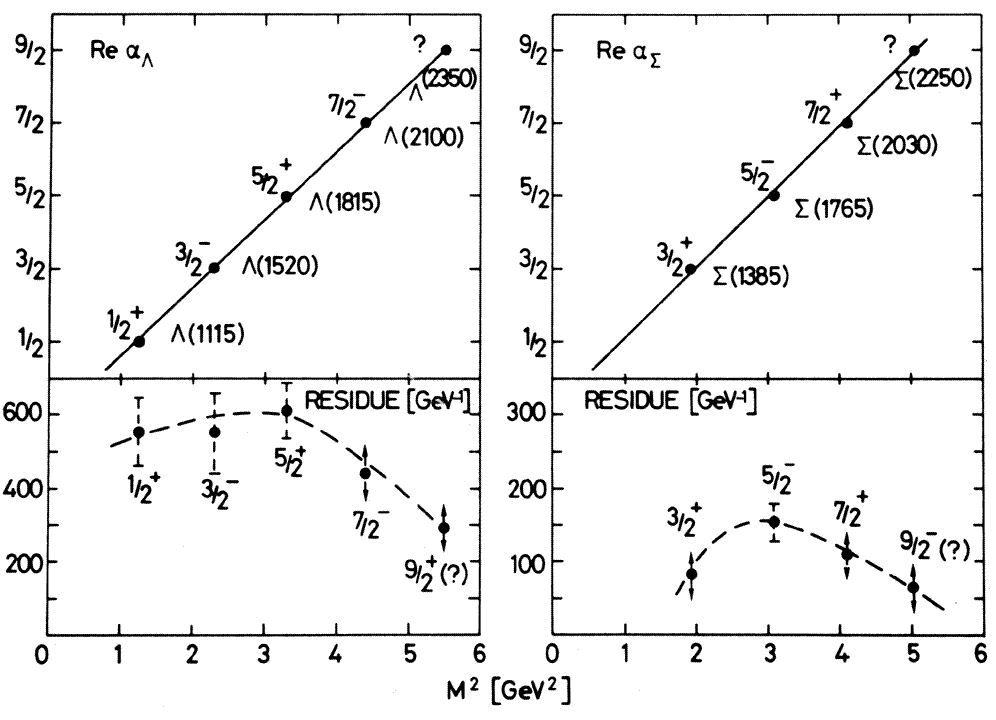}
\caption{\label{intro:regge}
Chew-Frautschi plot of the Regge trajectories $\alpha(M^2)$ for the $\Lambda$ and $\Sigma$ hadrons.  Poles at half-integer values of $\alpha$ correspond to the masses of the narrow resonances.  Figure adapted from \cite{Irving:1977ea}.}
\end{figure}

Flush with hadrons from the Bevatron, \agscomma, and \pscomma, plots of angular momentum vs.\ center-of-mass energy showed a stunning linear dependence with a common slope
\be
\alpha(s) = \alpha(0) + \alpha' s, \qquad \qquad \alpha' \sim 1.0 \; \textrm{(GeV)}^{-2};
\ee
see \fig{intro:regge} for one example.  It was argued from the crossing symmetry of analytically continued scattering amplitudes that the $t$-channel exchange of Regge poles dominated the high-energy, fixed momentum transfer, asymptotic behavior of physical amplitudes.  In this way 
\be
A(s,t)\sim \beta(t) (s/s_0)^{\alpha(s)}; \qquad s\rightarrow\infinity, t<0.
\ee
The partial wave expansion of this amplitude revealed a tower of Breit-Wigner resonances at physical values of $l$, integer for mesons and half-integer for baryons.

In order to fully characterize the S Matrix theory Chew proposed the bootstrap model, in which the hadrons themselves are exchanged to create the binding strong force; in a self-consistent way the hadrons create the force which allows their existence.  This model implied a duality: the poles in the $s$ and $t$ channels would be identical.  Unlike a Feynman-like calculation, including all channels would in this case result in double counting.

Then in 1968 Veneziano \cite{Veneziano:1968yb} proposed an actual formula for scattering amplitudes with all the required analytic and duality properties that also had linear Regge trajectories:
\bea
T & = & A(s,t) + A(s,u) + A(t,u) \\
\label{intro:open}
A(s,t) & = & \frac{\Gamma\bigl(-\alpha(s)\bigr)\Gamma\bigl(-\alpha(t)\bigr)}{\Gamma\bigl(-\alpha(s)-\alpha(t)\bigr)}.
\eea
Soon afterward Virasoro \cite{Virasoro:1969me} published an alternative form with full $s$, $t$, and $u$ symmetry:
\be
\label{intro:closed}
T = \frac{\Gamma\bigl(-\frac{1}{2}\alpha(s)\bigr)\Gamma\bigl(-\frac{1}{2}\alpha(t)\bigr)\Gamma\bigl(-\frac{1}{2}\alpha(u)\bigr)}{\Gamma\bigl(-\frac{1}{2}\alpha(s)-\frac{1}{2}\alpha(t)\bigr)\Gamma\bigl(-\frac{1}{2}\alpha(s)-\frac{1}{2}\alpha(u)\bigr)\Gamma\bigl(-\frac{1}{2}\alpha(t)-\frac{1}{2}\alpha(u)\bigr)}.
\ee
There were $N$-particle generalizations of these formulae, but most surprisingly they were found to be factorizable into a spectrum of single particle states of an infinite number of harmonic oscillators $\{\alpha^\mu_m\}$, $m=1,2,\ldots$.  This suggested a tantalizing interpretation: these were not just phenomenological fitting functions but rather a tree approximation of a full-fledged quantum field theory.

Unfortunately the Lorentz transformation properties of these operators implied the existence of negative norm states.  By imposing algebraic constraints, now named after him, on the oscillator operators Virasoro removed these states \cite{Virasoro:1969zu}.  Solving one problem led to another, though: this algebra implied that the leading trajectory associated with \eq{intro:open} had an intercept at $\alpha(0)=1$, which meant in turn that in addition to a massless vector the theory admitted a tachyonic ground state as well.  Additionally, unitarity required not the usual $d=4$ dimensions of hadronic physics but rather $d=26$.  Finally the intercepts for the trajectories of the $\pi$ and $\rho$ mesons gave them unphysically large masses.

Although there were attempts to resolve these issues and advances were made, with the advent of QCD and the Standard Model---and the convincing experimental evidence for them---string theory rapidly went out of favor.  Undaunted the remaining string practitioners abandoned it as a theory of hadrons and shot for the moon in 1974 by proposing it as a theory of everything.  By doing so the serious flaws of string theory as a theory of hadrons became assets: the massless spin-2 particle is the graviton and an inevitable consequence of strings; the theory is free of the usual UV divergences associated with point-particle quantum gravity; 
the massles spin-1 particles of open strings are Yang-Mills gauge bosons; and the extra dimension dynamics are set by the geometry of gravity.  
It is interesting to note that this promotion of string theory changed the string tension by 38 orders of magnitude, from $\alpha'\sim 1$ GeV$^{-2}$ to $\alpha'\sim 1/M_p^2$, where the Planck mass in $4D$ is $M_p\simeq 1.22 \times 10^{18}$ GeV.

The string theory story until the 1990's is of less interest to the nuclear physicist: \susy and \sugra emerged in the late 1970's; superstrings, anomaly cancellation in the early 1980's; and the discovery of heterotic strings and the compactification on Calabi-Yau manifolds of the mid-80's, the latter of which produced the first set of multiplets reminiscent of Standard Model particle classification.  
This second superstring revolution caused a frenzy of activity that continues today.  While appreciably closer to unifying the known forces, string theory still faces the daunting issues of the landscape and hierarchy problems.  
The more recent work of the 90's is of greater pertinence for heavy ion collisions.  It is during this time that string theory ideas were applied to black holes.  
\subsection{AdS/CFT Background}
\subsubsection{Holography}
The most important idea to emerge from attempting to consistently treat quantum mechanics and black holes is that of holography (for a review of holography, and especially its connection with AdS/CFT, see \cite{Bigatti:1999dp}).  A hologram in the vernacular is a 2D object that projects a 3D image of the recorded subject; in this way the full information of the 3D subject is encoded on a 2D surface.  Holography in string theory has the same connotation: given some volume the physics of the bulk is determined by the physics of the boundary.  A standard example of this principle is the derivation of the maximum entropy of a volume of space, $V$.  Imagine $V$ filled with mass and just enough energy that, should the system collapse into a black hole, its event horizon would be exactly the boundary of $V$, $\partial V$.  We know that the black hole entropy associated with this horizon is $A/4G$, where $A$ is the surface area of $\partial V$.  By the second law of thermodynamics this entropy must be greater than or equal to the entropy of the system before the collapse; therefore it gives a maximum entropy bound for $V$.  (There is some subtlety skimmed over in this argument.  As it will evaporate, the black hole is not really in equilibrium.  Nonetheless the entropy of the final state of radiation is $\mathcal{O}(1)$ times the entropy of the black hole).  Clearly AdS/CFT is another example of the holographic principle; the correspondence claims the boundary $d$-dimensional gauge field theory physics is the same as the bulk $d+1$ dimensional physics of supergravity on an \adsfive background.  

\subsubsection{$n$-dimensional GR}
Another important notion is that of a $p$ brane.  These are just black hole solutions to classical supergravity, found by extremizing the action.  Recall the physics of ordinary gravity from general relativity.  In $n$ dimensions the action is
\be
\label{ads:EH}
S = \frac{c^3}{16\pi \eqnnewtn}S_H + S_M = \int d^n x \sqrt{-g} \left[ \frac{c^3}{16\pi \eqnnewtn} R + \mathcal{L}_M \right],
\ee
where $S_H$ is the Einstein-Hilbert action, $\mathcal{L}_M$ is the Lagrange density for the matter, and $g=\det g_{\mu\nu}$ (in this AdS/CFT introduction a mostly plus metric is used), and $R$ is the Ricci scalar.  In these `unnatural' units $[\eqnnewtn] = L^{n-1} M^{-1} T^{-2}$, which leads to a length scale of a quantum theory of gravity, the Planck length,
\be
\frac{l_p^{n-2}}{2\pi} = \frac{\hbar\eqnnewtn}{c^3} \Rightarrow l_p = \left( \frac{h\eqnnewtn}{c^3} \right)^{\frac{1}{n-2}},
\ee
and a Planck mass,
\be
M_p = \left( \frac{\hbar^{n-3}}{\eqnnewtn c^{n-5}} \right)^{\frac{1}{n-2}}.
\ee
Unlike particle physics for which all quantities are in units of energy, or equivalently length, quantum gravity is dimensionless.  For completeness we note that a particle of mass $M$ has a Compton wavelength of $\lambda_c = h/Mc$ and an $n$-dimensional Schwarzschild radius
\be
R_s^{(n)} = \left( \frac{\kappa^2 M}{(n-2)\omega_{(n-2)}c^2} \right)^{\frac{1}{n-3}},
\ee
where $2 \kappa^2=16\pi\eqnnewtn/c^3$.  Then for $M\sim M_p$, $\lambda_c \sim R_s \sim l_p$ and quantum effects must be taken into account of a graviational treatment.

Variation of the action \eq{ads:EH} with respect to the metric \cite{Carroll:2004} yields the usual Einstein equations (modulo a subtlety due to varying the derivative of the metric on the boundary; see \cite{Ortin:2004}) 
\be
R_{\mu\nu}-\frac{1}{2}Rg_{\mu\nu} = \kappa^2 T_{\mu\nu},
\ee
where the symmetric, gauge-invariant energy-momentum tensor is
\be
T_{\mu\nu} = -2 \frac{1}{\sqrt{-g}}\frac{\delta \mathcal{L}_M}{\delta g^{\mu\nu}}.
\ee

With these (clearly poor) conventions the gravitational potential for a point particle is \cite{Ortin:2004}
\be
\label{ads:pot}
\phi(\vec{r}) = -\frac{\kappa^2 M c^3}{2(n-2)\omega_{(n-2)}}\frac{1}{|\vec{r}|^{n-3}},
\ee
where $\omega_d$ is the volume of a $d$-dimensional sphere of radius 1 ($\omega_1=2\pi$, $\omega_2 = 4\pi$, etc.).  We see that $n=4$ gives the usual $1/r$ potential.  \eq{ads:pot} yields the $n$-dimensional gravitational force on a test mass $m$,
\be
\vec{F}(\vec{r}) = -m\vec{\nabla}\phi(\vec{r}) = -\frac{8(n-3)\pi\eqnnewtn mM}{(n-2)\omega_{(n-2)}}\frac{\vec{r}}{|\vec{r}|^{n-1}}.
\ee

Taking $\mathcal{L}_M=0$ and a mass $M$ at the origin leads to the usual Schwarzschild solution in $4D$ (with $G\equiv G_N^{(4)}$, and $\hbar=c=1$ for the rest of the thesis),
\be
ds^2 = -\left( 1-\frac{2GM}{r} \right) dt^2+ \left( 1-\frac{2GM}{r} \right)^{-1} dr^2 + r^2 d\Omega_2.
\ee
There is a nonsingular event horizon at $r=2GM$ corresponding to $g_{rr}$ switching sign and a true singularity at $r=0$ (which can be seen from the invariant scalar $R^{\mu\nu\rho\sigma}R_{\mu\nu\rho\sigma}=48(GM)^2/r^6$; other scalars that can be examined for singularities are the Ricci scalar $R=R^\mu_{\phantom{\mu}\mu}$ and the Ricci tensor squared, $R^{\mu\nu}R_{\mu\nu}$).  

Now consider an $n=4$ gravity action including the E\&M Lagrangian, which is simply a two form field strength squared with a proportionality constant, $\mathcal{L}_\textrm{E\&M}=-(1/4)F^{\mu\nu}F_{\mu\nu}$:
\be
\label{ads:eandm}
S = \int d^4 x \sqrt{-g} \left( \frac{1}{16\pi G_N^{(4)}}R -\frac{1}{4}F^2_2 \right).
\ee
For a point particle of mass $M$ and charge $Q$ located at the origin variation of this action leads to the Reissner-Nordstr\"om metric,
\bea
ds^2 & = & -\Delta dt^2 + \Delta^{-1}dr^2+r^2d\Omega^2_2, \\
\Delta & = & 1-\frac{2GM}{r}+\frac{GQ^2}{r^2}.
\eea
There are now two special radial coordinates at which $g_{rr}=0$,
\be
r_\pm = GM \pm \sqrt{G^2M^2-GQ^2};
\ee
there can be zero, one, and two solutions to this equation.  For $GM^2<Q^2$ there are no solutions, which leads to a naked singularity at the origin (under very general conditions, the cosmic censorship conjecture prevents these from forming from gravitational collapse).  
For $GM^2>Q^2$ there are two event horizons, both of which are removable coordinate singularities.  The $r_+=r_-$ single solution case defines an \emph{extremal} black hole.

\subsubsection{$p$ branes}
Following \cite{Horowitz:1991cd,Aharony:1999ti} consider the bosonic part of a supergravity action in 10 dimensions, a slight generalization of \eq{ads:eandm}:
\be
S = \frac{1}{(2\pi)^7 l_s^8}\int d^{10}x\sqrt{-g} \left( e^{-2\phi}(R+4(\nabla \phi)^2) + \frac{2}{(8-p)!}F^2_{p+2} \right),
\ee
where $\phi$ is the dilaton field, $F_{p+2}$ is a $p+2$ form field strength, and $l_s$ is the string length; the string length $l_s$ is related to the string tension $(2\pi \alpha')^{-1}$ by $l_s^2=\alpha'$.  This form is suggestive of the result $G_N^{(10)}=8\pi^6 g_s^2 l_s^8$ \cite{Ortin:2004}, where $g_s$ is given by the vacuum expectation value of the dilaton field, $g_s = \exp(\phi)$.  Just as a 0 dimensional point charge sources a 1 form potential $A$ which creates a 2 form field strength $F=dA$ in E\&M, this $p+2$ form is sourced by a $p$ dimensional object, a $p$ brane.  Similarly to E\&M, setting a spherically symmetric charge at the origin the integral of the flux of the $p+2$ form yields the charge:
\be
\int_{S^{8-p}}* F_{p+2} = N.
\ee
Extremizing the above action for a $p$ brane with the $p$ parallel dimensions denoted by $x^i$, the radial distance away from the $p$ brane denoted by $\rho$, and the angular dimensions encoded in $d\Omega$, yields the metric
\bea
\label{ads:pmetric}
ds^2 & = & -\frac{f_+(\rho)}{\sqrt{f_-(\rho)}}dt^2 + \sqrt{f_-(\rho)} \sum_{i=1}^p dx^i dx^i + \frac{f_-(\rho)^{-\frac{1}{2}-\frac{5-p}{7-p}}}{f_+(\rho)}d\rho^2 \nonumber\\
& &  + \rho^2 f_-(\rho)^{\frac{1}{2}-\frac{5-p}{7-p}}d\Omega_{8-p}^2;\\
e^{-2\phi} & = & \frac{1}{g_s^2}f_-(\rho)^{-\frac{p-3}{2}},\\
f_\pm(\rho) & = & 1-\left(\frac{r_\pm}{\rho}\right)^{7-p}.
\eea
$r_\pm$ characterize the mass and charge of the $p$ brane by
\bea
M & = & \frac{1}{(7-p)(2\pi)^7d_pg_s^2 l_s^8}\left( (8-p)r_+^{7-p}-r_-^{7-p} \right) \\
N & = & \frac{1}{d_p g_s l_s^{7-p}}(r_+ r_-)^{\frac{7-p}{2}},
\eea
where $d_p$ is a numerical factor given by
\be
d_p = 2^{5-p} \pi^\frac{5-p}{2} \Gamma \left( \frac{7-p}{2} \right).
\ee
The interpretation for $p$ branes is exactly the same as for charged black holes: $r_->r_+$ gives a naked singularity at $r=r_-$; $r_+>r_-$ a horizon and removable coordinate singularity at $r=r_+$.  

For an extremal $p$ brane $r_+=r_-$ and \eq{ads:pmetric} becomes
\bea
ds^2 & = & \sqrt{f_+(\rho)} \left( -dt^2 + \sum_{i=1}^p dx^i dx^i \right) + f_+(\rho)^{\frac{3}{2}-\frac{5-p}{7-p}}d\rho^2 \nonumber\\
& &  + \rho^2 f_+(\rho)^{\frac{1}{2}-\frac{5-p}{7-p}}d\Omega_{8-p}^2.
\eea
Changing coordinates to a distance beyond the horizon
\be
r^{7-p} = \rho^{7-p}-r_+^{7-p}
\ee
we find the metric takes the form
\bea
\label{ads:rmetric}
ds^2 & = & \frac{1}{\sqrt{H(r)}} \left( -dt^2 + \sum_{i=1}^p dx^i dx^i \right) \nonumber\\
& & + \sqrt{H(r)}\left( dr^2 + r^2 d\Omega_{8-p}^2 \right); \\
e^{-2\phi} & = & \frac{1}{g_s^2} H(r)^\frac{3-p}{4}, \\
H(r) & = & \frac{1}{f_+(\rho)} = 1+\left(\frac{r_+}{r}\right)^{7-p},\\
r_+^{7-p} & = & g_s d_p N l_s^{7-p},
\eea
which we will find useful later on.

\subsubsection{$D$ branes and Chan-Paton Factors}
\begin{figure}[!htb]
\centering
\includegraphics[width=4 in]{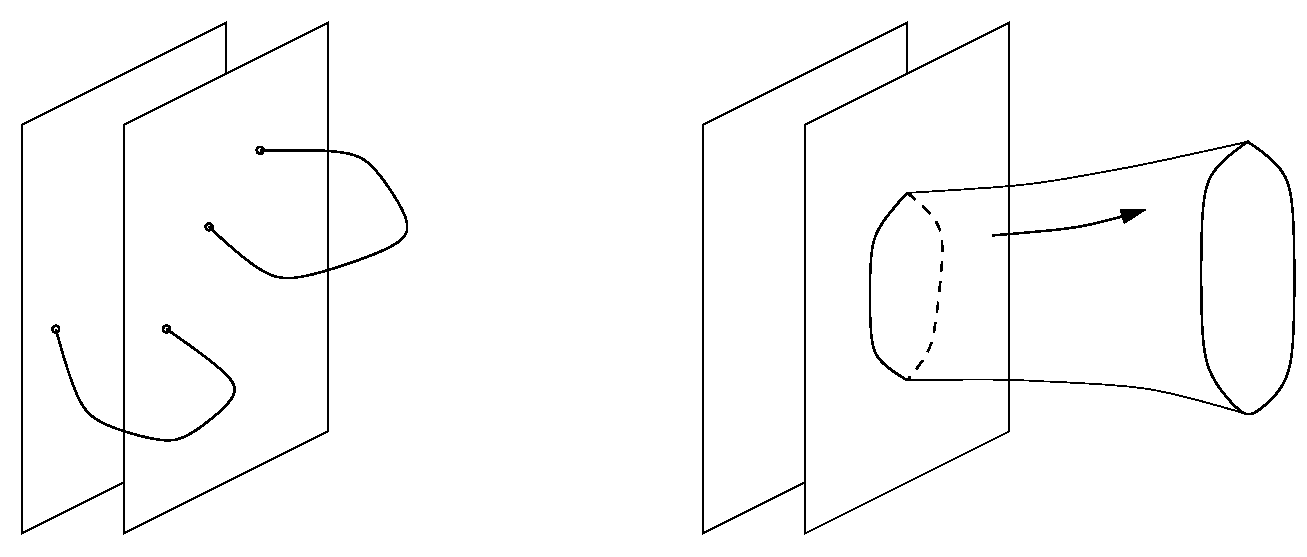}
$\begin{array}{cc}
\textrm{{\figsize(a)}} \qquad\qquad\qquad\qquad\qquad\qquad & \qquad\textrm{{\figsize(b)}}
\end{array}$
\caption{\label{ads:dbranes}(a) Open strings ending on a $D$ brane.  (b) $D$ branes as the source of closed strings.  Figure adapted from \cite{Aharony:1999ti}.}
\end{figure}

\begin{figure}[!htb]
\centering
$\begin{array}{cc}
\includegraphics[width=2.3in]{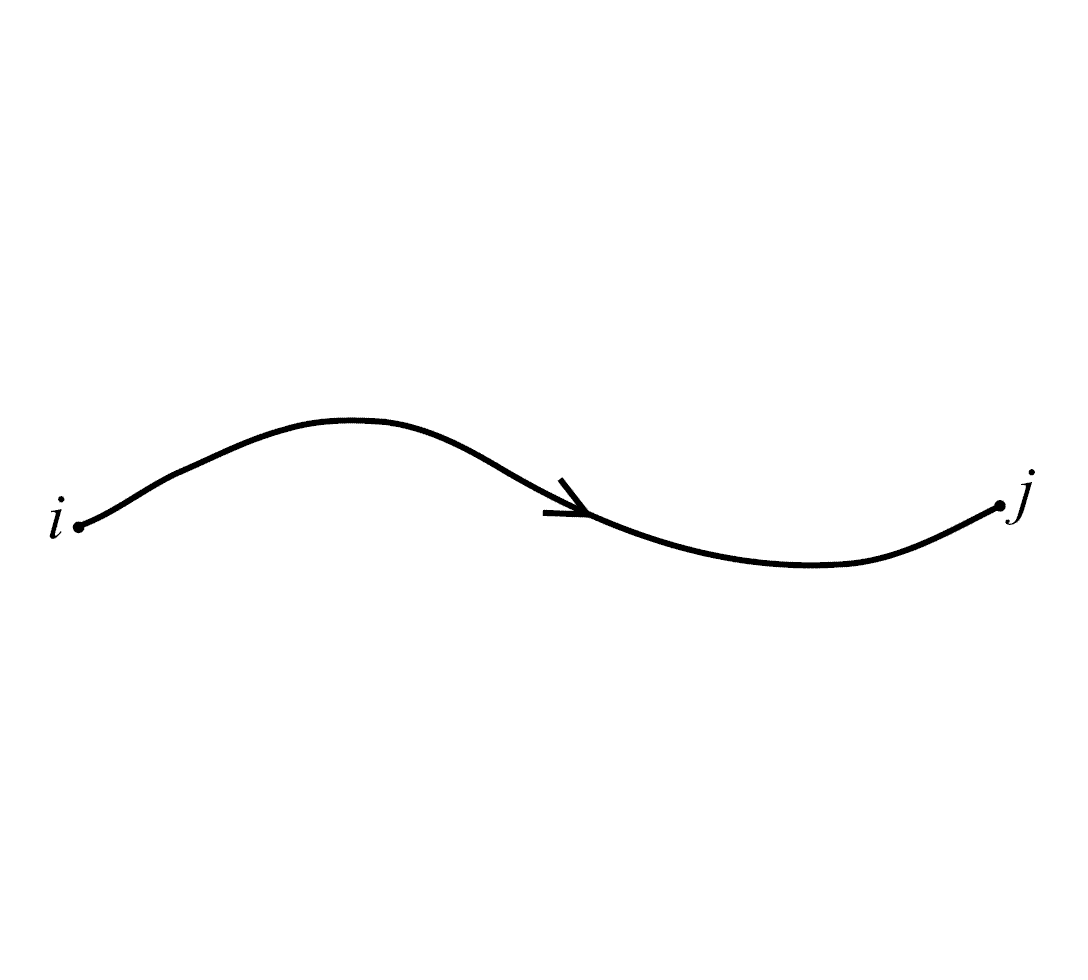} &
\includegraphics[width=2.3in]{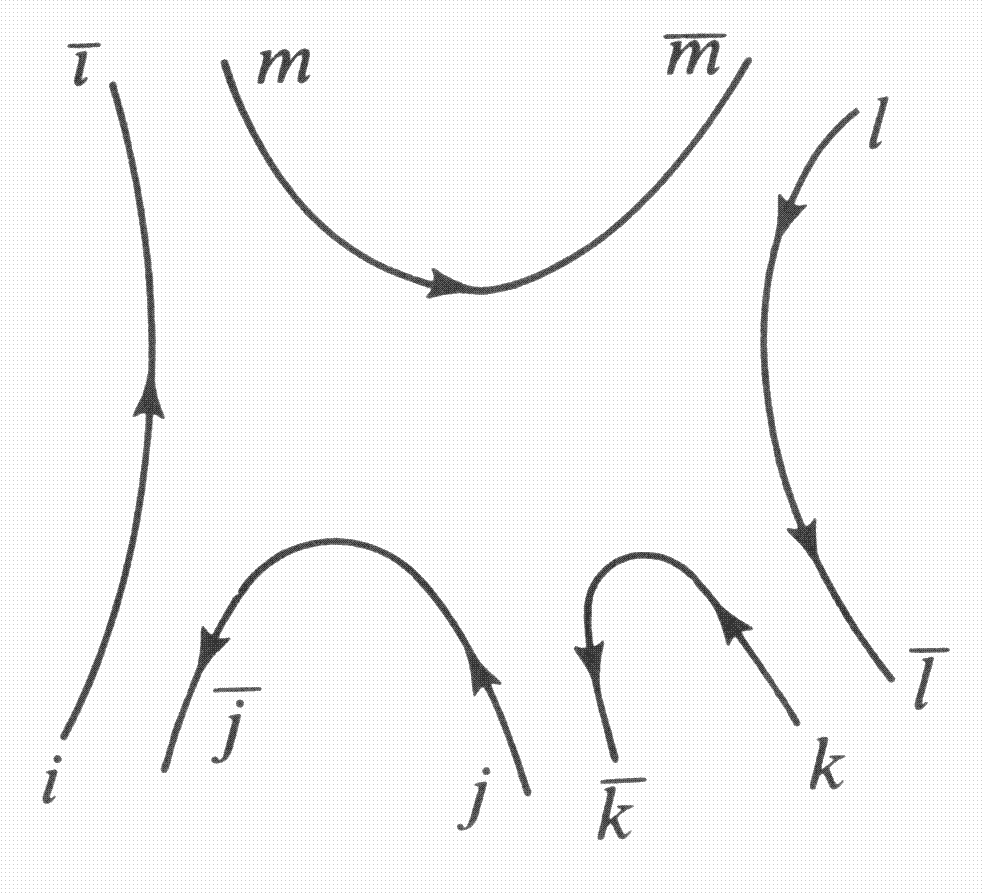} \\[-.05in]
\textrm{{\figsize(a)}} & \textrm{{\figsize(b)}}
\end{array}$
\caption{\label{ads:openstring}(a) Open string with $i$ and $j$ Chan-Paton degrees of freedom.  (b) Open string scattering amplitude; the Chan-Paton index for the right end of each string becomes that of the left end of the next.  In the hadronic string theory this corresponds to 3 meson to 2 meson scattering.  Figures adapted from \cite{Polchinski:1996na,Green:1987}.}
\end{figure}

Besides strings, string theory necessarily involves nonperturbative objects upon which open strings may end; these are $D$ branes, $D$ for Dirichlet, or sometimes $Dp$ branes (for a review of $D$ branes see \cite{Polchinski:1996na}).  It is thought that $p$ branes and $Dp$ branes are different descriptions of the same object, with the $p$ brane description of the backreaction on the geometry valid for small curvature in Planck units ($r_-/l_p\equiv R/l_p\gg1$).  The endpoints of open strings can be charged with a non-dynamical degree of freedom.  Each end is indexed by $i$ or $j$ running from 1 to $N$, where $N$ is the number of $D$ branes (only Type I string theory has open strings not ending on $D$ branes, but we are not interested in this string theory); see \fig{ads:openstring} (a).  The $N\times N$ matrices $\lambda_{ij}^a$ form a basis for this part of the string wavefunction, and in a scattering amplitude each vertex carries one of these matrices.  Consider a scattering amplitude in which the $N$ $D$ branes are coincident.  In this case one must sum over all possible Chan-Paton indices and, in addition, since the right endpoint of a string is the same as the left endpoint of another string---see \fig{ads:openstring} (b)---the total scattering amplitude results in a trace over these matrices $\Tr(\lambda^{a_1}\cdots\lambda^{a_n})$.  Therefore the amplitude is invariant under $U(N)$ transformations of these matrices, $\lambda \rightarrow U\lambda U^{-1}$.  One may then elevate this global symmetry to a local one in spacetime by defining the vertex operator $V^{a\mu}_{ij} = \lambda_{ij}^a \partial_t X^\mu \exp(ik\cdot X)$ that transforms under the adjoint of $U(N)$.  Since $U(N)=U(1)\times SU(N)$---$U\in U(N)$ is an $N\times N$ unitary matrix.  Unitarity implies that its determinant is a pure phase: $UU^\dag = 1 \Rightarrow |det(U)|=1$.  Therefore any $U(N)$ matrix may be written as $U=\exp(i\phi)SU$, where $SU\in SU(N)$---we now have a $p+1$ dimensional $SU(N)$ field theory defined on the stack of $Dp$ branes.  

Specifically, the low energy effective action for a $D3$ brane in $AdS$ space can be reliably approximated by the Born-Infeld action \cite{Leigh:1989jq,Aharony:1999ti} for the massless modes of the string theory.  For Type IIB the boson part of the action on the $D$ brane worldvolume reads:
\bea
\label{ads:daction}
S_{D3} & = & \frac{1}{(2\pi)^3g_s\alpha'^2}\int d^4 xf^{-1}\left[\phantom{\sqrt{\sqrt{f}}}\right. \nonumber\\
& & \left.\sqrt{-\det(\eta_{\alpha\beta}+f\partial_\alpha r\partial_\beta r+r^2fg_{ij}\partial_\alpha \theta^i \partial_\beta \theta^j+2\pi\alpha'\sqrt{f}F_{\alpha\beta})}-1\right], \nonumber\\
& & \\
f & = & \frac{4\pi g_s\alpha'^2N}{r^4},
\eea
where the $\theta^i$ are the angular coordinates of the 5-sphere.  The $r$ and $\theta^i$ are then six scalar fields that are functions of the brane coordinates.  Notice that by changing coordinates to $U=r/\alpha'$ the $\alpha'$ dependence drops out of $S_{D3}$.  The controlling parameter for \eq{ads:daction} then turns out to be $g_s N (\partial U)^2/U^4$ \cite{Aharony:1999ti}.  The lowest order term from the action yields $\mathcal{N}=4$ $SU(N)$ super-Yang-Mills from the $F_{\alpha\beta}$ term.  The quadratic term has no quantum corrections and the quartic term has only a one-loop correction, consistent with $\mathcal{N}=4$ SYM \cite{Seiberg:1994aj,Dine:1997nq}.  This loop correction has been calculated from the gauge theory and string theory, and the two agree \cite{Douglas:1996yp}.  Moreover, it can be argued that all higher order terms in \eq{ads:daction} are determined from the fourth-order term \cite{Maldacena:1997re}.  In order for this series to make sense $g_s N (\partial U)^2/U^4\ll 1$.  In particular in the supergravity regime of the \ads duality $g_s N\gg1$, and the higher order terms beyond $SU(N)$ may become important.

\subsection{Motivation for the AdS/CFT Correspondence}
In this section we will follow \cite{Aharony:1999ti} closely for a schematic motivation of the \ads conjecture (see the same for a very good review of the correspondence as well as an extensive bibliography).  Consider from a field theoretic standpoint Type IIB string theory in a 10 dimensional spacetime with a stack of $N$ coincident extremal ($r_+=r_-\equiv R$) $D3$ branes at the origin.  At energies small compared to the string scale $1/l_s=1/\sqrt{\alpha'}$ only the massless string modes can be excited.  
The full low energy effective action is then
\be
S = S_\textrm{bulk} + S_\textrm{branes} + S_\textrm{int}.
\ee
$S_\textrm{bulk}$ is the $10D$ supergravity action with higher derivative corrections.  $S_\textrm{branes}$ is the $3+1D$ SYM action with higher derivative corrections.  Finally $S_\textrm{int}$ describes the interaction between the brane and bulk modes.  It turns out that all of these actions are controlled by $\alpha'$: the bulk action is controlled by Newton's constant $\kappa$, the brane action by $\alpha'^2$, and $S_\textrm{int}$ also by $\kappa$.  Then taking the low energy limit $\kappa\sim g_s \alpha'^2\rightarrow0$ (specifically $\alpha'\rightarrow0$ with $U=r/\alpha'$ fixed), the higher derivative and interaction terms all drop out.  We are left with noninteracting supergravity in the bulk and $\mathcal{N}=4$ $SU(N)$ SYM on the branes, which are decoupled from each other.

On the other hand consider the geometrical interpretation of the low energy limit of Type IIB string theory in 10 dimensions.  From \eq{ads:rmetric} we have that
\bea
\label{ads:dmetric}
ds^2 & = & f^{-1/2}(-dt^2+dx_1^2+dx_2^2+dx_3^2)+f^{1/2}(dr^2+r^2 d\Omega_5^2),\\
\label{ads:coupling}
f & = & 1+\frac{R^4}{r^4}, \quad R^4 = 4\pi g_s l_s^4 N.
\eea
There are two regions of importance: the near horizon limit, $r\rightarrow0$, and the asymptotically flat limit, $r\rightarrow\infinity$.  In the flat limit the metric is the usual Minkowski one corresponding to, because of $\alpha'\rightarrow0$, free supergravity.  In the low energy limit these massless excitations decouple from the brane physics as their cross section goes $\sim \omega^3 R^8$ \cite{Klebanov:1997kc} (note that cross sections in $n$ spatial dimensions for problems with $p$ translational symmetries have dimension $L^{n-1-p}$).  On the other hand the nontrivial $g_{tt}$ component of \eq{ads:dmetric} means that excitations in the near horizon throat region appear red shifted to an observer at infinity, $E_\textrm{inf} = f^{-1/4} E_r$.  Since it is the energy as measured by an observer at infinity that is important, in the limit $r\rightarrow0$ the full Type IIB string theory must be kept.   
Nevertheless the higher energy modes from the string theory cannot escape the throat region without being redshifted away.  We are thus left with supergravity in flat asymptopia and IIB string theory compactified on the near-horizon geometry, and the two are decoupled.

Stepping back we find that we have two pictures of the same low energy limit of one theory, Type IIB string theory.  In the field theory picture we have two decoupled theories: supergravity in the far region and $\mathcal{N}=4$ $SU(N)$ SYM on the $D$ branes.  In the geometry picture we also have two decoupled theories: supergravity in asymptotically flat space and Type IIB string theory in the throat region.  Noticing that both pictures have identical asymptotic supergravity in them we boldly propose, in the spirit of Maldacena, that the other decoupled theories are also identical: Type IIB string theory compactified on the near horizon background of \eq{ads:dmetric} is dual to $3+1D$ $\mathcal{N}=4$ $SU(N)$ SYM.

The regions for which analytic tools exist for these two different pictures turn out to be completely incompatible.  Comparing the Born-Infeld action of $N$ coincident $D$ $p$ branes and that for an $SU(N)$ two form field strength yields
\be
g_{SYM}^2 = 2 g_s (2\pi)^{p-2}l_s^{3-p} \; \Rightarrow \; g_{SYM}^2 = 4\pi g_s\textrm{ for }p=3.
\ee
Perturbative field theory is a consistent approach for calculations when
\be
g_{SYM}^2 N = 4\pi g_s N = \frac{R^4}{l_s^4} \ll 1,
\ee
where the final relation came from \eq{ads:coupling}.  

Contrariwise string theory is tractable in the classical supergravity limit, in which case the characteristic length scale of the problem, $R$, is large compared to the two length scales of quantum string theory: the string length, $l_s$, and the Planck length, $l_p$; this ensures no string corrections and no loop corrections, respectively.  Demanding this implies
\be
\frac{R^4}{l_s^4} = 4\pi g_s N = g_{SYM}^2 N \gg 1; \quad \frac{R^4}{l_p^4} \sim N \gg 1.
\ee

There is an extra subtlety due to the $SL(2,\mathbb{Z})$ self-duality symmetry of both the gauge and string theory.  For the coupling $\tau = 4\pi i/g_{SYM}^2 = i/g_s$ the theories are invariant under $\tau\rightarrow (a\tau+b)/(c\tau+d)$ where $ad-bc=1$.  We can see that one can take $g_s\rightarrow1/g_s$ with $a=d=0$ and $b=1$, $c=-1$; a strong string coupling can be swapped for a weak one and vice-versa, but the number of colors must always remain large.  This is the origin of the $\lambda$ large and fixed while $N\rightarrow\infinity$ limit; this guarantees an $S$ transformation leaves $R$ large compared to both $l_s$ and $l_p$.  

Now we will show that the near horizon limit that is the background for the dual string theory is \adsfivecomma.  To see this, taking $r\rightarrow0$ in \eq{ads:dmetric} yields
\be
\label{ads:nhmetric}
ds^2 = \frac{r^2}{R^2}(-dt^2+dx_1^2+dx_2^2+dx_3^2)+\frac{R^2}{r^2}dr^2+R^2d\Omega_5^2.
\ee
Compare this to the metric of a $p+2$ dimensional hyperboloid
\be
\left.X^0\right.^2+\left.X^{p+2}\right.^2-\sum_{i=1}^{p+1}\left.X^i\right.^2 = R^2
\ee
embedded in a flat space with SO(2,p+1) isometries
\be
ds^2 = -\left.dX^0\right.^2 - \left.dX^{p+2}\right.^2 + \sum_{i=1}^{p+1}\left.dX^i\right.^2.
\ee
If we reparameterize the hypersurface with coordinates $y^\alpha = (t,r,\vec{x})$, $r>0$ such that
\bea
X^0 & = & \frac{1}{2r}\left( R^2+\frac{r^2}{R^2}(R^2+\vec{x}^2-t^2) \right), \\
X^i & = & \frac{rx^i}{R}, \quad (i=1\ldots p)\\
X^{p+1} & = & \frac{1}{2r}\left( R^2 - \frac{r^2}{R^2} (R^2-\vec{x}^2+t^2)\right), \\
X^{p+2} & = & \frac{rt}{R},
\eea
then the induced metric is obtained from the pullback \cite{Carroll:2004}
\bea
(y^* g)_{\alpha\beta} & = & \frac{dx^\mu}{dy^\alpha}\frac{dx^\nu}{dy^\beta}g_{\mu\nu} \\
& = & \frac{r^2}{R^2}(-dt^2 + \sum_{i=1}^{p+1} dx^i) + \frac{R^2}{r^2}dr^2.
\eea
Therefore \eq{ads:nhmetric} is exactly \adsfivecomma.

While there is no proof yet of the AdS/CFT conjecture a number of tests have found quantities calculated from the field theory side and the string theory side agree.  These are nontrivial in the sense that they must be independent of coupling.  These tests include checking the equivalence of symmetries in both theories; some correlation functions, usually associated with anomalies, that are independent of quantum corrections and $\lambda = g_{SYM}^2N$; the spectrum of chiral operators match for those that are currently calculable; see \cite{Aharony:1999ti} and references therein for a more complete list.

Having used the above picture to motivate the correspondence we now throw away that geometry and simply take our whole space to be \adsfivecomma.  It is useful to know that the boundary at asymptotic $r\rightarrow\infinity$ is the usual $3+1D$ Minkowski.  To see this first change coordinates to $y = R/r$.  Then \eq{ads:nhmetric} becomes
\bea
ds^2 & = & \frac{R^2}{y^2}(-dt^2+d\vec{x}^2) + \frac{R^2}{y^2}dy^2 + R^2 d\Omega_5^2 \\
& \rightarrow & -dt^2+d\vec{x}^2 + dy^2 + y^2 d\Omega_5^2,
\eea
where for the second step we Weyl rescaled the metric by an overall factor of $y^2/R^2$.  Now $r\rightarrow\infinity$ corresponds to $y\rightarrow0$, and we see that at the boundary the radius of the sphere shrinks to 0 and usual $3+1D$ Minkowski is recovered.

\subsection{Applications of \ads to Heavy Ion Collisions}
To make contact with thermal QCD physics we extend the conjecture to nonextremal $D3$ branes, in which case conformality is broken.  Starting with the nonextremal 3-brane metric \eq{ads:pmetric}, take $R=r_-$, $r_0=r_+^4-r_-^4$, and change coordinates to $r^4=\rho^4-R^4$; it then becomes
\bea
ds^2 & = & \frac{1}{\sqrt{H(r)}}\left[ -h(r)dt^2 + d\vec{x}^2 \right] + \sqrt{H(r)}\left[ \frac{dr^2}{h(r)} + r^2 d\Omega_5^2 \right],\\
H(r) & = & 1 + \frac{R^4}{r^4}; \quad h(r) = 1 - \frac{r_0^4}{r^4}.
\eea
This yields in the near-horizon, or decoupling, limit of $r\ll R$
\be
\label{ads:thermal}
ds^2 = \frac{r^2}{R^2} \left[ -h(r)dt^2 + d\vec{x}^2 \right] + \frac{R^2}{r^2 h(r)}dr^2 + R^2d\Omega_5^2.
\ee
We find the temperature associated with the horizon by requiring that the $\tau=it$ coordinate after Wick rotation has the correct periodicity for a path integral partition function.  We could integrate $\tau$ in a circle around $r=r_0$ \cite{Klebanov:2000me} but will rather follow \cite{Johnson:2003}: for a metric of the form $ds^2 = Vd\tau^2 + V^{-1}dr^2 + \ldots$ the temperature is $T=V'|_{r=r_0}/4\pi$, where $r_0$ is the radial position of the horizon.  Both methods yield
\be
\label{ads:temp}
T = \frac{r_0}{\pi R^2}.
\ee

Using this equation for $T$ one may follow \cite{Gubser:1996de,Klebanov:2000me} to compare the entropies of weakly coupled $\mathcal{N}=4$ $U(N)$ and strongly coupled gravity at the same temperature.  The first comes from the usual counting of states with $6N^2$ massless scalars and $4N^2$ massless Weyl fermions.  Taking the spatial volume of the 3-brane to be $V_3$ we have
\be
S = \frac{2\pi^3}{3}N^2 V_3 T^3.
\ee
The Bekenstein-Hawking relation, $S_{BH}=A/4G_N$, gives the entropy of the black $p$ brane in the supergravity picture.  The 8-dimensional area of the horizon may be read off of \eq{ads:thermal} as
\be
A = \left( \frac{r_0}{R} \right)^3 V_3 R^5 \Omega_5 = \pi^6 V_3 R^8 T^3,
\ee
where the second equality comes from \eq{ads:temp}.  By \eq{ads:coupling} $G_N^{(10)}=8\pi^6g_s^2l_s^8=\pi^4 R^8/(2N^2)$, so
\be
\label{ads:bhent}
S_{BH} = \frac{A}{4G_N} = \frac{\pi^2}{2}N^2 V_3 T^3.
\ee
The factor of $4/3$ difference between the weakly-coupled and black hole entropy limits is a prediction for the change in the number of degrees of freedom from weakly-coupled $SU(N)$ to strongly-coupled $SU(N)$.  This provides a natural explanation for the discrepancy of about 20\% between lattice results at a few times $T_c$, still at strong coupling, and the expected weak-coupling Stefan-Boltzmann limit as seen in \fig{intro:latticePT} (b).  

One may next find the strong coupling limit for the entropy to viscosity ratio, $\eta/s$.  By comparing the correlation function for the low energy scattering cross section used in the motivation for the correspondence to the Kubo formula for the viscosity of a strongly-coupled SYM plasma \cite{Kovtun:2003wp} one finds that
\be
\eta = \frac{\sigma_\textrm{abs}(0)}{16\pi G_N} = \frac{A}{16\pi G_N},
\ee
where \cite{Das:1996we} found that the cross section for extended $p$-branes has a universal low energy limit of the area of the horizon.  Using this with \eq{ads:bhent}, one quite trivially arrives at the famous strong coupling limit of viscosity to entropy
\be
\frac{\eta}{s} = \frac{1}{4\pi}.
\ee
The result led to another conjecture \cite{Kovtun:2004de}, that this is in fact a universal lower bound.  
As noted in Section \ref{bulk}, qualitative estimates of the viscosity at \rhic \cite{Teaney:2003kp} appear to require just such values of $\eta/s\sim.1\sim 1/4\pi$.  This gives great impetus to look for additional observables calculable in \adscomma.

\begin{figure}[!htb]
\centering
$\begin{array}{cc}
\includegraphics[width=2.35in]{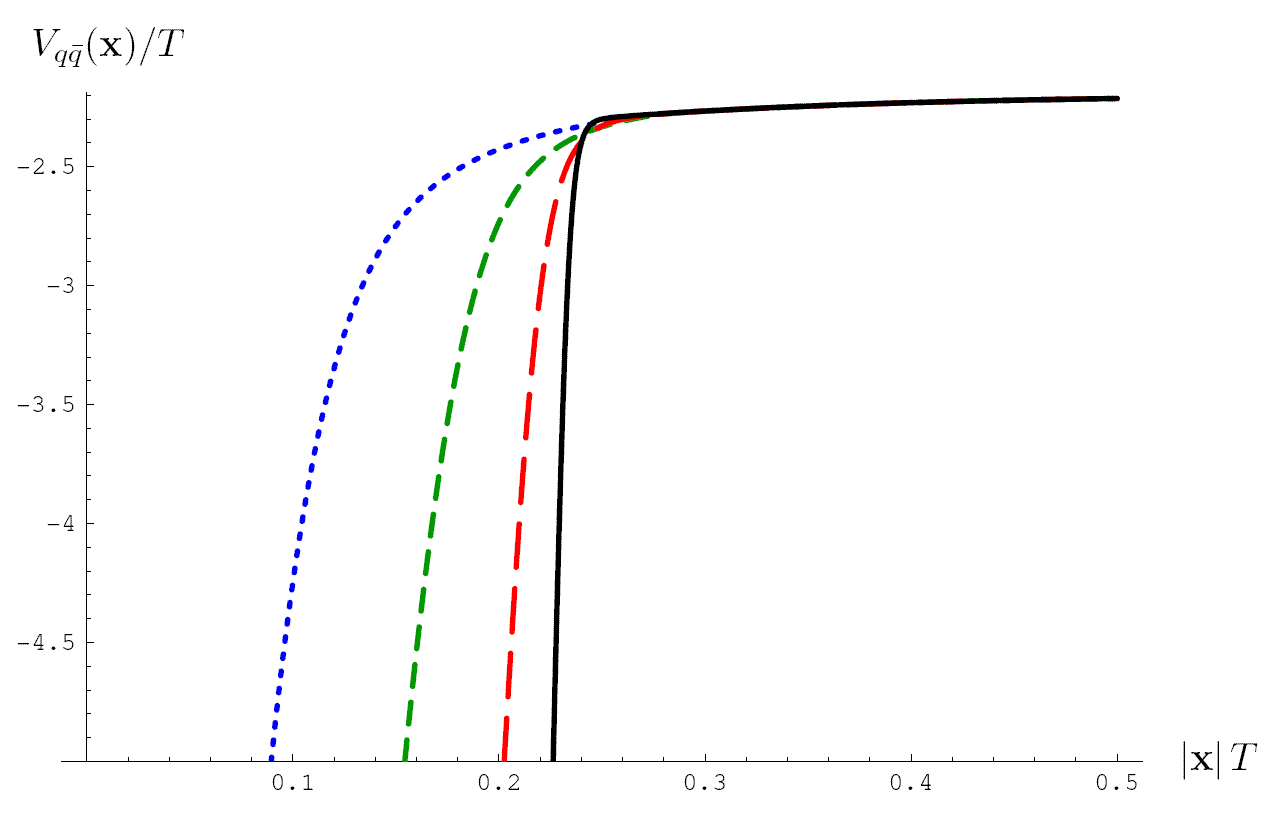} &
\includegraphics[width=2.75in]{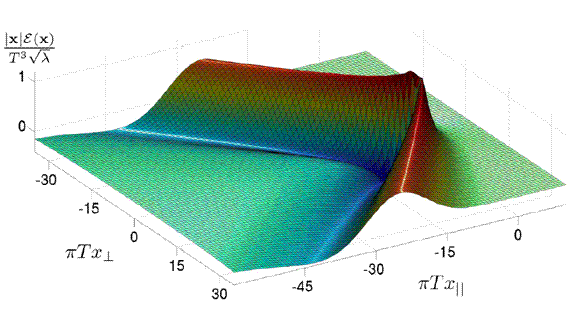} \\[-.05in]
\textrm{{\figsize(a)}} & \textrm{{\figsize(b)}}
\end{array}$
\caption{\label{ads:debye}
(a) Heavy $q\bar{q}$ potential for $N_c=3$ and $\lambda = 10$ (blue, dotted), $10^2$ (green, short dash), $10^3$ (red, long dash),
and $10^4$ (black, solid).  As $\lambda\rightarrow\infinity$ the potential develops a kink at $|\wv{x}|=.24/T$; for larger distances the potential continues to rise, but very slowly.  Figure adapted from \cite{Bak:2007fk}.  (b) The \ads scaled energy density of the SYM plasma with Mach cone for a heavy quark with $v=3/4$.  Figure adapted from \cite{Chesler:2007an}.}
\end{figure}

\begin{figure}[!htb]
\centering
\includegraphics[width=3.75 in]{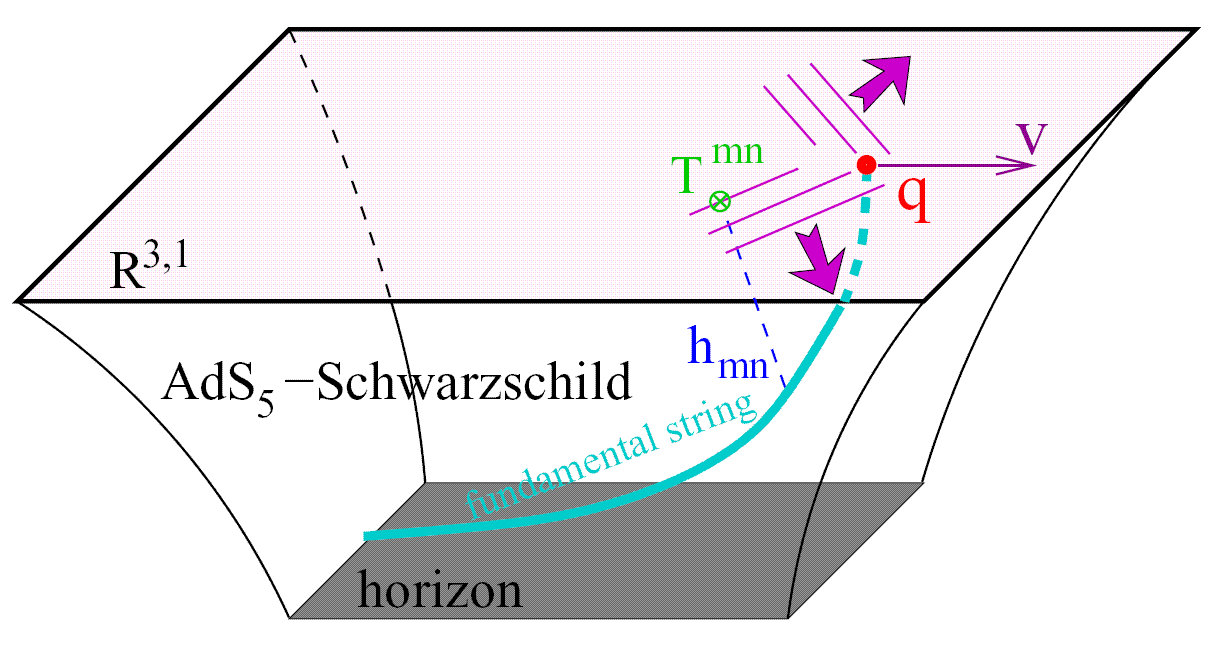}
\caption{\label{ads:Tmunu}
Visualization of a heavy quark propagating through a thermal plasma.  The fundamental quark is represented by a string hanging from a $D7$ brane (in this case at $r=\infinity$) down through the horizon onto the stack of coincident$D3$ branes.  Its motion causes a metric disturbance on the Mikowski boundary dual to a wake in the SYM plasma and a loss of momentum down the length of the string.  Figure adapted from \cite{Friess:2006fk}.}
\end{figure}

In fact we can learn a great deal by extending the \ads conjecture further by breaking some of the field theory supersymmetry encoded in the $S^5$ part of the geometry.  Specifically the introduction of $M\ll N$ $D7$ branes that fill space from the asymptotic $3+1D$ Minkowski boundary to some finite distance above $r=0$ and wrapping an equatorial $S^3\subset S^5$ corresponds to introducing $M$ fundamental flavors in the gauge theory \cite{Karch:2002sh}.  The $M\ll N$ requirement allows the $D3$ branes to backreact and warp spacetime in the large $N$ limit while the $D7$ branes act only as probes.  An open string ending on the $r<\infinity$ boundary of one of these $D7$ branes corresponds to a heavy fundamental quark.  At zero temperature the geometry supports mesons, with an open string having both its endpoints on $D7$ branes.  At finite temperature $T$ a string may hang from the $D7$ brane and terminate on one of the nonextremal $D3$ branes.  This is identified with an open heavy quark in a deconfined SYM plasma held at $T$.  One may calculate the Debye screening length for this process, see \fig{ads:debye} (a), 
\cite{Bak:2007fk} (and references therein) and also find a small enhancement to $q\bar{q}$ dissociation for moving heavy mesons such as $J/\Psi$ \cite{Liu:2006nn,Ejaz:2007hg}.  Given some initial velocity a hanging string will both perturb the Minkowski boundary metric, leaving a wake behind its path in the SYM plasma complete with Mach cones \cite{Friess:2006fk,Chesler:2007an} (see \fig{ads:debye} (b)), and also lose momentum from its $D7$ open heavy flavor endpoint into the $D3$ brane horizon \cite{Gubser:2006bz,Herzog:2006gh,VazquezPoritz:2008nw}; see \fig{ads:Tmunu}.  The big payoff from introducing this new complexity is analytic formulae for these intuitively familiar effects.  

Other applications of the conjecture include calculating the light-like Wilson loop \cite{Liu:2006ug,Liu:2006nn} conjectured to be connected with the jet quenching parameter $\hat{q}$ \cite{Kovner:2001vi} via \adscomma, although there is controversy over its ultimate magnitude \cite{Argyres:2006yz,Gubser:2006nz,Liu:2006he,Bertoldi:2007sf,Argyres:2008eg}.  Also the conjecture was used to investigate the diffusion coefficient in a Langevin formalism \cite{CasalderreySolana:2006rq,CasalderreySolana:2007qw}; this is only valid in the $v\rightarrow0$ limit \cite{Gubser:2006nz}, and it is not clear what the corrections for nonzero velocity are.

\subsection{Testing the Conjecture}
The most well known result of using this complicated machinery is the famous viscosity to entropy bound \cite{Gubser:1996de,Kovtun:2004de}
\be
\label{ads:etaovers}
\frac{\eta}{s} = \frac{1}{4\pi}\left( 1 + \mathcal{O}(\frac{1}{\sqrt{\lambda}}) \right).
\ee
As noted earlier, qualitative estimates of the viscosity at \rhic \cite{Teaney:2003kp} appear to require such small values of $\eta/s$.  This encourages continued investigation of applications of the \ads conjecture, but the big question is: how much can we trust the \ads results?  The duality of $\mathcal{N}=4$ SYM and Type IIB string theory on a pure \adsfive background has strong motivation and passed a number of theoretical tests, although it remains unproven.  But maximally supersymmetric Yang-Mills theory is very far from QCD.  As symmetries are broken and the field theory dual to string theory becomes more and more like QCD, the further the conjecture must be extended and the less evidence exists for its holding.  And while QCD is far from a quantitative description of heavy ion collisions there is a well known procedure for estimating the errors involved.  On the other hand, calculating quantities in a theory that is similar but not the same as QCD is an uncontrolled approximation, and it is far from clear how much those results will be modified by convergence to QCD.  

We are thus motivated to find qualitative features of heavy ion collisions for which \ads predictions are not and can not be reproduced by weakly-coupled QCD.  In this sense the entropy to viscosity ratio is the worst observable to use: the Frankfurt parton cascade group finds pQCD also gives $\eta/s\sim.1$; and the value of the ratio, due to the uncontrolled heavy ion initial conditions that so strongly affect hydrodynamics results and the unknown importance of viscous effects in hydrodynamics predictions, is very poorly determined.  
Nevertheless \ads can be applied to heavy flavor jets and jet-bulk interactions, and we will find that these are falsifiable predictions qualitatively different from pQCD.

Specifically \ads can be applied to the energy loss of heavy quarks propagating through a thermalized plasma \cite{Gubser:2006bz,Herzog:2006gh}.  It turns out that the drag experienced by these quarks is independent of their momentum and inversely proportional to their mass; this is entirely different from the logarithmic dependence on both mass and momentum found in pQCD.  A major result of this thesis is to propose an experimental observable, the ratio of charm to bottom quark $R_{AA}(\eqnpt)$, as a robust observable for differentiating between these two theories; see Chapter \ref{chapter:pqcdvsadscft}.  

\begin{figure}[!htb]
\centering
\includegraphics[width=5.25 in]{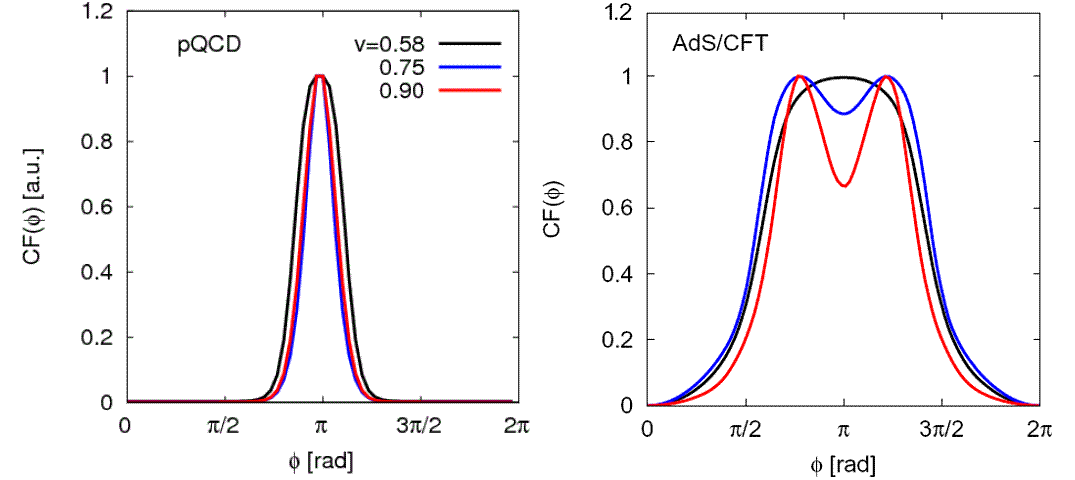}
\caption{\label{ads:dijets}
The away side correlation functions of particles produced by Cooper-Frye hadronization associated with a \highpt jet propagating through a thermalized bulk plasma.  The hydrodynamic source on the left comes from pQCD calculations \cite{Neufeld:2008fi}; the right from \ads \cite{Friess:2006fk}.  Figure adapted from \cite{Betz:2008wy}.}
\end{figure}

Another success of the \ads paradigm came from the consistency check of Mach cones produced in the bulk matter by heavy quarks propagating faster than the speed of sound, $v=1/\sqrt{3}$.  It turns out though that connecting this phenomenon to the observed conical flow in dijet correlations at \rhic is highly nontrivial \cite{Noronha:2007xe}.  
In fact one finds that hadronization removes all traces of a signal from any Mach wake in the plasma; the temperature variations are simply too small.  Recent work has surprisingly shown that the observable consequences from the so-called `neck' region, the bulk plasma trailing behind the quark jet, are qualitatively different for thermodynamic sources from pQCD and from \ads \cite{Betz:2008wy}; see \fig{ads:dijets}. 

\section{Outline of Thesis}
In the next chapter a simultaneous qualitative description of \intermediatept \raapt and \vtwopt is developed predicated on a nonperturbative momentum punch at the boundary of the QGP and the vacuum.  Another interesting source of this focusing of jets could come from the reduction of the Debye mass $m_D\sim gT$ as the parton propagates through a cooling plasma.  
This unexpected and heretofore inexplicably large \vtwo given the overall suppression is argued as an experimental signal for deconfinement.

In Chapter \ref{chapter:WHDG} we reduce the qualitative disagreement between pQCD calculations of heavy quark energy loss and observations of light and heavy meson suppression from \rhiccomma.  The inclusion of the elastic energy loss channel, whose neglect disproportionately affects the heavy quarks, increases the quenching of all parton flavors.  For single fixed pathlength calculations with the usual $L\sim5$ fm, pion predictions are falsified by data.  The realistic treatment of geometry is crucial for a simultaneous description of both pions and nonphotonic electrons: just as fluctuations in the number of emitted gluons softened the energy loss so does allowing fluctuations in the in-medium distances traveled by jets.  

The proposal of taking the ratio of charm to bottom quarks as a robust test of pQCD and AdS/CFT ideas is the focus of Chapter \ref{chapter:pqcdvsadscft}.  Most clearly seen at \lhccomma, scheduled to run heavy ions soon, the ratio goes to 1 in all pQCD predictions as the mass scale of the quark becomes irrelevant compared to the momentum scale of the particle.  On the other hand the ratio is flat and much smaller than one for \ads drag calculations.  The critical issue comes from the momentum regimes of applicability: the momentum above which the drag approximations fail is not clearly known, nor is the momenta below which perturbative predictions break down entirely understood, either.  Upgrades at \rhic will allow for discrimination between charm and bottom jets, possibly ending the controversy over the contributions to heavy quark energy loss; the advantage there will come from lower multiplicities so that the drag regime will extend to higher momenta.  

Finally we discuss heavy quark production radiation in Chapter \ref{HQPB}.  By including the influence of the away-side jet we find corrections to previously published Ter-Mikayelian effect results.  This is also the first step in computing the photon radiation associated not only with $2\rightarrow2$ processes but also from momentum changes due to gluon bremsstrahlung.

\clearpage
\mychapter{\texorpdfstring{$v_2$ vs.\ $R_{AA}$}{v2 vs.\ RAA} as a Constraint on sQGP Dynamics}{punch}
\section{Introduction}
A complete theoretical description of \rhic mid- to high-\pt particles must reproduce \raaphicomma, the ratio of observed high momentum leading hadrons in \aplusa collisions to binary scaled \pp collisions as a function of the angle made with respect to the reaction plane, $\phi$.  A useful means of representing this quantity is via Fourier expansion:
\be
\label{fourier}
R_{AA}(\phi,\eqnpt,y) = R_{AA}(\eqnpt,y) \left( 1+2\eqnvtwo(\eqnpt,y)\cos(2\phi)+\dots \right).
\ee
The goal of theorists, then, is to correctly replicate these moments, the most important being the first two: the normalization and the azimuthal anisotropy.

This is actually quite a difficult problem due to the anticorrelated nature of \raa and \vtwocomma.  In order to reproduce the anisotropy of the jet data, the partons must be made sensitive to the medium.  Then the anisotropy of the background created by noncentral \rhic collisions translates into anisotropy of the jets.  However, this sensitivity to the medium also leads to jet attenuation and a decrease in \raacomma.  As we will see, the very high \vtwo values measured at \rhic mean that, at best, previous models either oversuppressed \raa or underpredicted \vtwocomma.

Hydrodynamics has been applied to the \lowpt particles with great success.  Unfortunately, the lack of thermal equilibrium precludes the use of hydrodynamics in the momentum regime we are interested in.  Moreover, a na\"ive application would highly oversuppress \raa due to the Boltzmann factors.

Parton transport theory attempts to extend hydrodynamics' range of applicability to higher transverse momenta \cite{Zhang:1999rs,Molnar:2001ux}.  The M\'olnar parton cascade (MPC) succeeded in describing the low- and intermediate-\pt \vtwo results of \rhic only by taking the parton elastic cross sections to be extreme, $\sigma_t\sim45$ mb \cite{Molnar:2001ux}.  Even more damning, we will show that the MPC approach reaches the reported \vtwo values at the expense of reproducing the experimental \raa data. 

pQCD becomes valid for moderate and higher \pt partons, and models based on pQCD calculations of radiative energy loss have had success in reproducing the experimental \raapt data \cite{Vitev:2005ch}. These models use a single, representative pathlength; as such, they give $\eqnvtwo\equiv0$.

Compelling but not convincing, past geometric approximations of energy loss sought to demonstrate the inability of pure jet quenching to match the \rhic \vtwo findings \cite{Shuryak:2001me,Drees:2003zh,d'Enterria:2005cs}.  In \cite{Shuryak:2001me}, the author found that \vtwo reached a maximum value that was smaller than the data at the time.  Our formulation achieves large enough values of \vtwocomma, but over-quenches \raa to do so.  In \cite{Drees:2003zh}, the authors found that when they fit their model to the most central \raa data, they could not replicate the large \vtwo observations.  Instead of modeling \raa or \vtwo separately, we ask a different question: when considering the experimental \raa and \vtwo on equal footing, is there any way a pure geometric energy loss model can be consistent with the data within the new, 2D error ellipses?  We find the answer is no.  

\cite{Drees:2003zh,Gyulassy:2000gk} demonstrated the powerful influence of the nuclear density geometry used in calculations.  Sharp-edged approximations such as the hard sphere (HS) and especially the hard cylinder (HC) greatly enhance model \vtwo results.  Thus past works that used these geometries and could reproduce both \raa and \vtwocomma, such as energy loss with hydrodynamics \cite{Gyulassy:2000gk}, energy loss with thermal absorption \cite{Wang:2003mm}, and energy loss with \lowpt flow effects \cite{Armesto:2004vz} will no longer produce large enough \vtwo values in realistic, diffuse medium densities.  We will investigate the case of energy loss with thermal absorption more thoroughly later in this paper. 

\section{Plot of \texorpdfstring{\vtwovsraa}{v2 vs. RAA} and Failures of Previous Models}
In \fig{modelfailure}, we combine \star charged hadron $\eqnraa(\eqnpt)$ \cite{Adams:2003kv} and $\eqnvtwo(\eqnpt)$ (4 particle correlations only) \cite{Adams:2004bi}, \phenix charged hadron $\eqnraa(\eqnpt)$ \cite{Adler:2003au} and \vtwo\cite{Adler:2003kt} vs.\ centrality, and \phenix \pizero ($\eqnpt>4$ GeV) \raaphi\cite{Cole:2005yv} centrality data.  We na\"ively averaged the \star and \phenix $\eqnraa(\eqnpt)$ results to approximately match the \pt bins of their corresponding \vtwo measurements.  We report the \raa and \vtwo modes for the \phenix \pizero \raaphi data.  
The error estimates are unfortunately schematic only. 

The charged hadron data from both \star and \phenix appear to fall on a single line extending from low \raa and \vtwo to high \raa and \vtwocomma. The \vtwo values are larger in general than one would expect {\em a priori}, and the absence of a decline in \vtwo as $\eqnraa\rightarrow1$ is especially surprising. For the \pizero data, the possibly flat in \raa trend of \vtwo is also quite strange.  Taken as a conglomerate with the charged hadrons, however, the \pizero \raaphi data appears consistent within error.

The MPC has a single free parameter: the opacity, $\chi=\int dz\sigma_t\rho_g$ \cite{Molnar:2001ux}.  Allowing the MPC to run at an approximate 30\% centrality and varying the opacity gives the dashed curve in \fig{modelfailure}.  We see that this curve does not pass near the area in the \vtwovsraa diagram corresponding to the experimental results for the same centrality.  While in \cite{Molnar:2001ux} a single $\chi$ predicts \vtwopt well, the MPC fails at simultaneously reproducing the first two moments of \raaphicomma.

In order to make a rough calculation of both \raa and \vtwo resulting from partonic energy loss to the medium, we approximate the first order in radiative energy loss \cite{Gyulassy:2000er,Wiedemann:2000tf,Guo:2000nz} with a purely geometric model that neglects number fluctuations.  The asymptotic approximation of this loss in a static medium for a parton of energy $E$ traversing a path length $L$ in the limit of $EL\gg1$ and $E\gg\mu$ is
\be\label{radloss}
\frac{\Delta E^{(1)}_{rad}}{E} \approx \frac{\eqnalphas C_R\mu^2L^2}{4\lambda E}\left[\ln\frac{2E}{\mu^2L}-0.048\right],
\ee
where $\mu$ is the Debye screening mass, and $\lambda$ is the mean-free-path of the parton \cite{Wang:2001cs}. 
Taking the bracketed term to be constant (this is not such a good approximation: it changes by a factor of 2 for the range of parameters considered here) and
$\eqnrhopart\propto\eqndnslashdy$ (note that $\mu^2/\lambda\propto\eqndnslashdy$), we use an energy loss scheme similar to \cite{Drees:2003zh}: 
\be
\label{geomloss}
\epsilon = \frac{\Delta E_{rad}}{E} = \kappa I.
\ee
$\kappa$ is a free parameter encapsulating the $E$ dependence, etc.~of \eq{radloss} and the proportionality constant between \dnslashdy and \rhopartcomma. $I$ represents the integral through the Bjorken expanding medium, taken to be 
\be 
I=\int_0^\infinity\!\!\!\!\intd l\;\, l \frac{l_0}{l+l_0}\eqnrhopart(x+(l+l_0)n_x,\,y+(l+l_0)n_y),
\ee
where we use $l_0=.2$ fm as the formation time of the medium.  We consider only 1D expansion here because \cite{Gyulassy:2000gk} showed that including the transverse expansion of the medium had a negligible effect on the total energy loss.

Since the partonic production spectrum observed at \rhic can be approximated by a power law, we use the momentum Jacobian ($\eqnpt^f=(1-\epsilon)\eqnpt^i$) as the survival probability of hard partons; see Appendix \ref{punchappendix}. Following the factorization theorem and Glauber approach \cite{Glauber:1970jm}, we distribute partons in the overlap region according to $\eqnrhocoll=\eqntaa$ and isotropically in azimuth; hence 
\be
\label{GLVraaphi}
\eqnraa(\phi;\,b) = \frac{\iint\intd x\intd y\;\,\eqntaa(x,\,y;\,b)\left(1-\epsilon(x,\,y,\,\phi;\,b)\right)^n}{\eqnncoll (b)},
\ee
where $4\lesssim n\lesssim5$. The difference between $n=4$ and $n=5$ is less than 10\%, and in this paper we will always use the former.  

We evaluate \raaphi at 24 values of $\phi$ from 0-$2 \pi$ and then find the Fourier modes \raa and \vtwo of this distribution.  We label the results of this model in the plots as GREL for geometric radiative energy loss.  A different method for finding \vtwocomma, not used here, assumes the final parton distribution is given exactly by \raa and \vtwocomma, and then determines \vtwo from the ratio of $\eqnraa(0)$ and $\eqnraa(\pi/2)$; this systematically enhances \vtwocomma, especially at large centralities.

Choosing $40-50\%$ centrality as a good representative, we calculate the line in $(\eqnraa,\eqnpt)$ space corresponding to $0\lte\kappa < \infinity$, \fig{modelfailure}.  One sees in the figure that the curve does not cross the error ellipse for the relevant \phenix \pizero data.  This is despite the simplifying use of hard sphere geometry, where we took $R_{HS}=6.78$ fm to ensure that $\langle\eqnrperpWS^2\rangle=\langle\eqnrperpHS^2\rangle$, that amplifies the \vtwocomma.  
Since the experimental errors are the most basic estimate---it would be very interesting to see the results of a careful analysis---and the theoretical models do not currently have any error estimates associated with them prevent a statistical quantification of the disagreement, one is strongly persuaded that neither of these models encompasses all the relevant physics.
\bfig[!htb]
\begin{center}
\leavevmode
\includegraphics[width=.75 \columnwidth]{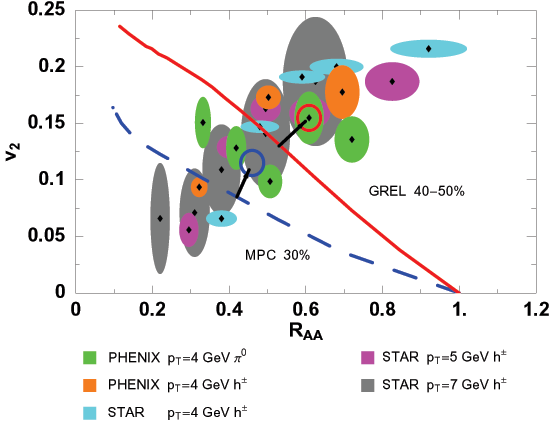}
\caption{\label{modelfailure}\captionsize{\star $h^\pm$ data for 0-5\%, 10-20\%, 20-30\%, 30-40\%, and 40-60\%, \phenix $h^\pm$ data for 0-20\%, 20-40\%, and 40-60\%, and \phenix \pizero data for 10-20\%, 20-30\%, \ldots, 50-60\% centralities.  MPC results for increasing $\chi$ that should go between the 20-30\% and 30-40\% \phenix \pizero results.  GREL results for increasing $\kappa$ that misses the 40-50\% \phenix \pizero data.}}
\end{center}
\efig

\section{Exclusion of Detailed Balance}
In \cite{Wang:2001cs}, Wang and Wang derived the first order in opacity formula for stimulated emission and thermal absorption associated with the multiple scattering of a propagating parton in a hot, dense QCD medium.  They found that this effects a net reduction of the energy loss.  Since their energy gain has a linear length dependence in a static medium, as opposed to the quadratic dependence for radiative energy loss, one might hope to reproduce the experimental \raa and \vtwo data by increasing the opacity enough that radiative energy loss creates the \vtwo and the energy gain properly reduces the oversuppression of \raacomma.  

From \cite{Wang:2001cs}, we have that in the limit of $E L\gg1$ and $E\gg\mu$, the asymptotic approximation of the effect in a static medium to the first order in opacity is:
\be
\label{abs}
\frac{\Delta E_{abs}^{(1)}}{E} \approx \frac{\pi\eqnalphas C_R L T^2}{3\lambda E^2} \left[\ln\frac{\mu^2 L}{T}-1+\gamma_E-\frac{6\zeta'(2)}{\pi^2}\right].
\ee
Since $\mu^2=4\pi\eqnalphas T^2$ and the bracketed term is actually approximately constant for our range of inputs, we use
\be
\label{wangwang}
\epsilon = \frac{\Delta E_{rad}}{E}-\frac{\Delta E_{abs}}{E} = \kappa I-k I_2,
\ee
where $\kappa I$ is the same as in \eq{geomloss}, k is a free parameter encapsulating the proportionality constants in \eq{abs}, and $I_2$ represents an integral through the 1D expanding medium:
\be
I_2 = \int_0^\infinity\!\!\!\!\intd l\;\, \frac{l_0}{l+l_0}\eqnrhopart(x+(l+l_0)n_x,\,y+(l+l_0)n_y).
\ee
Note that $I_2$ has one less power of $l$ in the integrand; this allows a unique determination of the two free parameters ($\kappa,\,k$) when fitting the model to a particular $(\eqnraa,\,\eqnvtwo)$ point.

Using the same hard sphere geometry as before, we are able to fit the 20-30\% centrality \phenix \pizero \raaphi data point with $\kappa=.5$ and $k=.25$.  
Keeping $\kappa$ and $k$ fixed, \fig{wangandpunch} shows the results when the impact parameter is varied (long-dashed, blue curve); one can see that this matches the \rhic \raa and \vtwo trends quite well.  

Taking \eq{wangwang} seriously, we invert the relationships and solve for \dnslashdycomma.  Recalling that $\mu=gT$, $\lambda^{-1}\approx 9 \pi \eqnalphas^2 (\rho/2)^{1/3}/4$, and the density for an expanding medium is $\rho=\eqndndy/\left(\eqnAperp (\frac{L}{2})\right)$, we find that, as a function of parton energy $E$, path length $L$, formation time $l_0$, and \alphascomma,
\bea
\label{raddndy}
\eqndndyrad & \sim & \kappa \frac{4E}{9\pi C_R \eqnalphas^3\tilde{v}_1} \frac{l_0 L}{l_0+L} \eqnnpart \\
\label{absdndy}
\eqndndyabs & \sim & k \frac{4E^2}{3\pi C_R \eqnalphas^2\tilde{v}_2} \frac{l_0 L}{l_0+L} \eqnnpart,
\eea
where $\tilde{v}_1$ and $\tilde{v}_2$ correspond to the bracketed terms in the energy loss and energy gain approximations, Eqs.~(\ref{radloss}) and (\ref{abs}), respectively.  

For our fitted values of $\kappa$ and $k$, the choice of $E=6$ GeV, $L=5$ fm, and $\eqnalphas=.4$ gives $\eqndnslashdyrad\sim1000$ from \eq{raddndy} and $\eqndnslashdyabs\sim3000$ from \eq{absdndy} in 0-5\% most central collisions.  Not only are the two values inconsistent, but the \dnslashdy value needed to create the necessary absorption is rather large.  Keeping the other quantities fixed but changing $E$ to 10 GeV, one finds $\eqndnslashdy\sim1000$ from \eq{raddndy} and $\eqndnslashdyabs\sim9000$ from \eq{absdndy}.  The huge increase of \dnslashdy to values inconsistent with the \rhic entropy data reflects the $E^2$ dependence of the Detailed Balance absorption.  It seems the only way to have a large enough energy gain while maintaining a $\eqndnslashdy\sim1000$ is to increase \alphas above 1.  Moreover, these calculations were performed using a hard sphere nuclear geometry profile, which naturally enhances \vtwocomma. For the more realistic Woods-Saxon profile, $\kappa$ would have to increase significantly.  Correspondingly, $k$ would rise to counter the quenching of \raacomma, aggravating the problem of unseemingly large $\eqndnslashdyabs$.
\bfig[!htb]
\begin{center}
\leavevmode
\includegraphics[width=.75 \columnwidth]{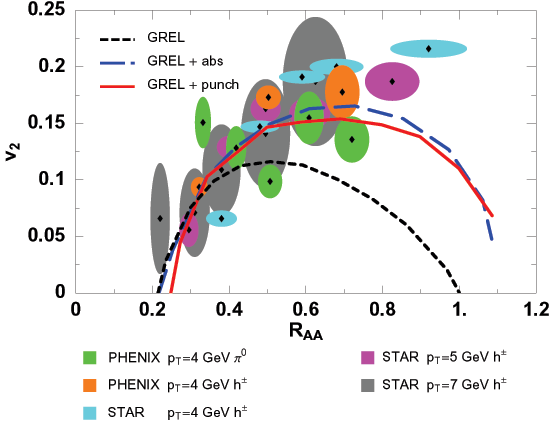}
\caption{
\label{wangandpunch}
\captionsize{Addition of thermal absorption (dashed blue curve) or momentum punch (solid red curve) to GREL; models' free parameters fit to \phenix 20-30\% centrality \pizero \raaphi.}
}
\end{center}
\efig

\section{Success of the Punch}
Building on the success of radiative energy loss in reproducing \raaptcomma, and supposing that latent heat, the bag constant, the screening mass, or other deconfinement effects might provide a small ($\sim1$ GeV) momentum boost to partons in the direction normal to the surface of emission, we created a new model based on the GREL model that includes a momentum ``punch,'' $\Delta\eqnpt$. After propagating to the edge of the medium with GREL, the parton's final, ``punched-up'' momentum and angle of emission are recomputed, giving a new probability of escape. Fitting to a single $(\eqnraa,\eqnvtwo)$ point provides a unique specification of $\kappa$ and punch magnitude. The results are astonishing: one sees from \fig{wangandpunch} that a tiny, .5 GeV, punch on a 10 GeV parton reproduces the data quite well over all centralities. Fitting the \phenix 20-30\% \pizero data sets $\kappa=.18$ and the aforementioned $\Delta\eqnpt=.5$ GeV. The size of the representative parton's initial momentum is on the high side for the displayed \rhic data; however, the important quantity is the ratio $\Delta\eqnpt/E$. Moreover, although the geometry used artificially enlarges the \vtwocomma, we feel confident that when this model is implemented for a Woods-Saxon geometry (a nontrivial task because the ``edge'' of the medium is then poorly defined), the necessarily larger final punch magnitude will still be of order 1 GeV. The magnitude of this deconfinement-caused momentum boost must be independent of the parton's momentum; hence $\eqnvtwo(\eqnpt)$ will decrease like $1/\eqnpt$. Moreover, since $\epsilon$ is larger out of plane than in, a fixed $\Delta\eqnpt$ enhances $\eqnraa(\pi/2)$ more than $\eqnraa(0)$. These are precisely the trends seen by \phenix \cite{Adler:2006bw}. Keeping the same values for $\kappa$, $k$, $\Delta\eqnpt$, etc.\ as for \auaucomma, we show in \fig{cucuresults} the centrality-binned \raa and \vtwo results for \cucu in the three geometric energy loss models. 
\bfig[!htb]
\begin{center}
\leavevmode
\includegraphics[width=.75 \columnwidth]{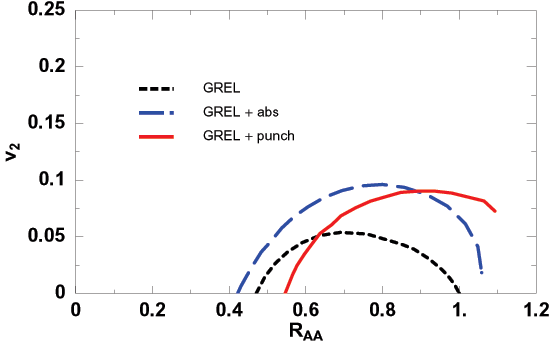}
\caption{
\label{cucuresults}
\captionsize{\cucu results for GREL (black, dotted), GREL with the punch (red, solid), and GREL with absorption (blue, dashed).}
}
\end{center}
\efig

We can in fact go further in thinking about the origin of this deconfinement mechanism.  Specifically we will find the change in the Debye mass of the jet $\mu_i(\vec{x},t)=gT(\vec{x}_i,t_i)$ as it propagates through the medium provides the right order of magnitude momentum punch.  In moving from position $i$ to $i+1$ one calculates the deflection of the jet following a slight generalization the usual means.  At $i$ the parton has momentum $k^\mu_i=(\omega,k_{||},k_{\perp,i})$, where the parallel direction is defined by the local line of constant temperature in $2D$ and the perpendicular direction is its normal; the normal is in the direction of the $2D$ gradient of the medium temperature, equivalently density.  By symmetry crossing the plane cannot affect the parallel component of the momentum so at $i+1$ we have $k^\mu_{i+1}=(\omega,k_{||},k_{\perp,i+1})$.  Then
\be
k_{\perp,i+1}^2 = k_{\perp,i}^2 + \mu_i^2-\mu_{i+1}^2.
\ee
As the jet moves to the less dense edges of the medium and the whole fireball cools as it Bjorken expands $\mu_{i+1}<\mu_i$, and the jet receives a continuous punch in the local normal direction.  This will tend to bend the jets, focusing them into higher \vtwo and simultaneously increase the \raacomma.  For a quark (gluon) produced in the center of a 20-40\% centrality collision at $\tau_0=.2$ fm and followed until $T=T_c=160$ MeV the change in Debye mass is 0.38 (0.55) GeV which is slightly smaller (larger) than the nominal 0.5 GeV investigated earlier.  It will be interesting to see if a more quantitative calculation in a realistic diffuse medium will meet the expectation that the data can be understood from this mechanism.

\section{Conclusions}

By failing to simultaneously match the \raa and \vtwo values seen at \rhic we discounted the MPC and pure GREL models. We showed that while including medium-induced absorption reproduces the \raaphi phenomena, it does so at the expense of inconsistent and huge \dnslashdycomma. Recent parton cascade results including $2\leftrightarrow3$ processes were claimed \cite{Fochler:2008ts} to simultaneously fit both \raapt and \vtwoptcomma.  Careful examination of the paper reveals that their large \vtwo comes from using huge $\alpha_s=.6$ and gluon jets only, but the claims of consistency with \raa use only $\alpha_s=.3$.  In this chapter we showed that the addition of a mere 5\% punch created a \rhiccomma-following trend.  The reduction in Debye mass from the hot, dense center of the medium to the cool, diffuse edge provides just such an order of magnitude momentum change.

\clearpage
\lhead{Chapter \arabic{chapter}: Fluctuations in Jet Tomography}
\mychapter{Elastic, Inelastic, and Path Length Fluctuations in Jet Tomography at RHIC and LHC}{chapter:WHDG}
\section{Introduction}
\begin{figure}[!hbt] 
\centering
\includegraphics[width=3.75 in]{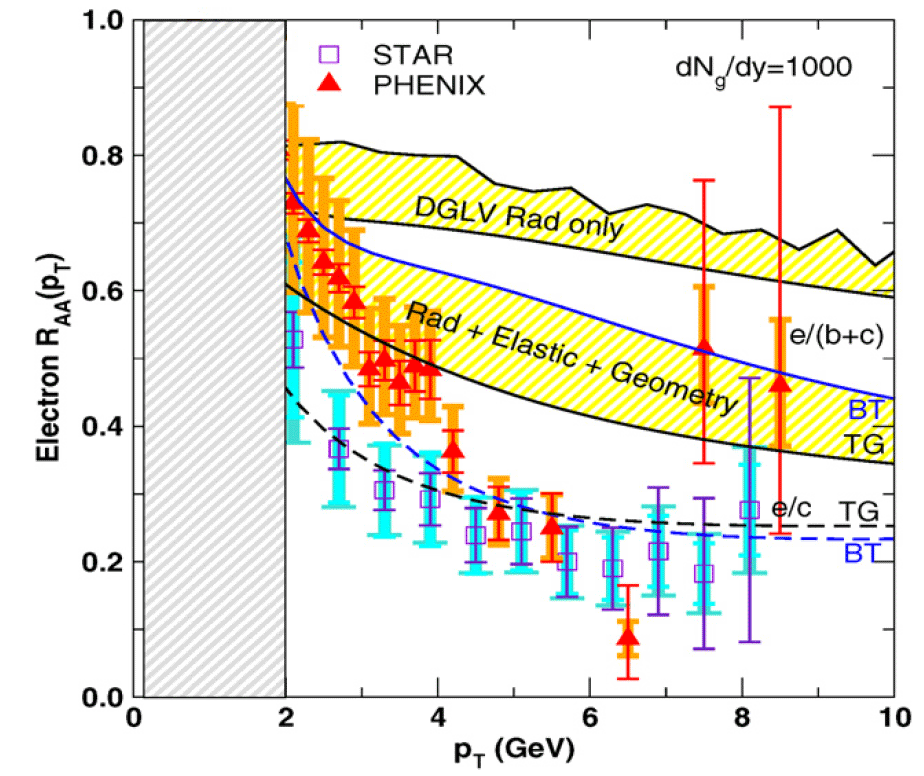}
\caption{\label{fig:eRAA} 
The suppression factor, $R_{AA}(p_\perp)$, for non-photonic electrons
  from the decays of quenched heavy quark (c+b) jets is compared to
  \phenix \cite{Adler:2005xv,Akiba:2005bs} and preliminary STAR
  data \cite{Bielcik:2005wu,Dong:2005nm} data in central Au+Au reactions at
  200~AGeV.  Shaded bars indicate systematic errors, while
  thin error lines indicate statistical ones.
All calculations assume initial $dN_g/dy=1000$.
  The upper yellow band from~\protect{\cite{Djordjevic:2005db}}
  takes into account radiative energy loss only, using a fixed $L=6$ fm; the lower yellow band is our new prediction, including both
  elastic and inelastic losses as well as jet path length fluctuations. 
The bands provide an estimate of current
theoretical uncertainties. The dashed curve shows the electron suppression using inelastic and TG elastic loss with bottom quark jets neglected.
}
\end{figure}

Light quark and gluon jet quenching observed via $\pi,\eta$
suppression \cite{Isobe:2005mh,Shimomura:2005en,Adler:2003qi} in Cu+Cu and Au+Au collisions at
$\sqrt{s} = 62-200$~AGeV at the Relativistic Heavy Ion Collider (\rhiccomma)
has been remarkably consistent thus far with
predictions \cite{Gyulassy:2003mc,Kovner:2003zj,Jacobs:2004qv,Wang:1991xy,Gyulassy:2000er,Gyulassy:2001nm,Vitev:2002pf}. However,
recent non-photonic single electron data \cite{Adler:2005xv,Akiba:2005bs,Bielcik:2005wu,Dong:2005nm} (which
present an indirect probe of heavy quark energy loss) have
significantly challenged the underlying assumptions of the jet
tomography theory (see \cite{Djordjevic:2005db}).  A much larger
suppression of electrons than predicted was observed in the
$p_T\sim 4-8 $ GeV region
{(see \fig{fig:eRAA})}. These data falsify the assumption that heavy quark
quenching is dominated by radiative energy loss when the bulk QCD
matter parton density is constrained by the observed $dN/dy\approx
1000$ rapidity density of produced hadrons.

The observed ``perfect fluidity'' \cite{Arsene:2004fa,Adcox:2004mh,Back:2004je,Adams:2005dq,BNLPress} of the sQGP
at long wavelengths ($p_T <  2$ GeV/c) provides direct evidence for highly 
nonperturbative dynamics leading to rapid equilibration and $PdV$ work 
consistent with the QCD equation of state \cite{Gyulassy:2004zy,Hirano:2005wx}.
Above $p_T>5-7$ GeV \cite{Winter:2005nw,Arsene:2004fa,Adcox:2004mh,Back:2004je,Adams:2005dq}, 
pQCD appears to provide increasingly reliable predictions, at least for the nuclear modification of light parton jets  \cite{Gyulassy:2003mc,Kovner:2003zj,Jacobs:2004qv,Wang:1991xy,Gyulassy:2000er,Gyulassy:2001nm,Vitev:2002pf}. So the question raised by the electron data is: to how short wavelengths, $1/p_T$, 
is the novel nonperturbative physics dominant in
the strongly interacting Quark Gluon Plasma (sQGP) \cite{Gyulassy:2004zy}?
This question is important because hard jets can be utilized as effective 
``external'' tomographic probes of the bulk sQGP matter only if their 
dynamics can be predicted reliably. Otherwise, jet quenching 
can only be an additional signal of new physics 
instead of providing a calibrated diagnostic probe of that physics. 

The upper band of \fig{fig:eRAA} shows that the 
predictions from  \cite{Djordjevic:2005db} 
considerably underestimated the electron nuclear modification even out
to $p_T\sim 8$ GeV.  This discrepancy points to either (1) missing
perturbative QCD physics, (2) incomplete understanding of the initial
heavy quark production and/or (3) novel non-perturbative mechanisms
affecting partonic physics out to $p_T > 10$ GeV. We note that
$p_T\sim 8$ GeV (single non-photonic) electrons originate in our
calculations from the fragmentation and decay of both charm and bottom
quarks with transverse momenta $p_T\sim 12\pm 4$ GeV (see \fig{fig:ptDist2}
in \cite{Djordjevic:2005db}).

Possibility (3) is of course the most radical and would imply the
persistence of non-perturbative physics in the sQGP down to extremely
short wavelengths, i.e. $1/p_T \sim 1/50 fm$.  These processes can be
postulated to improve the fit to the data\cite{Rapp:2005at}, but at
the price of losing theoretical control of the tomographic information
from jet quenching data. DGVW \cite{Djordjevic:2005db} showed that by
arbitrarily increasing the initial sQGP densities to unphysical $dN/dy
\gsim 4000$, the non-photonic electrons from heavy quarks can be
artificially suppressed to $R_{AA}\sim 0.5$.  Thus conventional
radiative energy loss requires either violation of bulk entropy bounds
or nonperturbatively large $\alpha_s$ extrapolations of the theory.
Even when ignoring the bottom contribution,
Ref. \cite{Armesto:2005iq}, found that a similarly excessive transport
coefficient \cite{Baier:2002tc}, $\hat{q}_{\eff}\sim 14$ GeV$^2$/fm,
was necessary to approach the measured suppression.

The main theoretical problem is that bottom quark jets are too weakly
quenched by radiative energy loss. Their contribution significantly reduces
the single electron suppression \cite{Djordjevic:2005db} compared to that of the charm jets alone.  While the
current data cannot rule out the possibility (2) that the bottom
production in this kinematic range is overestimated in NLO
theory \cite{Cacciari:2005rk}, in this paper we pursue the more
conservative approach of examining the inclusion of a previously
neglected pQCD component of the physics.

The recent work by Mustafa \cite{Mustafa:2003vh,Mustafa:2004dr} and others \cite{DuttMazumder:2004xk} motivated us to
revisit the assumption that pQCD elastic energy
loss \cite{Bjorken:1982tu} is negligible compared to radiative. In
earlier studies, the elastic energy
loss \cite{Bjorken:1982tu,Thoma:1990fm,Braaten:1991jj,Braaten:1991we,Wang:1994fx,Mustafa:1997pm,Lin:1997cn}
was found to be $dE^{el}/dx\sim 0.3-0.5 $ GeV/fm, which was
erroneously considered to be small compared to the several GeV/fm expected
from radiative energy loss.  The apparent weakness of conventional
pQCD collisional energy loss mechanisms was also supported by parton
transport theory results \cite{Molnar:2001ux,Moore:2004tg},
which showed that the typical thermal pQCD elastic cross section,
$\sigma_{el}\sim 3 $mb, is too small to explain the differential
elliptic flow at high $p_T> 2 $ GeV and also underestimates the high
$p_T$ quenching of pions.

\begin{figure}[!htb] 
\centering
\includegraphics[width=3.25 in]{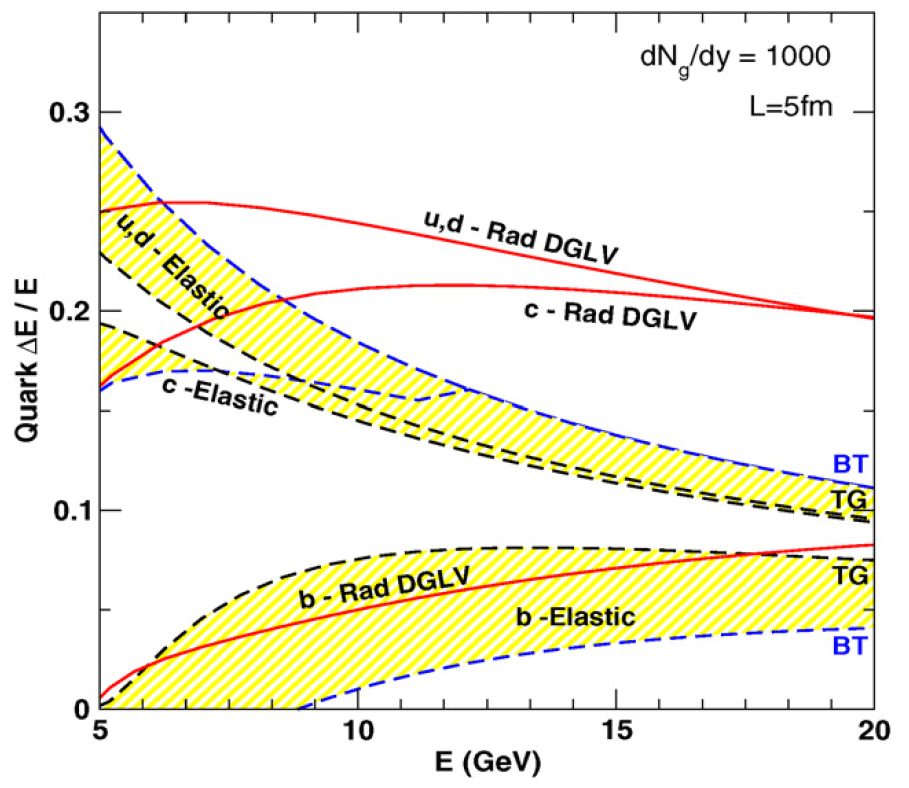}
\caption{\label{fig:ptDist} 
  $\Delta E/E$ for $u,c,b$ quarks as a function of $E$. A Bjorken
  expanding QGP with path length $L=5$~fm and initial density fixed
  by $dN_g/dy=1000$ is assumed. The curves are computed with the
  coupling $\alpha_{S}=0.3$ held fixed. For Debye mass $\mu_D\propto(dN_g/dy)^{(1/3)}$,
    the gluon mass is $\mu_D/\surd 2$, the light quark mass is
    $\mu_D/2$, the charm mass is $1.2$ GeV, and
    the bottom mass is $4.75$ GeV.
  Radiative DGLV first order energy loss is compared to elastic parton
  energy loss (in TG or BT approximations). The yellow bands provide
  an indication of current theoretical uncertainties in the elastic
  energy loss for bottom quarks.}
\end{figure} 

In contrast, Mustafa found that radiative and 
elastic energy losses for heavy quarks were in fact comparable over a very 
wide kinematic range accessible at \rhiccomma. In \fig{fig:ptDist}, we confirm 
Mustafa's finding \cite{Mustafa:2003vh,Mustafa:2004dr} and extend it to the light quark 
sector as well; the fractional energy loss,
$\Delta E/E$, from DGLV radiative for $u,c,b$ quarks (solid
curves) is compared to Thoma-Gyulassy (TG) \cite{Thoma:1990fm} and 
Braaten-Thoma (BT) \cite{Braaten:1991jj,Braaten:1991we} estimates of elastic (dashed
curves). For light quarks, the elastic energy loss 
decreases more rapidly with energy than radiative energy loss, but even at 20 GeV the elastic
is only 50\% smaller than the radiative. 

From \fig{fig:ptDist}, we see that above $E>10$ GeV, the light and charm quarks have 
similar elastic and inelastic energy losses. But due to the 
large mass effect \cite{Dokshitzer:2001zm,Djordjevic:2003be,Djordjevic:2003qk,Djordjevic:2004nq,Zhang:2003wk,Armesto:2005iq}, 
both radiative and elastic energy losses remain significantly smaller for
bottom quarks than for light and charm quarks. We present both TG and BT as a measure of the theoretical uncertainties of the Coulomb log. These are largest for the heaviest b quark, for which the leading log approximation \cite{Thoma:1990fm,Braaten:1991jj,Braaten:1991we} breaks down
in the kinematic range accessible at \rhic as they are not ultrarelativistic.
With advanced numerical covariant transport techniques \cite{Molnar:2001ux}
the theoretical errors on the elastic energy loss effects can be reduced 
considerably in the future.

\section{Theoretical Framework}
The quenched spectra of partons, hadrons, and leptons are calculated
as in \cite{Djordjevic:2005db} from the generic pQCD convolution
\be
\label{schem}
\frac{E d^3\sigma(e)}{dp^3} = \frac{E_i d^3\sigma(Q)}{dp^3_i} \otimes {P(E_i \rightarrow E_f )} \otimes D(Q \to H_Q) \otimes f(H_Q \to e),
\ee
where $Q$ denotes quarks and gluons.
For charm and bottom, the initial quark spectrum, 
$E d^3\sigma(Q)/dp^3$, is computed at next-to-leading order
using the code from \cite{Cacciari:2005rk,Mangano:1991jk};
for gluons and light quarks, the initial distributions are computed 
at leading order as in \cite{Vitev:2002pf}. $P(E_i \rightarrow E_f )$ is the energy loss probability, $D(Q \to
H_Q)$ is the fragmentation function of quark $Q$ to hadron $H_Q$, and
$f(H_Q \to e)$ is the decay function of hadron $H_Q$ into the observed
single electron. We use the same mass and factorization scales as
in \cite{Vogt:2001nh} and employ the CTEQ5M parton densities \cite{Pumplin:2002vw,Stump:2003yu}
with no intrinsic $k_T$. As in \cite{Vogt:2001nh} we neglect shadowing of
the nuclear parton distribution in this application.

We assume that the final quenched energy, $E_f$, is large enough that 
the Eikonal approximation can be employed. We also assume that in Au+Au 
collisions, the jet fragmentation function into hadrons is the same as in 
$e^+e^-$ collisions. This is expected to be valid in the deconfined 
pQCD medium case, where hadronization of $Q\rightarrow H_Q$ cannot occur 
until the quark emerges from the deconfined sQGP medium. 

The main difference between our previous 
calculation \cite{Djordjevic:2005db} 
and the present one is the inclusion of two new physics components in the  
energy loss probability $P(E_i \rightarrow E_f )$.
First, $P(E_i \rightarrow E_f)$ is generalized to include for the first time 
both elastic and inelastic energy loss and their fluctuations. We note that Vitev \cite{Vitev:2003xu,Vitev:2004kd,Qiu:2005ki} was the first to generalize GLV
to include {\it initial state} elastic energy loss effects
in D+Au. In this work, Eq. (\ref{fullconv}) extends the formalism to include
{\it final state} elastic loss effects  in $A+A$.

The second major
change relative to our previous applications is that we now take into account 
geometric path length fluctuations as follows:
\begin{eqnarray}
P(E_i \rightarrow E_f) & =& \int\frac{d\phi}{2\pi} \int \frac{d^2 \vx_\perp}{N_{bin}(b)} \; T_{AA}(\vx_\perp,\vb)\otimes P_{rad}(E_i \rightarrow E; L(\vx_\perp,\phi))  \nonumber \\
\label{fullconv}
& & \otimes P_{el}(E\rightarrow E_f; L(\vx_\perp,\phi)); \\
\label{Leff}
L(\vx_\perp,\phi) & = & \int d\tau \rho_p(\vx_\perp+\tau\hat{n}(\phi))/\langle\rho_p\rangle,
\end{eqnarray}
where $L$ is the locally determined effective path length of the jet given its initial 
production point $\vx_\perp$ and its initial azimuthal direction $\phi$
relative to the impact parameter plane $(x,y)$. The geometric path 
averaging used here is similar to that used in \cite{Gyulassy:2000gk}
and by Eskola et al. \cite{Eskola:2004cr}. However, 
the inclusion of elastic losses together with path fluctuations
in more realistic geometries was not considered before.

We consider a diffuse Woods-Saxon nuclear density profile \cite{Hahn}, which creates a participant transverse density,
$\rho_p(\vx_\perp)$, computed using the Glauber profiles,
$T_A(\vx)$, with inelastic cross section $\sigma_{NN}=42$ mb. The
bulk sQGP transverse density is assumed to be proportional to this
participant density, and its form is shown (for the $y=0$
slice) in \fig{fig:Survival} by the curve labeled $\rho_{{\rm QGP}}$. The distribution
of initial hard jet production points, $\vx_\perp$, is assumed on the
other hand to be proportional to the binary collision density,
$T_{AA}=T_A(\vx+\vb/2)T_A(\vx-\vb/2)$. This is illustrated in \fig{fig:Survival} by the narrower curve labeled $\rho_{\rm Jet}$.

\begin{figure}[!htb] 
\centering
\includegraphics[width=3.25 in]{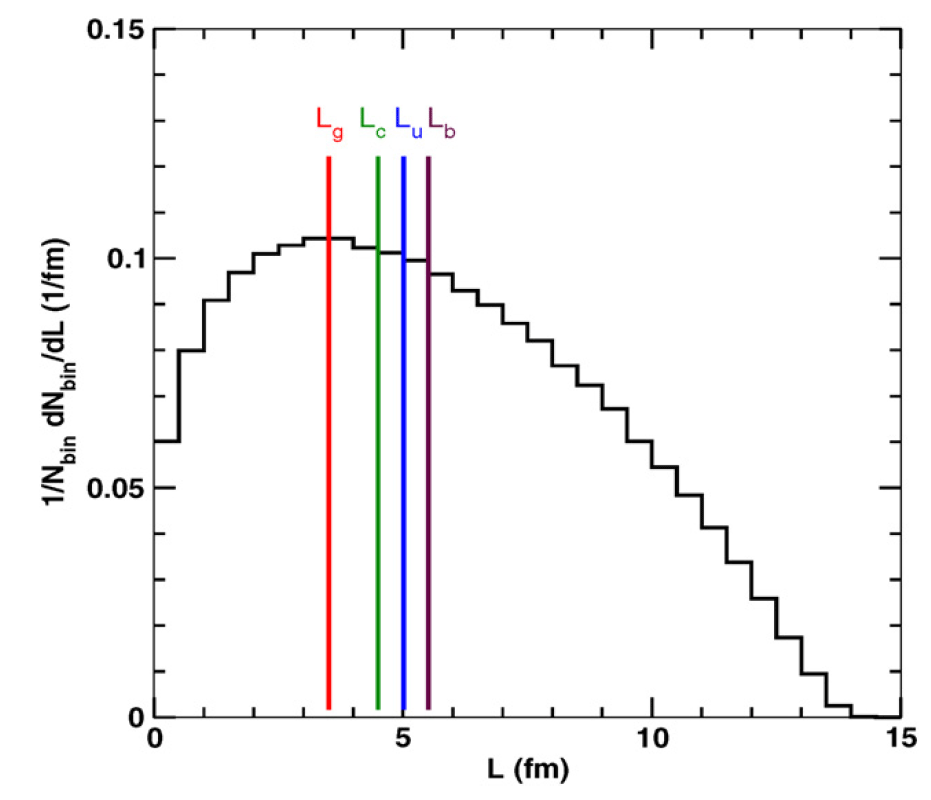}
\caption{\label{fig:pofl} Distribution of path lengths (given by \eq{Leff}) traversed by hard scatterers; the lengths, $L(\vx_\perp,\phi)$, are weighted by the probability of production and averaged over azimuth. An equivalent formulation of \eq{power} is $R_Q^I=\int dL 1/N_{bin} dN_{bin}/dL \int d\epsilon (1-\epsilon)^n P_Q^I(\epsilon;L)$. Since this is a purely geometric quantity, it is the same for all jet varieties. Also displayed are the single, representative pathlengths, $L_Q$, used as input in approach II. Note the hierarchy of scales with glue requiring the shortest, then charm, light quarks, and bottom the longest effective pathlength.}
\end{figure}

The combination of fluctuating DGLV radiative \cite{Djordjevic:2003zk}
with the new elastic energy losses and fluctuating path lengths (via 
the extra $d^2\vx_\perp d\phi$ integrations) adds a high
computational cost to the extended theory specified by
Eqs.~(\ref{schem},\ref{fullconv}). In this first study with the extended 
theory, we explore the relative order of magnitude of the competing effects
by combining two simpler approaches.

In approach I, we parameterize the heavy quark pQCD spectra by a simpler 
power law, $E d^3\sigma_Q/d^3k \propto 1/p_\perp^{n+2}$, with a slowly 
varying logarithmic index $n\equiv n(p_T)$. For the pure power
law case, the {\em partonic}  modification factor, 
$R_Q=d\sigma_Q^{final}/d\sigma_Q^{initial}$, (prior to fragmentation) 
is greatly simplified. This enables us to perform the path length 
fluctuations numerically via
\be
\label{power}
R^I_Q = \int\frac{d\phi}{2\pi}\int \frac{d^2 \vx_\perp}{N_{bin}(b)} \; T_{AA}(\vx_\perp,\vb) \; \int d\epsilon (1-\epsilon)^n P_Q^I(\epsilon; L(\vx_\perp,\phi)),
\ee
where
\be
P^I_Q(\epsilon; L) = \int dx P_{Q,rad}(x; L) P_{Q,el}(\epsilon-x; L).
\ee
Both the mean and width of those fractional 
energy losses depend on the local path length. We emphasize 
that no externally specified {\it a priori} path length, $L$, appears in 
Eq.~(\ref{power}); the path lengths explore the whole geometry. \fig{fig:pofl} shows the broad distribution of lengths traversed by hard partons in approach I.

\begin{figure}[!htb] 
\centering
\includegraphics[width=3.25 in]{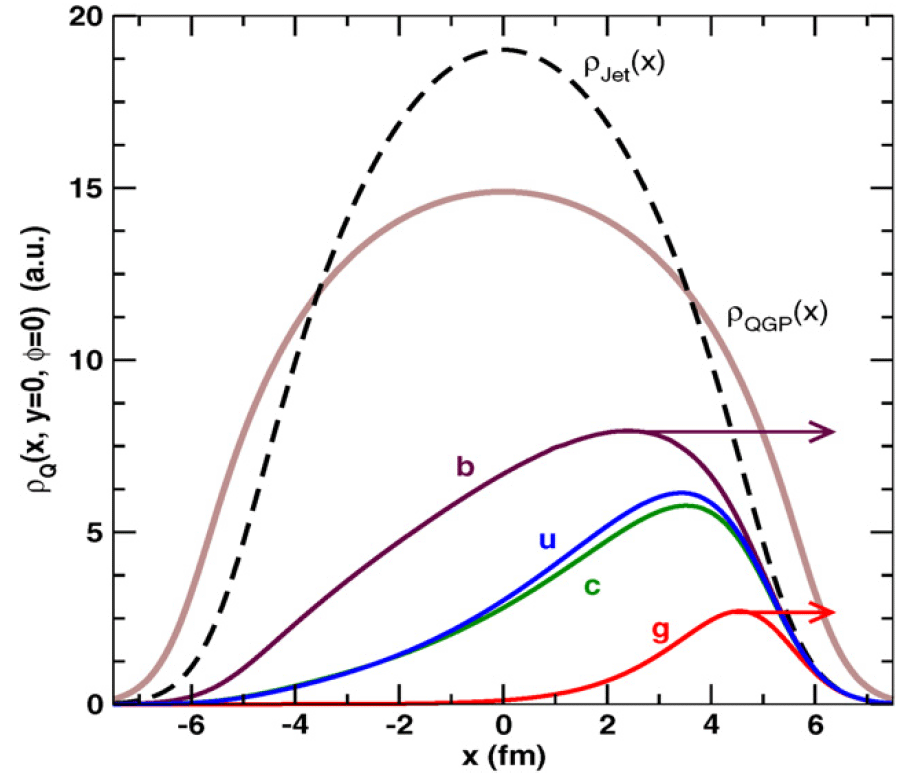}
\caption{\label{fig:Survival} Transverse coordinate
$(x,0)$ distribution of surviving
$p_T=15$ GeV, $Q=g,u,c,b$ jets moving in direction $\phi=0$
as indicated by the arrows. Units are arbitrary for illustration.
The transverse (binary collision) distribution
of initial jet production points, $\rho_{\rm Jet}(x,0)$, is shown
at midrapidity for Au+Au collisions at $b=2.1$ fm.
The ratio $\rho_Q/\rho_{\rm Jet}$ (see Eq.(\ref{rhoQ}))
gives the local quenching factor
including elastic and inelastic energy loss though
the bulk QGP matter distributed as $\rho_{\rm QGP}(x,0)$.}
\end{figure}

In the second approach, we determine effective path lengths, $L_Q$, 
for each parton flavor, $Q$, by varying fixed $L$ predictions 
${R}^{II}_Q(p_\perp, L)$ and comparing to approach I; see \fig{fig:pofl}. In approach II,  
${R}^{II}_Q(p_\perp, L)$ is calculated directly from Eq.~(1) with 
$P(E_i \rightarrow E_f)$ in Eq.~(2) replaced by the fixed $L$ approximation
\be
\label{fixedL}
P(E_i \rightarrow E_f, L) \approx  P_{rad}(E_i \rightarrow E; L) \otimes P_{el}(E\rightarrow E_f; L).
\ee
Here, jet quenching is performed via two independent branching 
processes in contrast to the additive convolution in Eq.~(\ref{power}).
For small energy losses the two approaches are similar. They differ
however in the case of long path lengths when the probability of complete
stopping approaches unity. In the convolution method,
the probability of $\epsilon>1$ is interpreted as complete stopping,
whereas in the branching algorithm the long path length case is just highly
suppressed. In both cases we take into account the
small finite probability that the fractional energy loss $\epsilon \leq 0$ due to fluctuations.

To illustrate the difference in approach II, consider the case of power law
initial $Q$ distributions as in Eq.~(\ref{power}). In this case
\be
\label{power2}
R^{II}_Q(p_\perp,L_Q) \equiv \langle(1-\epsilon_{Q}^r(L_{Q}))^n (1-\epsilon_{Q}^e(L_{Q}))^n \rangle_{\Delta E}.
\ee
The branching implementation is seen via the product of two distinct 
factors in contrast to the one quenching factor in Eq.~(\ref{power}).
For small $\langle \epsilon_Q^{r,e} \rangle $ both approaches 
obviously give rise to the same $R_Q=1- n \langle \epsilon_Q\rangle$.

In both approaches, fluctuations of the radiative energy loss due to gluon number fluctuations are computed as discussed in detail in
Ref. \cite{Djordjevic:2005db,Djordjevic:2004nq}. This involves using
the DGLV generalization \cite{Djordjevic:2003zk} of the GLV opacity
expansion \cite{Gyulassy:2000er} to heavy quarks. Bjorken longitudinal expansion is taken into account by evaluating the bulk density at an average time $\tau = L/2$ \cite{Djordjevic:2005db,Djordjevic:2004nq}. For elastic energy loss fluctuations, the full
Fokker-Planck solution applicable for small fractional energy
loss is approximated by a Gaussian centered at the average loss with
variance $\sigma_{el}^2 = 2 T \langle \Delta E^{el}(p_T,L)\rangle $
\cite{Moore:2004tg}. In approach I the correct, numerically intensive integration through the Bjorken expanding medium provides $\Delta E^{el}(p_T,L)$. In approach II the $\tau = L/2$ approximation is again used; numerical comparisons show that for $L\sim2-7$ fm this reproduces the full calculation well. Finally, we note that we use the additional numerical simplification of keeping the strong coupling constant $\alpha_S$ fixed at $0.3$.

\section{Numerical Results: Parton Level}
\begin{figure}[!htb] 
\centering
\includegraphics[width=3.27in]{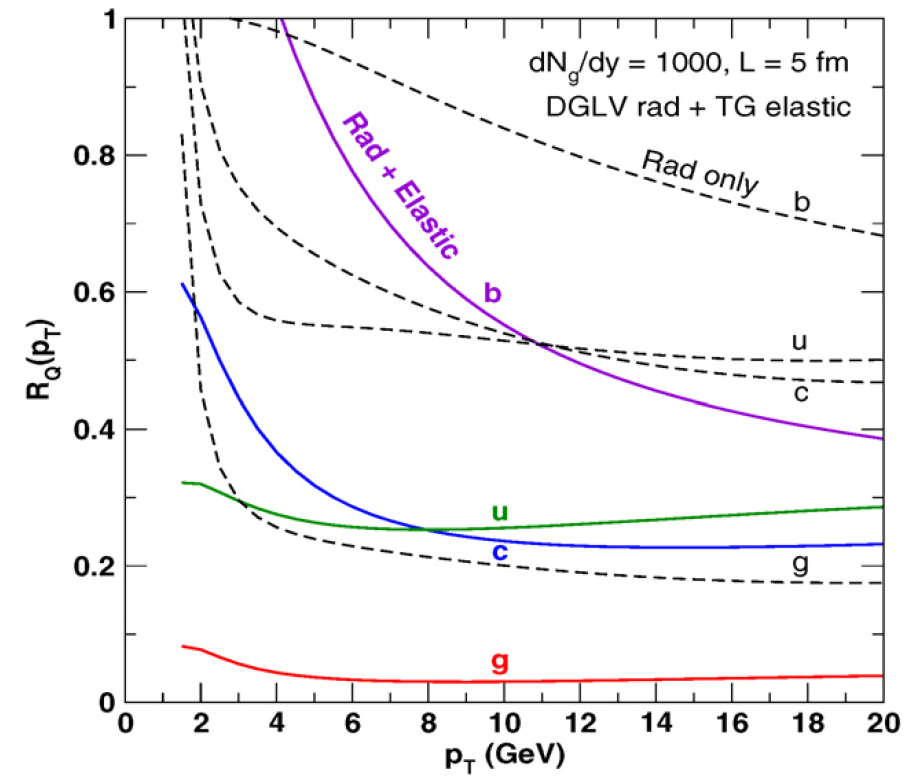}
\caption{\label{fig:ptDist2} 
Partonic nuclear modification, $R_{Q}^{II}(p_T)$ via Eq.(\ref{power2}),
for $g,u,c,b$ as a function
of $p_T$ for fixed L=5 fm path length and $dN_g/dy=1000$.
 Dashed curves
include only radiative energy loss,
while solid curves include elastic energy loss as well. 
}
\end{figure}

In \fig{fig:ptDist2}, we show the quenching pattern of $Q$ from the second approach
for a ``typical'' path length scale $L=5$ fm, similar to that used in previous calculations \cite{Djordjevic:2005db}. 
The curves show $R_Q(p_T)$, prior to hadronization, for
$Q=g,u,c,b$. The dashed curves show the quenching arising from only the 
DGLV radiative energy loss. The solid curves show the full results
after including TG elastic as well as DGLV radiative energy loss. Adding 
elastic loss is seen to increase the quenching of all flavors for fixed 
path length. Note especially the strong 
increase of the gluon suppression and the factor $\sim 2$ 
increase of the bottom suppression. The curious switch of 
the $u$ and the $c$ quenching reflects the
extra valence (smaller index $n_u$) contribution to light quarks.

\fig{fig:ptDist2} emphasizes the unavoidable result of using a fixed, ``typical'' 
path length scale, $L$, in jet tomography: the pion 
and single electron quenching can never be similar. If pions were composed only of light quarks and electrons only of charm, 
then we would expect comparable quenching
for both. However, contributions from highly quenched gluons 
decrease the pion $R_{AA}$ while weakly quenched bottom quarks 
increase the electron $R_{AA}$. Therefore, in the fixed length 
scenario, we expect a noticeable difference between pion 
and single electron suppression patterns.

In \fig{fig:Survival} we use fluctuating path lengths. The solid curves labeled by the parton flavor $Q=g,u,c,b$
show the relative transverse coordinate density of surviving jets defined by
\begin{eqnarray}
\rho_Q(\vx,\phi)\equiv \rho_{\mathrm{Jet}}(\vx)\int d\epsilon (1-\epsilon)^n P_Q^I(\epsilon; L(\vx, \phi)).
\label{rhoQ}
\end{eqnarray}
$\rho_Q$ is given by the initial transverse $\vx$ production distribution 
times the quenching factor from that position in direction $\phi$
with final momentum $p_T$. The case shown is for a $p_T=15$~GeV 
jet produced initially at $(x,0)$ and moving in the direction $\phi=0$
along the positive x axis. The quenching is determined by the participant 
bulk matter along its path $\rho_{\rm QGP}(x+ vt,0,t)$, and varies with $x$ because the local path length $L=L(x,0,0)$ changes according to Eq.~(\ref{Leff}).

What is most striking in \fig{fig:Survival} is the hierarchy of $Q$-dependent length 
scales. No single, representative path length can account for the distribution of all 
flavors. In general heavier flavors are less biased toward the surface 
(in direction $\phi$) than lighter flavors since the energy loss decreases 
with the parton mass. Gluons are more surface biased than light 
quarks due to their color Casimir enhanced energy loss. In addition, note
the surprising reversal of the $u$ and $ c$ suppressions, also seen in \fig{fig:ptDist2}. 
\fig{fig:ptDist} shows that the energy loss of $c$ is somewhat less than for $u$; 
however, the higher $p_T$ power index, $n$,
of $c$ relative to $u$ -- as predicted by pQCD and due to 
the valence component of $u$ -- compensates by amplifying its quenching. 

However, it is clear that none of the distributions can be usefully
categorized as surface emission. The characteristic widths of these
distributions range from $\Delta x \approx 3-6$ fm. Such a large
dynamic range of path length fluctuations allows for the consistency
of simultaneously reproducing both the electron and pion data.

\begin{figure}[!hbt] 
\centering
\includegraphics[width=3.25in]{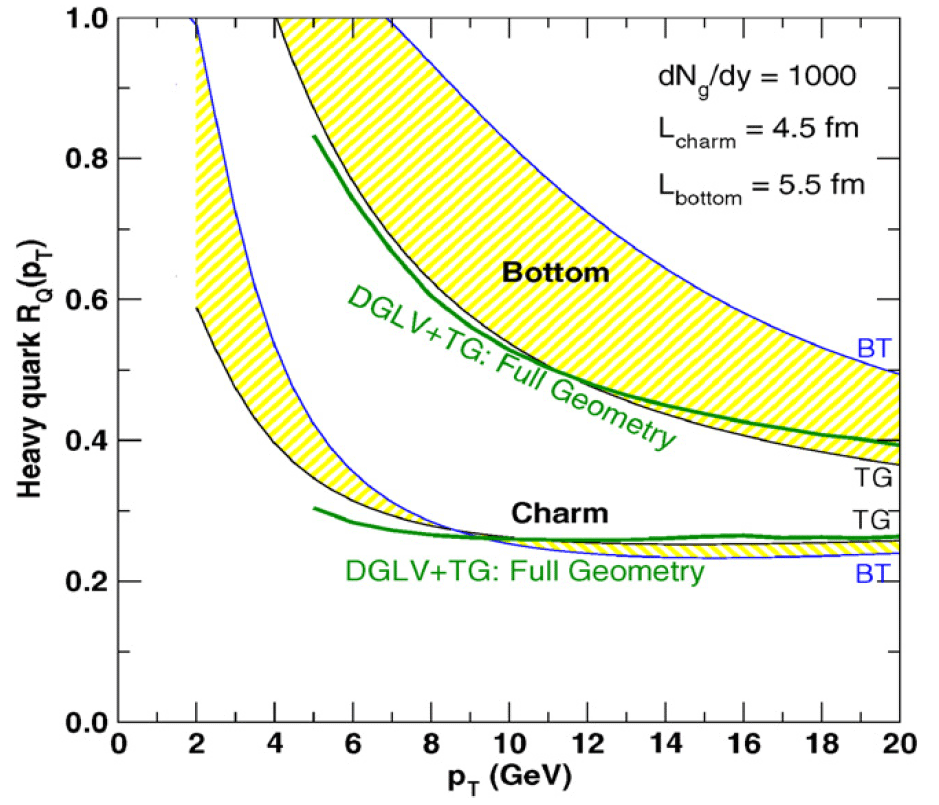}
\caption{\label{fig:cbRAA} Heavy quark jet quenching before fragmentation into mesons for $dN_g/dy=1000$. Solid green curves show the results of approach I based on Eq.~(\ref{power}) including full geometric path length fluctuations and DGLV radiative and TG elastic energy loss for c and b quarks. Upper and lower 
yellow bands show predictions using approach II via Eq.~(\ref{power2}) with effective path lengths taken as $L_{c}= 4.5$ and $L_{b}= 6.5$ fm. As previously noted in \fig{fig:ptDist}, the difference between TG and BT curves indicates the magnitude of theoretical uncertainties in the elastic energy loss.}
\end{figure}

\begin{figure}[!htb] 
\centering
\includegraphics[width=3.25in]{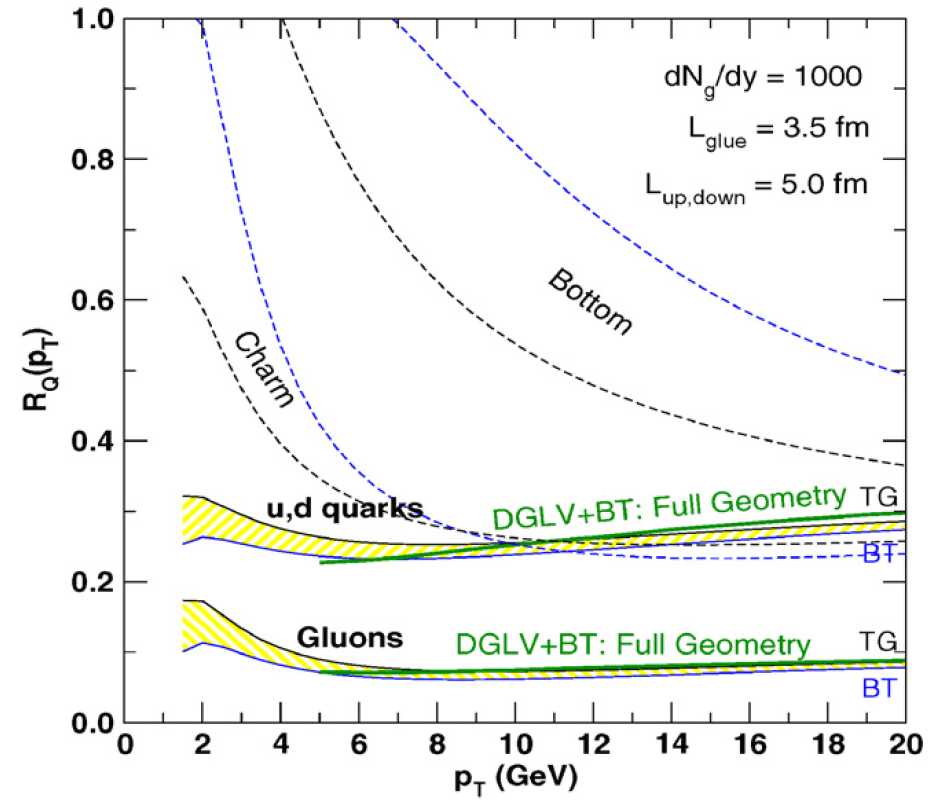}
\caption{\label{fig:qRAA} As in \fig{fig:cbRAA} 
but for light $u,d$ quarks and gluons. The yellow bands
are computed in this case with 
effective $g,u$ path lengths $L_{g}= 4.0$ and 
$L_{u}= 5.0$ fm based on Eq.~(\ref{power2}).
Note that  charm and light quark quenching
are similar in this $p_T$ range. }
\end{figure}

We turn next to Figs.~\ref{fig:cbRAA} and~\ref{fig:qRAA} to show the interplay between dynamical
geometry seen in \fig{fig:Survival} and the elastic-enhanced quenching of
partons.  In Figs.~\ref{fig:cbRAA} and~\ref{fig:qRAA} the solid green curves labeled ``DGLV+TG/BT: Full Geometry''
are the results using approach I based on Eq.~(\ref{power}). The
curves labeled TG and BT are from approach II based on
Eq.~(\ref{power2}). The effective fixed $L_Q$ in II were taken to
match approximately the green curves in which full path length
fluctuations are taken into account. This procedure is not exact because
of the different numerical approximations involved, but the trends are
well reproduced. The $L_Q$ are determined only to $\sim 0.5$ fm accuracy, as this suffices for our purposes here.
We show the comparison between approaches I and II for heavy quarks in \fig{fig:cbRAA} using $L_{c}= 4.5$ and $L_{b}= 6.5$ fm, and for gluons and light quarks in \fig{fig:qRAA} using  $L_{g}= 4.0$ and $L_{u}= 5.0$ fm; see \fig{fig:pofl} for a visual comparison of the input length distributions used. This hierarchy of $Q$-dependent length scales is in accord with that expected from \fig{fig:Survival}.

Note that geometric fluctuations reduce charm quark quenching in \fig{fig:cbRAA} 
relative to the fixed $L=5$ fm case in \fig{fig:ptDist2}. This is seen even more 
dramatically for gluons in \fig{fig:qRAA}: the overquenching of gluons in \fig{fig:ptDist2}
is reduced by a factor $\sim 2$ when geometric fluctuations are taken 
into account. Nevertheless, even with path fluctuations the gluons are still
quenched by a factor of 10 when elastic energy loss is included in addition to 
radiative. 
  
The amplified role of elastic energy loss is due to its smaller width
for fluctuations relative to radiative fluctuations.  Even in
moderately opaque media with $L/\lambda\sim 10$, inelastic energy loss
fluctuations are large because only a few, 2-3, extra gluons are
radiated \cite{Gyulassy:2001nm}. Thus, gluon number fluctuations, $O(1/\surd
N_g)$ lead to substantial reduction in the effect of radiative
energy loss. On the other hand, elastic energy loss
fluctuations are controlled by collision number fluctuations,
$O(\sqrt{\lambda/L})$, which are small in comparison. Therefore,
fluctuations of the elastic energy loss do not dilute the suppression of the nuclear
modification factor as much as $N_g$ fluctuations. The
increase in the sensitivity of the final quenching level to the
opacity is a novel and useful byproduct of including the elastic
channel.
The inclusion of elastic energy loss significantly reduces the 
fragility of pure radiative quenching \cite{Eskola:2004cr} and therefore increases 
the sensitivity of jet quenching to the opacity of the bulk medium.

\section{Numerical Results: Pions and Electrons}
We now return to \fig{fig:eRAA} to discuss the consequence of including
elastic energy loss of $c$ and $b$ on the electron spectrum. The
largest source of uncertainty, represented
by the lower yellow band, is the modest
but poorly determined elastic energy loss, $\Delta E/E\approx 0.0
-0.1$, of bottom quarks (see \fig{fig:ptDist}). Nevertheless, it is
remarkable that the lower yellow band can reach $R_{AA}\sim
0.5$ in spite of keeping $dN_g/dy=1000$, consistent with measured multiplicity, and
using a conservative $\alpha_S=0.3$. While the preliminary data suggest
that $R_{AA}^e< 0.4$, the Rad+Elastic band is not inconsistent with the
preliminary data above $p_T\gsim 7$ GeV within present experimental
and theoretical errors. Our tentative conclusion, therefore, is that
the combined elastic and inelastic pQCD approach may 
solve a substantial part of the heavy quark quenching puzzle.

\begin{figure}[!htb] 
\centering
\includegraphics[width=3.25in]{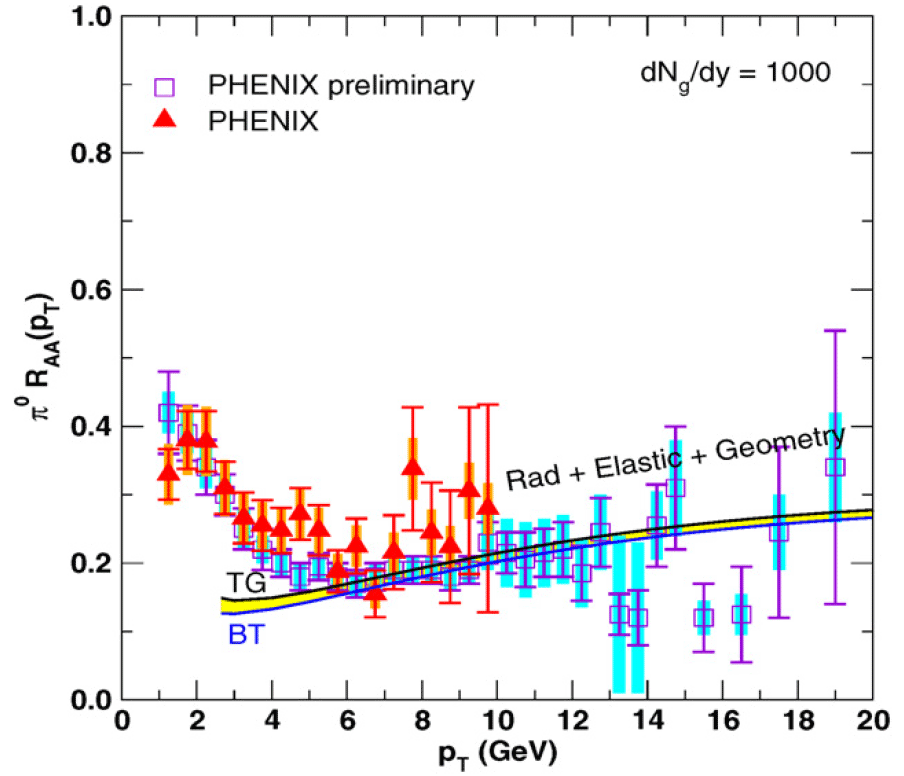}
\caption{\label{fig:piRAA} 
The consistency of the extended jet quenching theory
is tested by comparing its prediction to the nuclear modification
of the $\pi^0$ spectra observed by
\phenix \cite{Isobe:2005mh,Shimomura:2005en,Adler:2003qi}.
}
\end{figure}

However, as emphasized in \cite{Djordjevic:2005db}, any proposed solution
of this puzzle must also be checked for consistency with the extensive pion
quenching data \cite{Isobe:2005mh,Shimomura:2005en,Adler:2003qi}, for which  preliminary data
now extend out to $p_T\sim 20$ GeV. 
That this is a challenge is clearly 
seen in \fig{fig:ptDist2}, where for fixed $L=5$ fm, the addition of elastic loss 
would significantly overpredict the quenching of pions. However, the simultaneous 
inclusion of path fluctuations leads to a decrease of the mean $g$ and
$u,\; d$ path lengths that partially 
offsets the increased energy loss. 
Therefore, the combined three effects considered here makes it 
possible to satisfy $R_{AA}^e<0.5$ without violating the bulk $dN_g/dy=1000$ 
entropy constraint and without violating the pion quenching constraint 
$R_{AA}^{\pi^0}\approx 0.2\pm 0.1$ now observed out to 20 GeV; see \fig{fig:piRAA}.
We note that the slow rise of $R_{AA}^{\pi^0}$ with $p_T$ in the present
calculation is due in part to the neglect of initial $k_T$ smearing
that raises the low $p_T$ region and the EMC effect that lowers
the high $p_T$ region (see \cite{Vitev:2002pf}).

\section{Sensitivity and Imprecision in WHDG}
The ultimate goal of hard probe physics is jet tomography, the use of the high-\pt hadronic attenuation pattern to learn about the medium density.  One needs then a theoretical description of the processes involved that describes the data, has a mapping from its parameter(s) to the matter density, and is under calibrated control.  

For these tools to be truly tomographic, however, there is the additional requirement that the resulting medium measurement be reasonably precise.  The overall precision of density determination is set by a combination of experimental precision and theoretical sensitivity: if the experimental data are infinitely precise, any model with the previous paragraph's characteristics could asses the medium density with similar precision; but experimental error exists and must be taken seriously, limiting the precision of medium determination by the innate sensitivity of the theoretical model to changes in its input parameter.  
Additionally there will also be theoretical error.  Sources for this include, but are not limited to: neglect of higher order terms; treatment of the running coupling, especially at low $q^2$; and the translation of the realistic medium density to the brick problem for which the analytic theoretical derivation was applied.

\cite{Eskola:2004cr} claimed that BDMPS-based radiative energy loss models are a fragile probe of the medium; i.e.\ large changes in \qhatcomma, the input parameter for the theory, are not reflected in \raa predictions incompatible with data.  Surface emission, the process in which only the partons produced very close to the medium edge escape for observation, was posited as the reason for their observed insensitivity to increased medium opacity.  This is an appealing argument, and \cite{Dainese:2004te,Dainese:2005kb} provided some evidence for this idea.  The very pessimistic conclusion, then, was that jet tomography is not possible at \rhic and will not be possible at the \lhccomma.  

We counter the dire claims in \cite{Eskola:2004cr} by first noting that several approximations were made in their description of the medium produced in heavy ion collisions.  In \cite{Eskola:2004cr}, the authors used a hard cylinder initial nuclear density and did not include Bjorken expansion.  \cite{Dainese:2004te,Dainese:2005kb} used the realistic Woods-Saxon nuclear density distribution but again neglected Bjorken expansion.  These certainly aided in alleviating some of the high numerical cost involved in the calculations.  However we feel these geometrical simplifications strongly biased their results to the surface of the medium, resulting in a significant loss in model sensitivity.  Moreover, experimental error bars have shrunk considerably in the time since \cite{Eskola:2004cr} were published. These issues raise serious doubts about the validity of their conclusions.  

\begin{figure}[!htb] 
\centering
\includegraphics[width=3.25in]{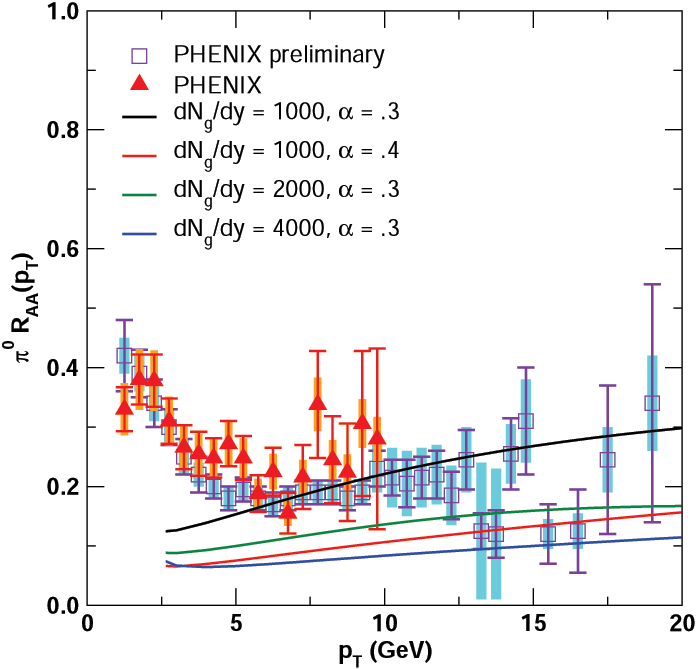}
\caption{\label{nofragility} Model sensitivity to varying \dnslashdycomma, which is much greater than observed in \cite{Eskola:2004cr}.  The susceptibility to changes in the fixed \alphas suggests the need to investigate the theoretical error stemming from the coupling.
}
\end{figure}

\begin{figure}[!htb] 
\centering
\includegraphics[width=3.25in]{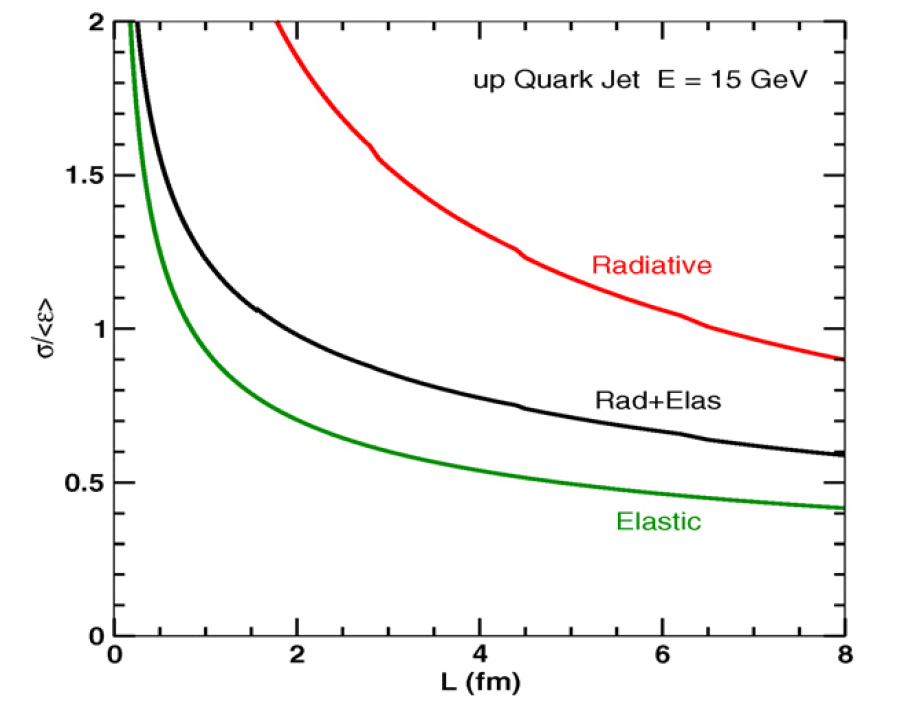}
\caption{\label{widths} 
The ratio of rms width, $\sigma(L)$, to the mean fractional
energy loss $ \langle \epsilon\rangle$ for radiative, elastic and convoluted 
energy
loss distributions is shown as a function of the path length, $L$,
for a Bjorken expanding plasma with $dN_g/dy=1000$.
The case of an up quark jet with $E=15$ GeV is shown.
Notice that the elastic distribution is significantly narrower
than the radiative one. This amplifies the effect of elastic energy loss on $R_{AA}$ relative to radiative.
}
\end{figure}

One can readily see in \fig{nofragility} that the use of fluctuating path lengths in a realistic and expanding background medium geometry results in a predicted $R_{AA}(\eqnpt)$ inconsistent with data when the medium density is artificially increased by a factor of 2.  Sensitivity to medium density changes is enhanced over pure radiative loss models by the inclusion of elastic energy loss, due to its smaller width for fluctuations relative to radiative \fig{widths}.  We note, however, that models using GLV-type radiative loss or higher-twist loss only are similarly not fragile \cite{Vitev:2005ch}.  More recent work in which jets were propagated through a medium evolved through hydrodynamics found that energy loss using the ASW quenching weights in conjunction with up-to-date experimental results are in fact not fragile, either \cite{Renk:2006pk}.  The pion \raapt then has the potential to be a quite good experimental signal for jet tomographic studies at \rhiccomma.  However the large systematic theoretical error seen in \fig{nofragility} associated only with changes in the coupling raise serious doubts about the theoretical imprecision of calculated results.  Before any firm conclusions may be made regarding the possibility of tomography these issues must be more rigorously investigated.

\section{LHC Predictions}
We show in \fig{lhcraa} the large qualitative difference in \pt dependence for predicted \lhc pion \raa from the WHDG model and from two different implementations of the ASW model; one easily sees the dramatic rise in \raa with increasing \pt from the first as opposed to the flat in \pt results of the latter two.  The consistency of Vitev's curve \cite{Vitev:2005he} with ours over a range of $dN_g/dy$ and the consistency of the two ASW calculations \cite{Eskola:2004cr,Dainese:2004te} suggest that this will be a robust test of the energy loss mechanisms.  The origin of the difference can be easily understood given our discussion of \rhic results.  For the case of WHDG, the pion \raa at \rhic does not require much overabsorption.  The modest increase in medium density, $\sim2-3$ based on either an extrapolation of \phobos results \cite{Back:2001ae,Adcox:2000sp} or predictions from the CGC \cite{Kharzeev:2004if,McLerran:2004fg}, for the \lhc leads to small energy losses at high momenta that can be well approximated by the pocket asymptotic energy loss formulae
\bea
\label{whdg:pocket1}
\epsilon_{rad}=\Delta E_{rad}/E & \sim & \eqnalphas^3\log(E/\mu^2L)/E\\
\label{whdg:pocket2}
\epsilon_{el}=\Delta E_{el}/E & \sim & \eqnalphas^2\log(\sqrt{ET}/m_g)/E;
\eea
see \fig{whdg:lhcasymp}.  As \pt increases the $\log(E)/E$ reduction in energy loss is not compensated by the slow (in comparison to \rhiccomma) increase in the power law partonic production spectrum, $dN/d\eqnpt\propto1/\eqnpt^{n(\eqnpt)}$ (see Appendix \ref{punchappendix}); thus \raa increases with \ptcomma.  On the other hand, the ASW models mimic the small normalization of the \rhic data by highly suppressing their jets; as discussed in the subsequent paragraphs, the significant quenching leads to a loss of information on the details of the energy loss process, flattening the results.  The two ASW models represented in \fig{lhcraa} used EKRT-type medium density scaling \cite{Eskola:1999fc} making the \lhc medium $\sim7$ times more dense than at \rhiccomma.  

\bfig[!htb]
\centering
$\begin{array}{@{\hspace{0in}}c@{\hspace{0in}}c@{\hspace{0in}}}
\includegraphics[width=2.75in]{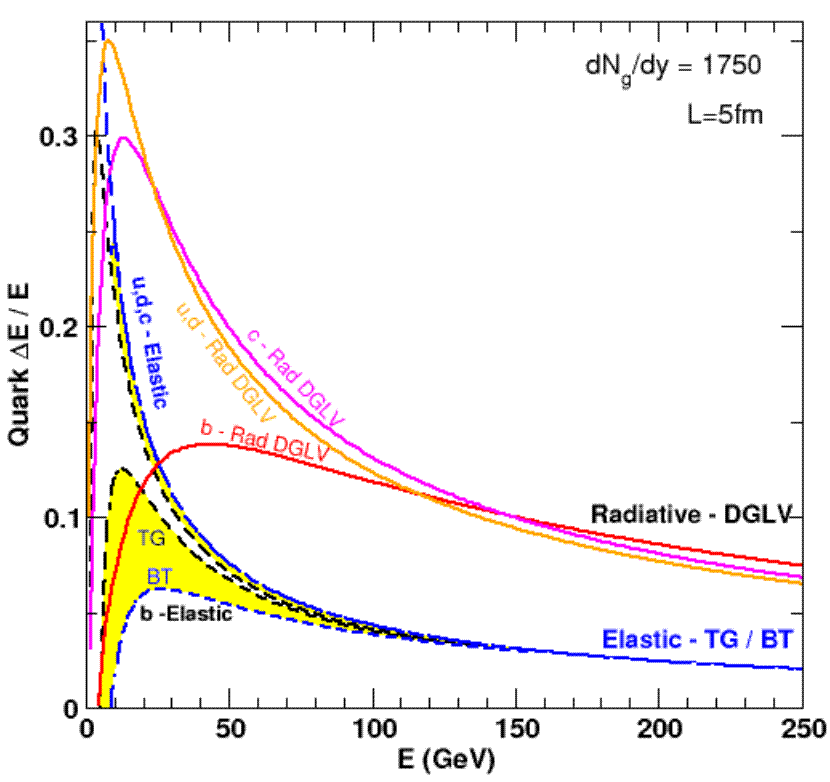} & 
\includegraphics[width=2.75in]{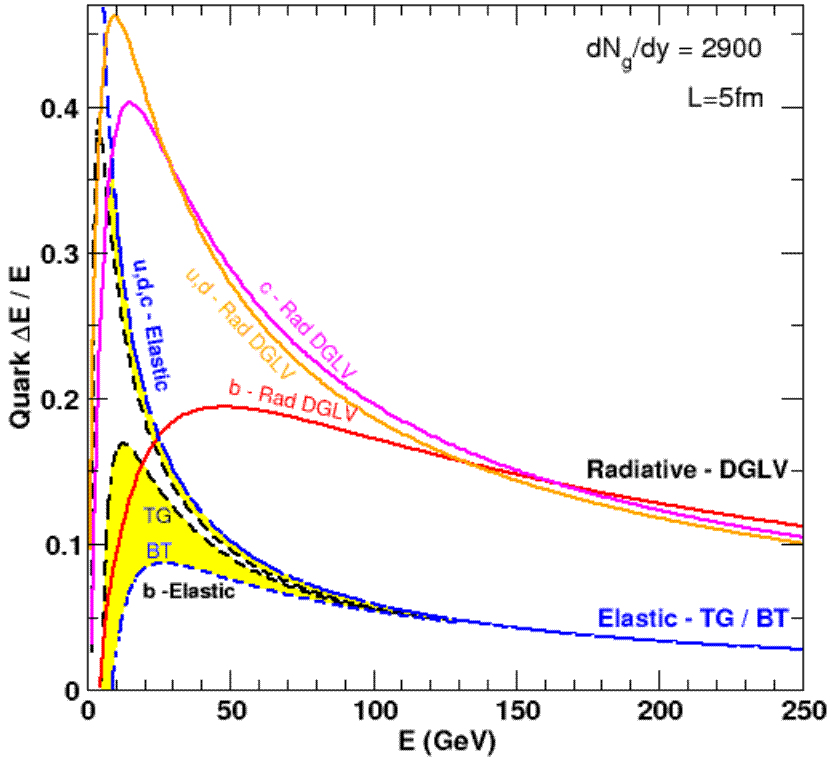} \\ [-.05in]
{\mbox {\scriptsize {\bf (a)}}} & {\mbox {\scriptsize {\bf (b)}}}
\end{array}$
\caption{\label{whdg:lhcasymp}
Radiative DGLV and TG and BT elastic partonic fractional energy loss as implemented in WHDG \cite{Wicks:2005gt} at \lhc momenta for all jet species at fixed $L=5$ fm and (a) $dN_g/dy=1750$ and (b) $dN_g/dy=2900$; the former density comes from the \phobos extrapolation \cite{Back:2004je,Busza:2007ke}, the latter from the KLN model of the color glass condensate (CGC) \cite{Kharzeev:2004if}.  $\Delta E/E$ exhibits asymptotic behavior for both energy loss channels as given by the analytic pocket formulae \eq{whdg:pocket1} and \eq{whdg:pocket2}.  
}
\efig

The important quantity in any of these calculations is $P(\epsilon)$, the probability that the final momentum is some fraction of the initial momentum, $p_T^f=(1-\epsilon)p_T^i$.  For calculations that include only radiative processes, $P(\epsilon)=P_{rad}(\epsilon)$, where
\be
\label{prad}
P_{rad}(x)=\left\{\begin{array}{l}
P_0^g\delta(x)+\widetilde{P}_{rad}(x)\\
e^{-N_g}\sum\limits_n\frac{N_g^n}{n!}\widetilde{P}_{rad,n}(x) = 
e^{-N_g}\delta(x)+ e^{-N_g} N_g \widetilde{P}_{rad,1}(x)+\ldots.
\end{array}\right.
\ee
The possibility that no gluons are emitted (subsequent to the original gluon radiation created by the initial hard scatter) is encapsulated in the coefficient, $P_0^g$, of the $\delta(x)$ term.  

Due to the approximations used in the radiative calculations overabsorption, $P(\epsilon>1)$, has a large support for highly suppressed jets.  Usually, one of two prescriptions is applied to remove this unphysical artifact.  Either the integrated excess probability weighs an explicit delta function,
\be
\label{norw}
P(\epsilon)=P_{old}(\epsilon)\theta(1-\epsilon) + \int_1^\infinity \!dxP_{old}(x) \delta(1-\epsilon),
\ee
or reweighs (rw) the original distribution,
\be
\label{rw}
P_{rw}(\epsilon)=\frac{1}{\int_1^\infinity \!dxP_{old}(x)}P_{old}(\epsilon)\theta(1-\epsilon).
\ee
Clearly the latter approach leads to larger \raa values for the two.  For large overabsorption energy loss details are lost to the removal of the unphysical $\epsilon>1$ region.  

\bfig[!htb]
\centering
\includegraphics[width=3.25in]{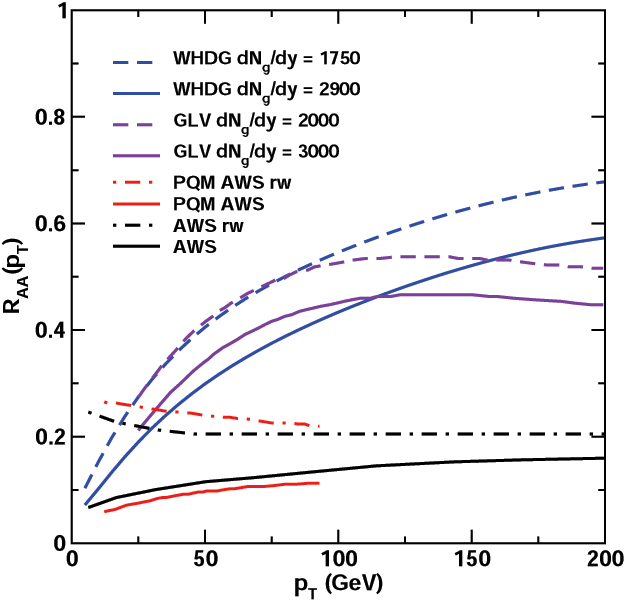}
\caption{\lhc predictions of pion \raapt for several energy loss models.  Curves correspond to WHDG \cite{Wicks:2005gt}, GLV \cite{Vitev:2005he}, PQM \cite{Dainese:2004te}, and ASW \cite{Eskola:2004cr}.  For the latter two, rw indicates the use of \eq{rw} for overabsorption; otherwise \eq{norw} was employed.  Also for the latter two, $\eqnqhat = 100$ and $\eqnqhat = 68$ were the values of the ASW input parameter, respectively.  Notice the sharp rise in \pt for the WHDG and GLV curves as opposed to the flatness of the ASW and PQM ASW results.  For heavy quark \raapt predictions from WHDG see Chapter \ref{chapter:pqcdvsadscft}.}
\label{lhcraa}
\efig 

\section{Conclusions}
While the results presented in this paper are encouraging, further improvements of the jet quenching theory will be required before stronger conclusions can be drawn. More work is needed to sort out coherence and correlation effects between elastic and inelastic processes that occur in a finite time and with multiple collisions. Classical electrodynamics calculations presented in \cite{Peigne:2005rk} suggested that radiative and elastic processes could destructively interfere over lengths far longer than previously thought.  However, the formalism used does not disentangle known radiative Ter-Mikayelin effects \cite{Djordjevic:2003be} from expected finite size collisional effects.  More important, the authors inconsistently treat their current as conserved (using $j^0=kj_L/\omega$) while simultaneously dropping the backwards-moving partner parton term from the current (thus violating current conservation by producing, at a finite time, a charge without a corresponding opposite charge); this results in a spurious subtraction of the (negative) binding energy of the pair and a correspondingly drastic reduction in the elastic energy loss.  As seen in \cite{Adil:2006ei}, a proper accounting of the current shows finite size effects persist out only to the expected lengths of order the screening scale, $1/\mu_D \lsim 1$ fm.  Recent work on the quantum mechanical treatment of elastic loss in a finite-sized medium \cite{Djordjevic:2006tw,Wang:2006qr} also concluded that these effects are small.  At the same time, the authors of \cite{Djordjevic:2006tw,Wang:2006qr} disagree on the effects of interference between the radiative and elastic contributions; a solution of this problem may involve the proper inclusion of transition radiation \cite{Djordjevic:2005nh}.

In addition, the
radiative and elastic energy losses depend sensitively on the coupling,
$\Delta E^{rad}\propto\alpha_S^3$ and $\Delta
E^{el}\propto\alpha_S^2$. Future calculations
will have to relax the fixed $\alpha_S$ approximation
and allow it to run. The energy loss involves integrals that probe momentum scales that are
certainly nonperturbative. Therefore it will be important to study the
irreducible uncertainty associated with the 
different maximum $\alpha_S$ cutoff prescriptions commonly used.
Further, improved numerical techniques will have to be devised
to bypass the approximations  employed here to incorporate
essential path length fluctuations. 
Nevertheless qualitatively different high momentum behaviors are predicted for \pizero suppression at the \lhc from the BDMPS-based ASW models and the GLV radiative and DGLV radiative convolved with elastic loss models: for the former \raapt is approximately flat in \ptcomma; for the latter it increases significantly with \ptcomma.  From an experimental
perspective, direct measurement of $D$ spectra would be valuable to
separate the different bottom and charm jet quark dynamics
that is at present convoluted in the electron spectra.

\clearpage
\lhead{Chapter \arabic{chapter}: pQCD vs.\ AdS/CFT}
\mychapter{pQCD vs.\ AdS/CFT Tested by Heavy Quark Energy Loss}{chapter:pqcdvsadscft}

\section{Introduction}
Recent discoveries
 at \rhic \cite{Adcox:2004mh,Gyulassy:2004zy,Riordan:2006df,BNLPress}
have led to suggestions \cite{Kovtun:2004de,Kovtun:2006pf,Nakamura:2006ih,Sin:2006pv,Janik:2005zt,Janik:2006gp,Kajantie:2006ya,Friess:2006kw,Shuryak:2005ia,Heller:2007qt} that
the properties of strongly coupled 
quark gluon plasmas (sQGP) produced in ultra-relativistic
nuclear collisions 
may be better approximated by
string theoretic models inspired by the AdS/CFT gravity-gauge theory
correspondence \cite{Maldacena:1997re,Witten:1998qj,Witten:1998zw} 
than conventional Standard Model perturbative QCD (pQCD). 
Four  main classes of observables have attacted
the most attention: (1) Entropy production
as probed by multiplicity distributions \cite{Back:2001ae}, 
(2) ``Perfect'' Fluidity \cite{Riordan:2006df} as 
probed by collective elliptic flow
observables \cite{Ackermann:2000tr,Adler:2001nb,Adler:2002ct,Adare:2006ti,Teaney:2000cw,Huovinen:2001cy,Molnar:2001ux,Teaney:2003kp}, 
(3) Jet Quenching and Tomography as probed by \highpt
hadrons \cite{Adcox:2001jp,Adler:2006hu,Wang:1991xy,Vitev:2002pf,Gyulassy:2003mc} and nonphotonic leptons
\cite{Abelev:2006db,Adler:2005xv,Adare:2006nq}, 
and (4) Dijet-Bulk Correlations
as probed by two and three particle correlations \cite{Adler:2002tq,Abelev:2006jr,Adams:2006yt,Wang:2005cg,Ulery:2006iw,
Adler:2005ee,Adare:2006nr}. 

\begin{figure}[!tbh]
\centering
\includegraphics[width=4.25 in]{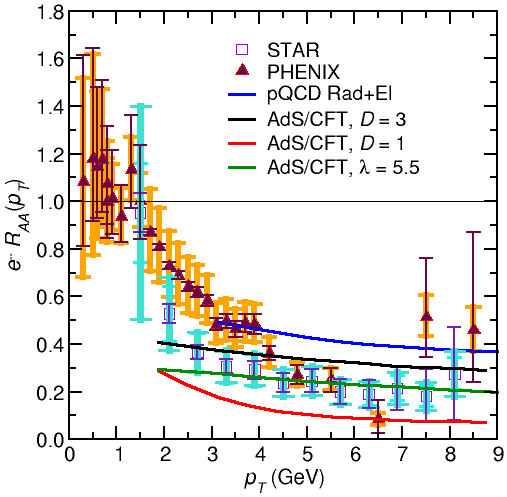}
\caption{\label{pqcd:elec}
Nonphotonic electron predictions from the AdS/CFT drag and pQCD WHDG \cite{Wicks:2005gt} models compared to \rhic data from \phenix \cite{Adler:2005xv,Akiba:2005bs} and \star \cite{Bielcik:2005wu,Dong:2005nm} in central 200 GeV $Au+Au$ collisions.  For a density of $dN_g/dy=1000$ the perturbative estimate is in qualitative disagreement; the drag calculation is in agreement in its range of reasonable choices of input parameters.  See text for details of the drag calculation; the electron fragmentation was done as in \cite{Wicks:2005gt}
}
\end{figure}

Qualitative  successes of recent \ads applications \cite{Kovtun:2004de,Kovtun:2006pf,Nakamura:2006ih,Sin:2006pv,Janik:2005zt,Janik:2006gp,Kajantie:2006ya,Friess:2006kw,Shuryak:2005ia,Heller:2007qt}
to nuclear collision phenomenology 
include the analytic account for (1) 
the surprisingly small ($\sim$ 3/4) drop \cite{Gubser:1996de,Klebanov:2005mh}
 of the entropy density in lattice QCD calculations
relative to Stefan-Boltzmann, 
(2) the order of magnitude reduction of the viscosity to entropy ratio
$\eta/s$ predicted relative to pQCD needed
to explain the seemingly near perfect
fluid flow of the sQGP observed at \rhiccomma, (3) the 
unexpectedly large stopping power of high transverse momenta
heavy quarks as inferred from heavy quark jet 
quenching, see \fig{pqcd:elec}, and elliptic flow, and (4) the possible occurrence of 
conical ``Mach'' wave-like correlations
of hadrons associated with tagged jets.

While quantitative and systematic comparisons of \ads gravity dual models
with  nuclear collision data are still incomplete  and 
while the conjectured
{\em double} %
Type IIB string theory$\leftrightarrow$conformal
 Supersymmetric Yang-Mills (SYM) gauge theory$\leftrightarrow$non-conformal, non-supersymmetric QCD correspondence
remains under debate (see,
e.g. \cite{McLerran:2007hz}),
the current successes provide strong  motivation
to seek more sensitive experimental tests 
that could help guide the theoretical development
of novel theoretical approaches that may be needed to explain 
recent \rhic and soon \lhc heavy ion data.

The aim of this chapter is twofold.  First is to propose a robust observable that can 
more readily reveal the kinematic boundaries and reaction conditions
where specific Standard Model  weak coupling pQCD approximations may fail
and where specific strong coupling AdS/CFT approximations may fail (or if they are applicable at all). 
Asymptotic freedom and the factorization theorems of QCD ensure that pQCD
should apply above some hard scale $Q(A,\surd s)$ that depends in general on both
atomic number, $A$, and center of mass energy, $\surd s$.
For jet production this scale is the greater of the gluon saturation scale $Q_{s}^2\sim Q_0^2 \log(A\surd s/p_T)
$, that defines a scale below which strong
{\em initial state} nuclear modification of the 
 parton distributions must be taken into account,
and $\hat{q}L \equiv \int dz d\sigma \rho(z) q^2 
\propto Q_{s}^2 \log^n(Q_s)$, below which strong
{\em final state} nuclear modifications 
 must be taken into account. For finite $A<238$ and finite
$\surd s <10$ ATeV the numerical values of these scales remains 
uncertain. 
Second is to suggest this observable will discriminate between current pQCD and  AdS/CFT dynamical models.
 
We start with the predicted nuclear modification, $R_{AA}^Q(\eqnpt)$,
of  the transverse momentum distribution
of identified heavy quark jets 
produced in central Pb+Pb reactions at 5.5 ATeV at \lhccomma.  
Specifically, we
propose that the double ratio of identified charm and bottom jet
nuclear modification factors $R_{AA}^c/R_{AA}^b$ is a remarkably 
robust observable that can distinguish between a rather 
wide class of pQCD energy loss mechanisms 
and a recently proposed class of gravity ``drag''
models \cite{Karch:2002sh,Herzog:2006gh,Gubser:2006bz,
Gubser:2006nz,Gubser:2007nd} of heavy quark dynamics.  
Similar tests can be performed at \rhic when identified jet flavor
detector upgrades
are completed. The main advantage of \lhc in comparison to \rhic is of course the much higher $p_T$ kinematical range that will be accessible. The main advantage of \rhic is a better control of the initial state saturation physics because
of extensive $d+A$ and $p+p$ control data at the same $\surd s$. 

The current failure of
pQCD based energy loss models \cite{Dokshitzer:2001zm,Djordjevic:2005db,
Wicks:2005gt,Armesto:2005mz} to account \underline{\em quantitatively} for the recent \rhic data from STAR
\cite{Abelev:2006db} and \phenix \cite{Adler:2005xv,Adare:2006nq} on the
nonphotonic electron spectrum
provides additional motivation to focus on heavy quark jet observables.
Unlike for 
light quark and gluon jet observables, where pQCD predictions
were found to be remarkably quantitative
\cite{Wang:1991xy,Vitev:2002pf,Gyulassy:2003mc}, 
heavy quark jet quenching, especially as inferred indirectly for bottom quarks,
appears to be significantly underpredicted \cite{Dokshitzer:2001zm,Djordjevic:2005db,
Wicks:2005gt,Armesto:2005mz}. Current issues that cloud the pQCD based energy loss predictions are
(1) the 
uncertainty in
the initial state nuclear production of bottom to charm
quarks, (2) the current controversy
over the relative magnitude of elastic versus radiative loss channels
\cite{Wicks:2005gt,Armesto:2005mz}, and (3)
the possibility that short formation time
nonperturbative hadronization effects may have to be taken into account
\cite{Adil:2006ra,vanHees:2005wb}.

The AdS/CFT correspondence has so far been applied to heavy ion jet physics 
in three ways. The first involves 
the calculation of the QCD Wilson line correlator that corresponds to
 the radiative transport coefficient \qhat \cite{Liu:2006ug,
Armesto:2003jh}. The second concentrates on estimating 
the heavy quark diffusion
coefficient $D$ \cite{CasalderreySolana:2006rq} that is an input 
to a Langevin model of drag 
\cite{Moore:2004tg}. The third is a prediction
of the heavy quark drag
coefficient based on the gravity dual dynamics of a 
classical string in an AdS black brane background \cite{Herzog:2006gh,Gubser:2006bz,Gubser:2006nz}.  
All three approaches remain under active debate (see, e.g. \cite{Gubser:2006nz,Liu:2006he,Bertoldi:2007sf}).

We focus in this chapter on the third proposed AdS/CFT application
that involves the most direct 
string theoretic inspired gravity ``realization'' 
of heavy quark dynamics \cite{Karch:2002sh,Herzog:2006gh,Gubser:2006bz}.
A heavy quark in the fundamental representation
is a bent Nambu-Goto string with 
one end attached to a probe brane 
and that trails back above the horizon
of a D3 black brane representing the 
uniform strongly coupled SYM plasma heat bath. 
This geometry
maps the drag force problem
into a modern string theoretic  
version of the 1696 Brachistochrone
problem and yields a remarkable, simple analytic solution for the string shape and momentum loss
per unit time.

\section{AdS/CFT compared to pQCD}\label{pqcdvsadscft}
Exploiting the AdS/CFT correspondence, the drag coefficient for
a massive quark moving through a strongly-coupled SYM plasma in the $\lambda =\eqngym^2N_c\gg 1$, $N_c\gg 1$, $M_Q\gg T^*$
limit is given in \cite{Herzog:2006gh,Gubser:2006bz,Gubser:2006nz} 
as  
\be
\label{mu}
\frac{d\eqnpt}{dt} = -\mu_Q \eqnpt = -
\frac{\pi\sqrt{\lambda}(T^*)^2}{2M_Q}\eqnpt, \ee
where $T^*$ is the temperature of the SYM plasma as fixed by the Hawking temperature of the dual D3 black brane.  This is parametrically quite similar to the Bethe-Heitler limit of incoherent radiative energy loss, in which case
\be
\frac{d\eqnpt}{dt} \sim -\frac{T^3}{M^2}\eqnpt,
\ee
differing from \eq{mu} only by a factor of $T/M$, and quite dissimilar to the usual LPM-dominated asymptotic pQCD result,
\be
\frac{d\eqnpt}{dt} \sim -LT^3 \log (\eqnpt/M).
\ee

There exists a  maximum momentum, or $\gamma_{c}\approx p_T^{crit}/M_Q$
beyond which \eq{mu} cannot be applied.
 Self consistency within the classical string picture 
requires a time-like boundary for the string worldsheet
\cite{Gubser:2006nz}. For constant velocity this limits
heavy quark ``speeds'' to $\gamma < \gamma_c$, where \be
\label{speedlimit}
\gamma_c=\left(1+\frac{2M}{\sqrt{\lambda}T^*}\right)^2\approx\frac{4M^2}{\lambda(T^*)^{2}}.
\ee 

The work of \cite{Herzog:2006gh} relaxed the assumptions of infinite quark mass and constant velocity.  The analytic form of \eq{mu} was found to well reproduce the full numerical results (most importantly $\mu_Q$ remained independent of \ptcomma) but with $M_Q$ no longer the bare quark rest mass.  All the calculations in this paper were based upon the infinite bare mass approximation of \cite{Gubser:2006bz,Gubser:2006nz} with $M_Q$ replaced with realistic quark masses.

We note that one may arrive at this same value through a different line of reasoning.  Assume that the quark's constant velocity is maintained by an electric field.  The largest electric field sustainable from the Born-Infeld action limits the magnitude of the momentum loss; the critical speed after which the field cannot be strong enough to keep the quark velocity constant is given by \eq{speedlimit} \cite{CasalderreySolana:2007qw}.  

While in the infinitely strongly coupled plasma dual
the quasiparticle picture is not applicable, a similar ``speed limit'' 
arises for ordinary incoherent Bethe-Heitler (BH) radiative energy loss (See Appendix \ref{BHtoLPMapp}).  
Landau-Pomeranchuck coherence effects invalidate 
the linear in $E\approx p_T$ rise of the energy loss when 
the formation time, $\tau_E \sim E/M^2$, exceeds the mean free path, 
$\bar{\lambda} =1/\rho\sigma \sim 1/\eqnalphas T$ 
(in a Debye-screened ultrarelativistic plasma).  Just as in the form of loss itself the requirement that $\tau_E < \bar{\lambda}$ therefore limits the applicability of BH to $\gamma<\gamma^{BH}_c = M_Q/\eqnalphas T$, similar to the AdS/CFT speed limit, \eq{speedlimit}, 
but with one less power of $T^*/M_Q$.  

The speed limit \eq{speedlimit} was found assuming a constant plasma temperature; this is not the case in experiment.  Nevertheless, to get a sense of
 the \pt scale where the \ads approximation could break down, we will plot
 the momentum cutoffs from \eq{speedlimit} for different
 SYM input parameters with two different assumptions for the mapping of QCD to \ads parameters, described in detail below.  There is still additional ambiguity in the $\gamma_c$ due to the time evolution of the QGP temperature.  The smallest $\gamma_c$ corresponds to the largest temperature; we take as the generous lower bound the extreme $T(\vec{x}=\vec{0};\tau=\tau_0)$, shown as a `(' in the figures.  For the largest $\gamma_c$ we use $T_c$, show as a ``].''  To further emphasize the possibility of corrections in the drag model we gradually fade the curves from the ``('' to the ``].''  Surprisingly the introduction of a thermal plasma in \ads results in an effective mass \emph{smaller} than the bare mass; this will result in a reduction of the momentum reach of the drag formalism.

\bfig[!htb]
\centering
\includegraphics[width=.9 \textwidth]{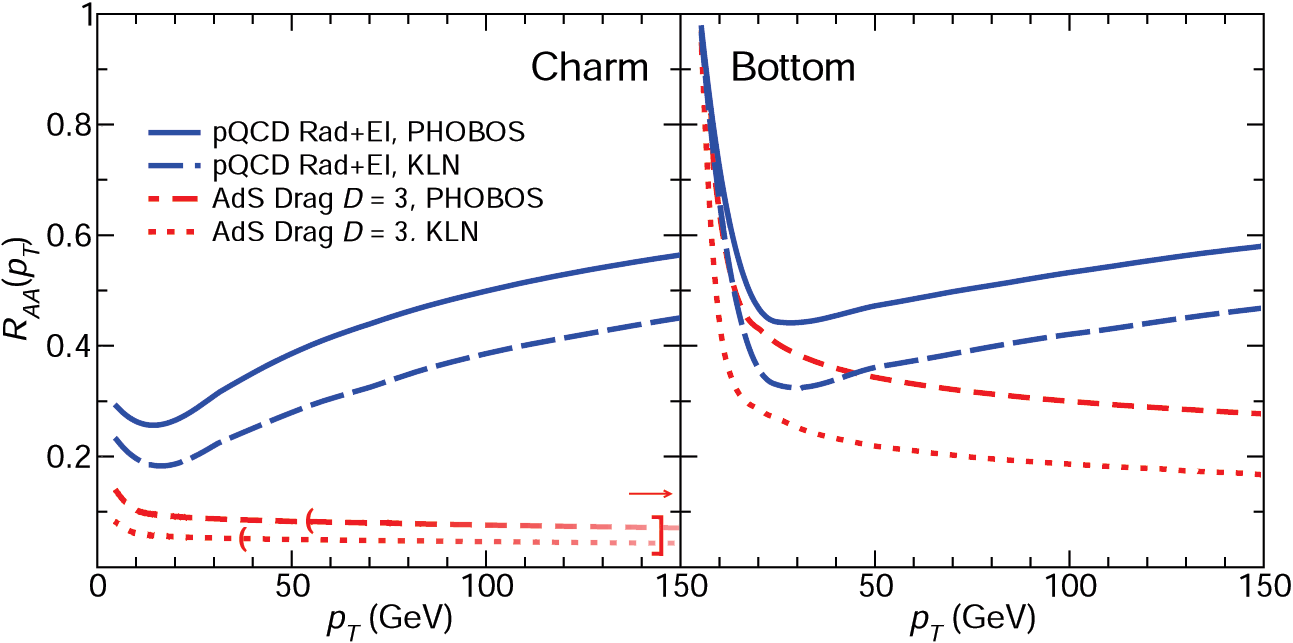}
\caption{\label{LHCcandbRAA}\captionsize{
\raacpt and \raabpt predicted for central $Pb+Pb$ at \lhc comparing \ads
\eq{mu} and pQCD using the WHDG model \cite{Wicks:2005gt}
convolving elastic and inelastic parton energy loss.  Possible initial gluon rapidity densities at \lhc 
are given by $dN_g/dy=1750$, from a \phobos \cite{Back:2004je,Busza:2007ke} extrapolation, or $dN_g/dy=2900$, from the KLN model of the color glass condensate (CGC) \cite{Kharzeev:2004if}.
The top two curves from pQCD increase with \pt while the bottom two curves from \ads slowly decrease with \ptcomma.  The \ads parameters here were found using the ``obvious'' prescription 
with $\eqnasym=.05$, $\tau_0=1$ fm/c, giving $D=3/2\pi T$ (abbreviated to $D=3$ in the figure). 
Similar trends were seen for the other input parameter possibilities discussed
in the text.  The ``('' and ``]'' denote momenta after which possible string theoretic corrections may need to be considered; the curves' increasing transparency from ``('' to ``]'' is meant to additionally emphasize this, see text.
}}  \efig

Applying \eq{mu} to heavy ion collisions requires an additional assumption
about how to map the QCD temperature and coupling to the gedanken SYM world
and its SUGRA dual.  The ``obvious''  first prescription \cite{Gubser:2006qh} takes
$\eqngym=\eqngs$ constant, $T^{*}=T^{QCD}$, and
$N_c=3$.
However it was suggested in \cite{Gubser:2006qh} that
a more physical ``alternative'' might be to equate energy
densities, giving $T^{*}\simeq T^{QCD}/3^{1/4}$, and
to fit the coupling $\lambda=\eqngym^2N_c\approx5.5$ in order to reproduce
the static quark-antiquark forces calculated via lattice QCD.

The string theoretic result for the diffusion coefficient
used in the Langevin model is $D=2/\sqrt{\lambda}\pi T^*$ \cite{CasalderreySolana:2006rq}.  This illustrates well the problem of connecting the $T^*$ and $\lambda$ of SYM
to ``our'' QCD world. Using the
``obvious'' prescription with $\eqnalphas=.3$, $N_c=3$, one finds
$D\sim1.2/2\pi T$.  However, $D=3/2\pi T$ was claimed in
\cite{Adare:2006nq,CasalderreySolana:2006rq} to fit \phenix data somewhat better. Note that
$D=3/2\pi T$
requires an unnaturally small $\eqnalphas\sim 0.05$ that is very far 
from the assumed $\lambda \gg 1$ 't Hooft limit. 

We proceed by computing the charm and bottom nuclear modification factors,
neglecting initial state shadowing or saturation effects.
In order to correctly 
deconvolute such effects from the final state effects
that we compute below, it will be necessary to measure nuclear
modification factors in $p+A$ as a function of 
$(y,\eqnpt)$ at \lhc just as $d+A$ was the critical
control experiment \cite{Adcox:2004mh} at \rhic \cite{Gyulassy:2004zy}.

Final state suppression of \highpt jets due to a fractional energy loss
$\epsilon$, $\eqnptf=(1-\epsilon)\eqnpti$, can be computed knowing 
the $Q$-flavor dependent spectral indices
$n_Q+1=-\frac{d}{d\log \eqnpt}\log\left(\frac{d\sigma_Q}{dyd\eqnpt}\right)$
from pQCD or directly from $p+p\rightarrow Q+X$ data. 
The nuclear modification factor is then
$R_{AA}^Q(\eqnpt)=<(1-\epsilon)^{n_Q}>$, where the average
is over the distribution $P(\epsilon;M_Q,\eqnpt,\ell)$ that depends
in general on the quark mass, \ptcomma, and the path length $\ell$ of the
jet through the sQGP. As in \cite{Wicks:2005gt} we find $n_Q$ from FONLL production cross sections \cite{Cacciari:2005rk,Mangano:1991jk}, average over jets produced according to the binary distribution
geometry, and compute $\ell$ through a participant
transverse density distribution taking into account the nuclear diffuseness.
Given $dN/dy$ of midrapidity partons, the temperature is computed
assuming isentropic Bjorken $1D$ Hubble flow (see Appendix \ref{appendix:whdg}) with $N_f=3$.  
As emphasized in \cite{Wicks:2005gt}, detailed geometric path length averaging
plays a crucial role in allowing consistency between $\pi^0,\eta$ and heavy quark
quenching in pQCD.

For \ads drag, \eq{mu} gives the average fractional
energy loss as $\bar{\epsilon}=1-\exp(-\mu_Q\ell)$.  Energy loss 
is assumed to start at thermalization, $\tau_0\sim0.6-1.0$ fm/c, and stops when the
confinement  temperature, $T_c\sim160$ MeV, is reached.  The exponentiated $T^2$ dependence in $\mu_Q$ leads to a significant sensitivity to the opacity of the medium, as well as to $\tau_0$ and $T_c$.

To understand the generic qualitative 
features of our numerical results
it is instructive to consider the simplest case of a geometric
path average over a static, finite, uniform plasma of thickness $L$; then
\be
R_{AA}^Q(\eqnpt)=\frac{1-e^{n_Q\mu_Q L}}{n_Q \mu_Q L}\approx\frac{1}{n_Q \mu_Q L},
\label{aprx}
\ee
where the \pt dependence is carried entirely by the spectral index $n_Q(\eqnpt)$.

\bfig[!htb]
\centering
\includegraphics[width=.95 \textwidth]{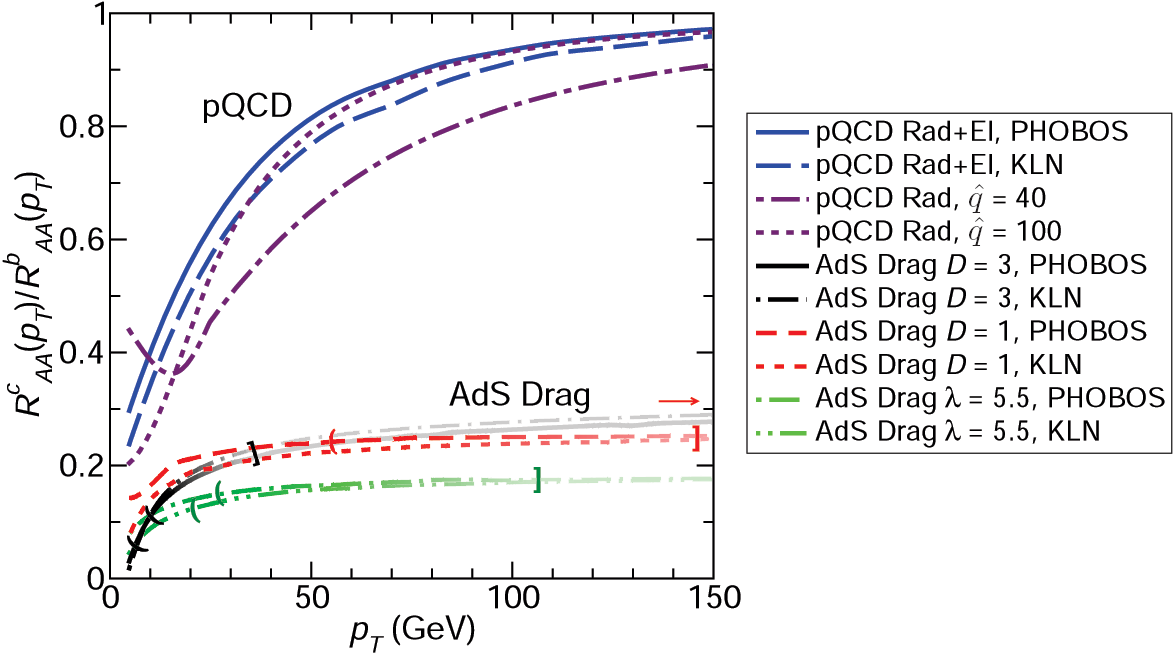}
\caption{\label{Ratio}\captionsize{The double ratio
of \raacpt to \raabpt predictions for \lhc using \eq{mu} for
\ads and WHDG \cite{Wicks:2005gt} for pQCD with a wide range
of input parameters. The generic difference between the pQCD results
tending to unity contrasted to
the much smaller and nearly \ptcomma-independent results from \ads
can be easily distinguished at \lhccomma.  The ``('' and ``]'' denote momenta after which possible string theoretic corrections may need to be considered; the curves' increasing transparency from ``('' to ``]'' is meant to additionally emphasize this, see text.}}
\efig

Two implementations of pQCD energy loss are used in this paper.  The
first is the full WHDG model convolving fluctuating
 elastic and inelastic loss with fluctuating path geometry
\cite{Wicks:2005gt}.  The second restricts WHDG to include
only radiative loss in order to facilitate comparison to 
\cite{Armesto:2003jh}.
Note that 
when
realistic nuclear geometries with Bjorken expansion are used, the
``fragility'' of \raa for large \qhat reported in \cite{Eskola:2004cr}
is absent in both implementations of WHDG. 

\begin{figure}[htb!]
\centering
\includegraphics[viewport = 20 25 423 416, clip = true, width=3 in]{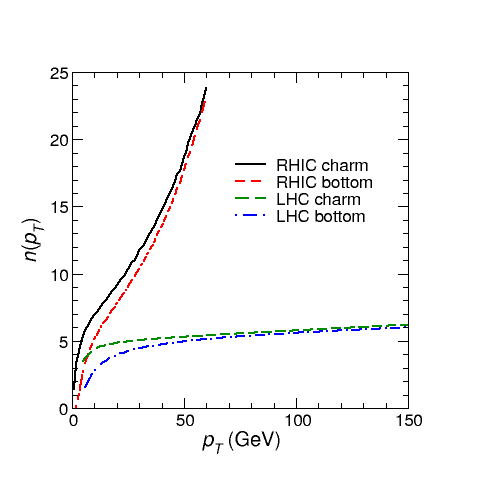}
\caption{
\label{prodindex} The power law production index $n_Q(\eqnpt)$ for \rhic and \lhccomma.  While it is quite flat--but slowly increasing--over a large momentum range at \lhccomma, at \rhic $n_Q(\eqnpt)$ hardens appreciably as momentum increases.
}
\end{figure}

Unlike the \ads dynamics, pQCD predicts \cite{Dokshitzer:2001zm,
Djordjevic:2005db,Wicks:2005gt} 
that the average energy loss fraction
in a static uniform plasma is approximately
$\bar{\epsilon} \approx 
\kappa L^2 \hat{q}
 \log(\eqnpt/M_Q)/\eqnpt$, 
with $\kappa$ a proportionality constant and $\hat{q}=\mu_D^2/\lambda_g$.
The most important feature in pQCD relative to \ads is that
$\bar{\epsilon}_{pQCD}\rightarrow 0$ asymptotically at \highpt while
$\bar{\epsilon}_{AdS}$ remains constant.  $n_Q(\eqnpt)$
is a slowly increasing function of momentum, \fig{prodindex}; thus $R_{AA}^{pQCD}$
increases with \pt whereas $R_{AA}^{AdS}$ decreases.  This generic difference can be observed in
\fig{LHCcandbRAA}, which shows representative predictions from the full
numerical calculations of charm and bottom \raapt at \lhccomma.

\section{Double Ratio of charm to bottom \texorpdfstring{$R_{AA}^Q$}{RAA}} 
A disadvantage of the $R_{AA}^Q(\eqnpt)$ observable alone
is that its normalization and slow 
\pt dependence can be fit with different model assumptions
compensated by using very different medium parameters. 
In particular, high value extrapolations
of the \qhat parameter
proposed in \cite{Armesto:2005mz} could simulate
the flat $p_T$-independent prediction from AdS/CFT. 

We propose to use the double ratio of charm to bottom \raa to amplify
the observable difference between the mass and \pt dependencies of the
\ads drag and pQCD-inspired energy loss models.  One can see in
\fig{Ratio} that not only are most overall normalization differences
canceled, but also that the curves remarkably bunch to either AdS/CFT-like 
or pQCD-like generic
results regardless of the input parameters used.

The numerical value of $R^{cb}$ shown in \fig{Ratio} for \ads
 can be roughly understood analytically from
\eq{aprx}
as, 
\be
\label{AdSRcb}
R^{cb}_{AdS} 
\approx \frac{M_c}{M_b} \frac{n^b(\eqnpt)}{n^c(\eqnpt)}
\approx\frac{M_c}{M_b}\approx 0.26,
\ee
where in this approximation all $\lambda$, $T^*$, $L$, and 
$n_c(\eqnpt)\approx n_b(\eqnpt)$ dependences drop out. 

The pQCD trend in \fig{Ratio} can be understood qualitatively
from the expected behavior of
$\bar{\epsilon}_{pQCD}$ noted above
giving (with $n_c\approx n_b=n$)
\be
\label{pQCDratio}
R^{cb}_{pQCD} \approx 1 - \frac{p_{cb}}{\eqnpt},
\ee
where $p_{cb}=\kappa
n(\eqnpt) L^2 \log(M_b/M_c) \eqnqhat$ sets the relevant
momentum scale.  Thus $R^{cb}\rightarrow1$ more slowly for higher opacity.
One can see this behavior reflected in the full numerical results shown in \fig{Ratio} for moderate suppression, but that the extreme opacity
 $\eqnqhat=100$ case deviates from \eq{pQCDratio}.  
 
Supposing that \lhc data are similar to the pQCD predictions one might be able to distinguish between convolved elastic and inelastic loss, for which $R^{cb}$ monotonically increases, and purely radiative energy loss, for which $R^{cb}$ dips to a minimum near $\eqnpt \sim 10$ GeV; one must be cautious, though, as many of the assumptions in the elastic energy loss derivations break down for $\eqnpt\not\gg M_Q$.

\begin{figure}[htb!]
\centering
\includegraphics[viewport = 2 0 950 466, clip = true, width=.95 \textwidth]{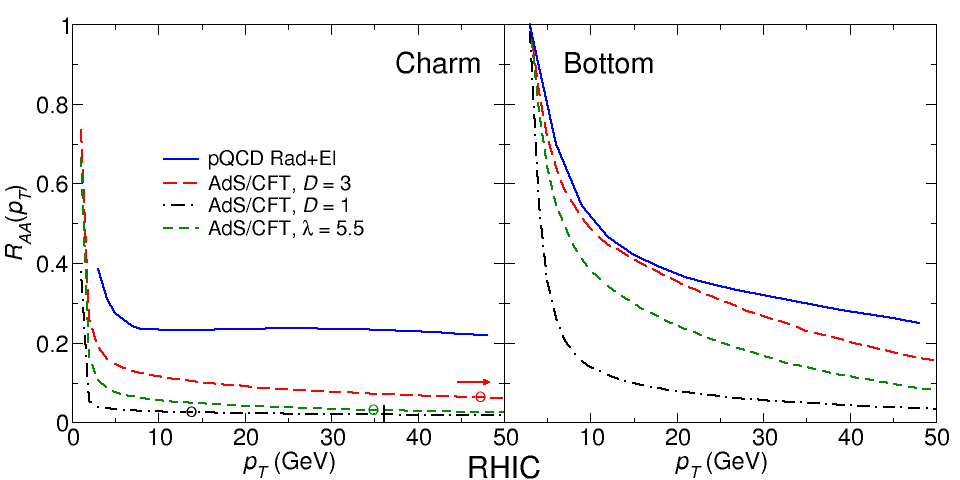}
\caption{
\label{rhiccandb} \raacpt and \raabpt for \rhic using \ads drag and the pQCD-based WHDG model.  Note how the rapid increase in $n_Q$ for \rhic as seen in \fig{prodindex} overcomes the decrease in the fractional energy loss for the pQCD predictions for charm and bottom \raapt so that \emph{both} \ads drag and pQCD results decrease as a function of momentum at \rhiccomma.}
\end{figure}

\begin{figure}[htb!]
\centering
\includegraphics[width=5.4 in]{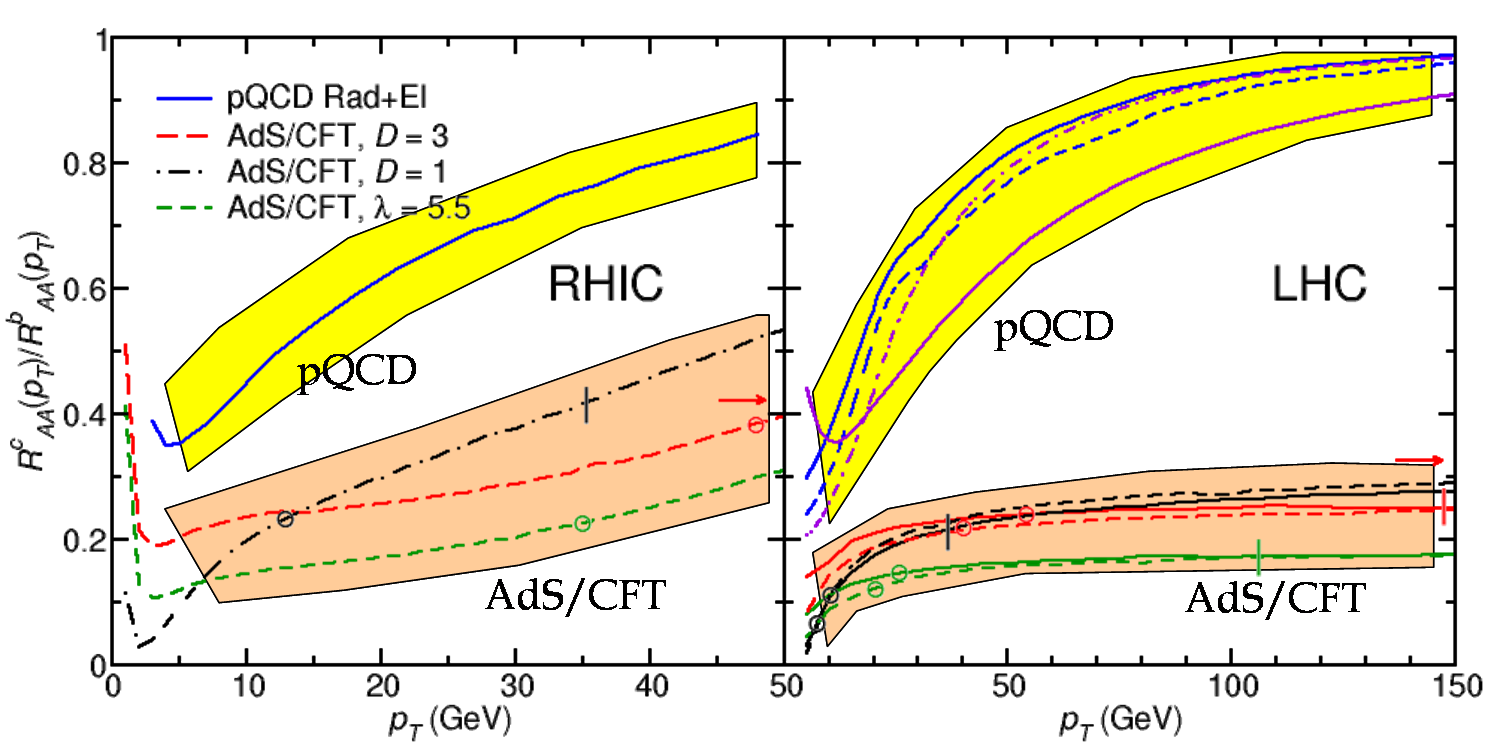}
\caption{
\label{rhicratio}
The double ratio of \raacpt to \raabpt predictions for \rhic using \eq{mu} for \ads and WHDG \cite{Wicks:2005gt} for pQCD with a range of input parameters.  While the hardening of the production spectrum reduces the dramatic bunching at \rhic as compared to \lhccomma, the lower temperature at \rhic means the \ads drag formalism is applicable to higher momenta.  Note that $R^{cb}$ is plotted to only 50 GeV for \rhiccomma.
}
\end{figure}

Future \rhic detector upgrades will allow for individual charm and bottom quark detection.  Predictions for \raacpt and \raabpt at \rhic from \eq{mu} and pQCD are shown in \fig{rhiccandb}.  As seen in \fig{prodindex} the power law production index grows quickly at \rhiccomma; we now expect the rapid increase in $n_Q$ to overcome the (relatively in comparison) slow decrease in $\epsilon_{pQCD}$ so that, unlike at \lhccomma, $dR_{AA}/d\eqnpt \ngtr 0$. In fact \fig{rhiccandb} shows that the full numerical results for $R_{AA}^Q$ from pQCD and \ads drag \emph{both} decrease with \ptcomma.  Nonetheless one may still examine the double ratio \rcbcomma, \fig{rhicratio}.  While the larger index makes the grouping less dramatic at \rhic one may still differentiate between pQCD and \ads drag.  Due to its smaller multiplicities, the temperature of the medium at \rhic is smaller than will be seen at \lhccomma; hence the \ads drag ``speed limit'', \eq{speedlimit}, is higher at \rhic than \lhccomma.  

\section{Conclusions}
Possible strong coupling alternatives to pQCD in nuclear collisions were studied based on a recent \ads  model of charm and bottom energy loss.  The predicted nuclear modification factors, $R_{AA}^Q$, were found to be decreasing as a function of \ptcomma, as compared to increasing as predicted from pQCD.  We showed that the momentum dependence differences in the individual $R_{AA}^Q$ can be masked by taking extreme energy loss extrapolations to \lhccomma.  However the double ratio \Rcb revealed very generic behavior, insensitive to the input parameters and radically different for the two coupling limits.  Of crucial importance is the momentum range over which pQCD and \ads drag are self-consistent.  Certainly pQCD must apply for $\eqnpt\rightarrow\infinity$, but the scale below which nonperturbative effects become important is not yet well understood.  Supposing that the AdS/CFT correspondence is relevant for heavy ion collisions, drag calculation momentum validity is limited from above.  Since $\gamma_c$ depends on $M_Q$ but the hard pQCD scale does not there is likely a large region of \pt for which both approximations are applicable for bottom quarks.  The disadvantage of studying the more robust \Rcb observable is that the AdS momentum reach is limited by the much smaller charm quark mass; an overlapping momentum region of validity for both coupling limits may not exist for this observable.  Further careful study of the ``speed limit'' and higher order corrections to the \ads result must be done in order to fruitfully compare $R_{AA}^c$ and \Rcb results to experiment.  Or, turning this around, experiment might inform theory as to the correct scales for which these theoretical predictions give reliable results.

\clearpage
\lhead{Chapter \arabic{chapter}: \leftmark}
\mychapter{Heavy Quark Photon Production Radiation}{HQPB}
\section{Introduction}\label{intro}

The confidence in the application of perturbative QCD methods to jet energy loss in heavy ion collisions gained from the early quantitative understanding of $\pi$ and $\eta$ suppression with null direct-$\gamma$ control \cite{Akiba:2005bs,Vitev:2002pf} has recently come into serious doubt \cite{ColeTalk}.  Evidence from measurements such as \highpt correlations \cite{Ackermann:2000tr,Winter:2006iz} and nonphotonic electrons \cite{Adare:2006nq,Abelev:2006db} demonstrates clear disagreement with perturbative models \cite{Shuryak:2001me,Horowitz:2005ja,Djordjevic:2005db,Wicks:2005gt,Armesto:2005mz}.  Several papers postulate alternative nonperturbative energy loss mechanisms \cite{vanHees:2005wb,Adil:2006ra,Gubser:2006bz,Herzog:2006gh}, and a new measurement, the double ratio of charm to bottom nuclear modification factors, has been suggested as a robust observable for testing some of these novel ideas \cite{Horowitz:2007su}.  Electromagnetic radiation from quark jets, as it is transparent to the partonic medium, holds enormous promise as a new tool for investigating jet energy loss mechanisms; naively one expects the spectrum of photons emitted from a jet that underwent a smooth, exponential slowdown will differ greatly from one that suffered the emission of a few hard gluons will differ from one that experienced a large number of soft scatterings.  Nascent data of $\gamma$-hadron correlations from $p+p$ collisions additionally motivates the theoretical exploration of photon bremsstrahlung in heavy ion collisions \cite{Hanks:2007qx}.  

In this paper we calculate as a warmup problem, and ultimately as an interesting problem in its own right, the \zeroth order in opacity energy loss of a heavy quark jet, the radiation associated with the production of a hard parton.  We will generalize the problem of a zero mass quark emitting a zero mass photon to a massive quark emitting a massive photon.  The lack of theoretical consistency in the understanding of light and heavy flavor jet suppression makes massive quark calculations of especial interest \cite{Horowitz:2005ja,Djordjevic:2005db,Wicks:2005gt}; moreover heavy quark predictions from pQCD will be necessary for comparison to AdS/CFT heavy quark drag results \cite{Horowitz:2007su}.  
For the case of \zeroth order emission QCD and QED are identical but for the replacement of $\alpha_{EM}$ with $\alpha_s$ and a color Casimir.  Using a massive photon will allow a comparison to already published results on the QCD Ter-Mikayelian effect \cite{Djordjevic:2003be}, whose main results were: (1) the Ter-Mikayelian effect leads to a large reduction in \zeroth order energy loss ($\sim30\%$ for charm quarks), (2) the full 1-loop HTL gluon propagator can be well approximated by using a fixed gluon mass $m_g=m_\infinity=\mu/\sqrt{2}$, and (3) the small-$x$, soft gluon, number distribution for \zeroth order in opacity is
\be
\label{magda}
\frac{dN_\mathrm{pQCD}^{(0)}}{d^3k} = \frac{Q^2\alpha}{\pi^2\omega} \frac{\wv{k}^2}{[\wv{k}^2+\mg^2+x^2\mmq^2]^2},
\ee
where, as usual, a bold variable represents a transverse two-vector.

\section{Calculation}
As a first step to compare to previously published results \cite{Djordjevic:2003be} we found the number distribution of emitted photons when simply plugging in massive 4-vectors into a standard classical E\&M calculation:
\bea
\label{eem}
E_\mathrm{EM} & = & \int\frac{d^3k}{(2\pi)^3}\frac{e^2}{2}\sum_{\lambda=1,2}\left|\vec{\epsilon}_\lambda(\vec{k})\cdot\left( \frac{\vec{p}'}{k\cdot p'} - \frac{\vec{p}}{k\cdot p} \right)\right|^2 \\
\label{preem}
& = & \int\frac{d^3k}{(2\pi)^3}\frac{e^2}{2} \left( \frac{2p\cdot p'}{(k\cdot p')(k\cdot p)}-\frac{\mmq^2}{(k\cdot p')^2}-\frac{\mmq^2}{(k\cdot p)^2} \right) \\
\Rightarrow \frac{dN_\mathrm{E\&M}^{(0)}}{d^3k} & = & \frac{Q^2\alpha}{\pi^2\omega} (1-x)^2 \frac{ \wv{k}^2 + (1-x)^2 \mg^2 }{[ \wv{k}^2 + (1-x)^2 \mg^2 + x^2\mmq^2 ]^2}+\mathcal{O}\left(1/E^+\right),\nonumber\\
\label{em}
\eea
where to get from the first to the second line we used the completeness relation $\sum_{\lambda=1,2}\epsilon_\lambda^\mu\epsilon_\lambda^{\nu*}\rightarrow -g_{\mu\nu}$, and where we took
\bea
p & = & [(1-x)E^+,\frac{\mmq^2+\wv{k}^2}{(1-x)E^+},-\wv{k}] \\
p' & = & [\frac{\mmq}{E^+},E^+,0] \\
k & = & [xE^+,\frac{\mg^2+\wv{k}^2}{xE^+},\wv{k}],
\eea
with brackets indicating lightcone coordinates.  

There are two elements seen in \eq{em} and not in  \eq{magda}: several factors of $(1-x)^2$, and an $\mg^2$ in the numerator.  
The first makes no difference in the limit of small $x$; however energy loss calculations integrate over all $x$, and it turns out that neglecting these factors is a large effect.  The second simply cannot be reconciled with \eq{magda}.  Interestingly this extra mass term in the numerator fills in the ``dead cone,'' the region of small angles with respect to the jet axis for which $dN_g/dx\rightarrow0$ in \eq{magda} as $\wv{k}\rightarrow0$ when $\mmq\ne0$; this motivates additional study because naively the dead cone leads to a reduction in heavy quark energy loss, inconsistent with the observation of similar suppression patterns for pions, decay fragments from gluons and light quarks, and nonphotonic electrons, decay fragments from heavy charm and bottom quarks.

One may rightly object that the results of \eq{em} were derived using the usual massless photon E\&M formulae.  Surprisingly the only modification of \eq{eem} when using the Proca Lagrangian is to change the polarization sum to include the longitudinal mode.  It turns out that the extra terms generated by the application of the identity $\sum_{\lambda=1,2,3}\newline \epsilon_\lambda^\mu \epsilon_\lambda^{\nu*} = -g_{\mu\nu} + k_\mu k_\nu/\mg^2$ exactly cancel, and \eq{em} is also valid for massive photon radiation.

\bfig[!ht]
\centering
\includegraphics[width=.75 \columnwidth]{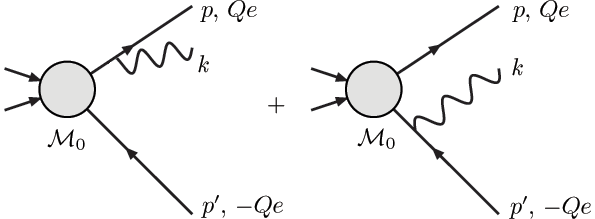}
\caption{\label{diagrams}\captionsize{The two diagrams contributing to the \zeroth order in opacity photon/gluon radiation spectrum.  Note the inclusion of the radiation from the away side jet, which is usually ignored in pQCD calculations.}}
\efig

In order to understand the discrepancy from the field theory perspective, consider the diagrams contributing to the \zeroth order shown in \fig{diagrams}.  Evaluation of these leads to
\be
\label{feyn}
i\mathcal{M} = Qe \bar{u}(p)\left[ \frac{2p\cdot\epsilon^*+\FMslash{\epsilon}^*\FMslash{k}}{2p\cdot k + m_\gamma^2}\mathcal{M}_0 - \mathcal{M}_0 \frac{2p'\cdot\epsilon^*+\FMslash{\epsilon}^*\FMslash{k}}{2p'\cdot k + m_\gamma^2} \right] v(p'),
\ee
where we have taken $\mathcal{M}_0(p+k,p')\approx\mathcal{M}_0(p,p'+k)\approx\mathcal{M}_0$ in the small $x$, soft radiation limit.  Most pQCD calculations ignore the away-side jet; one can easily see that the second term in \eq{feyn}, corresponding to the inclusion of the second diagram in \fig{diagrams}, is crucial for preserving the Ward identity \cite{Wang:1994fx}.  Simultaneously dropping the $\FMslash{k}$ in the numerator and the $m_\gamma$ in the denominator (consistent with the soft photon limit) exactly reproduces the classical Proca result.  We note that assuming $\mathcal{M}_0$ commutes with $\FMslash{\epsilon}^*\FMslash{k}$ and retaining $\mg\ne0$ in the denominator of \eq{feyn} results in a $dN_\mathrm{pQCD}/d^3k$ with leading order identical to \eq{em} but with $(1-x)^2\rightarrow(1-x/2)^2$ as the prefactor of the $\mg$ in the numerator and $(1-x)^2\rightarrow(1-x)$ as the prefactor of the $\mg$ in the denominator. 

\section{Size of Effects}
\begin{figure}[htb!]
\centering
$\begin{array}{c@{\hspace{.05in}}c}
\includegraphics[width=2.65in]{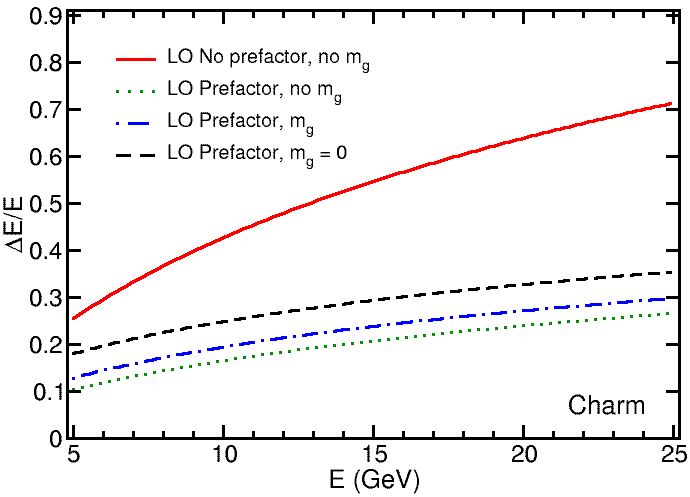} & 
\includegraphics[width=2.65in]{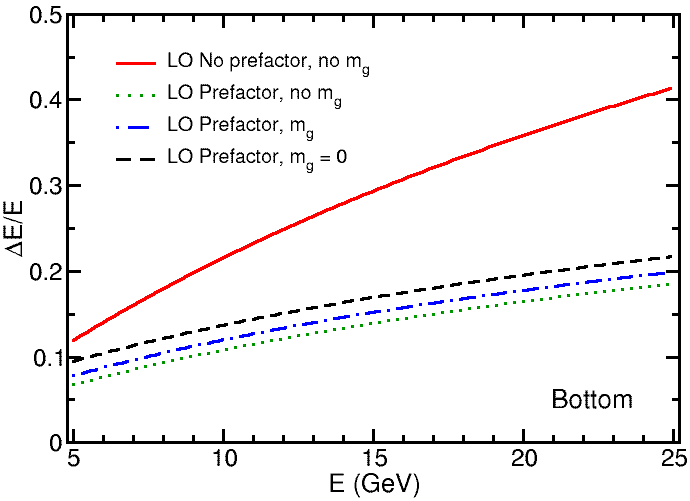} \\ [-.05in]
{\mbox {\scriptsize {\bf (a)}}} & {\mbox {\scriptsize {\bf (b)}}}
\end{array}$
\caption{
\label{effectsone}\captionsize{(Color online) \zeroth order radiative energy loss for (a) charm and (b) bottom quarks.  All results are to leader order (LO) in $1/E^+$.  One sees that the largest effect (50-150\%) comes from including the $(1-x)^2$ prefactor and that filling in the ``dead cone'' with the massive photon is a rather small one (5-20\%).  Comparison with $\mg=0$ yields the magnitude of the LO Ter-Mikayelian effect (10-40\%).}
}
\end{figure}

We wish to investigate quantitatively the effect of these extra terms on the \zeroth order energy loss.  To do so we enforce physicality by restricting the $x$ and $\wv{k}$ integration limits so that the emitted photon has $E_\gamma \gte m_g$ and leaves the jet with $E_{jet}\gte\mmq$.  For ease of comparison with \cite{Djordjevic:2003be} we set $\mu=.5$ GeV and $\alpha=.5$ fixed.  One can see from \fig{effectsone} the large (50-150\%) effect on $\Delta E/E$ of including the overall prefactor of $(1-x)^2$.  
Filling in the ``dead cone'' makes only a small difference to the energy lost (5-20\%); this is a surprise as the ``dead cone'' is the usual naive justification for heavy quarks having smaller radiative energy loss than light quarks.

\begin{figure}[htb!]
\centering
$\begin{array}{c@{\hspace{.05in}}c}
\includegraphics[width=2.65in]{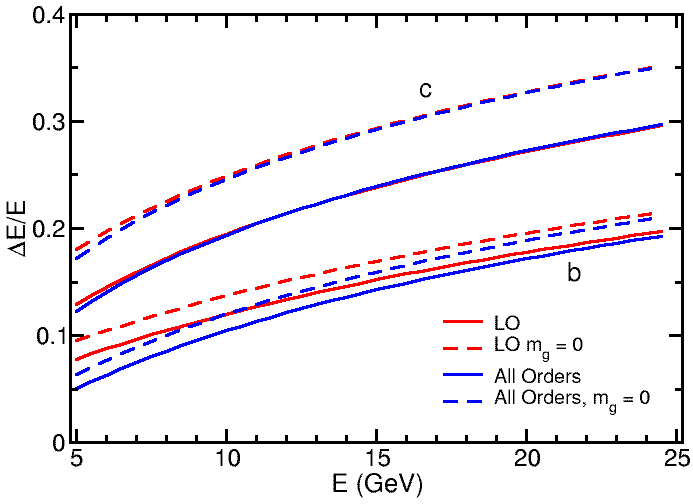} & 
\includegraphics[width=2.65in]{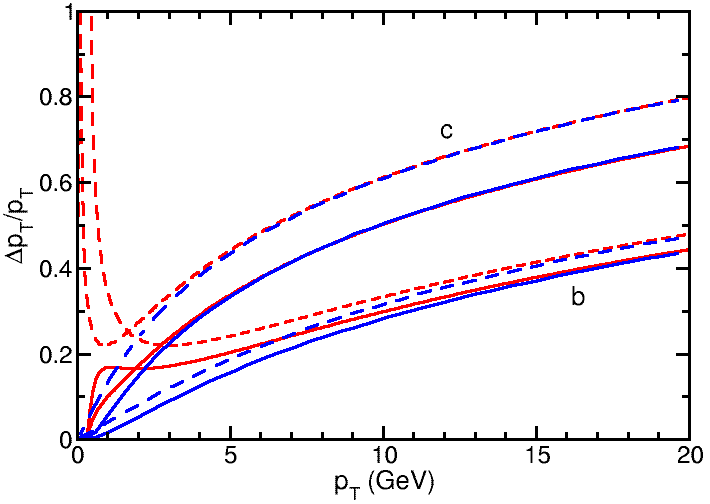} \\ [-.05in]
{\mbox {\scriptsize {\bf (a)}}} & {\mbox {\scriptsize {\bf (b)}}}
\end{array}$
\caption{
\label{effectstwo}\captionsize{The effect of including all terms from \eq{em} instead of just the leading order (LO) terms in $1/E^+$ for (a) $\Delta E/E$ and (b) $\Delta\eqnpt/\eqnpt$ (the legend in (a) applies to both plots).  For $\Delta E/E$, the size of the relative difference in magnitude---the Ter-Mik\-ay\-el\-i\-an effect---is changed little while the overall normalization is significantly altered at low energies.  For $\Delta\eqnpt/\eqnpt$ both the relative and overall normalizations change quite a bit, with the inclusion of all terms regulating the $\eqnpt\rightarrow0$ divergences in the vacuum production radiation spectrum.
}
}
\end{figure}

\fig{effectstwo} demonstrates the effect of including all the terms generated by \eq{em}, not just the LO in $1/E^+$.  Of course at higher $E$ and $\eqnpt$ the additional terms make little difference, but they regulate the otherwise divergent results in $\Delta \eqnpt/\eqnpt$ as $\eqnpt\rightarrow0$.  The Ter-Mikayelian effect, given by the difference between the $\mg\ne0$ and $\mg=0$ plots in \fig{effectstwo}, varies from 10-40\% for charm and bottom energy loss.

\section{Conclusions}\label{concl}
Unfortunately after many years of effort there is still no single satisfactory energy loss model for heavy ion collisions at \rhiccomma.  This leads to the need for basic experimental tests of the gross features of the underlying energy loss mechanism, whether it be more like pQCD, AdS/CFT, or some other approximation.  
Medium induced photon bremsstrahlung has the potential to provide unprecedented insight into the modes of jet energy loss, and in this paper we took the intermediate step of analyzing the \zeroth order in opacity production radiation energy loss.  
While enforcing gauge invariance by not neglecting the away side jet fills in the ``dead cone,'' this ultimately has only a small effect on the radiation spectrum.  On the other hand neglecting the overall factor of $(1-x)^2$ in the emitted photon distribution makes a surprisingly large difference.  This prefactor, also neglected in medium-induced gluon radiation derivations \cite{Gyulassy:2000er,Djordjevic:2003zk}, may significantly alter \raapt calculations, especially at smaller momenta.
\clearpage
\mychapter{Conclusions and Outlook}{chapter:conclusions}
Experiment drives physics.  Our description of natural laws evolved not from aesthetics but falsifiable predictions compared to data.  Scientific advance led us to the Standard Model of particle physics, with $SU(3)$ the gauge group of the strong force coupling to fundamental quarks of fractional charge.  Every tractable theory of nuclear physics points to a new paradigm at scales on the order of $\Lambda_\textrm{QCD}$, at which point quarks and gluons are presumably liberated from their hadronic prisons.  It is clear that the prerequisite densities are reached at \rhiccomma, but many questions remain, theoretically and experimentally.

If one were to sum up the discoveries at \rhic in two words they would be flow and jets.  The \lowpt azimuthal anisotropy appears to saturate the ideal hydrodynamics bound.  High-\pt jets are quenched, and, shockingly, heavy quarks are suppressed at a level similar to light quarks and gluons.  Additionally the \intermediatept \vtwo is surprisingly high for both light hadrons and nonphotonic electrons.  

Hydrodynamics results hold the tantalizing possibility for direct evidence of the degrees of freedom and the equation of state of the medium.  These are of course the golden measurements for demonstrating deconfinement, thermalization, and a QCD phase transition.  Heavy ion collisions may also form the most perfect fluid and provide the most stunning example of the application of string theoretic techniques to phenomenology.  Hydrodynamics is a tough business for even qualitative discoveries, though, as so little of the process is under theoretical or experimental control.  The initial state, so crucial in determining the final state, is up for grabs.  Viscous corrections during the fluid evolution phase are not well understood.  And the transition to detected particles through low-momentum hadronization is nonperturbative and not independently measurable.

High-\pt physics has the advantage of an initial state well under control theoretically and experimentally, 
although current techniques at best have significant error for heavy quarks.  Similarly the fragmentation into hadrons is an independently measured vacuum process that necessarily translates to light meson production, although this may not hold for heavy quarks.  The challenge comes in deriving a confident theoretical description of the in-medium energy loss mechanisms.  With this in hand one has the possibility of inverting the energy loss process using the theory to quantify properties of the medium, yielding jet tomography.  The original hope that a single uncontroversial theory of final state energy loss would emerge from the theoretical community and immediately agree quantitatively with data has been dashed.  There are currently four different major pQCD models of radiative energy loss and a number of elastic energy loss approaches.  Even more disturbing, none of these gives even a consistent qualitative agreement with data.  

In this thesis we confronted the two major areas of disagreement: the simultaneous description of (1) \intermediatept \raa and \vtwo and (2) nonphotonic electron and light meson \raacomma.  For the first we proposed a focusing mechanism that provides the necessary increase in \vtwo and also decreases as a function of momenta.  Future \vtwo results, should they persist at such high percentages, would cast serious doubt on the applicability of perturbative methods to light mesons.  On the other hand this might point to a medium geometry significantly different from our expectations, with radical consequences for hydrodynamics results.

For the second we included both radiative and collisional energy loss processes with realistic geometrical modeling.  This decreased the discrepancy between the presumed heavy quark and light quark and gluon energy loss as inferred from measurements at \rhiccomma.  We also showed that this model is sensitive to changes in its input parameter, the density of scattering centers.  However the significant \alphas dependence strongly suggests that the irreducible theoretical error will be large, which would reduce jets to mere qualitative probes of the medium.  It is clear that future work in estimating the magnitude and possible minimization of major sources of theoretical error is of critical importance.  

Working under the assumption of jets as qualitative probe we looked for a qualitative test of energy loss mechanisms and found one in the momentum dependence of heavy quarks.  Specifically at \lhc $dR^{c,b}_{AA}/d\eqnpt>0$ implies perturbative loss whereas $dR^{c,b}_{AA}/d\eqnpt<0$ implies AdS/CFT drag loss.  While the substantial suppression of heavy quarks at \lhc predicted by some models could make this difficult to see experimentally the ratio of the two nuclear modification factors, $R^{cb}(\eqnpt)=R^c_{AA}(\eqnpt)/R^b_{AA}(\eqnpt)$ is extremely robust to changes in the theoretical input parameters.  Perturbative predictions take $R^{cb}\rightarrow1$ for large momenta while the double ratio is momentum independent and significantly below one, $\sim .2$, from AdS/CFT drag.  The only issue is the regions of self-consistent applicability of the two approaches.  pQCD is valid at large momenta, but the momentum below which its approximations begin to break down is not quantitatively known.  On the other hand AdS/CFT drag is valid at small momenta, but the momentum above which its approximations begin to break down is not quantitatively known.  It is likely that for bottom quarks there is a momentum range for which both are applicable; for charm it is less clear.  In this sense the \pt behavior of the $R^{cb}$ ratio may indicate either the falsification of one or both sets or ideas or may simply give an experimental indication of these aforementioned momentum cutoffs.  Predictions for \rhiccomma, where the lower multiplicity will increase the momentum cutoff from the AdS/CFT drag derivation, were given.

Continuing to pursue qualitatively different predictions from energy loss formalisms we began to investigate heavy quark photon bremsstrahlung.  Even the intermediate results from production radiation were quite interesting: including the interference terms from the away-side quark jet filled in the radiative `dead cone.'  Not neglecting factors small in $x$ led to a significant reduction in the production radiation, although the Ter-Mikayelian effect remained similar in magnitude to previous computations.  The importance of including these factors in gluon bremsstrahlung radiation calculations is not yet known but is of course of crucial interest.  

The future of jet tomography in heavy ion collisions is far from rosy.  We demonstrated in this thesis that very interesting physics can be explored in a qualitative way using \highpt probes.  At the same time, making the not insignificant assumption of a self-consistent model of energy loss in qualitative agreement with data, the current, admittedly rough, estimates of theoretical precision are low.  Future experimental results will be invaluable in guiding the focus of theoretical research on the pertinent approximation regimes in nuclear physics; future work in quantifying, then reducing theoretical error may yet prove jet tomography possible.

\newpage
\phantomsection
\addcontentsline{toc}{chapter}{Bibliography}
\lhead{Bibliography}

\begin{spacing}{1}

\begin{thebibliography}{100}
\providecommand{\url}[1]{#1}
\providecommand{\urlprefix}{URL }
\providecommand{\eprint}[2][]{[\url{#2}]}

\bibitem{aristotle}
Aristotle, \emph{Metaphysica}.

\bibitem{aristotle:zeno}
Aristotle, \emph{Physica}.

\bibitem{popper}
K.~Popper, \emph{Logik der Forschung,} (1934).

\bibitem{newyorker}
B.~Wallace-Wells, \emph{Surfing the Universe}, The New Yorker \textbf{July 21},
  32 (2008).

\bibitem{pais}
A.~Pais, \emph{Inward Bound} (Oxford University Press, 1986), ISBN
  0-19-851971-0.

\bibitem{neeman}
Y.~Ne'eman and Y.~Kirsh, \emph{{The Particle Hunters}}, 2nd ed. ({Cambridge
  University Press}, 1996), ISBN 0-521-47676-0.

\bibitem{Griffiths:1987tjB}
D.~J. Griffiths, \emph{{Introduction to Elementary Particles}} (John Wiley \&
  Sons, Inc., 1987), ISBN 0-471-60386-4.

\bibitem{aristotle:democritus}
Aristotle, \emph{De Generatione et Corruptione}.

\bibitem{lavoisier}
A.-L. Lavoisier, \emph{{Trait\'e \'el\'ementaire de chimie}} (1789), see R.\
  D.\ Whitaker, \emph{An historical note on the conservation of mass}, JCE {\bf
  52 10}, 658 (1975) for a discussion of this historical attribution.

\bibitem{proust}
J.~L. Proust, \emph{{Reserches sur le bleu de Prusse}}, Journal de Physique
  \textbf{45}, 334 (1794).

\bibitem{dalton1}
J.~Dalton, \emph{On the Absorption of Gases by Water and Other Liquids},
  Memoirs of the Literary and Philosophical Society of Manchester, Second
  Series \textbf{I}, 271 (1805).

\bibitem{dalton2}
J.~Dalton, \emph{A New System of Chemical Philosophy} (1808).

\bibitem{Thomson:1897cm}
J.~J. Thomson, \emph{{Cathode rays}}, Phil. Mag. \textbf{44}, 293 (1897).

\bibitem{Rutherford:1911zz}
E.~Rutherford, \emph{{The scattering of $\alpha$ and $\beta$ particles by
  matter and the structure of the atom}}, Phil. Mag. \textbf{21}, 669 (1911).

\bibitem{rutherford:proton}
\emph{Physics at the British Association}, Nature \textbf{106}, 357 (1920).

\bibitem{Chadwick:1932ma}
J.~Chadwick, \emph{{POSSIBLE EXISTENCE OF A NEUTRON}}, Nature \textbf{129}, 312
  (1932).

\bibitem{Yukawa:1935xg}
H.~Yukawa, \emph{{On the interaction of elementary particles}}, Proc. Phys.
  Math. Soc. Jap. \textbf{17}, 48 (1935).

\bibitem{Lattes:1947mw}
C.~M.~G. Lattes, H.~Muirhead, G.~P.~S. Occhialini, and C.~F. Powell,
  \emph{{PROCESSES INVOLVING CHARGED MESONS}}, Nature \textbf{159}, 694 (1947).

\bibitem{Lattes:1947mx}
C.~M.~G. Lattes, G.~P.~S. Occhialini, and C.~F. Powell, \emph{{OBSERVATIONS ON
  THE TRACKS OF SLOW MESONS IN PHOTOGRAPHIC EMULSIONS. 1}}, Nature
  \textbf{160}, 453 (1947).

\bibitem{Rochester:1947mi}
G.~D. Rochester and C.~C. Butler, \emph{{EVIDENCE FOR THE EXISTENCE OF NEW
  UNSTABLE ELEMENTARY PARTICLES}}, Nature \textbf{160}, 855 (1947).

\bibitem{Pais:1952}
A.~Pais, \emph{{Some Remarks on the \emph{V}-Particles}}, Phys. Rev.
  \textbf{86}, 663 (1952).

\bibitem{Gell-Mann:1953}
M.~Gell-Mann, \emph{{Isotopic Spin and New Unstable Particles}}, Phys. Rev.
  \textbf{92}, 833 (1953).

\bibitem{Gell-Mann:1956}
M.~Gell-Mann, \emph{The Interpretation of the New Particles as Displaced
  Charged Multiplets}, Nuovo Cimento \textbf{4 Suppl.\ 2}, 848 (1956).

\bibitem{Nakano:1953}
T.~Nakano and K.~Nishijima, \emph{{Charge Independence for
  \emph{V}-particles}}, Prog. Theor. Phys. \textbf{10 5}, 581 (1953).

\bibitem{Gell-Mann:1964}
M.~Gell-Mann and Y.~Ne'eman, \emph{{The Eightfold Way}} (Benjamin, 1964).

\bibitem{Barnes:1964pd}
V.~E. Barnes et~al., \emph{{OBSERVATION OF A HYPERON WITH STRANGE\-NESS -3}},
  Phys. Rev. Lett. \textbf{12}, 204 (1964).

\bibitem{Lichtenberg:1980}
D.~B. Lichtenberg and S.~P. Rosen (Editors), \emph{{Developments in the Quark
  Theory of Hadrons}} (Nonantum: Hadronic Press, 1980).

\bibitem{Greenberg:1980}
O.~W. Greenberg, \emph{{Resource Letter Q-1: Quarks}}, Am. J. Phys.
  \textbf{50}, 1982 (1074-1089).

\bibitem{Bjorken:1968dy}
J.~D. Bjorken, \emph{{Asymptotic Sum Rules at Infinite Momentum}}, Phys. Rev.
  \textbf{179}, 1547 (1969).

\bibitem{Callan:1969uq}
J.~Callan, Curtis~G. and D.~J. Gross, \emph{{High-energy electroproduction and
  the constitution of the electric current}}, Phys. Rev. Lett. \textbf{22}, 156
  (1969).

\bibitem{Bloom:1969kc}
E.~D. Bloom et~al., \emph{{High-Energy Inelastic e p Scattering at 6-Degrees
  and 10- Degrees}}, Phys. Rev. Lett. \textbf{23}, 930 (1969).

\bibitem{Breidenbach:1969kd}
M.~Breidenbach et~al., \emph{{Observed Behavior of Highly Inelastic
  electron-Proton Scattering}}, Phys. Rev. Lett. \textbf{23}, 935 (1969).

\bibitem{Lichtenberg:1970}
D.~B. Lichtenberg, \emph{{Unitary Symmetry and Elementary Particles}} (Academic
  Press, 1970), ISBN 0124484506.

\bibitem{Han:1965pf}
M.~Y. Han and Y.~Nambu, \emph{{Three-triplet model with double SU(3)
  symmetry}}, Phys. Rev. \textbf{139}, B1006 (1965).

\bibitem{Feynman:1969ej}
R.~P. Feynman, \emph{{Very high-energy collisions of hadrons}}, Phys. Rev.
  Lett. \textbf{23}, 1415 (1969).

\bibitem{Yang:1969}
C.-N. Yang (Editor), \emph{{High Energy Collisions}} (Gordon and Breach, 1969).

\bibitem{Feynman:1973xcB}
R.~P. Feynman, \emph{{Photon-Hadron Interactions}} (W. A. Benjamin, 1972).

\bibitem{Kogut:1972di}
J.~B. Kogut and L.~Susskind, \emph{{The Parton picture of elementary
  particles}}, Phys. Rept. \textbf{8}, 75 (1973).

\bibitem{Bjorken:1969ja}
J.~D. Bjorken and E.~A. Paschos, \emph{{Inelastic Electron Proton and gamma
  Proton Scattering, and the Structure of the Nucleon}}, Phys. Rev.
  \textbf{185}, 1975 (1969).

\bibitem{Kuti:1971ph}
J.~Kuti and V.~F. Weisskopf, \emph{{Inelastic lepton - nucleon scattering and
  lepton pair production in the relativistic quark parton model}}, Phys. Rev.
  \textbf{D4}, 3418 (1971).

\bibitem{Hanson:1975fe}
G.~Hanson et~al., \emph{{Evidence for Jet Structure in Hadron Production by e+
  e- Annihilation}}, Phys. Rev. Lett. \textbf{35}, 1609 (1975).

\bibitem{Brandelik:1979bd}
R.~Brandelik et~al. (TASSO), \emph{{Evidence for Planar Events in e+ e-
  Annihilation at High- Energies}}, Phys. Lett. \textbf{B86}, 243 (1979).

\bibitem{Abreu:1990ce}
P.~Abreu et~al. (DELPHI), \emph{{Experimental study of the triple gluon
  vertex}}, Phys. Lett. \textbf{B255}, 466 (1991).

\bibitem{Decamp:1992ip}
D.~Decamp et~al. (ALEPH), \emph{{Evidence for the triple gluon vertex from
  measurements of the QCD color factors in Z decay into four jets}}, Phys.
  Lett. \textbf{B284}, 151 (1992).

\bibitem{Yang:1954ek}
C.-N. Yang and R.~L. Mills, \emph{{Conservation of isotopic spin and isotopic
  gauge invariance}}, Phys. Rev. \textbf{96}, 191 (1954).

\bibitem{Faddeev:1967fc}
L.~D. Faddeev and V.~N. Popov, \emph{{Feynman diagrams for the Yang-Mills
  field}}, Phys. Lett. \textbf{B25}, 29 (1967).

\bibitem{'tHooft:1971fh}
G.~'t~Hooft, \emph{{Renormalization of Massless Yang-Mills Fields}}, Nucl.
  Phys. \textbf{B33}, 173 (1971).

\bibitem{Gross:1973id}
D.~J. Gross and F.~Wilczek, \emph{{ULTRAVIOLET BEHAVIOR OF NON-ABELIAN GAUGE
  THEORIES}}, Phys. Rev. Lett. \textbf{30}, 1343 (1973).

\bibitem{Politzer:1973fx}
H.~D. Politzer, \emph{{RELIABLE PERTURBATIVE RESULTS FOR STRONG
  INTERACTIONS?}}, Phys. Rev. Lett. \textbf{30}, 1346 (1973).

\bibitem{Coleman:1973sx}
S.~R. Coleman and D.~J. Gross, \emph{{Price of asymptotic freedom}}, Phys. Rev.
  Lett. \textbf{31}, 851 (1973).

\bibitem{Weinberg:1973un}
S.~Weinberg, \emph{{Nonabelian Gauge Theories of the Strong Interactions}},
  Phys. Rev. Lett. \textbf{31}, 494 (1973).

\bibitem{Gross:1973ju}
D.~J. Gross and F.~Wilczek, \emph{{Asymptotically Free Gauge Theories. 1}},
  Phys. Rev. \textbf{D8}, 3633 (1973).

\bibitem{Gross:1974cs}
D.~J. Gross and F.~Wilczek, \emph{{ASYMPTOTICALLY FREE GAUGE THEORIES. 2}},
  Phys. Rev. \textbf{D9}, 980 (1974).

\bibitem{Georgi:1951sr}
H.~Georgi and H.~D. Politzer, \emph{{Electroproduction scaling in an
  asymptotically free theory of strong interactions}}, Phys. Rev. \textbf{D9},
  416 (1974).

\bibitem{Watanabe:1975su}
Y.~Watanabe et~al., \emph{{Test of Scale Invariance in Ratios of Muon
  Scattering Cross-Sections at 150-GeV and 56-GeV}}, Phys. Rev. Lett.
  \textbf{35}, 898 (1975).

\bibitem{Chang:1975sv}
C.~Chang et~al., \emph{{Observed Deviations from Scale Invariance in
  High-Energy Muon Scattering}}, Phys. Rev. Lett. \textbf{35}, 901 (1975).

\bibitem{Aurenche:2006vj}
P.~Aurenche, M.~Fontannaz, J.-P. Guillet, E.~Pilon, and M.~Werlen, \emph{{A new
  critical study of photon production in hadronic collisions}}, Phys. Rev.
  \textbf{D73}, 094007 (2006) \eprint{hep-ph/0602133}.

\bibitem{Cacciari:2007hn}
M.~Cacciari, \emph{{Heavy quark production: Theory}}, Nucl. Phys.
  \textbf{A783}, 189 (2007).

\bibitem{Kluth:2006bw}
S.~Kluth, \emph{{Tests of quantum chromo dynamics at e+ e- colliders}}, Rept.
  Prog. Phys. \textbf{69}, 1771 (2006) \eprint{hep-ex/0603011}.

\bibitem{Wilson:1974sk}
K.~G. Wilson, \emph{{CONFINEMENT OF QUARKS}}, Phys. Rev. \textbf{D10}, 2445
  (1974).

\bibitem{Karsch:2000kv}
F.~Karsch, E.~Laermann, and A.~Peikert, \emph{{Quark mass and flavor dependence
  of the QCD phase transition}}, Nucl. Phys. \textbf{B605}, 579 (2001)
  \eprint{hep-lat/0012023}.

\bibitem{Kaczmarek:2005ui}
O.~Kaczmarek and F.~Zantow, \emph{{Static quark anti-quark interactions in zero
  and finite temperature QCD. I: Heavy quark free energies, running coupling
  and quarkonium binding}}, Phys. Rev. \textbf{D71}, 114510 (2005)
  \eprint{hep-lat/0503017}.

\bibitem{Davies:2003ik}
C.~T.~H. Davies et~al. (HPQCD), \emph{{High-precision lattice QCD confronts
  experiment}}, Phys. Rev. Lett. \textbf{92}, 022001 (2004)
  \eprint{hep-lat/0304004}.

\bibitem{Narison:2004}
S.~Narison, \emph{{QCD as a Theory of Hadrons from Partons to Confinement}}
  (2004).

\bibitem{Halzen:1984}
F.~Halzen and A.~D. Martin, \emph{Quarks and Leptons} (John Wiley \& Sons,
  Inc., 1984), ISBN 0-471-88741-2.

\bibitem{Tung:2001cv}
W.~K. Tung, \emph{{Perturbative QCD and the parton structure of the nucleon}}
  In *Shifman, M. (ed.): At the frontier of particle physics, vol. 2* 887-971.

\bibitem{Baym:2001in}
G.~Baym, \emph{{RHIC: From dreams to beams in two decades}}, Nucl. Phys.
  \textbf{A698}, XXIII (2002) \eprint{hep-ph/0104138}.

\bibitem{Lee:1974ma}
T.~D. Lee and G.~C. Wick, \emph{{Vacuum Stability and Vacuum Excitation in a
  Spin 0 Field Theory}}, Phys. Rev. \textbf{D9}, 2291 (1974).

\bibitem{Hagedorn:1965st}
R.~Hagedorn, \emph{{Statistical thermodynamics of strong interactions at high-
  energies}}, Nuovo Cim. Suppl. \textbf{3}, 147 (1965).

\bibitem{Frautschi:1971ij}
S.~C. Frautschi, \emph{{Statistical bootstrap model of hadrons}}, Phys. Rev.
  \textbf{D3}, 2821 (1971).

\bibitem{Walecka:1974qa}
J.~D. Walecka, \emph{{A Theory of highly condensed matter}}, Annals Phys.
  \textbf{83}, 491 (1974).

\bibitem{Chin:1974sa}
S.~A. Chin and J.~D. Walecka, \emph{{An Equation of State for Nuclear and
  Higher-Density Matter Based on a Relativistic Mean-Field Theory}}, Phys.
  Lett. \textbf{B52}, 24 (1974).

\bibitem{Baym:1975mf}
G.~Baym and C.~Pethick, \emph{{Neutron Stars}}, Ann. Rev. Nucl. Part. Sci.
  \textbf{25}, 27 (1975).

\bibitem{Itoh:1970uw}
N.~Itoh, \emph{{Hydrostatic Equilibrium of Hypothetical Quark Stars}}, Prog.
  Theor. Phys. \textbf{44}, 291 (1970).

\bibitem{Collins:1974ky}
J.~C. Collins and M.~J. Perry, \emph{{Superdense Matter: Neutrons Or
  Asymptotically Free Quarks?}}, Phys. Rev. Lett. \textbf{34}, 1353 (1975).

\bibitem{Shuryak:1980tp}
E.~V. Shuryak, \emph{{Quantum Chromodynamics and the Theory of Superdense
  Matter}}, Phys. Rept. \textbf{61}, 71 (1980).

\bibitem{Adcox:2004mh}
K.~Adcox et~al. (PHENIX), \emph{{Formation of dense partonic matter in
  relativistic nucleus nucleus collisions at RHIC: Experimental evaluation by
  the PHENIX collaboration}}, Nucl. Phys. \textbf{A757}, 184 (2005)
  \eprint{nucl-ex/0410003}.

\bibitem{Cheng:2007jq}
M.~Cheng et~al., \emph{{The QCD Equation of State with almost Physical Quark
  Masses}}, Phys. Rev. \textbf{D77}, 014511 (2008) \eprint{0710.0354}.

\bibitem{Walecka:2004}
J.~D. Walecka, \emph{Theoretical Nuclear and Subnuclear Physics}, 2nd ed.
  (World Scientific Publishing Company, Inc., 2004), ISBN 9-812-38795-1.

\bibitem{Kaczmarek:1999mm}
O.~Kaczmarek, F.~Karsch, E.~Laermann, and M.~Lutgemeier, \emph{{Heavy quark
  potentials in quenched QCD at high temperature}}, Phys. Rev. \textbf{D62},
  034021 (2000) \eprint{hep-lat/9908010}.

\bibitem{Karsch:2001vs}
F.~Karsch, \emph{{Lattice results on QCD thermodynamics}}, Nucl. Phys.
  \textbf{A698}, 199 (2002) \eprint{hep-ph/0103314}.

\bibitem{Doring:2005mt}
M.~Doring, S.~Ejiri, O.~Kaczmarek, F.~Karsch, and E.~Laermann, \emph{{Heavy
  quark free energies and screening at finite temperature and density}}, PoS
  \textbf{LAT2005}, 193 (2006) \eprint{hep-lat/0509150}.

\bibitem{Adil:2007}
A.~Adil, \emph{{Testing the case for the creation of a strongly interacting
  quark gluon plasma at RHIC}}, Ph.D. thesis, Columbia University (2007),
  [\href{http://proquest.umi.com/pqdweb?did=1414129031&sid=2&Fmt=2&clientId=15%
403&RQT=309&VName=PQD}{UMI-32-85033}].

\bibitem{Aoki:2006br}
Y.~Aoki, Z.~Fodor, S.~D. Katz, and K.~K. Szabo, \emph{{The QCD transition
  temperature: Results with physical masses in the continuum limit}}, Phys.
  Lett. \textbf{B643}, 46 (2006) \eprint{hep-lat/0609068}.

\bibitem{Arsene:2004fa}
I.~Arsene et~al. (BRAHMS), \emph{{Quark gluon plasma and color glass condensate
  at RHIC? The perspective from the BRAHMS experiment}}, Nucl. Phys.
  \textbf{A757}, 1 (2005) \eprint{nucl-ex/0410020}.

\bibitem{Adams:2005dq}
J.~Adams et~al. (STAR), \emph{{Experimental and theoretical challenges in the
  search for the quark gluon plasma: The STAR collaboration's critical
  assessment of the evidence from RHIC collisions}}, Nucl. Phys. \textbf{A757},
  102 (2005) \eprint{nucl-ex/0501009}.

\bibitem{Back:2004je}
B.~B. Back et~al., \emph{{The PHOBOS perspective on discoveries at RHIC}},
  Nucl. Phys. \textbf{A757}, 28 (2005) \eprint{nucl-ex/0410022}.

\bibitem{Rafelski:1982pu}
J.~Rafelski and B.~Muller, \emph{{Strangeness Production in the Quark - Gluon
  Plasma}}, Phys. Rev. Lett. \textbf{48}, 1066 (1982).

\bibitem{Koch:1986ud}
P.~Koch, B.~Muller, and J.~Rafelski, \emph{{Strangeness in Relativistic Heavy
  Ion Collisions}}, Phys. Rept. \textbf{142}, 167 (1986).

\bibitem{Becattini:1995xt}
F.~Becattini, A.~Giovannini, and S.~Lupia, \emph{{Multiplicity distributions in
  a thermodynamical model of hadron production in e+ e- collisions}}, Z. Phys.
  \textbf{C72}, 491 (1996) \eprint{hep-ph/9511203}.

\bibitem{Becattini:1997rv}
F.~Becattini and U.~W. Heinz, \emph{{Thermal hadron production in p p and p
  anti-p collisions}}, Z. Phys. \textbf{C76}, 269 (1997)
  \eprint{hep-ph/9702274}.

\bibitem{Cleymans:1992hy}
J.~Cleymans, K.~Redlich, H.~Satz, and E.~Suhonen, \emph{{The Hadronization of a
  quark - gluon plasma}}, Z. Phys. \textbf{C58}, 347 (1993).

\bibitem{Kaneta:2004zr}
M.~Kaneta and N.~Xu, \emph{{Centrality dependence of chemical freeze-out in Au
  + Au collisions at RHIC}}  (2004) \eprint{nucl-th/0405068}.

\bibitem{Letessier:2005qe}
J.~Letessier and J.~Rafelski, \emph{{Hadron production and phase changes in
  relativistic heavy ion collisions}}, Eur. Phys. J. \textbf{A35}, 221 (2008)
  \eprint{nucl-th/0504028}.

\bibitem{Castorina:2007eb}
P.~Castorina, D.~Kharzeev, and H.~Satz, \emph{{Thermal Hadronization and
  Hawking-Unruh Radiation in QCD}}, Eur. Phys. J. \textbf{C52}, 187 (2007)
  \eprint{0704.1426}.

\bibitem{HanburyBrown:1956pf}
R.~Hanbury~Brown and R.~Q. Twiss, \emph{{A Test of a new type of stellar
  interferometer on Sirius}}, Nature \textbf{178}, 1046 (1956).

\bibitem{HanburyBrown:1954wr}
R.~Hanbury~Brown and R.~Q. Twiss, \emph{{A New type of interferometer for use
  in radio astronomy}}, Phil. Mag. \textbf{45}, 663 (1954).

\bibitem{Lisa:2005dd}
M.~A. Lisa, S.~Pratt, R.~Soltz, and U.~Wiedemann, \emph{{Femtoscopy in
  relativistic heavy ion collisions}}, Ann. Rev. Nucl. Part. Sci. \textbf{55},
  357 (2005) \eprint{nucl-ex/0505014}.

\bibitem{Rischke:1998fq}
D.~H. Rischke, \emph{{Fluid dynamics for relativistic nuclear collisions}}
  (1998) \eprint{nucl-th/9809044}.

\bibitem{Kolb:2003dz}
P.~F. Kolb and U.~W. Heinz, \emph{{Hydrodynamic description of
  ultrarelativistic heavy-ion collisions}}  (2003) \eprint{nucl-th/0305084}.

\bibitem{Huovinen:2006jp}
P.~Huovinen and P.~V. Ruuskanen, \emph{{Hydrodynamic models for heavy ion
  collisions}}, Ann. Rev. Nucl. Part. Sci. \textbf{56}, 163 (2006)
  \eprint{nucl-th/0605008}.

\bibitem{Hirano:2008hy}
T.~Hirano, N.~van~der Kolk, and A.~Bilandzic, \emph{{Hydrodynamics and Flow}}
  (2008) \eprint{0808.2684}.

\bibitem{Gyulassy:2004vg}
M.~Gyulassy, \emph{{The QGP discovered at RHIC}}  (2004)
  \eprint{nucl-th/0403032}.

\bibitem{Gyulassy:2004zy}
M.~Gyulassy and L.~McLerran, \emph{{New forms of QCD matter discovered at
  RHIC}}, Nucl. Phys. \textbf{A750}, 30 (2005) \eprint{nucl-th/0405013}.

\bibitem{Ludlam:2003rh}
T.~Ludlam and L.~McLerran, \emph{{What have we learned from the Relativistic
  Heavy Ion Collider?}}, Phys. Today \textbf{56N10}, 48 (2003).

\bibitem{Huovinen:2001cy}
P.~Huovinen, P.~F. Kolb, U.~W. Heinz, P.~V. Ruuskanen, and S.~A. Voloshin,
  \emph{{Radial and elliptic flow at RHIC: Further predictions}}, Phys. Lett.
  \textbf{B503}, 58 (2001) \eprint{hep-ph/0101136}.

\bibitem{Alt:2003ab}
C.~Alt et~al. (NA49), \emph{{Directed and elliptic flow of charged pions and
  protons in Pb + Pb collisions at 40-A-GeV and 158-A-GeV}}, Phys. Rev.
  \textbf{C68}, 034903 (2003) \eprint{nucl-ex/0303001}.

\bibitem{Adler:2003kt}
S.~S. Adler et~al. (PHENIX), \emph{{Elliptic flow of identified hadrons in Au +
  Au collisions at s(NN)**(1/2) = 200-GeV}}, Phys. Rev. Lett. \textbf{91},
  182301 (2003) \eprint{nucl-ex/0305013}.

\bibitem{Adler:2004rq}
S.~S. Adler et~al. (PHENIX), \emph{{Bose-Einstein correlations of charged pion
  pairs in Au + Au collisions at s(NN)**(1/2) = 200-GeV}}, Phys. Rev. Lett.
  \textbf{93}, 152302 (2004) \eprint{nucl-ex/0401003}.

\bibitem{Adams:2003ra}
J.~Adams et~al. (STAR), \emph{{Azimuthally sensitive HBT in Au + Au collisions
  at s(NN)**(1/2) = 200-GeV}}, Phys. Rev. Lett. \textbf{93}, 012301 (2004)
  \eprint{nucl-ex/0312009}.

\bibitem{Hirano:2002ds}
T.~Hirano and K.~Tsuda, \emph{{Collective flow and two pion correlations from a
  relativistic hydrodynamic model with early chemical freeze out}}, Phys. Rev.
  \textbf{C66}, 054905 (2002) \eprint{nucl-th/0205043}.

\bibitem{Heinz:2002un}
U.~W. Heinz and P.~F. Kolb, \emph{{Two RHIC puzzles: Early thermalization and
  the HBT problem}}  (2002) \eprint{hep-ph/0204061}.

\bibitem{Soff:2002qw}
S.~Soff, \emph{{HBT interferometry and the parton-hadron phase transition}}
  (2002) \eprint{hep-ph/0202240}.

\bibitem{Adil:2005qn}
A.~Adil and M.~Gyulassy, \emph{{3D jet tomography of twisted strongly coupled
  quark gluon plasmas}}, Phys. Rev. \textbf{C72}, 034907 (2005)
  \eprint{nucl-th/0505004}.

\bibitem{Adil:2005bb}
A.~Adil, M.~Gyulassy, and T.~Hirano, \emph{{3D jet tomography of the twisted
  color glass condensate}}, Phys. Rev. \textbf{D73}, 074006 (2006)
  \eprint{nucl-th/0509064}.

\bibitem{Hirano:2005xf}
T.~Hirano, U.~W. Heinz, D.~Kharzeev, R.~Lacey, and Y.~Nara, \emph{{Hadronic
  dissipative effects on elliptic flow in ultrarelativistic heavy-ion
  collisions}}, Phys. Lett. \textbf{B636}, 299 (2006) \eprint{nucl-th/0511046}.

\bibitem{Das:1977cp}
K.~P. Das and R.~C. Hwa, \emph{{Quark - anti-Quark Recombination in the
  Fragmentation Region}}, Phys. Lett. \textbf{B68}, 459 (1977).

\bibitem{Hwa:1979pn}
R.~C. Hwa, \emph{{Clustering and Hadronization of Quarks: A Treatment of the
  Low p(t) Problem}}, Phys. Rev. \textbf{D22}, 1593 (1980).

\bibitem{Biro:1994mp}
T.~S. Biro, P.~Levai, and J.~Zimanyi, \emph{{ALCOR: A Dynamic model for
  hadronization}}, Phys. Lett. \textbf{B347}, 6 (1995).

\bibitem{Csizmadia:1998vp}
P.~Csizmadia et~al., \emph{{Strange hyperon and antihyperon production from
  quark and string-rope matter}}, J. Phys. \textbf{G25}, 321 (1999)
  \eprint{hep-ph/9809456}.

\bibitem{Teaney:2003kp}
D.~Teaney, \emph{{Effect of shear viscosity on spectra, elliptic flow, and
  Hanbury Brown-Twiss radii}}, Phys. Rev. \textbf{C68}, 034913 (2003)
  \eprint{nucl-th/0301099}.

\bibitem{Hiscock:1983zz}
W.~A. Hiscock and L.~Lindblom, \emph{{Stability and causality in dissipative
  relativistic fluids}}, Annals Phys. \textbf{151}, 466 (1983).

\bibitem{PhysRevD.31.725}
W.~A. Hiscock and L.~Lindblom, \emph{Generic instabilities in first-order
  dissipative relativistic fluid theories}, Phys. Rev. D \textbf{31 4}, 725
  (1985).

\bibitem{PhysRevD.35.3723}
W.~A. Hiscock and L.~Lindblom, \emph{Linear plane waves in dissipative
  relativistic fluids}, Phys. Rev. D \textbf{35 12}, 3723 (1987).

\bibitem{Muller:1967}
M\ .

\bibitem{Israel:1976tn}
W.~Israel, \emph{{Nonstationary irreversible thermodynamics: A Causal
  relativistic theory}}, Ann. Phys. \textbf{100}, 310 (1976).

\bibitem{Stewart:1977}
J.~M. Stewart, \emph{On Transient Relativistic Thermodynamics and Kinetic
  Theory}, Proc. Roy. Soc. Lon. A \textbf{357}, 59 (1977).

\bibitem{Israel:1979wp}
W.~Israel and J.~M. Stewart, \emph{{Transient relativistic thermodynamics and
  kinetic theory}}, Ann. Phys. \textbf{118}, 341 (1979).

\bibitem{Romatschke:2007mq}
P.~Romatschke and U.~Romatschke, \emph{{Viscosity Information from Relativistic
  Nuclear Collisions: How Perfect is the Fluid Observed at RHIC?}}, Phys. Rev.
  Lett. \textbf{99}, 172301 (2007) \eprint{0706.1522}.

\bibitem{Song:2008si}
H.~Song and U.~W. Heinz, \emph{{Multiplicity scaling in ideal and viscous
  hydrodynamics}}  (2008) \eprint{0805.1756}.

\bibitem{Danielewicz:1984ww}
P.~Danielewicz and M.~Gyulassy, \emph{{Dissipative Phenomena in Quark Gluon
  Plasmas}}, Phys. Rev. \textbf{D31}, 53 (1985).

\bibitem{Baym:1990uj}
G.~Baym, H.~Monien, C.~J. Pethick, and D.~G. Ravenhall, \emph{{TRANSVERSE
  INTERACTIONS AND TRANSPORT IN RELATIVISTIC QUARK - GLUON AND ELECTROMAGNETIC
  PLASMAS}}, Phys. Rev. Lett. \textbf{64}, 1867 (1990).

\bibitem{Arnold:2000dr}
P.~Arnold, G.~D. Moore, and L.~G. Yaffe, \emph{{Transport coefficients in high
  temperature gauge theories. I: Leading-log results}}, JHEP \textbf{11}, 001
  (2000) \eprint{hep-ph/0010177}.

\bibitem{Arnold:2003zc}
P.~Arnold, G.~D. Moore, and L.~G. Yaffe, \emph{{Transport coefficients in high
  temperature gauge theories. II: Beyond leading log}}, JHEP \textbf{05}, 051
  (2003) \eprint{hep-ph/0302165}.

\bibitem{Zhang:1999rs}
B.~Zhang, M.~Gyulassy, and C.~M. Ko, \emph{{Elliptic flow from a parton
  cascade}}, Phys. Lett. \textbf{B455}, 45 (1999) \eprint{nucl-th/9902016}.

\bibitem{Molnar:2000jh}
D.~Molnar and M.~Gyulassy, \emph{{New solutions to covariant nonequilibrium
  dynamics}}, Phys. Rev. \textbf{C62}, 054907 (2000) \eprint{nucl-th/0005051}.

\bibitem{Xu:2004mz}
Z.~Xu and C.~Greiner, \emph{{Thermalization of gluons in ultrarelativistic
  heavy ion collisions by including three-body interactions in a parton
  cascade}}, Phys. Rev. \textbf{C71}, 064901 (2005) \eprint{hep-ph/0406278}.

\bibitem{Molnar:2001ux}
D.~Molnar and M.~Gyulassy, \emph{{Saturation of elliptic flow at RHIC: Results
  from the covariant elastic parton cascade model MPC}}, Nucl. Phys.
  \textbf{A697}, 495 (2002) \eprint{nucl-th/0104073}.

\bibitem{Molnar:2008xj}
D.~Molnar and P.~Huovinen, \emph{{Dissipative effects from transport and
  viscous hydrodynamics}}  (2008) \eprint{0806.1367}.

\bibitem{Xu:2007ns}
Z.~Xu and C.~Greiner, \emph{{Shear viscosity in a gluon gas}}, Phys. Rev. Lett.
  \textbf{100}, 172301 (2008) \eprint{0710.5719}.

\bibitem{Fochler:2008ts}
O.~Fochler, Z.~Xu, and C.~Greiner, \emph{{Towards a unified understanding of
  jet-quenching and elliptic flow within perturbative QCD parton transport}}
  (2008) \eprint{0806.1169}.

\bibitem{Jacob:1972ym}
M.~Jacob and S.~Nussinov, \emph{{Pion production in high-energy proton
  anti-proton annihilation}}  (1972), {FERMILAB-PUB-72-067-T}.

\bibitem{Jacob:1972yc}
M.~Jacob, R.~Slansky, and C.~C. Wu, \emph{{Systematics of single particle
  spectra in proton proton collisions}}  (1972), {FERMILAB-PUB-72-062-T}.

\bibitem{Busser:1973hs}
F.~W. Busser et~al., \emph{{Observation of pi0 mesons with large transverse
  momentum in high-energy proton proton collisions}}, Phys. Lett. \textbf{B46},
  471 (1973).

\bibitem{Alper:1973nv}
B.~Alper et~al. (British-Scandinavian ISR), \emph{{Production of high
  transverse momentum particles in p p collisions in the central region at the
  CERN ISR}}, Phys. Lett. \textbf{B44}, 521 (1973).

\bibitem{Banner:1973nu}
M.~Banner et~al., \emph{{Large transverse momentum particle production at 90
  degrees in proton-p proton collisions at the ISR}}, Phys. Lett. \textbf{B44},
  537 (1973).

\bibitem{Bjorken:1982tu}
J.~D. Bjorken, \emph{{Energy Loss of Energetic Partons in Quark - Gluon Plasma:
  Possible Extinction of High p(t) Jets in Hadron - Hadron Collisions}}
  (1982), {FERMILAB-PUB-82-059-THY}.

\bibitem{Aversa:1988vb}
F.~Aversa, P.~Chiappetta, M.~Greco, and J.~P. Guillet, \emph{{QCD Corrections
  to Parton-Parton Scattering Processes}}, Nucl. Phys. \textbf{B327}, 105
  (1989).

\bibitem{Adler:2003pb}
S.~S. Adler et~al. (PHENIX), \emph{{Mid-rapidity neutral pion production in
  proton proton collisions at s**(1/2) = 200-GeV}}, Phys. Rev. Lett.
  \textbf{91}, 241803 (2003) \eprint{hep-ex/0304038}.

\bibitem{Jager:2002xm}
B.~Jager, A.~Schafer, M.~Stratmann, and W.~Vogelsang, \emph{{Next-to-leading
  order QCD corrections to high-p(T) pion production in longitudinally
  polarized p p collisions}}, Phys. Rev. \textbf{D67}, 054005 (2003)
  \eprint{hep-ph/0211007}.

\bibitem{deFlorian:2002az}
D.~de~Florian, \emph{{Next-to-leading order QCD corrections to one hadron
  production in polarized p p collisions at RHIC}}, Phys. Rev. \textbf{D67},
  054004 (2003) \eprint{hep-ph/0210442}.

\bibitem{Wicks:2005gt}
S.~Wicks, W.~Horowitz, M.~Djordjevic, and M.~Gyulassy, \emph{{Elastic,
  Inelastic, and Path Length Fluctuations in Jet Tomography}}, Nucl. Phys.
  \textbf{A784}, 426 (2007) \eprint{nucl-th/0512076}.

\bibitem{Libby:1978qf}
S.~B. Libby and G.~Sterman, \emph{{Jet and Lepton Pair Production in
  High-Energy Lepton- Hadron and Hadron-Hadron Scattering}}, Phys. Rev.
  \textbf{D18}, 3252 (1978).

\bibitem{Ellis:1978sf}
R.~K. Ellis, H.~Georgi, M.~Machacek, H.~D. Politzer, and G.~G. Ross,
  \emph{{Factorization and the Parton Model in QCD}}, Phys. Lett. \textbf{B78},
  281 (1978).

\bibitem{Ellis:1978ty}
R.~K. Ellis, H.~Georgi, M.~Machacek, H.~D. Politzer, and G.~G. Ross,
  \emph{{Perturbation Theory and the Parton Model in QCD}}, Nucl. Phys.
  \textbf{B152}, 285 (1979).

\bibitem{Amati:1978aa}
D.~Amati, R.~Petronzio, and G.~Veneziano, \emph{{Relating hard QCD processes
  through universality of mass singularities}}, Nucl. Phys. \textbf{B140}, 54
  (1978).

\bibitem{Amati:1978bb}
D.~Amati, R.~Petronzio, and G.~Veneziano, \emph{{ Relating hard QCD processes
  through universality of mass singularities (II)}}, Nucl. Phys. \textbf{B146},
  29 (1978).

\bibitem{Curci:1980uw}
G.~Curci, W.~Furmanski, and R.~Petronzio, \emph{{Evolution of Parton Densities
  Beyond Leading Order: The Nonsinglet Case}}, Nucl. Phys. \textbf{B175}, 27
  (1980).

\bibitem{Collins:1983ju}
J.~C. Collins, D.~E. Soper, and G.~Sterman, \emph{{ALL ORDER FACTORIZATION FOR
  DRELL-YAN CROSS-SECTIONS}}, Phys. Lett. \textbf{B134}, 263 (1984).

\bibitem{Collins:1992xw}
J.~C. Collins, \emph{{Hard scattering in QCD with polarized beams}}, Nucl.
  Phys. \textbf{B394}, 169 (1993) \eprint{hep-ph/9207265}.

\bibitem{Pumplin:2002vw}
J.~Pumplin et~al., \emph{{New generation of parton distributions with
  uncertainties from global QCD analysis}}, JHEP \textbf{07}, 012 (2002)
  \eprint{hep-ph/0201195}.

\bibitem{Gluck:2007ck}
M.~Gluck, P.~Jimenez-Delgado, and E.~Reya, \emph{{Dynamical parton
  distributions of the nucleon and very small-x physics}}, Eur. Phys. J.
  \textbf{C53}, 355 (2008) \eprint{0709.0614}.

\bibitem{Martin:2007bv}
A.~D. Martin, W.~J. Stirling, R.~S. Thorne, and G.~Watt, \emph{{Update of
  Parton Distributions at NNLO}}, Phys. Lett. \textbf{B652}, 292 (2007)
  \eprint{0706.0459}.

\bibitem{Alekhin:2006zm}
S.~Alekhin, K.~Melnikov, and F.~Petriello, \emph{{Fixed target Drell-Yan data
  and NNLO QCD fits of parton distribution functions}}, Phys. Rev.
  \textbf{D74}, 054033 (2006) \eprint{hep-ph/0606237}.

\bibitem{deFlorian:2007aj}
D.~de~Florian, R.~Sassot, and M.~Stratmann, \emph{{Global analysis of
  fragmentation functions for pions and kaons and their uncertainties}}, Phys.
  Rev. \textbf{D75}, 114010 (2007) \eprint{hep-ph/0703242}.

\bibitem{Albino:2008fy}
S.~Albino, B.~A. Kniehl, and G.~Kramer, \emph{{AKK Update: Improvements from
  New Theoretical Input and Experimental Data}}, Nucl. Phys. \textbf{B803}, 42
  (2008) \eprint{0803.2768}.

\bibitem{Hirai:2007cx}
M.~Hirai, S.~Kumano, T.~H. Nagai, and K.~Sudoh, \emph{{Determination of
  fragmentation functions and their uncertainties}}, Phys. Rev. \textbf{D75},
  094009 (2007) \eprint{hep-ph/0702250}.

\bibitem{Kniehl:2000fe}
B.~A. Kniehl, G.~Kramer, and B.~Potter, \emph{{Fragmentation functions for
  pions, kaons, and protons at next-to-leading order}}, Nucl. Phys.
  \textbf{B582}, 514 (2000) \eprint{hep-ph/0010289}.

\bibitem{Bourhis:1997yu}
L.~Bourhis, M.~Fontannaz, and J.~P. Guillet, \emph{{Quark and gluon
  fragmentation functions into photons}}, Eur. Phys. J. \textbf{C2}, 529 (1998)
  \eprint{hep-ph/9704447}.

\bibitem{Horowitz:code}
\url{http://www.phy.uct.ac.za/people/horowitz}.

\bibitem{Dokshitzer:1977sg}
Y.~L. Dokshitzer, \emph{{Calculation of the Structure Functions for Deep
  Inelastic Scattering and e+ e- Annihilation by Perturbation Theory in Quantum
  Chromodynamics. (In Russian)}}, Sov. Phys. JETP \textbf{46}, 641 (1977).

\bibitem{Gribov:1972ri}
V.~N. Gribov and L.~N. Lipatov, \emph{{Deep inelastic e p scattering in
  perturbation theory}}, Sov. J. Nucl. Phys. \textbf{15}, 438 (1972).

\bibitem{Altarelli:1977zs}
G.~Altarelli and G.~Parisi, \emph{{Asymptotic Freedom in Parton Language}},
  Nucl. Phys. \textbf{B126}, 298 (1977).

\bibitem{Binoth:1999qq}
T.~Binoth, J.~P. Guillet, E.~Pilon, and M.~Werlen, \emph{{A full next to
  leading order study of direct photon pair production in hadronic
  collisions}}, Eur. Phys. J. \textbf{C16}, 311 (2000) \eprint{hep-ph/9911340}.

\bibitem{Catani:2002ny}
S.~Catani, M.~Fontannaz, J.~P. Guillet, and E.~Pilon, \emph{{Cross section of
  isolated prompt photons in hadron hadron collisions}}, JHEP \textbf{05}, 028
  (2002) \eprint{hep-ph/0204023}.

\bibitem{Cacciari:1998it}
M.~Cacciari, M.~Greco, and P.~Nason, \emph{{The p(T) spectrum in heavy-flavour
  hadroproduction}}, JHEP \textbf{05}, 007 (1998) \eprint{hep-ph/9803400}.

\bibitem{Cacciari:2001td}
M.~Cacciari, S.~Frixione, and P.~Nason, \emph{{The p(T) spectrum in
  heavy-flavor photoproduction}}, JHEP \textbf{03}, 006 (2001)
  \eprint{hep-ph/0102134}.

\bibitem{Mueller:1985wy}
A.~H. Mueller and J.-w. Qiu, \emph{{Gluon Recombination and Shadowing at Small
  Values of x}}, Nucl. Phys. \textbf{B268}, 427 (1986).

\bibitem{Eskola:1993mb}
K.~J. Eskola, J.-w. Qiu, and X.-N. Wang, \emph{{Perturbative gluon shadowing in
  heavy nuclei}}, Phys. Rev. Lett. \textbf{72}, 36 (1994)
  \eprint{nucl-th/9307025}.

\bibitem{Frankfurt:2002kd}
L.~Frankfurt, V.~Guzey, M.~McDermott, and M.~Strikman, \emph{{Nuclear shadowing
  in deep inelastic scattering on nuclei: Leading twist versus eikonal
  approaches}}, JHEP \textbf{02}, 027 (2002) \eprint{hep-ph/0201230}.

\bibitem{Nikolaev:1990yw}
N.~N. Nikolaev and B.~G. Zakharov, \emph{{SCALING PROPERTIES OF NUCLEAR
  SHADOWING IN DEEP INELASTIC SCATTERING}}, Phys. Lett. \textbf{B260}, 414
  (1991).

\bibitem{Brodsky:1989qz}
S.~J. Brodsky and H.~J. Lu, \emph{{Shadowing and Antishadowing of Nuclear
  Structure Functions}}, Phys. Rev. Lett. \textbf{64}, 1342 (1990).

\bibitem{Geesaman:1995yd}
D.~F. Geesaman, K.~Saito, and A.~W. Thomas, \emph{{The nuclear EMC effect}},
  Ann. Rev. Nucl. Part. Sci. \textbf{45}, 337 (1995).

\bibitem{Bodek:1980ar}
A.~Bodek and J.~L. Ritchie, \emph{{Fermi Motion Effects in Deep Inelastic
  Lepton Scattering from Nuclear Targets}}, Phys. Rev. \textbf{D23}, 1070
  (1981).

\bibitem{Bodek:1981wr}
A.~Bodek and J.~L. Ritchie, \emph{{Further Studies of Fermi Motion Effects in
  Lepton Scattering from Nuclear Targets}}, Phys. Rev. \textbf{D24}, 1400
  (1981).

\bibitem{Peilert:1992hv}
G.~Peilert et~al., \emph{{Dynamical treatment of Fermi motion in a microscopic
  description of heavy ion collisions}}, Phys. Rev. \textbf{C46}, 1457 (1992).

\bibitem{Eskola:1998df}
K.~J. Eskola, V.~J. Kolhinen, and C.~A. Salgado, \emph{{The scale dependent
  nuclear effects in parton distributions for practical applications}}, Eur.
  Phys. J. \textbf{C9}, 61 (1999) \eprint{hep-ph/9807297}.

\bibitem{Eskola:1998iy}
K.~J. Eskola, V.~J. Kolhinen, and P.~V. Ruuskanen, \emph{{Scale evolution of
  nuclear parton distributions}}, Nucl. Phys. \textbf{B535}, 351 (1998)
  \eprint{hep-ph/9802350}.

\bibitem{Hirai:2001np}
M.~Hirai, S.~Kumano, and M.~Miyama, \emph{{Determination of nuclear parton
  distributions}}, Phys. Rev. \textbf{D64}, 034003 (2001)
  \eprint{hep-ph/0103208}.

\bibitem{deFlorian:2003qf}
D.~de~Florian and R.~Sassot, \emph{{Nuclear parton distributions at next to
  leading order}}, Phys. Rev. \textbf{D69}, 074028 (2004)
  \eprint{hep-ph/0311227}.

\bibitem{Drees:2003zh}
A.~Drees, H.~Feng, and J.~Jia, \emph{{Medium induced jet absorption at RHIC}},
  Phys. Rev. \textbf{C71}, 034909 (2005) \eprint{nucl-th/0310044}.

\bibitem{Adcox:2001jp}
K.~Adcox et~al. (PHENIX), \emph{{Suppression of hadrons with large transverse
  momentum in central Au + Au collisions at s**(1/2)(N N) = 130-GeV}}, Phys.
  Rev. Lett. \textbf{88}, 022301 (2002) \eprint{nucl-ex/0109003}.

\bibitem{Datta:2003ww}
S.~Datta, F.~Karsch, P.~Petreczky, and I.~Wetzorke, \emph{{Behavior of
  charmonium systems after deconfinement}}, Phys. Rev. \textbf{D69}, 094507
  (2004) \eprint{hep-lat/0312037}.

\bibitem{Mocsy:2007jz}
A.~Mocsy and P.~Petreczky, \emph{{Color Screening Melts Quarkonium}}, Phys.
  Rev. Lett. \textbf{99}, 211602 (2007) \eprint{0706.2183}.

\bibitem{Gribov:1981kg}
L.~V. Gribov, E.~M. Levin, and M.~G. Ryskin, \emph{{HIGH P(T) HADRONS IN THE
  PIONIZATION REGION IN QCD}}, Phys. Lett. \textbf{B100}, 173 (1981).

\bibitem{Gribov:1984tu}
L.~V. Gribov, E.~M. Levin, and M.~G. Ryskin, \emph{{Semihard Processes in
  QCD}}, Phys. Rept. \textbf{100}, 1 (1983).

\bibitem{Laenen:1994gh}
E.~Laenen and E.~Levin, \emph{{Parton densities at high-energy}}, Ann. Rev.
  Nucl. Part. Sci. \textbf{44}, 199 (1994).

\bibitem{McLerran:1993ni}
L.~D. McLerran and R.~Venugopalan, \emph{{Computing quark and gluon
  distribution functions for very large nuclei}}, Phys. Rev. \textbf{D49}, 2233
  (1994) \eprint{hep-ph/9309289}.

\bibitem{McLerran:1993ka}
L.~D. McLerran and R.~Venugopalan, \emph{{Gluon distribution functions for very
  large nuclei at small transverse momentum}}, Phys. Rev. \textbf{D49}, 3352
  (1994) \eprint{hep-ph/9311205}.

\bibitem{McLerran:1994vd}
L.~D. McLerran and R.~Venugopalan, \emph{{Green's functions in the color field
  of a large nucleus}}, Phys. Rev. \textbf{D50}, 2225 (1994)
  \eprint{hep-ph/9402335}.

\bibitem{Balitsky:1995ub}
I.~Balitsky, \emph{{Operator expansion for high-energy scattering}}, Nucl.
  Phys. \textbf{B463}, 99 (1996) \eprint{hep-ph/9509348}.

\bibitem{Kovchegov:1999ua}
Y.~V. Kovchegov, \emph{{Unitarization of the BFKL pomeron on a nucleus}}, Phys.
  Rev. \textbf{D61}, 074018 (2000) \eprint{hep-ph/9905214}.

\bibitem{Iancu:2000hn}
E.~Iancu, A.~Leonidov, and L.~D. McLerran, \emph{{Nonlinear gluon evolution in
  the color glass condensate. I}}, Nucl. Phys. \textbf{A692}, 583 (2001)
  \eprint{hep-ph/0011241}.

\bibitem{JalilianMarian:1996xn}
J.~Jalilian-Marian, A.~Kovner, L.~D. McLerran, and H.~Weigert, \emph{{The
  intrinsic glue distribution at very small x}}, Phys. Rev. \textbf{D55}, 5414
  (1997) \eprint{hep-ph/9606337}.

\bibitem{JalilianMarian:1997jx}
J.~Jalilian-Marian, A.~Kovner, A.~Leonidov, and H.~Weigert, \emph{{The BFKL
  equation from the Wilson renormalization group}}, Nucl. Phys. \textbf{B504},
  415 (1997) \eprint{hep-ph/9701284}.

\bibitem{JalilianMarian:1997gr}
J.~Jalilian-Marian, A.~Kovner, A.~Leonidov, and H.~Weigert, \emph{{The Wilson
  renormalization group for low x physics: Towards the high density regime}},
  Phys. Rev. \textbf{D59}, 014014 (1999) \eprint{hep-ph/9706377}.

\bibitem{JalilianMarian:1997dw}
J.~Jalilian-Marian, A.~Kovner, and H.~Weigert, \emph{{The Wilson
  renormalization group for low x physics: Gluon evolution at finite parton
  density}}, Phys. Rev. \textbf{D59}, 014015 (1999) \eprint{hep-ph/9709432}.

\bibitem{JalilianMarian:1998cb}
J.~Jalilian-Marian, A.~Kovner, A.~Leonidov, and H.~Weigert,
  \emph{{Unitarization of gluon distribution in the doubly logarithmic regime
  at high density}}, Phys. Rev. \textbf{D59}, 034007 (1999)
  \eprint{hep-ph/9807462}.

\bibitem{Kovner:2000pt}
A.~Kovner, J.~G. Milhano, and H.~Weigert, \emph{{Relating different approaches
  to nonlinear QCD evolution at finite gluon density}}, Phys. Rev.
  \textbf{D62}, 114005 (2000) \eprint{hep-ph/0004014}.

\bibitem{Weigert:2000gi}
H.~Weigert, \emph{{Unitarity at small Bjorken x}}, Nucl. Phys. \textbf{A703},
  823 (2002) \eprint{hep-ph/0004044}.

\bibitem{Ferreiro:2001qy}
E.~Ferreiro, E.~Iancu, A.~Leonidov, and L.~McLerran, \emph{{Nonlinear gluon
  evolution in the color glass condensate. II}}, Nucl. Phys. \textbf{A703}, 489
  (2002) \eprint{hep-ph/0109115}.

\bibitem{Blaizot:2004px}
J.-P. Blaizot and F.~Gelis, \emph{{Searching evidence for the color glass
  condensate at RHIC}}, Nucl. Phys. \textbf{A750}, 148 (2005)
  \eprint{hep-ph/0405305}.

\bibitem{Andersson:2002cf}
B.~Andersson et~al. (Small x), \emph{{Small x phenomenology: Summary and
  status}}, Eur. Phys. J. \textbf{C25}, 77 (2002) \eprint{hep-ph/0204115}.

\bibitem{Szczurek:2003fu}
A.~Szczurek, \emph{{From unintegrated gluon distributions to particle
  production in nucleon nucleon collisions at RHIC energies}}, Acta Phys.
  Polon. \textbf{B34}, 3191 (2003) \eprint{hep-ph/0304129}.

\bibitem{Lonnblad:2004zp}
L.~Lonnblad and M.~Sjodahl, \emph{{Uncertainties on central exclusive scalar
  luminosities from the unintegrated gluon distributions}}, JHEP \textbf{05},
  038 (2005) \eprint{hep-ph/0412111}.

\bibitem{Nikolaev:2004cu}
N.~N. Nikolaev and W.~Schafer, \emph{{Breaking of k(T)-factorization for single
  jet production off nuclei}}, Phys. Rev. \textbf{D71}, 014023 (2005)
  \eprint{hep-ph/0411365}.

\bibitem{Cronin:1974zm}
J.~W. Cronin et~al., \emph{{Production of Hadrons with Large Transverse
  Momentum at 200-GeV, 300-GeV, and 400-GeV}}, Phys. Rev. \textbf{D11}, 3105
  (1975).

\bibitem{Adams:2003im}
J.~Adams et~al. (STAR), \emph{{Evidence from d + Au measurements for
  final-state suppression of high p(T) hadrons in Au + Au collisions at RHIC}},
  Phys. Rev. Lett. \textbf{91}, 072304 (2003) \eprint{nucl-ex/0306024}.

\bibitem{Aurenche:1983ws}
P.~Aurenche, A.~Douiri, R.~Baier, M.~Fontannaz, and D.~Schiff, \emph{{Prompt
  Photon Production at Large p(T) in QCD Beyond the Leading Order}}, Phys.
  Lett. \textbf{B140}, 87 (1984).

\bibitem{Aurenche:1987fs}
P.~Aurenche, R.~Baier, M.~Fontannaz, and D.~Schiff, \emph{{Prompt Photon
  Production at Large p(T) Scheme Invariant QCD Predictions and Comparison with
  Experiment}}, Nucl. Phys. \textbf{B297}, 661 (1988).

\bibitem{Baer:1990ra}
H.~Baer, J.~Ohnemus, and J.~F. Owens, \emph{{A NEXT-TO-LEADING LOGARITHM
  CALCULATION OF DIRECT PHOTON PRODUCTION}}, Phys. Rev. \textbf{D42}, 61
  (1990).

\bibitem{Gordon:1993qc}
L.~E. Gordon and W.~Vogelsang, \emph{{Polarized and unpolarized prompt photon
  production beyond the leading order}}, Phys. Rev. \textbf{D48}, 3136 (1993).

\bibitem{Adler:2005ig}
S.~S. Adler et~al. (PHENIX), \emph{{Centrality dependence of direct photon
  production in s(NN)**(1/2) = 200-GeV Au + Au collisions}}, Phys. Rev. Lett.
  \textbf{94}, 232301 (2005) \eprint{nucl-ex/0503003}.

\bibitem{Thoma:1990fm}
M.~H. Thoma and M.~Gyulassy, \emph{{QUARK DAMPING AND ENERGY LOSS IN THE HIGH
  TEMPERATURE QCD}}, Nucl. Phys. \textbf{B351}, 491 (1991).

\bibitem{Braaten:1989kk}
E.~Braaten and R.~D. Pisarski, \emph{{Resummation and Gauge Invariance of the
  Gluon Damping Rate in Hot QCD}}, Phys. Rev. Lett. \textbf{64}, 1338 (1990).

\bibitem{Braaten:1989mz}
E.~Braaten and R.~D. Pisarski, \emph{{Soft Amplitudes in Hot Gauge Theories: A
  General Analysis}}, Nucl. Phys. \textbf{B337}, 569 (1990).

\bibitem{Braaten:1990az}
E.~Braaten and R.~D. Pisarski, \emph{{DEDUCING HARD THERMAL LOOPS FROM WARD
  IDENTITIES}}, Nucl. Phys. \textbf{B339}, 310 (1990).

\bibitem{Blaizot:2001nr}
J.-P. Blaizot and E.~Iancu, \emph{{The quark-gluon plasma: Collective dynamics
  and hard thermal loops}}, Phys. Rept. \textbf{359}, 355 (2002)
  \eprint{hep-ph/0101103}.

\bibitem{Braaten:1991jj}
E.~Braaten and M.~H. Thoma, \emph{{Energy loss of a heavy fermion in a hot
  plasma}}, Phys. Rev. \textbf{D44}, 1298 (1991).

\bibitem{Braaten:1991we}
E.~Braaten and M.~H. Thoma, \emph{{Energy loss of a heavy quark in the quark -
  gluon plasma}}, Phys. Rev. \textbf{D44}, 2625 (1991).

\bibitem{Svetitsky:1987gq}
B.~Svetitsky, \emph{{DIFFUSION OF CHARMED QUARK IN THE QUARK - GLUON PLASMA}},
  Phys. Rev. \textbf{D37}, 2484 (1988).

\bibitem{Romatschke:2004au}
P.~Romatschke and M.~Strickland, \emph{{Collisional energy loss of a heavy
  quark in an anisotropic quark-gluon plasma}}, Phys. Rev. \textbf{D71}, 125008
  (2005) \eprint{hep-ph/0408275}.

\bibitem{Gyulassy:1990bh}
M.~Gyulassy and M.~Plumer, \emph{{Jet quenching as a probe of dense matter}},
  Nucl. Phys. \textbf{A527}, 641 (1991).

\bibitem{Wang:1994fx}
X.-N. Wang, M.~Gyulassy, and M.~Plumer, \emph{{The LPM effect in QCD and
  radiative energy loss in a quark gluon plasma}}, Phys. Rev. \textbf{D51},
  3436 (1995) \eprint{hep-ph/9408344}.

\bibitem{Jackson:1998}
J.~D. Jackson, \emph{Classical Electrodynamics}, 3rd ed. (John Wiley \& Sons,
  Inc., 1998), ISBN 0-471-30932-X.

\bibitem{Mustafa:2004dr}
M.~G. Mustafa, \emph{{Energy loss of charm quarks in the quark-gluon plasma:
  Collisional vs radiative}}, Phys. Rev. \textbf{C72}, 014905 (2005)
  \eprint{hep-ph/0412402}.

\bibitem{Adler:2005xv}
S.~S. Adler et~al. (PHENIX), \emph{{Nuclear modification of electron spectra
  and implications for heavy quark energy loss in Au + Au collisions at
  s(NN)**(1/2) = 200-GeV}}, Phys. Rev. Lett. \textbf{96}, 032301 (2006)
  \eprint{nucl-ex/0510047}.

\bibitem{Akiba:2005bs}
Y.~Akiba (PHENIX), \emph{{Probing the properties of dense partonic matter at
  RHIC}}, Nucl. Phys. \textbf{A774}, 403 (2006) \eprint{nucl-ex/0510008}.

\bibitem{Bielcik:2005wu}
J.~Bielcik (STAR), \emph{{Centrality dependence of heavy flavor production from
  single electron measurement in s(NN)**(1/2) = 200-GeV Au + Au collisions}},
  Nucl. Phys. \textbf{A774}, 697 (2006) \eprint{nucl-ex/0511005}.

\bibitem{Dong:2005nm}
X.~Dong, \emph{{Open charm production at RHIC}}, AIP Conf. Proc. \textbf{828},
  24 (2006) \eprint{nucl-ex/0509038}.

\bibitem{Peshier:2006hi}
A.~Peshier, \emph{{The QCD collisional energy loss revised}}, Phys. Rev. Lett.
  \textbf{97}, 212301 (2006) \eprint{hep-ph/0605294}.

\bibitem{Braun:2006vd}
J.~Braun and H.-J. Pirner, \emph{{Effects of the running of the QCD coupling on
  the energy loss in the quark-gluon plasma}}, Phys. Rev. \textbf{D75}, 054031
  (2007) \eprint{hep-ph/0610331}.

\bibitem{Wicks:2008}
S.~Wicks, \emph{Fluctuations with small numbers: Developing the perturbative
  paradigm for jet physics in the QGP at RHIC and LHC}, Ph.D. thesis, Columbia
  University (2008).

\bibitem{Peigne:2005rk}
S.~Peigne, P.-B. Gossiaux, and T.~Gousset, \emph{{Retardation effect for
  collisional energy loss of hard partons produced in a QGP}}, JHEP
  \textbf{04}, 011 (2006) \eprint{hep-ph/0509185}.

\bibitem{Adil:2006ei}
A.~Adil, M.~Gyulassy, W.~A. Horowitz, and S.~Wicks, \emph{{Collisional energy
  loss of non asymptotic jets in a QGP}}, Phys. Rev. \textbf{C75}, 044906
  (2007) \eprint{nucl-th/0606010}.

\bibitem{Gossiaux:2007gd}
P.~B. Gossiaux, J.~Aichelin, C.~Brandt, T.~Gousset, and S.~Peigne,
  \emph{{Energy loss of a heavy quark produced in a finite-size quark-gluon
  plasma}}, J. Phys. \textbf{G34}, S817 (2007) \eprint{hep-ph/0703095}.

\bibitem{Djordjevic:2006tw}
M.~Djordjevic, \emph{{Collisional energy loss in a finite size QCD matter}},
  Phys. Rev. \textbf{C74}, 064907 (2006) \eprint{nucl-th/0603066}.

\bibitem{Ayala:2007cq}
A.~Ayala, J.~Magnin, L.~M. Montano, and E.~Rojas, \emph{{Collisional parton
  energy loss in a finite size QCD medium revisited: Off mass-shell effects}},
  Phys. Rev. \textbf{C77}, 044904 (2008) \eprint{0706.2377}.

\bibitem{Moore:2004tg}
G.~D. Moore and D.~Teaney, \emph{{How much do heavy quarks thermalize in a
  heavy ion collision?}}, Phys. Rev. \textbf{C71}, 064904 (2005)
  \eprint{hep-ph/0412346}.

\bibitem{vanHees:2005wb}
H.~van Hees, V.~Greco, and R.~Rapp, \emph{{Heavy-quark probes of the
  quark-gluon plasma at RHIC}}, Phys. Rev. \textbf{C73}, 034913 (2006)
  \eprint{nucl-th/0508055}.

\bibitem{Peskin:1995}
M.~E. Peskin and D.~V. Schroeder, \emph{Introduction to Quantum Field Theory}
  (Westview Press, 1995), ISBN 0-201-50397-2.

\bibitem{Baier:1996vi}
R.~Baier, Y.~L. Dokshitzer, A.~H. Mueller, S.~Peigne, and D.~Schiff, \emph{{The
  Landau-Pomeranchuk-Migdal effect in QED}}, Nucl. Phys. \textbf{B478}, 577
  (1996) \eprint{hep-ph/9604327}.

\bibitem{Baier:1996kr}
R.~Baier, Y.~L. Dokshitzer, A.~H. Mueller, S.~Peigne, and D.~Schiff,
  \emph{{Radiative energy loss of high energy quarks and gluons in a
  finite-volume quark-gluon plasma}}, Nucl. Phys. \textbf{B483}, 291 (1997)
  \eprint{hep-ph/9607355}.

\bibitem{Baier:1996sk}
R.~Baier, Y.~L. Dokshitzer, A.~H. Mueller, S.~Peigne, and D.~Schiff,
  \emph{{Radiative energy loss and p(T)-broadening of high energy partons in
  nuclei}}, Nucl. Phys. \textbf{B484}, 265 (1997) \eprint{hep-ph/9608322}.

\bibitem{Baier:1998yf}
R.~Baier, Y.~L. Dokshitzer, A.~H. Mueller, and D.~Schiff, \emph{{Radiative
  energy loss of high energy partons traversing an expanding {QCD} plasma}},
  Phys. Rev. \textbf{C58}, 1706 (1998) \eprint{hep-ph/9803473}.

\bibitem{Baier:1998kq}
R.~Baier, Y.~L. Dokshitzer, A.~H. Mueller, and D.~Schiff, \emph{{Medium-induced
  radiative energy loss: Equivalence between the BDMPS and Zakharov
  formalisms}}, Nucl. Phys. \textbf{B531}, 403 (1998) \eprint{hep-ph/9804212}.

\bibitem{Baier:1999ds}
R.~Baier, Y.~L. Dokshitzer, A.~H. Mueller, and D.~Schiff, \emph{{Angular
  dependence of the radiative gluon spectrum and the energy loss of hard jets
  in QCD media}}, Phys. Rev. \textbf{C60}, 064902 (1999)
  \eprint{hep-ph/9907267}.

\bibitem{Baier:2000mf}
R.~Baier, D.~Schiff, and B.~G. Zakharov, \emph{{Energy loss in perturbative
  QCD}}, Ann. Rev. Nucl. Part. Sci. \textbf{50}, 37 (2000)
  \eprint{hep-ph/0002198}.

\bibitem{Baier:2001qw}
R.~Baier, Y.~L. Dokshitzer, A.~H. Mueller, and D.~Schiff, \emph{{On the angular
  dependence of the radiative gluon spectrum}}, Phys. Rev. \textbf{C64}, 057902
  (2001) \eprint{hep-ph/0105062}.

\bibitem{Baier:2001yt}
R.~Baier, Y.~L. Dokshitzer, A.~H. Mueller, and D.~Schiff, \emph{{Quenching of
  hadron spectra in media}}, JHEP \textbf{09}, 033 (2001)
  \eprint{hep-ph/0106347}.

\bibitem{Zakharov:1996fv}
B.~G. Zakharov, \emph{{Fully quantum treatment of the Landau-Pomeranchuk-Migdal
  effect in QED and QCD}}, JETP Lett. \textbf{63}, 952 (1996)
  \eprint{hep-ph/9607440}.

\bibitem{Zakharov:1996cm}
B.~G. Zakharov, \emph{{Landau-Pomeranchuk-Migdal effect for finite-size
  targets}}, Pisma Zh. Eksp. Teor. Fiz. \textbf{64}, 737 (1996)
  \eprint{hep-ph/9612431}.

\bibitem{Zakharov:1997uu}
B.~G. Zakharov, \emph{{Radiative energy loss of high energy quarks in
  finite-size nuclear matter and quark-gluon plasma}}, JETP Lett. \textbf{65},
  615 (1997) \eprint{hep-ph/9704255}.

\bibitem{Zakharov:2002ik}
B.~G. Zakharov, \emph{{Coherent final state interaction in jet production in
  nucleus nucleus collisions}}, JETP Lett. \textbf{76}, 201 (2002)
  \eprint{hep-ph/0207206}.

\bibitem{Kovner:2001vi}
A.~Kovner and U.~A. Wiedemann, \emph{{Eikonal evolution and gluon radiation}},
  Phys. Rev. \textbf{D64}, 114002 (2001) \eprint{hep-ph/0106240}.

\bibitem{Wiedemann:2000ez}
U.~A. Wiedemann, \emph{{Transverse dynamics of hard partons in nuclear media
  and the QCD dipole}}, Nucl. Phys. \textbf{B582}, 409 (2000)
  \eprint{hep-ph/0003021}.

\bibitem{Wiedemann:2000za}
U.~A. Wiedemann, \emph{{Gluon radiation off hard quarks in a nuclear
  environment: Opacity expansion}}, Nucl. Phys. \textbf{B588}, 303 (2000)
  \eprint{hep-ph/0005129}.

\bibitem{Wiedemann:2000tf}
U.~A. Wiedemann, \emph{{Jet quenching versus jet enhancement: A quantitative
  study of the BDMPS-Z gluon radiation spectrum}}, Nucl. Phys. \textbf{A690},
  731 (2001) \eprint{hep-ph/0008241}.

\bibitem{Salgado:2003gb}
C.~A. Salgado and U.~A. Wiedemann, \emph{{Calculating quenching weights}},
  Phys. Rev. \textbf{D68}, 014008 (2003) \eprint{hep-ph/0302184}.

\bibitem{Armesto:2003jh}
N.~Armesto, C.~A. Salgado, and U.~A. Wiedemann, \emph{{Medium-induced gluon
  radiation off massive quarks fills the dead cone}}, Phys. Rev. \textbf{D69},
  114003 (2004) \eprint{hep-ph/0312106}.

\bibitem{Gyulassy:1999ig}
M.~Gyulassy, P.~Levai, and I.~Vitev, \emph{{Jet quenching in thin plasmas}},
  Nucl. Phys. \textbf{A661}, 637 (1999) \eprint{hep-ph/9907343}.

\bibitem{Gyulassy:1999zd}
M.~Gyulassy, P.~Levai, and I.~Vitev, \emph{{Jet quenching in thin quark-gluon
  plasmas. I: Formalism}}, Nucl. Phys. \textbf{B571}, 197 (2000)
  \eprint{hep-ph/9907461}.

\bibitem{Gyulassy:2000fs}
M.~Gyulassy, P.~Levai, and I.~Vitev, \emph{{Non-Abelian energy loss at finite
  opacity}}, Phys. Rev. Lett. \textbf{85}, 5535 (2000)
  \eprint{nucl-th/0005032}.

\bibitem{Gyulassy:2000er}
M.~Gyulassy, P.~Levai, and I.~Vitev, \emph{{Reaction operator approach to
  non-Abelian energy loss}}, Nucl. Phys. \textbf{B594}, 371 (2001)
  \eprint{nucl-th/0006010}.

\bibitem{Gyulassy:2001nm}
M.~Gyulassy, P.~Levai, and I.~Vitev, \emph{{Jet tomography of Au + Au reactions
  including multi-gluon fluctuations}}, Phys. Lett. \textbf{B538}, 282 (2002)
  \eprint{nucl-th/0112071}.

\bibitem{Gyulassy:2002yv}
M.~Gyulassy, P.~Levai, and I.~Vitev, \emph{{Reaction operator approach to
  multiple elastic scatterings}}, Phys. Rev. \textbf{D66}, 014005 (2002)
  \eprint{nucl-th/0201078}.

\bibitem{Vitev:2002pf}
I.~Vitev and M.~Gyulassy, \emph{{High-p(T) tomography of d + Au and Au + Au at
  SPS, RHIC, and LHC}}, Phys. Rev. Lett. \textbf{89}, 252301 (2002)
  \eprint{hep-ph/0209161}.

\bibitem{Djordjevic:2003zk}
M.~Djordjevic and M.~Gyulassy, \emph{{Heavy quark radiative energy loss in QCD
  matter}}, Nucl. Phys. \textbf{A733}, 265 (2004) \eprint{nucl-th/0310076}.

\bibitem{Djordjevic:2008iz}
M.~Djordjevic and U.~W. Heinz, \emph{{Radiative energy loss in a finite
  dynamical QCD medium}}, Phys. Rev. Lett. \textbf{101}, 022302 (2008)
  \eprint{0802.1230}.

\bibitem{Wang:2000uj}
X.-N. Wang, \emph{{Dynamical screening and radiative parton energy loss in a
  quark-gluon plasma}}, Phys. Lett. \textbf{B485}, 157 (2000)
  \eprint{nucl-th/0003033}.

\bibitem{Guo:2000nz}
X.-f. Guo and X.-N. Wang, \emph{{Multiple scattering, parton energy loss and
  modified fragmentation functions in deeply inelastic e A scattering}}, Phys.
  Rev. Lett. \textbf{85}, 3591 (2000) \eprint{hep-ph/0005044}.

\bibitem{Wang:2001ifa}
X.-N. Wang and X.-f. Guo, \emph{{Multiple parton scattering in nuclei: Parton
  energy loss}}, Nucl. Phys. \textbf{A696}, 788 (2001) \eprint{hep-ph/0102230}.

\bibitem{Wang:2001cs}
E.~Wang and X.-N. Wang, \emph{{Parton energy loss with detailed balance}},
  Phys. Rev. Lett. \textbf{87}, 142301 (2001) \eprint{nucl-th/0106043}.

\bibitem{Osborne:2002dx}
J.~A. Osborne, E.~Wang, and X.-N. Wang, \emph{{Evolution of parton
  fragmentation functions at finite temperature}}, Phys. Rev. \textbf{D67},
  094022 (2003) \eprint{hep-ph/0212131}.

\bibitem{Zhang:2003wk}
B.-W. Zhang, E.~Wang, and X.-N. Wang, \emph{{Heavy quark energy loss in nuclear
  medium}}, Phys. Rev. Lett. \textbf{93}, 072301 (2004)
  \eprint{nucl-th/0309040}.

\bibitem{Arnold:2002ja}
P.~Arnold, G.~D. Moore, and L.~G. Yaffe, \emph{{Photon and gluon emission in
  relativistic plasmas}}, JHEP \textbf{06}, 030 (2002) \eprint{hep-ph/0204343}.

\bibitem{Turbide:2005fk}
S.~Turbide, C.~Gale, S.~Jeon, and G.~D. Moore, \emph{{Energy loss of leading
  hadrons and direct photon production in evolving quark-gluon plasma}}, Phys.
  Rev. \textbf{C72}, 014906 (2005) \eprint{hep-ph/0502248}.

\bibitem{Gunion:1981qs}
J.~F. Gunion and G.~Bertsch, \emph{{HADRONIZATION BY COLOR BREMSSTRAHLUNG}},
  Phys. Rev. \textbf{D25}, 746 (1982).

\bibitem{Landau:1953um}
L.~D. Landau and I.~Pomeranchuk, \emph{{Limits of applicability of the theory
  of bremsstrahlung electrons and pair production at high-energies}}, Dokl.
  Akad. Nauk Ser. Fiz. \textbf{92}, 535 (1953).

\bibitem{Migdal:1956tc}
A.~B. Migdal, \emph{{Bremsstrahlung and pair production in condensed media at
  high-energies}}, Phys. Rev. \textbf{103}, 1811 (1956).

\bibitem{Brodsky:1992nq}
S.~J. Brodsky and P.~Hoyer, \emph{{A Bound on the energy loss of partons in
  nuclei}}, Phys. Lett. \textbf{B298}, 165 (1993) \eprint{hep-ph/9210262}.

\bibitem{Gyulassy:1993hr}
M.~Gyulassy and X.-n. Wang, \emph{{Multiple collisions and induced gluon
  Bremsstrahlung in QCD}}, Nucl. Phys. \textbf{B420}, 583 (1994)
  \eprint{nucl-th/9306003}.

\bibitem{Moliere:1947}
G.~Z. Moli\`{e}re, \emph{{Theorie der Streuung schneller geladener Teilchen. I.
  Einzelstreuung am abgeschirmten Coulomb-Field}}, Z. Naturforsch \textbf{2a},
  133 (1947).

\bibitem{Moliere:1948}
G.~Z. Moli\`{e}re, \emph{{Theorie der Streuung schneller geladener Teilchen.
  II. Mehrfach- und Vielfachstreuung}}, Z. Naturforsch \textbf{3a}, 78 (1948).

\bibitem{Bethe:1953va}
H.~A. Bethe, \emph{{Moli\`{e}re's theory of multiple scattering}}, Phys. Rev.
  \textbf{89}, 1256 (1953).

\bibitem{Dainese:2004te}
A.~Dainese, C.~Loizides, and G.~Paic, \emph{{Leading-particle suppression in
  high energy nucleus nucleus collisions}}, Eur. Phys. J. \textbf{C38}, 461
  (2005) \eprint{hep-ph/0406201}.

\bibitem{Gyulassy:2003mc}
M.~Gyulassy, I.~Vitev, X.-N. Wang, and B.-W. Zhang, \emph{{Jet quenching and
  radiative energy loss in dense nuclear matter}}  (2003)
  \eprint{nucl-th/0302077}.

\bibitem{Dokshitzer:2001zm}
Y.~L. Dokshitzer and D.~E. Kharzeev, \emph{{Heavy quark colorimetry of QCD
  matter}}, Phys. Lett. \textbf{B519}, 199 (2001) \eprint{hep-ph/0106202}.

\bibitem{Jeon:2003gi}
S.~Jeon and G.~D. Moore, \emph{{Energy loss of leading partons in a thermal QCD
  medium}}, Phys. Rev. \textbf{C71}, 034901 (2005) \eprint{hep-ph/0309332}.

\bibitem{Baier:1998ej}
V.~N. Baier and V.~M. Katkov, \emph{{Quantum theory of transition radiation and
  transition pair creation}}, Phys. Lett. \textbf{A252}, 263 (1999)
  \eprint{hep-ph/9811302}.

\bibitem{Schildknecht:2005sc}
D.~Schildknecht and B.~G. Zakharov, \emph{{Transition radiation in quantum
  regime as a diffractive phenomenon}}  (2005) \eprint{hep-ph/0501241}.

\bibitem{Djordjevic:2005nh}
M.~Djordjevic, \emph{{Transition radiation in QCD matter}}, Phys. Rev.
  \textbf{C73}, 044912 (2006) \eprint{nucl-th/0512089}.

\bibitem{Ter-Mikayelian:1954}
M.~L. Ter-Mikayelian, Dokl. Akad. Nauk SSSR \textbf{94}, 1033 (1954).

\bibitem{Ter-Mikayelian:1972}
M.~L. Ter-Mikayelian, \emph{{High-Energy Electromagnetic Processes in Condensed
  Media}} (John Wiley \& Sons, 1972).

\bibitem{Djordjevic:2003be}
M.~Djordjevic and M.~Gyulassy, \emph{{The Ter-Mikayelian effect on QCD
  radiative energy loss}}, Phys. Rev. \textbf{C68}, 034914 (2003)
  \eprint{nucl-th/0305062}.

\bibitem{Eskola:2004cr}
K.~J. Eskola, H.~Honkanen, C.~A. Salgado, and U.~A. Wiedemann, \emph{{The
  fragility of high-p(T) hadron spectra as a hard probe}}, Nucl. Phys.
  \textbf{A747}, 511 (2005) \eprint{hep-ph/0406319}.

\bibitem{Lai:2008zp}
Y.-S. Lai and B.~A. Cole, \emph{{Jet reconstruction in hadronic collisions by
  Gaussian filtering}}  (2008) \eprint{0806.1499}.

\bibitem{Salur:2008}
S.~Salur, \emph{{Direct measurement of jets in $\sqrt{s_{NN}}=200$ GeV Heavy
  Ion Collisions by STAR}}, in \emph{{Hard Probes}} (2008).

\bibitem{Putschke:2008}
J.~Putschke, \emph{{Modified fragmentation measurements with full jet
  reconstruction in heavy ion collisions at $\sqrt{s_{NN}}=200$ GeV by STAR}},
  in \emph{{Hard Probes}} (2008).

\bibitem{Adare:2008cg}
A.~Adare et~al. (PHENIX), \emph{{Quantitative Constraints on the Opacity of Hot
  Partonic Matter from Semi-Inclusive Single High Transverse Momentum Pion
  Suppression in Au+Au collisions at sqrt(s\_NN) = 200 GeV}}, Phys. Rev.
  \textbf{C77}, 064907 (2008) \eprint{0801.1665}.

\bibitem{Adare:2008qa}
A.~Adare et~al. (PHENIX), \emph{{Suppression pattern of neutral pions at high
  transverse momentum in Au+Au collisions at sqrt(s\_NN) = 200 GeV and
  constraints on medium transport coefficients}}  (2008) \eprint{0801.4020}.

\bibitem{Maldacena:1997re}
J.~M. Maldacena, \emph{{The large N limit of superconformal field theories and
  supergravity}}, Adv. Theor. Math. Phys. \textbf{2}, 231 (1998)
  \eprint{hep-th/9711200}.

\bibitem{Schwarz:2007yc}
J.~H. Schwarz, \emph{{The Early Years of String Theory: A Personal
  Perspective}}  (2007) \eprint{0708.1917}.

\bibitem{Irving:1977ea}
A.~C. Irving and R.~P. Worden, \emph{{Regge Phenomenology}}, Phys. Rept.
  \textbf{34}, 117 (1977).

\bibitem{Veneziano:1968yb}
G.~Veneziano, \emph{{Construction of a crossing - symmetric, Regge behaved
  amplitude for linearly rising trajectories}}, Nuovo. Cim. \textbf{A57}, 190
  (1968).

\bibitem{Virasoro:1969me}
M.~A. Virasoro, \emph{{Alternative constructions of crossing-symmetric
  amplitudes with regge behavior}}, Phys. Rev. \textbf{177}, 2309 (1969).

\bibitem{Virasoro:1969zu}
M.~S. Virasoro, \emph{{Subsidiary conditions and ghosts in dual resonance
  models}}, Phys. Rev. \textbf{D1}, 2933 (1970).

\bibitem{Bigatti:1999dp}
D.~Bigatti and L.~Susskind, \emph{{TASI lectures on the holographic principle}}
   (1999) \eprint{hep-th/0002044}.

\bibitem{Carroll:2004}
S.~M. Carroll, \emph{Spacetime and Geometry: An Introduction to General
  Relativity} (Addison Wesley, 2004), ISBN 0-8053-8732-3.

\bibitem{Ortin:2004}
T.~Ort\'in, \emph{{Gravity and Strings}} (Cambridge University Press, 2004).

\bibitem{Horowitz:1991cd}
G.~T. Horowitz and A.~Strominger, \emph{{Black strings and P-branes}}, Nucl.
  Phys. \textbf{B360}, 197 (1991).

\bibitem{Aharony:1999ti}
O.~Aharony, S.~S. Gubser, J.~M. Maldacena, H.~Ooguri, and Y.~Oz, \emph{{Large N
  field theories, string theory and gravity}}, Phys. Rept. \textbf{323}, 183
  (2000) \eprint{hep-th/9905111}.

\bibitem{Polchinski:1996na}
J.~Polchinski, \emph{{Lectures on D-branes}}  (1996) \eprint{hep-th/9611050}.

\bibitem{Green:1987}
M.~B. Green, J.~H. Schwarz, and E.~Witten, \emph{{Superstring theory}}, vol.~1
  (Cambridge University Press, 1987).

\bibitem{Leigh:1989jq}
R.~G. Leigh, \emph{{Dirac-Born-Infeld Action from Dirichlet Sigma Model}}, Mod.
  Phys. Lett. \textbf{A4}, 2767 (1989).

\bibitem{Seiberg:1994aj}
N.~Seiberg and E.~Witten, \emph{{Monopoles, duality and chiral symmetry
  breaking in N=2 supersymmetric QCD}}, Nucl. Phys. \textbf{B431}, 484 (1994)
  \eprint{hep-th/9408099}.

\bibitem{Dine:1997nq}
M.~Dine and N.~Seiberg, \emph{{Comments on higher derivative operators in some
  SUSY field theories}}, Phys. Lett. \textbf{B409}, 239 (1997)
  \eprint{hep-th/9705057}.

\bibitem{Douglas:1996yp}
M.~R. Douglas, D.~N. Kabat, P.~Pouliot, and S.~H. Shenker, \emph{{D-branes and
  short distances in string theory}}, Nucl. Phys. \textbf{B485}, 85 (1997)
  \eprint{hep-th/9608024}.

\bibitem{Klebanov:1997kc}
I.~R. Klebanov, \emph{{World-volume approach to absorption by non-dilatonic
  branes}}, Nucl. Phys. \textbf{B496}, 231 (1997) \eprint{hep-th/9702076}.

\bibitem{Klebanov:2000me}
I.~R. Klebanov, \emph{{TASI lectures: Introduction to the AdS/CFT
  correspondence}}  (2000) \eprint{hep-th/0009139}.

\bibitem{Johnson:2003}
C.~V. Johnson, \emph{D-branes} (Cambridge University Press, 2003).

\bibitem{Gubser:1996de}
S.~S. Gubser, I.~R. Klebanov, and A.~W. Peet, \emph{{Entropy and Temperature of
  Black 3-Branes}}, Phys. Rev. \textbf{D54}, 3915 (1996)
  \eprint{hep-th/9602135}.

\bibitem{Kovtun:2003wp}
P.~Kovtun, D.~T. Son, and A.~O. Starinets, \emph{{Holography and hydrodynamics:
  Diffusion on stretched horizons}}, JHEP \textbf{10}, 064 (2003)
  \eprint{hep-th/0309213}.

\bibitem{Das:1996we}
S.~R. Das, G.~W. Gibbons, and S.~D. Mathur, \emph{{Universality of low energy
  absorption cross sections for black holes}}, Phys. Rev. Lett. \textbf{78},
  417 (1997) \eprint{hep-th/9609052}.

\bibitem{Kovtun:2004de}
P.~Kovtun, D.~T. Son, and A.~O. Starinets, \emph{{Viscosity in strongly
  interacting quantum field theories from black hole physics}}, Phys. Rev.
  Lett. \textbf{94}, 111601 (2005) \eprint{hep-th/0405231}.

\bibitem{Karch:2002sh}
A.~Karch and E.~Katz, \emph{{Adding flavor to AdS/CFT}}, JHEP \textbf{06}, 043
  (2002) \eprint{hep-th/0205236}.

\bibitem{Bak:2007fk}
D.~Bak, A.~Karch, and L.~G. Yaffe, \emph{{Debye screening in strongly coupled
  N=4 supersymmetric Yang-Mills plasma}}, JHEP \textbf{08}, 049 (2007)
  \eprint{0705.0994}.

\bibitem{Liu:2006nn}
H.~Liu, K.~Rajagopal, and U.~A. Wiedemann, \emph{{An AdS/CFT calculation of
  screening in a hot wind}}, Phys. Rev. Lett. \textbf{98}, 182301 (2007)
  \eprint{hep-ph/0607062}.

\bibitem{Ejaz:2007hg}
Q.~J. Ejaz, T.~Faulkner, H.~Liu, K.~Rajagopal, and U.~A. Wiedemann, \emph{{A
  limiting velocity for quarkonium propagation in a strongly coupled plasma via
  AdS/CFT}}, JHEP \textbf{04}, 089 (2008) \eprint{0712.0590}.

\bibitem{Friess:2006fk}
J.~J. Friess, S.~S. Gubser, G.~Michalogiorgakis, and S.~S. Pufu, \emph{{The
  stress tensor of a quark moving through N = 4 thermal plasma}}, Phys. Rev.
  \textbf{D75}, 106003 (2007) \eprint{hep-th/0607022}.

\bibitem{Chesler:2007an}
P.~M. Chesler and L.~G. Yaffe, \emph{{The wake of a quark moving through a
  strongly-coupled $\mathcal N=4$ supersymmetric Yang-Mills plasma}}, Phys.
  Rev. Lett. \textbf{99}, 152001 (2007) \eprint{0706.0368}.

\bibitem{Gubser:2006bz}
S.~S. Gubser, \emph{{Drag force in AdS/CFT}}, Phys. Rev. \textbf{D74}, 126005
  (2006) \eprint{hep-th/0605182}.

\bibitem{Herzog:2006gh}
C.~P. Herzog, A.~Karch, P.~Kovtun, C.~Kozcaz, and L.~G. Yaffe, \emph{{Energy
  loss of a heavy quark moving through N = 4 supersymmetric Yang-Mills
  plasma}}, JHEP \textbf{07}, 013 (2006) \eprint{hep-th/0605158}.

\bibitem{VazquezPoritz:2008nw}
J.~F. Vazquez-Poritz, \emph{{Drag force at finite 't Hooft coupling from
  AdS/CFT}}  (2008) \eprint{0803.2890}.

\bibitem{Liu:2006ug}
H.~Liu, K.~Rajagopal, and U.~A. Wiedemann, \emph{{Calculating the jet quenching
  parameter from AdS/CFT}}, Phys. Rev. Lett. \textbf{97}, 182301 (2006)
  \eprint{hep-ph/0605178}.

\bibitem{Argyres:2006yz}
P.~C. Argyres, M.~Edalati, and J.~F. Vazquez-Poritz, \emph{{Spacelike strings
  and jet quenching from a Wilson loop}}, JHEP \textbf{04}, 049 (2007)
  \eprint{hep-th/0612157}.

\bibitem{Gubser:2006nz}
S.~S. Gubser, \emph{{Momentum fluctuations of heavy quarks in the gauge-string
  duality}}, Nucl. Phys. \textbf{B790}, 175 (2008) \eprint{hep-th/0612143}.

\bibitem{Liu:2006he}
H.~Liu, K.~Rajagopal, and U.~A. Wiedemann, \emph{{Wilson loops in heavy ion
  collisions and their calculation in AdS/CFT}}, JHEP \textbf{03}, 066 (2007)
  \eprint{hep-ph/0612168}.

\bibitem{Bertoldi:2007sf}
G.~Bertoldi, F.~Bigazzi, A.~L. Cotrone, and J.~D. Edelstein, \emph{{Holography
  and Unquenched Quark-Gluon Plasmas}}, Phys. Rev. \textbf{D76}, 065007 (2007)
  \eprint{hep-th/0702225}.

\bibitem{Argyres:2008eg}
P.~C. Argyres, M.~Edalati, and J.~F. Vazquez-Poritz, \emph{{Lightlike Wilson
  loops from AdS/CFT}}, JHEP \textbf{03}, 071 (2008) \eprint{0801.4594}.

\bibitem{CasalderreySolana:2006rq}
J.~Casalderrey-Solana and D.~Teaney, \emph{{Heavy quark diffusion in strongly
  coupled N = 4 Yang Mills}}, Phys. Rev. \textbf{D74}, 085012 (2006)
  \eprint{hep-ph/0605199}.

\bibitem{CasalderreySolana:2007qw}
J.~Casalderrey-Solana and D.~Teaney, \emph{{Transverse momentum broadening of a
  fast quark in a N = 4 Yang Mills plasma}}, JHEP \textbf{04}, 039 (2007)
  \eprint{hep-th/0701123}.

\bibitem{Noronha:2007xe}
J.~Noronha, G.~Torrieri, and M.~Gyulassy, \emph{{Near Zone Navier-Stokes
  Analysis of Heavy Quark Jet Quenching in an $\mathcal{N}$ =4 SYM Plasma}}
  (2007) \eprint{0712.1053}.

\bibitem{Betz:2008wy}
B.~Betz, M.~Gyulassy, J.~Noronha, and G.~Torrieri, \emph{{Anomalous Conical
  Di-jet Correlations in pQCD vs AdS/CFT}}  (2008) \eprint{0807.4526}.

\bibitem{Neufeld:2008fi}
R.~B. Neufeld, B.~Muller, and J.~Ruppert, \emph{{Sonic Mach Cones Induced by
  Fast Partons in a Perturbative Quark-Gluon Plasma}}  (2008)
  \eprint{0802.2254}.

\bibitem{Vitev:2005ch}
I.~Vitev, \emph{{Jet quenching in relativistic heavy ion collisions}}, J. Phys.
  Conf. Ser. \textbf{50}, 119 (2006) \eprint{hep-ph/0503221}.

\bibitem{Shuryak:2001me}
E.~V. Shuryak, \emph{{The azimuthal asymmetry at large p(t) seems to be too
  large for a 'jet quenching'}}, Phys. Rev. \textbf{C66}, 027902 (2002)
  \eprint{nucl-th/0112042}.

\bibitem{d'Enterria:2005cs}
D.~G. d'Enterria, \emph{{High p(T) leading hadron suppression in nuclear
  collisions at s(NN)**(1/2) = 20-GeV - 200-GeV: Data versus parton energy loss
  models}}, Eur. Phys. J. \textbf{C43}, 295 (2005) \eprint{nucl-ex/0504001}.

\bibitem{Gyulassy:2000gk}
M.~Gyulassy, I.~Vitev, and X.~N. Wang, \emph{{High p(T) azimuthal asymmetry in
  noncentral A + A at RHIC}}, Phys. Rev. Lett. \textbf{86}, 2537 (2001)
  \eprint{nucl-th/0012092}.

\bibitem{Wang:2003mm}
X.-N. Wang, \emph{{High p(T) hadron spectra, azimuthal anisotropy and back-
  to-back correlations in high-energy heavy-ion collisions}}, Phys. Lett.
  \textbf{B595}, 165 (2004) \eprint{nucl-th/0305010}.

\bibitem{Armesto:2004vz}
N.~Armesto, C.~A. Salgado, and U.~A. Wiedemann, \emph{{Low-p(T) collective flow
  induces high-p(T) jet quenching}}, Phys. Rev. \textbf{C72}, 064910 (2005)
  \eprint{hep-ph/0411341}.

\bibitem{Adams:2003kv}
J.~Adams et~al. (STAR), \emph{{Transverse momentum and collision energy
  dependence of high p(T) hadron suppression in Au + Au collisions at
  ultrarelativistic energies}}, Phys. Rev. Lett. \textbf{91}, 172302 (2003)
  \eprint{nucl-ex/0305015}.

\bibitem{Adams:2004bi}
J.~Adams et~al. (STAR), \emph{{Azimuthal anisotropy in Au + Au collisions at
  s(NN)**(1/2) = 200-GeV}}, Phys. Rev. \textbf{C72}, 014904 (2005)
  \eprint{nucl-ex/0409033}.

\bibitem{Adler:2003au}
S.~S. Adler et~al. (PHENIX), \emph{{High-p(T) charged hadron suppression in Au
  + Au collisions at s(NN)**(1/2) = 200-GeV}}, Phys. Rev. \textbf{C69}, 034910
  (2004) \eprint{nucl-ex/0308006}.

\bibitem{Cole:2005yv}
B.~A. Cole et~al. (PHENIX), \emph{{Differential probes of medium-induced energy
  loss}}, Eur. Phys. J. \textbf{C43}, 271 (2005).

\bibitem{Glauber:1970jm}
R.~J. Glauber and G.~Matthiae, \emph{{High-energy scattering of protons by
  nuclei}}, Nucl. Phys. \textbf{B21}, 135 (1970).

\bibitem{Adler:2006bw}
S.~S. Adler et~al. (PHENIX), \emph{{A detailed study of high-p(T) neutral pion
  suppression and azimuthal anisotropy in Au + Au collisions at s(NN)**(1/2) =
  200-GeV}}, Phys. Rev. \textbf{C76}, 034904 (2007) \eprint{nucl-ex/0611007}.

\bibitem{Isobe:2005mh}
T.~Isobe, \emph{{Measurement of neutral pions in s(NN)**(1/2) = 200-GeV and
  62.4-GeV Au + Au collisions at RHIC-PHENIX}}, Acta Phys. Hung. \textbf{A27},
  227 (2006) \eprint{nucl-ex/0510085}.

\bibitem{Shimomura:2005en}
M.~Shimomura (PHENIX), \emph{{High-p(T) pi0, eta, identified and inclusive
  charged hadron spectra from PHENIX}}, Nucl. Phys. \textbf{A774}, 457 (2006)
  \eprint{nucl-ex/0510023}.

\bibitem{Adler:2003qi}
S.~S. Adler et~al. (PHENIX), \emph{{Suppressed pi0 production at large
  transverse momentum in central Au + Au collisions at s(NN)**(1/2) =
  200-GeV}}, Phys. Rev. Lett. \textbf{91}, 072301 (2003)
  \eprint{nucl-ex/0304022}.

\bibitem{Kovner:2003zj}
A.~Kovner and U.~A. Wiedemann, \emph{{Gluon radiation and parton energy loss}}
  (2003) \eprint{hep-ph/0304151}.

\bibitem{Jacobs:2004qv}
P.~Jacobs and X.-N. Wang, \emph{{Matter in extremis: Ultrarelativistic nuclear
  collisions at RHIC}}, Prog. Part. Nucl. Phys. \textbf{54}, 443 (2005)
  \eprint{hep-ph/0405125}.

\bibitem{Wang:1991xy}
X.-N. Wang and M.~Gyulassy, \emph{{Gluon shadowing and jet quenching in A + A
  collisions at s**(1/2) = 200-GeV}}, Phys. Rev. Lett. \textbf{68}, 1480
  (1992).

\bibitem{Djordjevic:2005db}
M.~Djordjevic, M.~Gyulassy, R.~Vogt, and S.~Wicks, \emph{{Influence of bottom
  quark jet quenching on single electron tomography of Au + Au}}, Phys. Lett.
  \textbf{B632}, 81 (2006) \eprint{nucl-th/0507019}.

\bibitem{BNLPress}
B.~P. Release, \emph{RHIC Scientists Serve Up ``Perfect'' Liquid},
  \url{http://www.bnl.gov/bnlweb/pubaf/pr/PR_display.asp?prID=05-38} (2005).

\bibitem{Hirano:2005wx}
T.~Hirano and M.~Gyulassy, \emph{{Perfect fluidity of the quark gluon plasma
  core as seen through its dissipative hadronic corona}}, Nucl. Phys.
  \textbf{A769}, 71 (2006) \eprint{nucl-th/0506049}.

\bibitem{Winter:2005nw}
D.~Winter (PHENIX), \emph{{PHENIX measurement of particle yields at high p(T)
  with respect to reaction plane in Au + Au collisions at s**(1/2) = 200-GeV}},
  Nucl. Phys. \textbf{A774}, 545 (2006) \eprint{nucl-ex/0511039}.

\bibitem{Rapp:2005at}
R.~Rapp, V.~Greco, and H.~van Hees, \emph{{Heavy-quark spectra at RHIC and
  resonances in the QGP}}, Nucl. Phys. \textbf{A774}, 685 (2006)
  \eprint{hep-ph/0510050}.

\bibitem{Armesto:2005iq}
N.~Armesto, A.~Dainese, C.~A. Salgado, and U.~A. Wiedemann, \emph{{Testing the
  color charge and mass dependence of parton energy loss with heavy-to-light
  ratios at RHIC and LHC}}, Phys. Rev. \textbf{D71}, 054027 (2005)
  \eprint{hep-ph/0501225}.

\bibitem{Baier:2002tc}
R.~Baier, \emph{{Jet quenching}}, Nucl. Phys. \textbf{A715}, 209 (2003)
  \eprint{hep-ph/0209038}.

\bibitem{Cacciari:2005rk}
M.~Cacciari, P.~Nason, and R.~Vogt, \emph{{QCD predictions for charm and bottom
  production at RHIC}}, Phys. Rev. Lett. \textbf{95}, 122001 (2005)
  \eprint{hep-ph/0502203}.

\bibitem{Mustafa:2003vh}
M.~G. Mustafa and M.~H. Thoma, \emph{{Quenching of hadron spectra due to the
  collisional energy loss of partons in the quark gluon plasma}}, Acta Phys.
  Hung. \textbf{A22}, 93 (2005) \eprint{hep-ph/0311168}.

\bibitem{DuttMazumder:2004xk}
A.~K. Dutt-Mazumder, J.-e. Alam, P.~Roy, and B.~Sinha, \emph{{Stopping power of
  hot QCD plasma}}, Phys. Rev. \textbf{D71}, 094016 (2005)
  \eprint{hep-ph/0411015}.

\bibitem{Mustafa:1997pm}
M.~G. Mustafa, D.~Pal, D.~K. Srivastava, and M.~Thoma, \emph{{Radiative
  energy-loss of heavy quarks in a quark-gluon plasma}}, Phys. Lett.
  \textbf{B428}, 234 (1998) \eprint{nucl-th/9711059}.

\bibitem{Lin:1997cn}
Z.-w. Lin, R.~Vogt, and X.-N. Wang, \emph{{Energy loss effects on charm and
  bottom production in high-energy heavy-ion collisions}}, Phys. Rev.
  \textbf{C57}, 899 (1998) \eprint{nucl-th/9705006}.

\bibitem{Djordjevic:2003qk}
M.~Djordjevic and M.~Gyulassy, \emph{{Where is the charm quark energy loss at
  RHIC?}}, Phys. Lett. \textbf{B560}, 37 (2003) \eprint{nucl-th/0302069}.

\bibitem{Djordjevic:2004nq}
M.~Djordjevic, M.~Gyulassy, and S.~Wicks, \emph{{The charm and beauty of RHIC
  and LHC}}, Phys. Rev. Lett. \textbf{94}, 112301 (2005)
  \eprint{hep-ph/0410372}.

\bibitem{Mangano:1991jk}
M.~L. Mangano, P.~Nason, and G.~Ridolfi, \emph{{Heavy quark correlations in
  hadron collisions at next-to- leading order}}, Nucl. Phys. \textbf{B373}, 295
  (1992).

\bibitem{Vogt:2001nh}
R.~Vogt (Hard Probe), \emph{{The A dependence of open charm and bottom
  production}}, Int. J. Mod. Phys. \textbf{E12}, 211 (2003)
  \eprint{hep-ph/0111271}.

\bibitem{Stump:2003yu}
D.~Stump et~al., \emph{{Inclusive jet production, parton distributions, and the
  search for new physics}}, JHEP \textbf{10}, 046 (2003)
  \eprint{hep-ph/0303013}.

\bibitem{Vitev:2003xu}
I.~Vitev, \emph{{Initial state parton broadening and energy loss probed in d +
  Au at RHIC}}, Phys. Lett. \textbf{B562}, 36 (2003) \eprint{nucl-th/0302002}.

\bibitem{Vitev:2004kd}
I.~Vitev, \emph{{The perturbative QCD factorization approach in high energy
  nuclear collisions}}, J. Phys. \textbf{G31}, S557 (2005)
  \eprint{hep-ph/0409297}.

\bibitem{Qiu:2005ki}
J.-W. Qiu, \emph{{Rescattering effects in hadron nucleus and heavy-ion
  collisions}}, Eur. Phys. J. \textbf{C43}, 239 (2005) \eprint{hep-ph/0507268}.

\bibitem{Hahn}
B.~Hahn, D.~G. Ravenhall, and R.~Hofstadter, \emph{{High-Energy Electron
  Scattering and the Charge Distributions of Selected Nuclei}}, Phys. Rev.
  \textbf{101}, 1131 (1956).

\bibitem{Dainese:2005kb}
A.~Dainese, C.~Loizides, and G.~Paic, \emph{{Leading-particle suppression and
  surface emission in nucleus nucleus collisions}}, Acta Phys. Hung.
  \textbf{A27}, 245 (2006) \eprint{hep-ph/0511045}.

\bibitem{Renk:2006pk}
T.~Renk and K.~Eskola, \emph{{Prospects of medium tomography using back-to-back
  hadron correlations}}, Phys. Rev. \textbf{C75}, 054910 (2007)
  \eprint{hep-ph/0610059}.

\bibitem{Vitev:2005he}
I.~Vitev, \emph{{Testing the theory of QGP-induced energy loss at RHIC and the
  LHC}}, Phys. Lett. \textbf{B639}, 38 (2006) \eprint{hep-ph/0603010}.

\bibitem{Back:2001ae}
B.~B. Back et~al. (PHOBOS), \emph{{Energy dependence of particle multiplicities
  in central Au + Au collisions}}, Phys. Rev. Lett. \textbf{88}, 022302 (2002)
  \eprint{nucl-ex/0108009}.

\bibitem{Adcox:2000sp}
K.~Adcox et~al. (PHENIX), \emph{{Centrality dependence of charged particle
  multiplicity in Au Au collisions at s(N N)**(1/2) = 130-GeV}}, Phys. Rev.
  Lett. \textbf{86}, 3500 (2001) \eprint{nucl-ex/0012008}.

\bibitem{Kharzeev:2004if}
D.~Kharzeev, E.~Levin, and M.~Nardi, \emph{{Color glass condensate at the LHC:
  Hadron multiplicities in p p, p A and A A collisions}}, Nucl. Phys.
  \textbf{A747}, 609 (2005) \eprint{hep-ph/0408050}.

\bibitem{McLerran:2004fg}
L.~McLerran, \emph{{What is the evidence for the color glass condensate?}}
  (2004) \eprint{hep-ph/0402137}.

\bibitem{Eskola:1999fc}
K.~J. Eskola, K.~Kajantie, P.~V. Ruuskanen, and K.~Tuominen, \emph{{Scaling of
  transverse energies and multiplicities with atomic number and energy in
  ultrarelativistic nuclear collisions}}, Nucl. Phys. \textbf{B570}, 379 (2000)
  \eprint{hep-ph/9909456}.

\bibitem{Busza:2007ke}
W.~Busza, \emph{{Trends in multiparticle production and some 'predictions' for
  pp and PbPb collisions at LHC}}, J. Phys. \textbf{G35}, 044040 (2008)
  \eprint{0710.2293}.

\bibitem{Wang:2006qr}
X.-N. Wang, \emph{{Interference effect in elastic parton energy loss in a
  finite medium}}, Phys. Lett. \textbf{B650}, 213 (2007)
  \eprint{nucl-th/0604040}.

\bibitem{Riordan:2006df}
M.~Riordan and W.~A. Zajc, \emph{{The first few microseconds}}, Sci. Am.
  \textbf{294N5}, 24 (2006).

\bibitem{Kovtun:2006pf}
P.~Kovtun and A.~Starinets, \emph{{Thermal spectral functions of strongly
  coupled N = 4 supersymmetric Yang-Mills theory}}, Phys. Rev. Lett.
  \textbf{96}, 131601 (2006) \eprint{hep-th/0602059}.

\bibitem{Nakamura:2006ih}
S.~Nakamura and S.-J. Sin, \emph{{A holographic dual of hydrodynamics}}, JHEP
  \textbf{09}, 020 (2006) \eprint{hep-th/0607123}.

\bibitem{Sin:2006pv}
S.-J. Sin, S.~Nakamura, and S.~P. Kim, \emph{{Elliptic flow, Kasner universe
  and holographic dual of RHIC fireball}}, JHEP \textbf{12}, 075 (2006)
  \eprint{hep-th/0610113}.

\bibitem{Janik:2005zt}
R.~A. Janik and R.~B. Peschanski, \emph{{Asymptotic perfect fluid dynamics as a
  consequence of AdS/CFT}}, Phys. Rev. \textbf{D73}, 045013 (2006)
  \eprint{hep-th/0512162}.

\bibitem{Janik:2006gp}
R.~A. Janik and R.~B. Peschanski, \emph{{Gauge / gravity duality and
  thermalization of a boost- invariant perfect fluid}}, Phys. Rev.
  \textbf{D74}, 046007 (2006) \eprint{hep-th/0606149}.

\bibitem{Kajantie:2006ya}
K.~Kajantie and T.~Tahkokallio, \emph{{Spherically expanding matter in
  AdS/CFT}}, Phys. Rev. \textbf{D75}, 066003 (2007) \eprint{hep-th/0612226}.

\bibitem{Friess:2006kw}
J.~J. Friess, S.~S. Gubser, G.~Michalogiorgakis, and S.~S. Pufu,
  \emph{{Expanding plasmas and quasinormal modes of anti-de Sitter black
  holes}}, JHEP \textbf{04}, 080 (2007) \eprint{hep-th/0611005}.

\bibitem{Shuryak:2005ia}
E.~Shuryak, S.-J. Sin, and I.~Zahed, \emph{{A gravity dual of RHIC
  collisions}}, J. Korean Phys. Soc. \textbf{50}, 384 (2007)
  \eprint{hep-th/0511199}.

\bibitem{Heller:2007qt}
M.~P. Heller and R.~A. Janik, \emph{{Viscous hydrodynamics relaxation time from
  AdS/CFT}}, Phys. Rev. \textbf{D76}, 025027 (2007) \eprint{hep-th/0703243}.

\bibitem{Witten:1998qj}
E.~Witten, \emph{{Anti-de Sitter space and holography}}, Adv. Theor. Math.
  Phys. \textbf{2}, 253 (1998) \eprint{hep-th/9802150}.

\bibitem{Witten:1998zw}
E.~Witten, \emph{{Anti-de Sitter space, thermal phase transition, and
  confinement in gauge theories}}, Adv. Theor. Math. Phys. \textbf{2}, 505
  (1998) \eprint{hep-th/9803131}.

\bibitem{Ackermann:2000tr}
K.~H. Ackermann et~al. (STAR), \emph{{Elliptic flow in Au + Au collisions at
  s(N N)**(1/2) = 130-GeV}}, Phys. Rev. Lett. \textbf{86}, 402 (2001)
  \eprint{nucl-ex/0009011}.

\bibitem{Adler:2001nb}
C.~Adler et~al. (STAR), \emph{{Identified particle elliptic flow in Au + Au
  collisions at s(NN)**(1/2) = 130-GeV}}, Phys. Rev. Lett. \textbf{87}, 182301
  (2001) \eprint{nucl-ex/0107003}.

\bibitem{Adler:2002ct}
C.~Adler et~al. (STAR), \emph{{Azimuthal anisotropy and correlations in the
  hard scattering regime at RHIC}}, Phys. Rev. Lett. \textbf{90}, 032301 (2003)
  \eprint{nucl-ex/0206006}.

\bibitem{Adare:2006ti}
A.~Adare et~al. (PHENIX), \emph{{Scaling properties of azimuthal anisotropy in
  Au + Au and Cu + Cu collisions at s(NN)**(1/2) = 200-GeV}}, Phys. Rev. Lett.
  \textbf{98}, 162301 (2007) \eprint{nucl-ex/0608033}.

\bibitem{Teaney:2000cw}
D.~Teaney, J.~Lauret, and E.~V. Shuryak, \emph{{Flow at the SPS and RHIC as a
  quark gluon plasma signature}}, Phys. Rev. Lett. \textbf{86}, 4783 (2001)
  \eprint{nucl-th/0011058}.

\bibitem{Adler:2006hu}
S.~S. Adler et~al. (PHENIX), \emph{{Common suppression pattern of eta and pi0
  mesons at high transverse momentum in Au + Au collisions at s(NN)**(1/2) =
  200-GeV}}, Phys. Rev. Lett. \textbf{96}, 202301 (2006)
  \eprint{nucl-ex/0601037}.

\bibitem{Abelev:2006db}
B.~I. Abelev et~al. (STAR), \emph{{Transverse momentum and centrality
  dependence of high-pt non-photonic electron suppression in Au+Au collisions
  at $\sqrt{s_{NN}}$ = 200 GeV}}, Phys. Rev. Lett. \textbf{98}, 192301 (2007)
  \eprint{nucl-ex/0607012}.

\bibitem{Adare:2006nq}
A.~Adare et~al. (PHENIX), \emph{{Energy Loss and Flow of Heavy Quarks in Au+Au
  Collisions at sqrt(s\_NN) = 200 GeV}}, Phys. Rev. Lett. \textbf{98}, 172301
  (2007) \eprint{nucl-ex/0611018}.

\bibitem{Adler:2002tq}
C.~Adler et~al. (STAR), \emph{{Disappearance of back-to-back high p(T) hadron
  correlations in central Au + Au collisions at s(NN)**(1/2) = 200-GeV}}, Phys.
  Rev. Lett. \textbf{90}, 082302 (2003) \eprint{nucl-ex/0210033}.

\bibitem{Abelev:2006jr}
B.~I. Abelev et~al. (STAR), \emph{{Identified baryon and meson distributions at
  large transverse momenta from Au + Au collisions at s(NN)**(1/2) = 200-GeV}},
  Phys. Rev. Lett. \textbf{97}, 152301 (2006) \eprint{nucl-ex/0606003}.

\bibitem{Adams:2006yt}
J.~Adams et~al. (STAR), \emph{{Direct observation of dijets in central Au + Au
  collisions at s(NN)**(1/2) = 200-GeV}}, Phys. Rev. Lett. \textbf{97}, 162301
  (2006) \eprint{nucl-ex/0604018}.

\bibitem{Wang:2005cg}
F.~Wang, \emph{{Distributions of charged hadrons associated with high p(T)
  particles}}, J. Phys. Conf. Ser. \textbf{27}, 32 (2005)
  \eprint{nucl-ex/0508021}.

\bibitem{Ulery:2006iw}
J.~G. Ulery and F.~Wang, \emph{{Analysis method for jet-like three-particle
  azimuthal correlations}}  (2006) \eprint{nucl-ex/0609016}.

\bibitem{Adler:2005ee}
S.~S. Adler et~al. (PHENIX), \emph{{Modifications to di-jet hadron pair
  correlations in Au + Au collisions at s(NN)**(1/2) = 200-GeV}}, Phys. Rev.
  Lett. \textbf{97}, 052301 (2006) \eprint{nucl-ex/0507004}.

\bibitem{Adare:2006nr}
A.~Adare et~al. (PHENIX), \emph{{System size and energy dependence of
  jet-induced hadron pair correlation shapes in Cu + Cu and Au + Au collisions
  at s(NN)**(1/2) = 200-GeV and 62.4-GeV}}, Phys. Rev. Lett. \textbf{98},
  232302 (2007) \eprint{nucl-ex/0611019}.

\bibitem{Klebanov:2005mh}
I.~R. Klebanov, \emph{{QCD and string theory}}, Int. J. Mod. Phys.
  \textbf{A21}, 1831 (2006) \eprint{hep-ph/0509087}.

\bibitem{McLerran:2007hz}
L.~McLerran, \emph{{Theory summary: Quark Matter 2006}}, J. Phys. \textbf{G34},
  S583 (2007) \eprint{hep-ph/0702004}.

\bibitem{Gubser:2007nd}
S.~S. Gubser and S.~S. Pufu, \emph{{Master field treatment of metric
  perturbations sourced by the trailing string}}, Nucl. Phys. \textbf{B790}, 42
  (2008) \eprint{hep-th/0703090}.

\bibitem{Armesto:2005mz}
N.~Armesto, M.~Cacciari, A.~Dainese, C.~A. Salgado, and U.~A. Wiedemann,
  \emph{{How sensitive are high-p(T) electron spectra at RHIC to heavy quark
  energy loss?}}, Phys. Lett. \textbf{B637}, 362 (2006)
  \eprint{hep-ph/0511257}.

\bibitem{Adil:2006ra}
A.~Adil and I.~Vitev, \emph{{Collisional dissociation of heavy mesons in dense
  QCD matter}}, Phys. Lett. \textbf{B649}, 139 (2007) \eprint{hep-ph/0611109}.

\bibitem{Gubser:2006qh}
S.~S. Gubser, \emph{{Comparing the drag force on heavy quarks in N = 4 super-
  Yang-Mills theory and QCD}}, Phys. Rev. \textbf{D76}, 126003 (2007)
  \eprint{hep-th/0611272}.

\bibitem{ColeTalk}
B.~Cole, \emph{Hard probes: past, present, and future},
  \url{http://www.veccal.ernet.in/~pmd/qm2008/webpage/Program/10Feb/bcole.pdf}
  (2008).

\bibitem{Winter:2006iz}
D.~L. Winter (PHENIX), \emph{{High-p(T) pi0 production with respect to the
  reaction plane using the PHENIX detector at RHIC}}, Eur. Phys. J.
  \textbf{C49}, 47 (2007) \eprint{nucl-ex/0609019}.

\bibitem{Horowitz:2005ja}
W.~A. Horowitz, \emph{{Large observed v(2) as a signature for deconfinement}},
  Acta Phys. Hung. \textbf{A27}, 221 (2006) \eprint{nucl-th/0511052}.

\bibitem{Horowitz:2007su}
W.~A. Horowitz and M.~Gyulassy, \emph{{Testing AdS/CFT Deviations from pQCD
  Heavy Quark Energy Loss with Pb+Pb at LHC}}  (2007) \eprint{0706.2336}.

\bibitem{Hanks:2007qx}
A.~Hanks (PHENIX), \emph{{Measuring bremsstrahlung photons in 200-GeV p p
  collisions}}, Int. J. Mod. Phys. \textbf{E16}, 2182 (2007)
  \eprint{0705.0526}.

\bibitem{Kaneta:2001}
M.~Kaneta, \emph{{Results from the Relativistic Heavy Ion Collider (Part II)}}
  (2001).

\bibitem{PhysRevC.18.1756}
L.~Ray, G.~W. Hoffmann, G.~S. Blanpied, W.~R. Coker, and R.~P. Liljestrand,
  \emph{Analysis of 0.8-GeV polarized-proton elastic scattering from $Pb208$,
  $Zr90$, $Ni58$, and $C12$}, Phys. Rev. C \textbf{18 4}, 1756 (1978).

\bibitem{PhysRev.101.1131}
B.~Hahn, D.~G. Ravenhall, and R.~Hofstadter, \emph{High-Energy Electron
  Scattering and the Charge Distributions of Selected Nuclei}, Phys. Rev.
  \textbf{101 3}, 1131 (1956).

\bibitem{DeVries:1987}
H.~De~Vries, C.~W. De~Jager, and C.~De~Vries, \emph{{Nuclear
  charge-density-distribution parameters from elastic electron scattering}},
  ADNDT \textbf{36}, 495 (1987).

\bibitem{Djordjevic:2005rr}
M.~Djordjevic, \emph{{Radiative heavy quark energy loss in a strongly
  interacting quark gluon plasma}}, Ph.D. thesis, Columbia University (2005),
  [\href{http://proquest.umi.com/pqdlink?did=954041911&Fmt=7&clientId=15403&RQ%
T=309&VName=PQD}{UMI-31-82937}].

\end{thebibliography}

\end{spacing}

\clearpage

\begin{figure}[!htb]
\centering
\includegraphics[width=3 in]{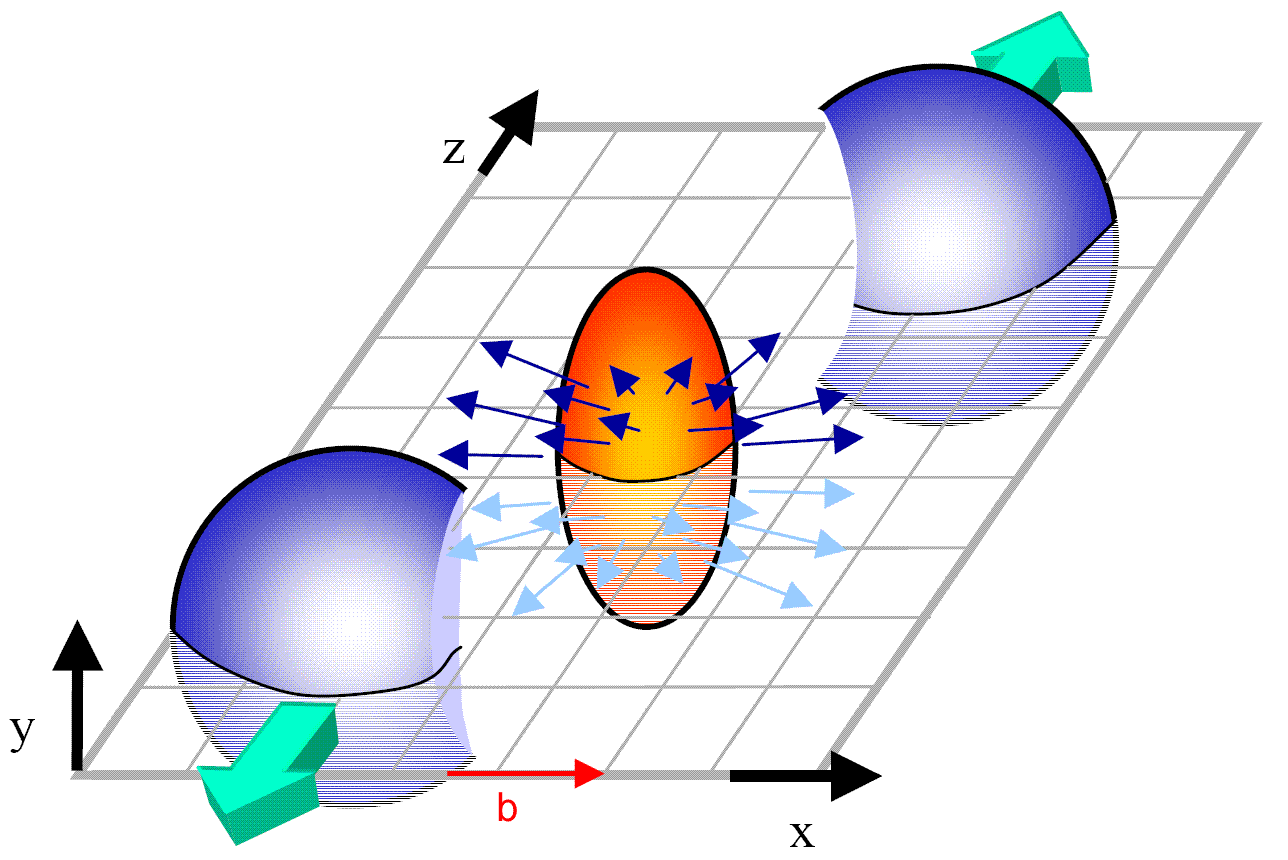}
\caption{\label{geom:collisiongeom}
An illustration of an $A+A$ heavy ion collision at impact parameter $b$.  The soft medium particles tend to be distributed similarly to the participant density, $\rho_\textrm{part}$ while the hard collisions scale like the binary density, $\rho_\textrm{coll}$.  Figure adapted from \cite{Kaneta:2001}.}
\end{figure}

\appendix
\lhead{Appendix \Alph{chapter}: \leftmark}
\mychapter{Nuclear Geometry}{app:geometry}
For a given nuclear density distribution $\rho(\vec{r}$, normalized such that $\int d^3x \rho = A$, one defines the nuclear thickness function $T_A(\vec{x}=\{x,y\}) = \int dz \rho(\vec{r})$; see \fig{geom:collisiongeom} for an illustration of the heavy ion collision geometry.  Then the nuclear overlap as a function of impact parameter $b$ is $T_{AA}(\vec{x};b) = \sigma_{NN} T_A(x+b/2,y) T_A(x-b/2,y)$; $\sigma_{NN}$ is the nucleon-nucleon inelastic cross section, taken here as 42 mb = 4.2 fm$^2$.  This gives the density of binary collisions in the reaction plane given $b$.  
The participant density is given by the probability that there is at least one interaction between a nucleon from nucleus $A$ and the entire stack of nucleons it passes through in nucleus $B$,
\bea
\rho_\textrm{part}(\vec{x};b) & = & T_A(x-b/2,y)\left( 1-e^{-\sigma_{NN}T_A(x+b/2,y)} \right) \nonumber\\
& & + T_A(x+b/2,y)\left( 1-e^{-\sigma_{NN}T_A(x-b/2,y)} \right).
\eea
The total number of participants is $N_\textrm{part}=\int d^2x\rho_\textrm{part}$.  

Centrality is an experimental measurement associated with the impact parameter of a collision.  Theoretically it is found from the inelastic cross section as a function of impact parameter,
\bea
\frac{d\sigma}{db} & = & 2\pi b \left( 1 - \left(1-\frac{N_\textrm{part}(b)}{2A}\right)^A\right),\\
\sigma_\textrm{tot} & = & \int db \frac{d\sigma}{db}.
\eea
The impact parameter $b_\textrm{cent}$ for a given centrality of $C\%$ is defined by
\be
\int_0^{b_\textrm{cent}} db \frac{d\sigma}{db} = C\sigma_\textrm{tot}.
\ee
Then the impact parameter for a given centrality class $C_1-C_2\%$ is found from the average centrality:
\be
\label{geom:cent}
\int_0^{b_\textrm{cent}} db \frac{d\sigma}{db} = \frac{1}{2}(C_1+C_2)\sigma_\textrm{tot}.
\ee

\begin{figure}[!htb]
\centering
$\begin{array}{cc}
\includegraphics[width=2.95in]{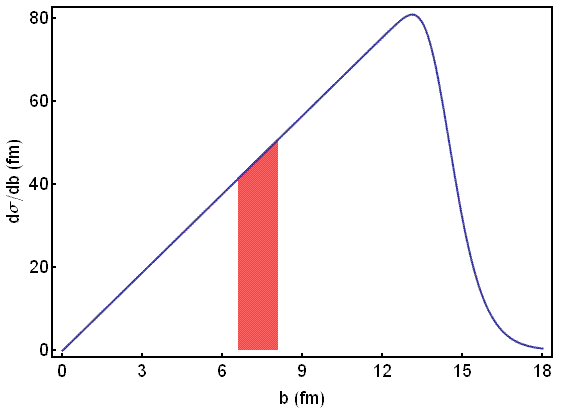} & 
\includegraphics[width=2.25in]{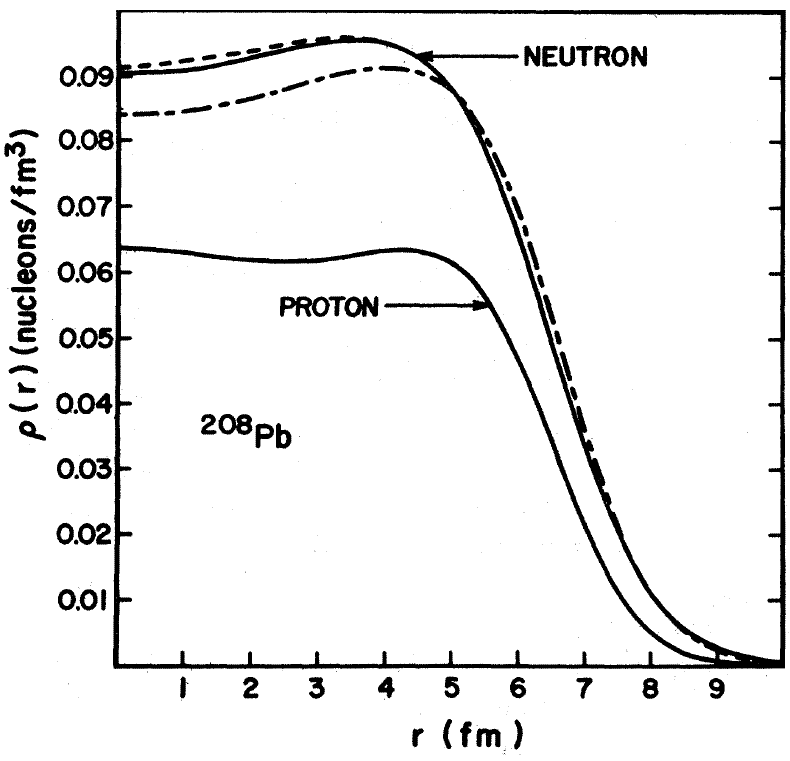} \\[-.05in]
\textrm{{\figsize(a)}} & \textrm{{\figsize(b)}}
\end{array}$
\caption{\label{geometry:pbdensity}
(a) Plot of $d\sigma/db$ as a function of $b$ for the Woods-Saxon $^{197}Au$ nucleus.  The shaded region corresponds to $20-30\%$ centrality with endpoints at $b=6.6$ and 8.1 fm; the single representative impact parameter for this centrality class, found by properly weighting from \eq{geom:cent}, is $b=7.4$ fm.  (b) Plot of the nuclear density as a function of radius separately for protons and neutrons in $^{208}Pb$ \cite{PhysRevC.18.1756}.}
\end{figure}

The nuclear density profile can take many forms.  The simplest numerically is the hard cylinder geometry, which is constant in density over the reaction plane:
\be
\rho_{HC}(x,y) = \rho_0 \theta(\sqrt{x^2+y^2}-R_{HC}).
\ee
This of course leads to participant and binary densities that are also constant across the reaction plane.  The next simplest is hard spheres:
\be
\rho_{HS}(\vec{r}) = \rho_0 \theta(|\vec{r}|-R_{HS}).
\ee
The Woods-Saxon distribution is a commonly used diffuse nuclear density, parameterized by
\be
\rho_{WS}(\vec{r}) = \rho_0 \left( e^{-\frac{|\vec{r}|-R}{z}} - 1\right)^{-1}.
\ee
For $^{197}Au$ \cite{PhysRev.101.1131} found $R=6.36$ fm and $z=.535$ fm by comparing the charge distribution as measured by high-energy electron scattering to predictions from the Dirac equation.  Values for $^{63}Cu$ can be found in \cite{DeVries:1987}.  These experimental determinations of the nuclear densities can become nearly arbitrarily complex; see \fig{geometry:pbdensity} for the results of fits of the proton and neutron distributions with 9 free parameters each \cite{PhysRevC.18.1756}.

\clearpage
\mychapter{Power Law \texorpdfstring{$R_{AA}$}{RAA}}{punchappendix}
\section{Fragmentation Functions}
The fragmentation function $D_q^h(z)$, where $z=2E_h/Q$, gives the probability of producing hadron $h$ of energy $E_h$ from parton $q$ in a process of total energy $Q$.  
If we take the problem as the production of a quark and an antiquark and ignore all masses then, in the center of mass frame, each parton carries half the total energy,
\be
p_q = E_q = \sum_{h} E_h = Q/2.
\ee
Therefore
\be
\label{Punch:z}
z = \frac{2E_h}{Q} = \frac{E_h}{p_q} = \frac{p_h}{p_q}.
\ee
Conservation of energy then demands
\be
\sum_{h}\int_0^1 zD_q^h(z) dz = \sum_h \frac{2\langle E_h \rangle}{Q} = 1.
\ee

From the definition of a fragmentation function we have that
\bea
\label{Punch:frag1}
dN^h (p_h) & = & dN^q (p_q) D(z) dz.\\
\Rightarrow \frac{dN^h}{dp_h}(p_h) & = & \int \frac{dN^q}{dp_h}(p_q) D(z) dz \nonumber\\
\label{Punch:frag2}
& = & \int \frac{dN^q}{dp_q}\left( \frac{p_h}{z} \right) \frac{1}{z} D(z) dz,
\eea
where we get to the last line by employing \eq{Punch:z} twice.
\section{Partonic \texorpdfstring{$R_{AA}^Q$}{RAA}}
For in-medium momentum loss in which the final momentum is related to the initial momentum by $p_f = (1-\epsilon) p_i$ we have the probability of fractional energy loss for parton $q$ with initial momentum $p_i$ is $P_q(\epsilon|p_i)$.  Then, exactly as for fragmentation functions (\eq{Punch:frag1}),
\be
dN^q(p_f) = dN^q(p_i)P(\epsilon|p_i)d\epsilon,
\ee
where we have suppressed the $q$ index on $P$, and hence (\eq{Punch:frag2})
\be
\frac{dN^q_\mathrm{final}}{dp_f}(p_f) = \int d\epsilon \frac{dN^q_\mathrm{prod}}{dp_i} \left( \frac{p_f}{1-\epsilon} \right) \frac{1}{1-\epsilon} P\left(\epsilon \left| \frac{p_f}{1-\epsilon}\right.\right).
\ee

If we assume that the production spectrum can be approximated by a power law,
\be
\frac{dN^q_\mathrm{prod}}{dp_i}(p_i) = \frac{A}{p_i^{n(p_i)}},
\ee
where $A$ is some normalization constant, then we may find a simple equation for the partonic nuclear modification factor,
\bea
R_{AA}^q(\eqnpt) & \equiv & \frac{\frac{dN^q_{AA}}{d\eqnpt}(\eqnpt)}{\eqnncoll\frac{dN^q_{pp}}{d\eqnpt}(\eqnpt)} \\
& = & \frac{\eqnncoll\int d\epsilon \frac{dN^q_\mathrm{prod}}{dp_i} \left( \frac{\eqnpt}{1-\epsilon} \right) \frac{1}{1-\epsilon} P(\epsilon|\frac{\eqnpt}{1-\epsilon})}{\eqnncoll\frac{dN^q_\mathrm{prod}}{d\eqnpt}(\eqnpt)} \nonumber\\
& = & \frac{\eqnncoll\int d\epsilon \frac{A}{\left( \eqnpt/1-\epsilon \right)^{n(\eqnpt/1-\epsilon)}} \frac{1}{1-\epsilon} P(\epsilon|\frac{\eqnpt}{1-\epsilon})}{\eqnncoll\frac{A}{\eqnpt^{n(\eqnpt)}}} \nonumber\\
& = & \int d\epsilon P(\epsilon|\frac{\eqnpt}{1-\epsilon})(1-\epsilon)^{n(\eqnpt/1-\epsilon)-1} \frac{\eqnpt^{n(\eqnpt)}}{\eqnpt^{n(\eqnpt/1-\epsilon)}}.
\eea
In the case of a slowly varying power law $n(\eqnpt)$ is approximately independent of $\eqnpt$; if we further assume a slow variation of $P$ with respect to $\eqnpt$ we have
\be
\label{Punch:RAA}
R_{AA}^q(\eqnpt) \simeq \int d\epsilon P(\epsilon)(1-\epsilon)^{n(\eqnpt)-1}.
\ee
We note that if the production spectra are $dN^q_\mathrm{prod}/d^2\eqnpt$ or $dN^q_\mathrm{prod}/d\eqnpt^2$ then the exponent in \eq{Punch:RAA} goes from $n(\eqnpt)-1$ to $n(\eqnpt)-2$.
\section{Hadronic \texorpdfstring{$R_{AA}^h$}{RAA} from Partonic \texorpdfstring{$R_{AA}^q$}{RAA}}
\be
R_{AA}^h(\eqnpt) = \frac{\frac{dN^h_{AA}}{d\eqnpt} \scriptstyle{(\eqnpt)}}{\eqnncoll\frac{dN^h_{pp}}{d\eqnpt} \scriptstyle{(\eqnpt)}} = \frac{\sum_q\int dz \frac{1}{z}D_q^h(z) \frac{dN_{AA}^q}{dp_q}\left( \frac{\eqnpt}{z} \right)}{\eqnncoll\sum_q\int dz \frac{1}{z}D_q^h(z) \frac{dN_{pp}^q}{dp_q}\left( \frac{\eqnpt}{z} \right)}.
\ee
Since
\bea
& \frac{dN^q_{AA}}{d\eqnpt}(\eqnpt) = \frac{dN^q_{pp}}{d\eqnpt}(\eqnpt) R_{AA}^q(\eqnpt) & \\
& \therefore R_{AA}^h(\eqnpt) = \frac{\sum_q\int dz \frac{1}{z}D_q^h(z) \frac{dN_{AA}^q}{dp_q}\left( \frac{\eqnpt}{z} \right)R_{AA}^q\left(\frac{\eqnpt}{z}\right)}{\sum_q\int dz \frac{1}{z}D_q^h(z) \frac{dN_{pp}^q}{dp_q}\left( \frac{\eqnpt}{z} \right)}. &
\eea

\clearpage
\mychapter{Statistical Mechanics Results}{appendix:whdg}
We wish to relate the plasma density to its temperature.  In general
\be
dN = g\, n_\varepsilon d\tau = g\, n_\varepsilon \frac{dVd^3p}{(2\pi\hbar)^3},
\ee
where $n_\varepsilon = [\exp \big( (\varepsilon-\mu)/T \big) \pm 1]^{-1}$ is the occupation number ($-$ taken for fermions, $+$ taken for bosons), and $g$ is the degeneracy.  In an ultrarelativistic gas the single-particle energy is $\varepsilon=cp$.  Therefore $d^3p = 4\pi p^2 dp = 4\pi\varepsilon^2 d\varepsilon/c^3$ and
\bea
N_F & = & \frac{4\pi g_F V}{(2\pi \hbar c)^3} \int_0^\infinity d\varepsilon \frac{\varepsilon^2}{e^{\varepsilon/T}-1} \\
& = & \frac{4\pi g_F V}{(2\pi \hbar c)^3} T^3 2 \zeta(3)
\eea
for a Fermi gas at zero chemical potential, $\mu=0$.  Then the density is
\be
\frac{N_F}{V} = \rho_F = \frac{g_F \zeta(3)}{\pi^2 (\hbar c)^3} T^3.
\ee
For a Bose gas at $\mu=0$ we have, as the $\varepsilon$ integral yields $T^3 (3/2) \zeta(3)$,
\be
\frac{N_B}{V} = \rho_B = \frac{g_B \zeta(3)}{4 \pi^2 (\hbar c)^3} T^3.
\ee
For a plasma of both fermions and bosons $N_{tot} = N_F + N_B$ and
\bea
\frac{N_{tot}}{V} = & \rho & = \frac{\zeta(3)}{\pi^2 (\hbar c)^3} \left[ g_F + \frac{3}{4} g_B \right] T^3 \\
\Rightarrow & T & = \left( \frac{\pi^2 (\hbar c)^3}{\zeta(3) \left[ g_F + \frac{3}{4} g_B \right]} \rho \right)^{1/3} \\
& & \simeq \left( \frac{\pi^2 (\hbar c)^3}{1.202 \left[ 2 (N_c^2-1) + 3 N_c N_f \right]} \rho \right)^{1/3},
\eea
where $g_F = 2 \cdot (N_c^2-1)$ for gluons in $SU(N_c)$ with 2 helicities, $g_B = 2\cdot 2\cdot N_c \cdot N_f$ for spin-1/2 quarks and anti-quarks of $N_f$ flavors and $N_c$ colors, and $\zeta(3)\simeq1.202$.  For a Bjorken-expanding QGP with a boost-invariant rapidity distribution $dN/dy$ of transverse area $A_T$ the temperature is then ($N_c=3$)
\be
\label{temperature}
T(\tau) = \left( \frac{\pi^2 (\hbar c)^3}{1.202 \left[ 16 + 9 N_f \right]} \frac{dN/dy}{A_T \tau} \right)^{1/3}.
\ee
\clearpage
\mychapter{Formation Time and Brick Estimates}{appendix:pqcdvsadscft}
\section{\texorpdfstring{$\gamma_c$}{\9003\263\9000\137c} for Bethe-Heitler to LPM Regime}\label{BHtoLPMapp}
\subsection{Formation Physics}
By the Heisenberg Uncertainty Principle, the formation time, $\tau_f'$, for a gluon (or photon) of energy $\left.k^0\right.'$ separating from its parent quark (or gluon)---without loss of generality taken as a quark from now on---in the rest frame of the quark, is
\be
\label{tauf}
\tau_f'\sim\frac{1}{\left.k^0\right.'},
\ee
where the primed coordinates are in the rest frame of the quark and unprimed coordinates will be in the lab frame.  To find $\tau_f$ in the lab frame, one must first boost the gluon energy in the lab frame into the rest frame of the quark, then boost back to the lab frame.  Take a quark with 4-momentum $p=(p^0,p^z,\wv{p})$ and gluon of momentum $k=(k^0,k^z,\wv{k})$.  The velocity of the quark is $\beta = p^z/p^0$.  The energy of the gluon in the quark's rest frame is then $\left.k^0\right.' = \gamma (k^0-\beta k^z)$.  Boosting the formation time, \eq{tauf}, back to the lab frame will dilate the time by $\gamma$; therefore in the lab
\bea
\label{tauflab}
\tau_f & = & \frac{1}{k^0-\beta k^z} \\
\label{migdal}
& \simeq & \frac{\lambdabar}{1-\beta},
\eea
where for the last line we have taken $k^0\simeq k^z$ so that \eq{migdal} reproduces \cite{Migdal:1956tc}.  

For a quark of mass $M$ and an emitted gluon of mass $m_g$ with momenta
\bea
\label{BHmomenta}
p & = & \left( \sqrt{M^2+(1-x)^2p_z^2+\wv{k}^2},(1-x)p_z,-\wv{k} \right) \\
k & = & \left( \sqrt{m_g^2+x^2p_z^2+\wv{k}^2},xp_z,\wv{k} \right),
\eea
the formation time to lowest order (assuming large $p_z$ and small $\wv{k}$) is
\be
\label{tau}
\tau_f \simeq \frac{2x(1-x)^2p_z}{\wv{k}^2 [1-2x(1-x)]+(1-x)^2m_g^2+x^2M^2}
\ee
this agrees with \cite{Djordjevic:2005rr} in the limit of small-$x$ considered there.  Note that using \eq{migdal} one finds (also for large $p_z$ and small $\wv{k}$)
\be
\tau_f \simeq \frac{2(1-x)^2p_z}{x(\wv{k}^2+M^2)},
\ee
which agrees with \eq{tau} in the large-$x$ Bethe-Heitler limit.

For the same momenta, \eq{BHmomenta}, the leading order lightcone coordinate representation,
\bea
p & \simeq & [(1-x)E^+,\frac{M^2+\wv{k}^2}{(1-x)E^+},-\wv{k} ] \\
k & \simeq & [xE^+,\frac{M^2+\wv{k}^2}{xE^+},-\wv{k} ]
\eea
yields exactly the same formation time, \eq{tau}, but with $p_z$ replaced by $E^+$.
\subsection{Bethe-Heitler to LPM}
The ``speed limit'' at which incoherent Bethe-Heitler turns over to coherent LPM, $\gamma_c$, is found when the formation length for the produced radiation is on the order of the radiation's mean free path:
\be
\label{BHtoLPM}
\tau_f \sim \lambda_g,
\ee
where $\tau_f$ is given by \eq{tauflab} and
\bea
\lambda_g & = & \frac{1}{\sigma\rho} \sim \frac{1}{\left( \frac{\alpha^2}{\mu^2} \right) T^3} \\
& \sim & \frac{1}{\alpha T}.
\eea

For large-$x$ Bethe-Heitler radiation one sees from \eq{tau} that
\be
\tau_f \sim \frac{E}{M^2} \sim \frac{\gamma}{M}.
\ee

\eq{BHtoLPM} then gives
\be
\gamma_c \sim \frac{M}{\alpha T}.
\ee
\section{AdS/CFT Brick Estimate}
We wish to more rigorously derive \eq{aprx}.  Recall that \eq{aprx} justifies \eq{AdSRcb}, the back-of-the-envelope explanation for the momentum independences and normalization for $R^{cb}(\eqnpt)=R_{AA}^c(\eqnpt)/R_{AA}^b(\eqnpt)$ for AdS/CFT Drag-type heavy quark energy loss.

From the simple trailing string model of \cite{Gubser:2006bz,Herzog:2006gh} we have by \eq{mu}
\be
\label{localmu}
\frac{d\eqnpt}{dt} = -\mu p,
\ee
where
\be
\mu = \frac{\pi\sqrt{\lambda}T^2}{2M},
\ee
with $T$ the temperature of the SYM plasma and $M$ the heavy quark mass.  Solving \eq{localmu} one finds that
\be
\label{adspt}
p(t) = p(0)e^{-\int dt' \mu(t')}.
\ee

Defining the fractional momentum loss $\epsilon$ as
\bea
p_f & = & (1-\epsilon)p_i \\
\label{AdSeps}
\Rightarrow \; \epsilon & = & 1-e^{-\int dt' \mu(t')}.
\eea

From 
Eq.[] we have that for a quark (or gluon) with slowly varying power law production index $n_Q(\eqnpt)+1$,
\be
\label{AdSRAA}
R_{AA}(\eqnpt) \simeq \langle(1-\epsilon)^n\rangle_{geom},
\ee
where $\langle\cdots\rangle_{geom}$ indicates averaging over the geometry of the collision.

\subsection{Static, Constant Medium}\label{NCM}
In a static, constant medium $\mu(\vec{x}+\vec{v}t)=\mu$.  For this case $R_{AA}$ for a single, fixed path of length $L$ is
\be
\label{AdSExp}
R_{AA}(\eqnpt) \simeq e^{-n_Q(\eqnpt)\mu L/v},
\ee
where we have assumed $v\approx\mathrm{constant}$ (a good approximation for ultrarelativistic motion with $v\simeq1$).  From now on we will simply take $v\simeq1$ and suppress it in our derivations.  One can see that \eq{AdSExp}, plotted in \fig{AdSBrickExp}, qualitatively reproduces the full numerical results of \fig{LHCcandbRAA} for the \emph{a posteriori} input parameter choice of $L=1$ fm.

\bfig[!htb]
\centering
\includegraphics[width=3.77 in]{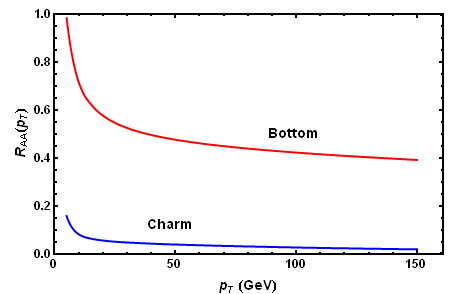}
\caption{
\label{AdSBrickExp}
$R_{AA}(\eqnpt)$ for charm and bottom quarks traversing a single pathlength of $L=1$ fm in a static medium of temperature $T=300$ MeV (from \eq{temperature} with the PHOBOS extrapolation to LHC gluon density $dN_g/dy=1750$ \cite{Back:2004je,Busza:2007ke}, $\tau_0=.2$ fm, $\tau = 1$ fm, and $N_f=3$), with $D=3/2\pi T$ from \eq{AdSExp}.   
The magnitude and momentum dependence are similar to the results of the full numerical simulation with Bjorken expansion and realistic medium geometry shown in \fig{LHCcandbRAA}.
}
\efig

One may also readily find the nuclear modification factor assuming a uniform, static brick of plasma.  Then \eq{AdSRAA} and \eq {AdSeps} give
\bea
R_{AA}(\eqnpt) & \simeq & \int_0^L\frac{d\ell}{L}e^{-n\int_\ell^L d\ell' \mu(\ell')} \nonumber\\
& = & \int_0^L\frac{d\ell}{L}e^{-n\mu(L-\ell)} \nonumber\\
& = & \frac{e^{-n\mu L}}{L} \left[ \frac{1}{n\mu} \left( e^{n\mu L}-1 \right) \right] \nonumber\\
\label{AdSOneOver}
& \simeq & \frac{1}{n\mu L}.
\eea
The results of \eq{AdSOneOver} are plotted in \fig{AdSOneOverPlot} with $L=5$ fm, the temperature $T\simeq 210$ MeV taken at $L/2=2.5$ fm, and $D/2\pi T=3$; they are again in qualitative agreement with the full numerics shown in \fig{LHCcandbRAA}, and the \emph{a posteriori} input parameters are somewhat less arbitrary.

\bfig[!htb]
\centering
\includegraphics[width=3.75in]{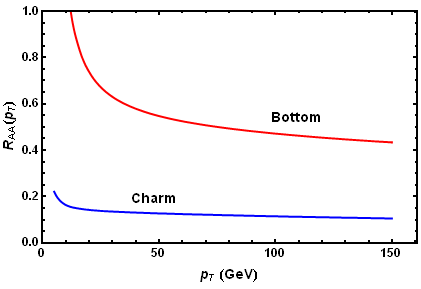}
\caption{
\label{AdSOneOverPlot}
$R_{AA}(\eqnpt)$ for charm and bottom quarks created uniformly along a static brick of SYM plasma of length $L=5$ fm and temperature $T\simeq210$ MeV (from \eq{temperature} with PHOBOS gluon density $dN_g/dy=1750$ \cite{Back:2004je,Busza:2007ke}, $\tau_0=.2$ fm, and $\tau = L/2 = 2.5$ fm) according to \eq{AdSOneOver} with $D=3/2\pi T$.   
The magnitude and momentum dependence are similar to the results of the full numerical simulation with Bjorken expansion and realistic medium geometry shown in \fig{LHCcandbRAA}.
}
\efig

\subsection{Static Nonconstant Medium}
We note in passing the analytically interesting case of a time-independent medium of nonconstant density, $\mu=\mu(\vec{x})$.  Then
\be
\label{AdS:NCM}
\frac{dp}{p} = -\mu(x)dt = -\mu \frac{1}{v}dx.
\ee
Since $p=\gamma mv$ we have that $dp = mvd\gamma + \gamma m dv = (1+\gamma^2v^2)dv$.  Then \eq{AdS:NCM} becomes
\be
\frac{v (1+\gamma^2v^2)\gamma mdv}{\gamma mv} = -\mu dx.
\ee
Moreover
\be
\int\frac{v^2}{1-v^2}dv = \frac{1}{2}\ln\frac{1+v}{1-v}-v,
\ee
so we have that
\be
\left.\frac{1}{2}\ln\frac{1+v}{1-v}\right|_{v=v_i}^{v_f} = - \int_{x_i}^{x_f}\mu dx,
\ee
where the integral on the RHS is really a line integral through the medium from the initial starting point $\vec{x}_i$ to the final point of interest, $\vec{x}_f$.  Therefore
\be
\label{AdS:v}
\frac{1+v_f}{1-v_f} = \frac{1+v_i}{1-v_i}e^{-2\int\mu dx}.
\ee
For the case of ultrarelativistic motion, $1+v_f\simeq2$, and \eq{AdS:v} simplifies to
\be
v_f = 1- (1-v_i)e^{-2\int\mu dx}.
\ee
The full solution to \eq{AdS:v} is
\be
v_f = \frac{\frac{1+v_i}{1-v_i}e^{-2\int\mu dx} - 1}{\frac{1+v_i}{1-v_i}e^{-2\int\mu dx} + 1}.
\ee

\subsection{Bjorken Expanding Medium}
One might reasonably wonder the consequences of including Bjorken expansion on our estimate of $R_{AA}$ from \ref{NCM}.  Defining 
\be
\nu = n \frac{\pi \sqrt{\lambda}}{2 M},
\ee
\eq{AdSRAA} becomes for a uniform brick of SYM plasma
\bea
\label{RAABj}
R_{AA} & \simeq & \int_0^L \frac{d\ell}{L}e^{-\nu \int_{\ell+\tau_0}^L d\ell' T^2(\ell')} \\
& = & \int_0^{L-\tau_0} \frac{d\ell}{L}e^{-\nu \int_{\ell+\tau_0}^L \left(\frac{\kappa}{\ell'-\ell} \right)^{2/3}} + \frac{\tau_0}{L},
\eea
where \eq{temperature} gives
\be
\kappa = \frac{dN_g/dy}{A_T} \frac{\pi^2}{\zeta(3)(16+9N_f)}.
\ee
The integral in the exponential of \eq{RAABj} is
\be
\int_{\ell+\tau_0}^L \left(\frac{\kappa}{\ell'-\ell} \right)^{2/3} = 3\kappa^{2/3}\left[ (L-\ell)^{1/3}-\tau_0^{1/3} \right].
\ee
Changing variables to $y=L-\ell \; \Rightarrow \; dy = -d\ell$ and defining $\omega = 3\kappa^{2/3}\nu$, the remaining integral in \eq{RAABj} becomes
\bea
-\frac{e^{\omega\tau_0^{1/3}}}{L}\int dy e^{-\omega y^{1/3}} 
& = & \frac{e^{\omega\tau_0^{1/3}}}{L} \left\{ \frac{3 e^{-\omega y^{1/3}}}{\omega^3} \left[ 2+2\omega y^{1/3}+\omega^2 y^{2/3} \right] \right\} \nonumber\\
& = & \frac{e^{\omega\tau_0^{1/3}}}{L} \Bigg\{ \frac{3 e^{-\omega (L-\ell)^{1/3}}}{\omega^3}\times \nonumber\\
& & \left[ 2+2\omega (L-\ell)^{1/3}+\omega^2 (L-\ell)^{2/3} \right] \Bigg\}_{\ell=0}^{\ell=L-\tau_0} \nonumber\\
& = & \frac{e^{\omega\tau_0^{1/3}}}{L} \left\{ \frac{3 e^{-\omega \tau_0^{1/3}}}{\omega^3} \left[ 2+2\omega \tau_0^{1/3}+\omega^2 \tau_0^{2/3} \right] \right. \nonumber\\
& & - \left. \frac{3 e^{-\omega L^{1/3}}}{\omega^3} \left[ 2+2\omega L^{1/3}+\omega^2 L^{2/3} \right] \right\}.
\eea
$\omega$ is in fact so large that for any reasonable $\tau_0\sim .2$ fm, $\omega\tau_0\gg1$.  Therefore we arrive at, to leading order,
\be
\label{AdSRAABj}
R_{AA} \simeq \frac{3 \tau_0^{1/3}}{\omega L}+\frac{\tau_0}{L} = \frac{3\tau_0^{1/3}+\omega\tau_0}{\omega L}.
\ee
Since $\omega \propto n_Q/M_Q$, with the proportionality constant the same for charm and bottom quarks, we still find that the ratio of charm to bottom \raa goes as
\be
R^{cb}_{AA}(\eqnpt) \simeq \frac{n_b}{n_c}\frac{M_c}{M_b}.
\ee
Note the essential importance of the plasma formation time, $\tau_0$, in \eq{AdSRAABj} in order to reproduce the linear in $M_c/M_b$ behavior.  The results of the full calculation (not just the leading order terms) for charm and bottom quarks are shown in \fig{AdSBrickBj}.  The momentum dependence is very similar to \fig{LHCcandbRAA}.  The normalization was made similar by \emph{a posteriori} setting the Bjorken expanding brick to the abbreviated length $L=3$ fm.  

\bfig[!htb]
\centering
\includegraphics[width=3.75in]{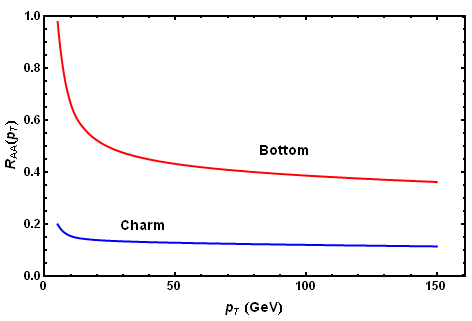}
\caption{
\label{AdSBrickBj}
$R_{AA}(\eqnpt)$ for charm and bottom quarks created uniformly along a Bjorken expanding brick of SYM plasma of length $L=3$ fm as given by the full analytic solution to \eq{RAABj} with $D=3/2\pi T$.  
The magnitude and momentum dependence are similar to the results of the full numerical simulation with Bjorken expansion and realistic medium geometry shown in \fig{LHCcandbRAA}.
}
\efig
\clearpage
\renewcommand{\mq}{M_Q}

\mychapter{Standard pQCD Results}{chapter:pqcd}
\section{Light Cone Coordinates}
\subsection{Definitions; Identities}
A four-vector is not bold-faced (e.g. $p$, $k$), a three-vector is bold-faced with a vector symbol (e.g. $\rv{p}$, $\rv{k}$), and a transverse two-vector is bold-faced without a vector symbol (e.g. $\wv{p}$, $\wv{k}$).  Minkowski four-vectors are written with parentheses, (); light-cone four-vectors with brackets, [].
\be
p=(p^0,p^z,\wv{p})=[p^+,p^-,\wv{p}].
\ee
We will use non-symmetrized lightcone coordinates:
\bea
p^+ & = & p^0+p^z \\
p^- & = & p^0-p^z \\
\wv{p} & = & \wv{p}.
\eea
The inverse transformation is then
\bea
p^0 & = & \frac{1}{2}(p^++p^-) \\
p^z & = & \frac{1}{2}(p^+-p^-) \\
\wv{p} & = & \wv{p}.
\eea
The Minkowski dot product in lightcone coordinates is:
\be
p\cdot k = p^0k^0-p^zk^z-\wv{p\cdot k} = \frac{1}{2}(p^+k^-+p^-k^+)-\wv{p\cdot k}.
\ee
The length of a vector using lightcone coordinates is then:
\be
p\cdot p = p^+p^--\wv{p\cdot p}
\ee.

\subsection{Eikonality in Light Cone Coordinates}
Note that the eikonal approximation requires collinear emission, which in turn requires $k^+\gg k^-$ (for emitted boson with 4-momentum $k$).  If this were otherwise, the $k^z$ emission would be in the opposite direction of $p^z$: $p^z=\frac{1}{2}(p^+-p^-)\simeq\frac{1}{2}p^+=\frac{1}{2}(1-x)E^+$; on the other hand $k^z=\frac{1}{2}(k^+-k^-)$ must be approximately $\frac{1}{2}k^+$ or else the sign is wrong.

A useful relation comes from eikonality:
\bea
& k^+ \gg k^- \Rightarrow xE^+ \gg \frac{\wv{k}^2}{xE^+} \Rightarrow (xE^+)^2 \gg \wv{k}^2 & \nonumber\\
\label{eikonal}
& \Rightarrow xE^+ \gg |\wv{k}|.&
\eea

\section{Monojet vs.\ Dijet \texorpdfstring{\zeroth}{0th} Order Radiation}
In order to motivate the work of Appendix \ref{zerothchapter} we will derive the \zeroth order bremsstrahlung radiation loss for massive quarks and gluons using two methods and find that the results are inconsistent.  We will find that the usual pQCD approach of calculating the radiation spectrum from a monojet only is consistent with a dijet calculation in the massless limit only.  For massive gluons the longitudinal polarization must be considered, and its contribution to the spectrum from the away-side jet will be $\mathcal{O}(1)$ in an expansion in powers of $1/E^+$.

\subsection{Peskin's Trick}
We will find the following relations useful in evaluating Feynman diagrams related to bremsstrahlung, which we will refer to as Peskin's Trick:
\be
\label{peskin}
\begin{split}
\bar{u}(p')\gamma^\mu\epsilon_\mu^*(\FMslash{p}'+m) & = \bar{u}(p')2p'^\mu\epsilon_\mu^*; \\
(\FMslash{P}+m)\gamma^\mu\epsilon_\mu^*u(P) & = 2P^\mu\epsilon_\mu^*u(P).
\end{split}
\ee
The first relation is proved as follows:
\bea
\bar{u}(p')\gamma\cdot\epsilon^*(\FMslash{p}'+m) & = & \bar{u}(p')\gamma^\mu\epsilon_\mu^*(\gamma^\nu p'_\nu+m) \nonumber\\
& = & \bar{u}(p')\big(2 \epsilon^*\cdot p'-(\FMslash{p}'+m)\gamma\cdot\epsilon^*\big) \nonumber\\
& = & \bar{u}(p')2p'^\mu\epsilon_\mu^*.
\eea
The second line comes from using $\{\gamma^\mu,\gamma^\nu\}=2g^{\mu\nu}$; the third line results from the Dirac equation, $\bar{u}(p)(\FMslash{p}+m)=0$.  The proof of the conjugate relation follows the exact same line of reasoning.

\subsection{Diagrammatic Calculation, Massless Photon and Quark Monojet}
\begin{figure}[!htb]
\begin{center}
\includegraphics[width=3. in]{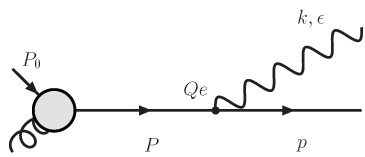}
\end{center}
\caption{\label{zerothorder}Zeroth order radiation diagram.  The blob on the left represents the factorized production of the hard quark.}
\end{figure}
There is no scattering off a medium center.  Hence
\bea
P & \simeq & (E,E,\wv{0}) = [E^+=2E,0,\wv{0}] \\
p & = & [(1-x)E^+,\frac{\wv{k}^2}{(1-x)E^+},-\wv{k}] \\
k & = & [xE^+,\frac{\wv{k}^2}{xE^+}=\frac{\wv{k}^2}{2xE}=\frac{\wv{k}^2}{2\omega}\equiv\omega_0,\wv{k}] \\
\epsilon & = & [0,\frac{2\wv{k\cdot}\wvg{\epsilon}}{xE^+},\wvg{\epsilon}].
\eea
Note that $P$ is off-shell so it cannot be exactly $P=(E,E,\wv{0})$.  $p^-$ is found by enforcing $p\cdot p=0$; similarly for $k\cdot k = 0$ and $k\cdot\epsilon = 0$.  Some useful relations used when deriving the radiation spectrum from this diagram are:
\bea
2p\cdot\epsilon & = & p^+\epsilon^--2\wv{p\cdot}\wvg{\epsilon} = (1-x)E^+\frac{2\wv{k\cdot}\wvg{\epsilon}}{xE^+}+2\wv{k\cdot}\wvg{\epsilon} \nonumber\\
& = & \frac{2\wv{k\cdot}\wvg{\epsilon}}{x}; \\
(p+k)^2 & = & [E^+,\frac{\wv{k}^2}{(1-x)E^+}+\frac{\wv{k}^2}{xE^+},\wv{0}]^2\approx[E^+,\frac{\wv{k}^2}{xE^+},0]^2 \nonumber\\
& = & \frac{\wv{k}^2}{x}.
\eea

We assume some kind of factorization so that we don't have to explicitly calculate the production process.  Therefore
\bea
i\mathcal{M} 
& = & \bar{u}(p) iQe\gamma^\mu\epsilon_\mu^* \frac{\FMslash{P}}{P^2+i\epsilon} \mathcal{M}_0(P,P_0) u(P_0) \\
& \simeq & \bar{u}(p) iQe\gamma^\mu\epsilon_\mu^* \frac{\FMslash{p}}{(p+k)^2+i\epsilon} \mathcal{M}_0(P,P_0) u(P_0) \\
& \simeq & iQe \frac{2p\cdot\epsilon^*}{2p\cdot k+i\epsilon} \bar{u}(P) \mathcal{M}_0(P,P_0) u(P_0) \\
& = & iQe \frac{2\wv{k}\cdot\wvg{\epsilon}^*}{\wv{k}^2} \bar{u}(P) \mathcal{M}_0(P,P_0) u(P_0)
\eea
The first line comes from writing down the result of the diagram, where $\mathcal{M}_0$ denotes the part of the amplitude corresponding to the hard production.  In the next line we approximate the numerator of the propagator $P$ with $p$ (dropping $k$ because it is small).  The third line results from exploiting Peskin's trick, \eq{peskin}, and approximating $u(p)$ by $u(P)$.  

We wish to calculate the distribution of emitted photons \emph{given} that production has already taken place.  Since production is represented by $\bar{u}\mathcal{M}_0u$, we will drop this term in our further analysis.

The quantity of interest is then
\be
\sum_{\lambda}|\mathcal{M_\beta}| = \frac{4(Qe)^2}{\wv{k}^4}\sum_{\lambda}|\wv{k}\cdot\wvg{\epsilon}_\lambda^*|^2.
\ee
We may choose our polarization vector however we may wish; without loss of generality take $\wvg{\epsilon}_1^* = \hat{x}$ and $\wvg{\epsilon}_2^* = \hat{y}$.  Then
\be
\sum_{\lambda}|\wv{k}\cdot\wvg{\epsilon}_\lambda^*|^2 = \wv{k}^2 \left(\sin(\phi)^2+\cos(\phi)^2\right).
\ee

Therefore
\be
\sum_{\lambda}|\mathcal{M_\beta}| = \frac{4(Qe)^2}{\wv{k}^2}.
\ee
\subsection{Soft Photon Approximation, Massless Dijet}
Compute electron (mass $m=0$) beta decay starting with \eq{softdijet},
\be
i\mathcal{M} = \bar{u}(p)\left[\mathcal{M}_0(p,P)\right]u(P)\cdot\left[Qe\left(\frac{p\cdot\epsilon^*}{p\cdot k} - \frac{P\cdot\epsilon^*}{P\cdot k}\right)\right].
\ee
We will want to exploit the replacement $\sum\epsilon_\mu\epsilon_\nu^*\rightarrow-g_{\mu\nu}$ so we will have to preserve the Ward identity (charge conservation),
\be
k_\mu\left(\frac{p^\mu}{k\cdot p}-\frac{P^\mu}{k\cdot P}\right) = 0.
\ee
To do so we will have to keep the second term explicitly with $P_M\simeq[M,M,0,0]$ being the momentum of the ``proton" after (as well as before) beta emission.  Then (essentially \eq{qftradEM}),
\bea
\sum_{\lambda}|\mathcal{M}|^2 & = & (Qe)^2(-g_{\mu\nu})\left(\frac{p^\mu}{k\cdot p}-\frac{P_M^\mu}{k\cdot P_M}\right)\left(\frac{p^\nu}{k\cdot p}-\frac{P_M^\nu}{k\cdot P_M}\right) \nonumber\\
& = & (Qe)^2\left(\frac{2p\cdot P_M}{(k\cdot p)(k\cdot P_M)} - \frac{M^2}{(k\cdot P_M)^2} - \frac{m^2}{(k\cdot p)^2}\right)
\label{classical}
\eea
We will need the following dot products:
\bea
p\cdot P & = & [(1-x)E^+,\frac{\wv{k}^2}{(1-x)E^+},-\wv{k}]\cdot[M,M,\wv{0}] \nonumber\\
& \simeq & \frac{1}{2}ME^+ \\
p\cdot k & = & \frac{1}{2}\left(\frac{\wv{k}^2}{x}+x\wv{k}^2\right)+\wv{k}^2 \simeq \frac{\wv{k}^2}{2x} \\
k\cdot P & = & \frac{1}{2}\left(MxE^++M\frac{\wv{k}^2}{xE^+}\right) \simeq \frac{1}{2}MxE^+
\eea
With these in hand we have that
\bea
\sum_{\lambda}|\mathcal{M}_\beta|^2 & = & (Qe)^2 \left\{ \frac{2\left(\frac{1}{2}ME^+\right)}{\left(\frac{\wv{k}^2}{2x}\right)\left(\frac{1}{2}MxE^+\right)} - \frac{M^2}{\left(\frac{1}{2}MxE^+)^2\right)} \right\} \\
& = & \frac{4(Qe)^2}{\wv{k}^2}-\frac{4(Qe)^2}{(xE^+)^2} \\
& \simeq & \frac{4(Qe)^2}{\wv{k}^2},
\eea
where the final relation comes from $\wv{k}\ll xE^+$ due to eikonality, \eq{eikonal}.

So
\bea
d^3N_\gamma^{(0)} & = & \frac{d^3\rv{k}}{(2\pi)^32\omega}\sum_{\lambda}|\mathcal{M}_\beta|^2 \\
\Rightarrow \frac{d^3N_\gamma^{(0)}}{d^3\rv{k}} & = & \frac{(Qe)^2}{4\pi^3\omega\wv{k}^2} = \frac{Q^2\alpha}{\pi^2\omega\wv{k}^2}\\
\eea
We wish to transform coordinates
\bea
k_x & = & k_\perp\cos(\phi) \\
k_y & = & k_\perp\sin(\phi) \\
k_z & = & \sqrt{k^2-k_\perp^2} \simeq k
\eea
Thus the Jacobian gives $d^3k = k_\perp dkdk_\perp d\phi = k_\perp d\omega dk_\perp d\phi$.  Hence
\bea
\frac{d^3N_\gamma^{(0)}}{k_\perp d\omega dk_\perp d\phi} & = & \frac{Q^2\alpha}{\pi^2\omega\wv{k}^2} \\
\frac{d^2N_\gamma^{(0)}}{k_\perp d\omega dk_\perp} & = & 
\frac{2 Q^2\alpha}{\pi\omega\wv{k}^2} \\
\frac{d^2N_\gamma^{(0)}}{d\omega d\wv{k}^2} & = & 
\frac{Q^2\alpha}{\pi\omega\wv{k}^2}.
\eea
Nicely, we recover the Sudakov double logarithm \cite{Peskin:1995}:
\be
N_\gamma^{(0)} = \frac{Q^2\alpha}{\pi} \log\left(\frac{\omega_{max}}{\omega_{min}}\right) \log\left(\frac{\wv{k}^2_{max}}{\wv{k}^2_{min}}\right).
\ee

\subsection{Diagrammatic Calculation, Massive Photon and Massive Quark Monojet}
The idea is to reproduce Djordjevic's calculation of zeroth order radiation in \cite{Djordjevic:2003be}.  The diagram is the same as \fig{zerothorder} except that the quark has a mass $\mq$ and the photon has mass $\mg$.  The relevant four-vectors are now
\bea
P & \simeq & [E^+,\frac{\mq^2}{E^+},\wv{0}] \\
p & = & [(1-x)E^+,\frac{\wv{k}^2+\mq^2}{(1-x)E^+},-\wv{k}] \\
k & = & [xE^+,\frac{\wv{k}^2+\mg^2}{xE^+},\wv{k}] \\
\epsilon & = & [0,\frac{\wvg{\epsilon}\cdot\wv{k}}{xE^+},\wvg{\epsilon}].
\eea
The full matrix element is then
\bea
i\mathcal{M} & = &
\bar{u}(p)iQe\gamma^\mu\epsilon_\mu^*\frac{i(\FMslash{P}+M)}{P^2-M^2+i\epsilon}\mathcal{M}_0(P,P_0)u(P_0) \\
& \simeq & i^2Qe\frac{2p\cdot\epsilon^*}{2p\cdot k+\mg^2+i\epsilon}\bar{u}(P)\mathcal{M}_0(P,P_0)u(P_0),
\eea
where Peskin's trick, \eq{peskin}, was used.  The latter amplitude corresponds to factorized production and will be dropped.  We need the following relations
\bea
2p\cdot k + \mg^2 & = & \left[ \frac{(1-x)(\wv{k}^2+\mg^2)}{x} + \frac{x(\wv{k}^2+\mq^2)}{(1-x)} \right] \nonumber\\
& & \quad + 2\wv{k}^2 + \mg^2 \\
& = & \frac{\wv{k}^2+(1-x)^2\mg^2+x^2\mq^2}{x(1-x)} + \mg^2 \\
& = & \frac{\wv{k}^2+(1-x)\mg^2+x^2\mq^2}{x(1-x)} \\
2p\cdot\epsilon & = & \left[ \frac{2(1-x)\wv{k}\cdot\wvg{\epsilon}}{x} + 2\wv{k}\cdot\wvg{\epsilon} \right] = \frac{2\wv{k}\cdot\wvg{\epsilon}}{x}.
\eea
Therefore
\bea
i\mathcal{M}_\beta
& = & Qe\frac{2\wv{k}\cdot\wvg{\epsilon}/x}{\wv{k}^2+(1-x)\mg^2+x^2\mq^2/x(1-x)} \\
& = & 2Qe\frac{(1-x)\wv{k}\cdot\wvg{\epsilon}}{\wv{k}^2+(1-x)\mg^2+x^2\mq^2} \\
& \simeq & 2Qe\frac{\wv{k}\cdot\wvg{\epsilon}}{\wv{k}^2+(1-x)\mg^2+x^2\mq^2}.
\eea
Hence the distribution of emitted photons is
\be
\label{massivemonojet}
\frac{dN_\gamma^{(0)}}{d^3k} = \frac{Q^2\alpha}{\pi^2\omega} \frac{\wv{k}^2}{[\wv{k}^2+(1-x)\mg^2+x^2\mq^2]^2}.
\ee
This does not agree with \cite{Djordjevic:2003be}, but does agree with the---presumably corrected version---in \cite{Djordjevic:2005rr}.  Note that we have not included radiation into the longitudinal mode.  

\subsection{Classical Calculation, Massive Quark Jet and \texorpdfstring{\\}{} Massive Photon}
Note that while the sum over polarizations changes to $\sum_{\lambda}\epsilon_\mu\epsilon^*_\nu = -g_{\mu\nu}+k_\mu k_\nu/m_g^2$, \eq{classical} is unchanged; the additional terms exactly cancel each other.  We need to calculate the relevant dot products using the four momenta from the previous section.  One finds that
\bea
p\cdot P_M & = &\frac{M}{2(1-x)E^+} \left[ \left( (1-x)E^+ \right)^2 + \wv{k}^2 + \mq^2 \right] \\
p\cdot k & = & \frac{1}{2x(1-x)} \left[ \wv{k}^2 + (1-x)^2\mg^2 + x^2\mq^2 \right] \\
P_M\cdot k & = & \frac{M}{2xE^+} \left[ (xE^+)^2 + \wv{k}^2 + \mg^2 \right].
\eea
The result for \eq{classical} expanded to first order in $1/E^+$ is
\be
\label{massivedijet}
\sum_{\lambda}|\mathcal{M}|^2 = \frac{4 (Qe)^2 (1-x)^2 \left( \wv{k}^2 + (1-x)^2\mg^2 \right) }{[ \wv{k}^2 + (1-x)^2\mg^2 + x^2\mq^2 ]^2} + \mathcal{O}\left( \left( 1/xE^+ \right)^2 \right).
\ee

There are three differences between \eq{massivemonojet} and \eq{massivedijet}: the additional overall factor of $(1-x)^2$ in \eq{massivedijet}, $(1-x)^2$ in \eq{massivedijet} instead of $1-x$ in \eq{massivemonojet} as the coefficient of $\mg^2$ in the denominator, and the additional $(1-x)^2\mg^2$ in the numerator of \eq{massivedijet}.  In the limit of small $x$ for which the calculations are valid, the first two disappear; however the third is a true inconsistency.  In the following Appendix we will find that the second is due to including the mass of the gluon from the $k^2$ term in the denominator of the quark propagator and that the first and third are due to the interference terms from the away-side jet.

\renewcommand{\mq}{$m_q$}

\clearpage
\mychapter{Heavy Quark Production Radiation Derivation}{zerothchapter}
\section{Classical Computation}
\subsection{Variation of the Determinant of the Metric}
\subsubsection{Proof of Identity}
First we must derive a very important relation: 
\be
\label{metricvariation}
\partial_\mu g = g\,g^{\alpha\beta}\,\partial_\mu g_{\alpha\beta}.
\ee
This derivation is due to Simon Judes.

Let $G\equiv g_{\mu\nu}$ be the metric matrix, and let $\det G \equiv g$.  The inverse matrix is $G^{-1}\equiv g^{\mu\nu}$; necessarily $\det G^{-1} = g^{-1}$ ($\det (M^{-1}) = (\det M)^{-1}$).

From the definition of the metric,
\be
ds^2 = g_{\mu\nu}dx^\mu dx^\nu
\ee
it follows that $G$ can be chosen to be symmetric without loss of generality.  It can therefore be diagonalized by a unitary transformation.  Define the diagonalized metric as
\be
\mathcal{G} = U G U^{-1},
\ee
where $U$ is the unitary matrix whose rows are the (normalized) eigenvectors of $G$.  Note that
\be
\det \mathcal{G} = \det(UGU^{-1}) = \det(U^{-1}UG) = \det G.
\ee

Consider
\be
\partial_\mu g = \partial_\mu e^{\ln \det G}.
\ee
Since $\det G = \det\mathcal{G}$, from above, and, trivially from the definition, $\ln \det \mathcal{G} = \Tr \ln \mathcal{G}$, we have that (where $\mathcal{G}_i$ is the $i$th eigenvalue)
\bea
\partial_\mu g & = & g \partial_\mu \Tr \ln \mathcal{G} \nonumber\\
& = & g \partial_\mu \sum_{i} \ln \mathcal{G}_i \nonumber\\
& = & g \sum_{i} \mathcal{G}^{-1}_i\partial_\mu\mathcal{G}_i \nonumber\\
& = & g \Tr \mathcal{G}^{-1} \partial_\mu \mathcal{G}.
\eea

Now we must prove that $g^{\alpha\beta}\partial_\mu g_{\alpha\beta}$ is equal to this last quantity:
\bea
g^{\alpha\beta}\partial_\mu g_{\alpha\beta} & = & \Tr G^{-1}\partial_\mu G \nonumber\\
& = & \Tr UG^{-1}U^{-1}U\partial_\mu GU^{-1} \nonumber\\
& = & \Tr \mathcal{G}^{-1} \left[ \partial_\mu \mathcal{G} - (\partial_\mu U)GU^{-1} - UG(\partial_\mu U^{-1}) \right] \nonumber\\
& = & \Tr \mathcal{G}^{-1} \partial_\mu \mathcal{G} - UG^{-1}U^{-1}(\partial_\mu U)GU^{-1} - UG^{-1}U^{-1}UG(\partial_\mu U^{-1}) \nonumber\\
& = & \Tr \mathcal{G}^{-1} \partial_\mu \mathcal{G} - U^{-1}(\partial_\mu U) - U(\partial_\mu U^{-1}) \nonumber\\
& = & \Tr \mathcal{G}^{-1} \partial_\mu \mathcal{G} - \partial_\mu (U^{-1}U) \nonumber\\
& = & \Tr \mathcal{G}^{-1} \partial_\mu \mathcal{G}.
\eea
\subsubsection{Use of Identity}
We will use an important example of the identity \eq{metricvariation}:
\be
\delta g = g g^{\mu\nu} \delta g_{\mu\nu} = - g g_{\mu\nu} \delta g^{\mu\nu}.
\ee
The last equality comes from taking the variation of $g_{\mu\nu}g^{\mu\nu} = \delta^\nu_\nu = \#$:
\bea
\delta(g_{\mu\nu}g^{\mu\nu}) & = & g^{\mu\nu}\delta g_{\mu\nu} + g_{\mu\nu}\delta g^{\mu\nu} \nonumber\\[.15cm]
\Rightarrow g^{\mu\nu}\delta g_{\mu\nu} & = & -g_{\mu\nu}\delta g^{\mu\nu}.
\eea

Therefore we find an identity we will use often in the following:
\be
\frac{\delta \sqrt{-g}}{\delta g^{\mu\nu}} = \frac{1}{2\sqrt{-g}}g g_{\mu\nu} = -\frac{\sqrt{-g}}{2} g_{\mu\nu};
\ee
similarly
\be
\frac{\delta \sqrt{-g}}{\delta g_{\mu\nu}} = \frac{\sqrt{-g}}{2} g^{\mu\nu}.
\ee

\subsection{\texorpdfstring{$T^{\mu\nu}$}{T\9003\274\9003\275} from Varying the Metric}
We will use variation with respect to the metric to find the (automatically) symmetric, gauge-invariant form of $T^{\mu\nu}$ \cite{Carroll:2004}:
\be
\label{tmunu}
T_{\mu\nu} = \frac{2}{\sqrt{-g}} \frac{\delta \sqrt{-g}\mathcal{L}}{\delta g^{\mu\nu}},
\ee
where $g$ is the determinant of the metric and the coefficient out front gives the proper normalization for the usual stress-energy tensor (assuming a $+---$ type metric).

\subsubsection{E\&M}
As a warmup exercise we use \eq{tmunu} to derive the usual E\&M energy density from its Lagrangian,
\be
\mathcal{L}_{\mathrm{EM}} = -\frac{1}{4}F_{\mu\nu}F^{\mu\nu} = -\frac{1}{4}g^{\alpha\beta}g^{\gamma\delta}F_{\alpha\gamma}F_{\beta\delta}.
\ee

Therefore
\bea
T^\mathrm{EM}_{\mu\nu} & = & \frac{2}{\sqrt{-g}} \frac{\delta}{\delta g^{\mu\nu}} \left[ -\frac{1}{4}\sqrt{-g}g^{\alpha\beta}g^{\gamma\delta}F_{\alpha\gamma}F_{\beta\delta} \right] \nonumber\\
& = & -\frac{1}{2\sqrt{-g}} \left[ -\frac{\sqrt{-g}}{2}g_{\mu\nu}F^2 + \sqrt{-g} \delta^\alpha_\mu\delta^\beta_\nu g^{\gamma\delta} F_{\alpha\gamma}F_{\beta\delta} \right. \nonumber\\
& & \left. \phantom{\frac{\sqrt{-g}}{2}}+ \sqrt{-g} g^{\alpha\beta}\delta^\gamma_\mu\delta^\delta_\nu F_{\alpha\gamma}F_{\beta\delta} \right] \nonumber\\
& = & -\frac{1}{2} \left[ F_{\mu\gamma}F_\nu^{\phantom{\nu}\gamma} + F^\beta_{\phantom{\beta}\mu} F_{\beta\nu} - \frac{1}{2}g_{\mu\nu}F^2 \right] \nonumber\\
& = & F_{\mu\lambda}F^\lambda_{\phantom{\lambda}\nu} + \frac{1}{4}g_{\mu\nu} F_{\alpha\beta}F^{\alpha\beta}.
\eea

We know that $F^2 = 2 (\vec{B}^2 + \vec{E}^2)$, and since $E = - F^{0i}$ we have that
\bea
T^{00}_\mathrm{EM} & = & F^{0\lambda}F_\lambda^{\phantom{\lambda}0} + \frac{1}{4} F^2 \nonumber\\
& = & F^{0i}F_i^{\phantom{i}0} + \frac{1}{2} (\vec{B}^2 - \vec{E}^2) \nonumber\\
& = & \frac{1}{2} (\vec{E}^2 + \vec{B}^2).
\eea

\subsubsection{Proca}
Let's now find the stress-energy tensor for the Proca Lagrangian,
\bea
\mathcal{L}_{\mathrm{Proc}} & = & \mathcal{L}_{\mathrm{EM}} + \frac{\mu^2}{2} A^\mu A_\mu \\
& = & -\frac{1}{4} F_{\mu\nu}F^{\mu\nu} + \frac{\mu^2}{2} A^\mu A_\mu \\
& = & -\frac{1}{4}g^{\alpha\beta}g^{\gamma\delta}F_{\alpha\gamma}F_{\beta\delta} + \frac{\mu^2}{2} g^{\alpha\beta}A_\alpha A_\beta,
\eea
where $\mu$ is the mass of the photon.

Consider the variation of the new, massive photon part of the Lagrangian:
\bea
\frac{2}{\sqrt{-g}} \frac{\delta}{\delta g^{\mu\nu}} \left[ \frac{\mu^2}{2} \sqrt{-g} g^{\alpha\beta}A_\alpha A_\beta \right]
& = & \frac{\mu^2}{\sqrt{-g}} \left[ -\frac{\sqrt{-g}}{2}g_{\mu\nu}A^2 + \sqrt{-g}A_\mu A_\nu \right] \nonumber\\
& = & \mu^2 \left( A_\mu A_\nu - \frac{1}{2}g_{\mu\nu} A^\lambda A_\lambda \right).
\eea

Therefore
\bea
T_\mathrm{Proca}^{\mu\nu} & = & T_{\mathrm{EM}}^{\mu\nu} + \mu^2 \left( A^\mu A^\nu - \frac{1}{2}g^{\mu\nu} A^\lambda A_\lambda \right) \nonumber\\
& = & F^{\mu\lambda}F_\lambda^{\phantom{\lambda}\nu} + \frac{1}{4}g^{\mu\nu} F_{\alpha\beta}F^{\alpha\beta} + \mu^2 \left( A^\mu A^\nu - \frac{1}{2}g^{\mu\nu} A^\lambda A_\lambda \right).\nonumber\\
\eea

And
\be
\label{prt}
T_\mathrm{Proca}^{00} = \frac{1}{2} \big[\vec{E}^2 + \vec{B}^2 + \mu^2(\left.A^0\right.^2 + \vec{A}^2)\big].
\ee
This agrees with Jackson problem 12.16 (a) \cite{Jackson:1998}.  Note that we recover $T_\mathrm{EM}^{\mu\nu}$ for $\mu\rightarrow0$.

\subsection{Classical Bremsstrahlung Radiation}
\subsubsection{E\&M}
The E\&M Lagrangian coupled to a classical source is
\be
\mathcal{L}_\mathrm{E\&M} = -\frac{1}{4}F^{\mu\nu}F_{\mu\nu} - j^\mu A_\mu.
\ee
Using the E-L equations ($\partial_\alpha \partial \mathcal{L}/\partial \partial_\alpha A_\beta - \partial \mathcal{L}/\partial A_\beta = 0$), we find that
\bea
\partial_\alpha F^{\alpha\beta} & = & j^\beta \\
\partial_\alpha \partial^\alpha A^\beta - \partial^\beta \partial_\alpha A^\alpha & = & j^\beta \nonumber\\
\label{emeom}
\partial_\alpha \partial^\alpha A^\beta & = & j^\beta,
\eea
where we get from the second to the third line be choosing to be in the Lorentz gauge ($\partial_\alpha A^\alpha = 0$).

For the Beta decay problem, following Peskin 6.1 (pgs.\ 177-180) \cite{Peskin:1995}, our external current takes the form
\be
\label{jmu}
j^\mu(x) = e \int d\tau \frac{dy^\mu(\tau)}{d\tau}\delta^{(4)}\big(x-y(\tau)\big),
\ee
where, for the trajectory of our particle, we take
\be
\label{path}
y^\mu(\tau) = \left\{ \begin{array}{ll}
(p^\mu/m)\tau \quad & \mathrm{for}\;\tau<0;\\
(p'^\mu/m)\tau \quad & \mathrm{for}\;\tau>0. \end{array}\right.
\ee

We will work with the Fourier transform of the pertinent quantities:
\bea
A^\mu(x) & = & \int \frac{d^4k}{(2\pi)^4}e^{-ik\cdot x}\tilde{A}^\mu(k) \\
j^\mu(x) & = & \int \frac{d^4k}{(2\pi)^4}e^{-ik\cdot x}\tilde{j}^\mu(k).
\eea
With this normalization for the transforms we have that
\bea
\tilde{A}^\mu(k) & = & \int d^4ke^{ik\cdot x}A^\mu(x) \\
\tilde{j}^\mu(k) & = & \int d^4ke^{ik\cdot x}j^\mu(x).
\eea

Solving the EOM (\eq{emeom}) in momentum space we find that
\be
\tilde{A}^\mu(k) = -\frac{1}{k^2}\tilde{j}^\mu(k).
\ee

We can find $\tilde{j}$ from the given trajectory (\eq{path}) and the definition of the Fourier transform:
\bea
j^\mu(x) & = & e\int_{0}^{\infinity}d\tau\frac{p'^\mu}{m}\delta^{(4)}\left(x-\frac{p'}{m}\tau\right) + e\int_{-\infinity}^{0}d\tau\frac{p^\mu}{m}\delta^{(4)}\left(x-\frac{p}{m}\tau\right) \nonumber\\
\\
\Rightarrow \tilde{j}^\mu(k) & = & \int d^4xe^{ik\cdot x)}j^\mu(x) \nonumber\\
& = & e\int_{0}^{\infinity}d\tau\frac{p'^\mu}{m}e^{i(k\cdot p'/m + i\epsilon)\tau} + e\int_{-\infinity}^{0}d\tau\frac{p^\mu}{m}e^{i(k\cdot p/m - i\epsilon)\tau} \nonumber\\
& = & ie \left( \frac{p'^\mu}{k\cdot p'+i\epsilon} - \frac{p^\mu}{k\cdot p - i\epsilon} \right).
\eea

Therefore we have the field in position space as an integral over momentum:
\be
\label{amu}
A^\mu(x) = \int \frac{d^4k}{(2\pi)^4}e^{-ik\cdot x}\frac{-ie}{k^2}\left[ \frac{p'^\mu}{k\cdot p'+i\epsilon} - \frac{p^\mu}{k\cdot p-i\epsilon} \right].
\ee

The integral on the right hand side (RHS) has the pole structure as shown in \fig{fig1}.
\bfig[!hbt]
\centering
\includegraphics[width=4 in]{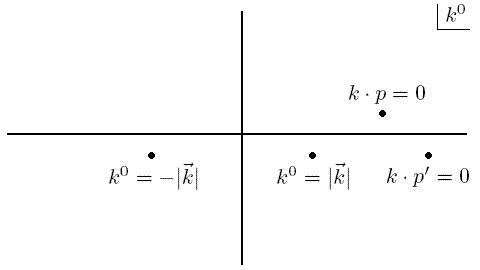}
\caption{\label{fig1}Pole structure of \eq{amu}.}
\efig

For $t<0$ we close the contour in the upper half of the plane, picking up the $k^0 = \vec{k}\cdot\vec{p}/p^0$ pole; the result is
\be
\label{amutgreater}
A^\mu(x) = \int \frac{d^3k}{(2\pi)^3} e^{i\vec{k}\cdot\vec{x}}e^{-i\left( \frac{\vec{k}\cdot\vec{p}}{p^0} \right) t}\frac{(2\pi i)}{2\pi} \frac{-ie}{k^2}\frac{-p^\mu}{p^0},
\ee
where the $k^0$ in the $k^2$ term is given by its value at the pole.  Note that the $1/p^0$ comes from correctly evaluating the contour integral using
\bea
\frac{p^\mu}{k\cdot p-i\epsilon} & = & \frac{p^\mu}{k^0p^0-\vec{k}\cdot\vec{p}-i\epsilon} \nonumber\\
& = & \frac{p^\mu}{p^0} \left[ \frac{1}{k^0-\frac{\vec{k}\cdot\vec{p}}{p^0}-i\epsilon} \right].
\eea

We can see that the $k\cdot p=0$ pole for $t<0$ corresponds to the Coulomb field of the particle by boosting to its rest frame; we have that $k^2=-|\vec{k}|^2$ (because $k\cdot p = 0 \; \Rightarrow  \; k^0 = 0$ for $p^\mu=(m,\vec{0})^\mu$), and \eq{amutgreater} becomes
\bea
A^\mu(x) & = & \int \frac{d^3k}{(2\pi)^3} e^{i\vec{k}\cdot\vec{x}} \frac{e}{|\vec{k}|^2}(1,\vec{0})^\mu \nonumber\\
& = & \frac{2\pi}{(2\pi)^3}e\underbrace{\int_{0}^{\infinity}dk\int_{0}^{\pi}\sin\theta d\theta e^{ikr\cos\theta}}_{\displaystyle{\pi/r}}(1,\vec{0})^\mu \nonumber\\
& = & \frac{1}{4\pi r}(1,\vec{0})^\mu.
\eea

For $t>0$ we will ignore the $k\cdot p' = 0$ pole because that merely gives the Coulomb field; radiation is produced by the $k^0=|\vec{k}|$ and $k^0 = -|\vec{k}|$ poles.  Since $k^2 = (k^0-|\vec{k}|)(k^0+|\vec{k}|)$ and $k\cdot p = k^0p^0 - \vec{k}\cdot\vec{p}$ we have that
\bea
A^\mu(x)
& = & \int \frac{d^3k}{(2\pi)^3}\frac{(-2\pi i)}{2\pi}(-ie)\left[ \phantom{\frac{p'^\mu}{|\vec{k}|p'^0-\vec{k}\cdot \vec{p}'}} \right. \nonumber\\
& & e^{-i|\vec{k}|t}e^{i\vec{k}\cdot\vec{x}}\frac{1}{2|\vec{k}|} \left( \frac{p'^\mu}{|\vec{k}|p'^0-\vec{k}\cdot \vec{p}'}-\frac{p^\mu}{|\vec{k}|p^0-\vec{k}\cdot \vec{p}} \right) \nonumber\\
& & + \left. e^{i|\vec{k}|t}e^{i\vec{k}\cdot\vec{x}}\frac{1}{-2|\vec{k}|} \left( \frac{p'^\mu}{-|\vec{k}|p'^0-\vec{k}\cdot \vec{p}'}-\frac{p^\mu}{-|\vec{k}|p^0-\vec{k}\cdot \vec{p}} \right) \right] \nonumber\\
& = & -e \int \frac{d^3k}{(2\pi)^3}\frac{1}{2|\vec{k}|}\left[ e^{-i|\vec{k}|t}e^{i\vec{k}\cdot\vec{x}} \left( \frac{p'^\mu}{|\vec{k}|p'^0-\vec{k}\cdot \vec{p}'} - \frac{p^\mu}{|\vec{k}|p^0-\vec{k}\cdot \vec{p}} \right) \right. \nonumber\\
& & + \left. e^{i|\vec{k}|t}e^{-i\vec{k}\cdot\vec{x}} \left( \frac{p'^\mu}{|\vec{k}|p'^0-\vec{k}\cdot \vec{p}'} - \frac{p^\mu}{|\vec{k}|p^0-\vec{k}\cdot \vec{p}} \right) \right],
\eea
where, to get to the last line, we were able to freely change $\vec{k}\rightarrow-\vec{k}$ because of the integration over all space.  Therefore we have that
\bea
\label{complexa}
A^\mu(x) & = & \Re \int \frac{d^3k}{(2\pi)^3}\mathcal{A}^\mu(k)e^{-ik\cdot x}, \\
\mathcal{A}^\mu(k) & = & -\frac{e}{|\vec{k}|} \left( \frac{p'^\mu}{k\cdot p'} - \frac{p^\mu}{k\cdot p} \right).
\eea
Note that we have implicitly taken $k^0 = |\vec{k}|$ and will continue to do so for the rest of the calculation.

If we define
\bea
E^i(x) & = & \Re \int \frac{d^3k}{(2\pi)^3}\mathcal{E}^i(x)e^{-ik\cdot x} \\
B^i(x) & = & \Re \int \frac{d^3k}{(2\pi)^3}\mathcal{B}^i(x)e^{-ik\cdot x}
\eea
and recall that
\bea
E^i & = & -F^{0i} \nonumber\\
& = & \partial^iA^0-\partial^0A^i \\
B & = & *F \nonumber\\
\Rightarrow B^i & = & \frac{1}{2}\epsilon^i_{\phantom{i}jk}F^{jk} \nonumber\\
& = & \frac{1}{2}\epsilon^i_{\phantom{i}jk} \left( \partial^jA^k - \partial^kA^j \right) = \epsilon^i_{\phantom{i}jk} \partial^jA^k \nonumber\\
& = & (\vec{\nabla}\times\vec{A})^i.
\eea
we then have
\bea
\label{complexe}
\vec{\mathcal{E}}(\vec{k}) & = & ik^0\vec{\mathcal{A}}(\vec{k}) - i\vec{k}\mathcal{A}^0(\vec{k}) \\
\label{complexb}
\vec{\mathcal{B}}(\vec{k}) & = & i\vec{k}\times\vec{\mathcal{A}}(\vec{k}).
\eea

Let's now prove some nice relations regarding $\mathcal{E}$ and $\mathcal{B}$ that will be useful later on: transversality and orthogonality.  To prove that $\vec{\mathcal{E}}$ is orthogonal to $\vec{k}$ recall that $\mathcal{A}^\mu\propto(p'^\mu/k\cdot p') - (p^\mu/k\cdot p)$.  Then
\bea
\vec{k}\cdot\vec{\mathcal{E}}(\vec{k}) & \propto & -i|\vec{k}|^2(\hat{k}\cdot\vec{\mathcal{A}}-\mathcal{A}^0) \nonumber\\
& \propto & -i|\vec{k}| \left[ \left( \frac{\vec{k}\cdot\vec{p}'}{k\cdot p'} - \frac{\vec{k}\cdot\vec{p}}{k\cdot p} \right) \right. \nonumber\\
& & \left. - \left( \frac{k^0p'^0}{k\cdot p'} - \frac{k^0p^0}{k\cdot p} \right) \right] \nonumber\\
& \propto & -i|\vec{k}| \left[ \frac{k\cdot p'}{k\cdot p'} - \frac{k\cdot p}{k\cdot p} \right] = 0.
\eea
Also
\bea
\vec{k}\times\vec{\mathcal{E}} & = & \vec{k}\times \left[ ik^0\vec{\mathcal{A}}-i\vec{k}\mathcal{A}^0 \right] \nonumber\\
& = & ik^0\vec{k}\times\vec{\mathcal{A}};
\eea
since $k^0 = \pm |\vec{k}|$, and we won't be interested in the sign, we have that
\be
\vec{\mathcal{B}} = \hat{k}\times\vec{\mathcal{E}}.
\ee

Let's use all our knowledge to find the energy radiated by this motion.  Recall that
\bea
E_\mathrm{EM} & = & \int d^3x T^{00}_\mathrm{EM} \nonumber\\
& = & \frac{1}{2}\int d^3x \left( |\vec{E}(\vec{x})|^2 + |\vec{B}(\vec{x})|^2 \right).
\eea
Consider first the electric piece:
\bea
E_\mathrm{E} & = & \frac{1}{2} \int d^3x |\vec{E}(\vec{x})|^2 \nonumber\\
& = & \frac{1}{8} \int d^3x \int\frac{d^3k}{(2\pi)^3} \int\frac{d^3k'}{(2\pi)^3} \left[ \ek e^{-ik\cdot x} + \estark e^{ik\cdot x}\right] \nonumber\\
& & \cdot \left[ \ekp e^{-ik'\cdot x} + \estarkp e^{ik'\cdot x}\right] \nonumber\\
& = & \frac{1}{8} \int\frac{d^3k}{(2\pi)^3} \left[ \ek\cdot\emk e^{-2ik^0t} + \estark\cdot\estarmk e^{2ik^0t} \right.\nonumber\\
& & \left. + 2 \ek\cdot\estark \right].
\eea
Now consider the magnetic piece:
\bea
E_\mathrm{B} & = & \frac{1}{2} \int d^3x |\vec{B}(\vec{x})|^2 \nonumber\\
& = & \frac{1}{8} \int d^3x \int\frac{d^3k}{(2\pi)^3} \int\frac{d^3k'}{(2\pi)^3} \left[ \hat{k}\times\ek e^{-ik\cdot x} + \hat{k}\times\estark e^{ik\cdot x}\right] \nonumber\\
& & \cdot \left[ \hat{k}'\times\ekp e^{-ik'\cdot x} + \hat{k}'\times\estarkp e^{ik'\cdot x}\right] \nonumber\\
& = & \frac{1}{8} \int\frac{d^3k}{(2\pi)^3} \bigg\{ \left[ \hat{k}\times\ek \right]  \cdot \left[(-\hat{k})\times\emk \right] e^{-2ik^0t} \nonumber\\
& & + \left[\hat{k}\times\estark\right] \cdot \left[(-\hat{k})\times\estarmk\right] e^{2ik^0t} \nonumber\\
& & + 2 \left[\hat{k}\times\ek\right] \cdot \left[\hat{k}\times\estark\right]\bigg\}.
\eea
We will now prove that $(\hat{k}\times\vec{\mathcal{E}})\cdot(\hat{k}\times\vec{\mathcal{E}}) = \vec{\mathcal{E}}\cdot\vec{\mathcal{E}}$.
\bea
\epsilon^{ijk}k^j\mathcal{E}^k\epsilon^{ilm}k^l\mathcal{E}^m & = & (\delta^{jl}\delta^{km}-\delta^{jm}\delta^{kl})k^j\mathcal{E}^kk^l\mathcal{E}^m \nonumber\\
& = & (\hat{k}\cdot\hat{k})\big(\vec{\mathcal{E}}\cdot\vec{\mathcal{E}}\big) - (\hat{k}\cdot\vec{\mathcal{E}})(\hat{k}\cdot\vec{\mathcal{E}}) \nonumber\\
& = & \vec{\mathcal{E}}\cdot\vec{\mathcal{E}}.
\eea
Therefore we find that the energy radiated is
\be
E_\mathrm{EM} = \frac{1}{2}\int\frac{d^3k}{(2\pi)^3}\ek\cdot\estark.
\ee

With two transverse polarizations $\vec{\epsilon}_\lambda(\vec{k})$ we have that
\bea
\ek\cdot\estark & = & \fe_1\fe^*_1+\fe_2\fe^*_2 = \sum_{\lambda=1,2}\fe_\lambda\fe^*_\lambda = \sum_{\lambda=1,2}|\vec{\epsilon}_\lambda(\vec{k})\cdot\ek|^2,\nonumber\\
& = & \left.k^0\right.^2\sum_{\lambda=1,2}|\vec{\epsilon}_\lambda(\vec{k})\cdot\vec{\fa}(\vec{k})|^2 = |\vec{k}|^2\sum_{\lambda=1,2}|\vec{\epsilon}_\lambda(\vec{k})\cdot\vec{\fa}(\vec{k})|^2.\nonumber\\
\eea

This leads us finally to
\be
E_\mathrm{EM} = \int\frac{d^3k}{(2\pi)^3}\frac{e^2}{2}\sum_{\lambda=1,2}\left|\vec{\epsilon}_\lambda(\vec{k})\cdot\left( \frac{\vec{p}'}{k\cdot p'} - \frac{\vec{p}}{k\cdot p} \right)\right|^2.
\ee

We would now like to show that one may replace the three vector quantities with their corresponding four vectors, then take $\sum \epsilon_\mu\epsilon^*_\nu\rightarrow-g_{\mu\nu}$.  If we orient $k^\mu = (k,0,0,k)^\mu$ along the $z$-direction, then we may take our two transverse polarizations as $\epsilon_1^\mu=(0,1,0,0)^\mu$ and $\epsilon_2^\mu=(0,0,1,0)^\mu$.  It is now clear that we may freely take the three vectors to four, as the zeroth components of the polarization vectors is zero.  To turn the sum over polarizations into a metric we need that the Ward identity be satisfied, $k_\mu j^\mu = k_\mu \fa^\mu = 0$.  Then, with our $k$ oriented along $z$,
\bea
k^\mu A_\mu = k A_0 + k A_3 = 0 \nonumber\\
\Rightarrow A_0 = -A_3.
\eea
Then
\bea
\sum \epsilon_\mu\epsilon^*_\nu A^\mu A^{*\nu} & = & \left|A_1\right|^2 + \left|A_2\right|^2 \nonumber\\
& = & -\left|A_0\right|^2 + \left|A_1\right|^2 + \left|A_2\right|^2 + \left|A_3\right|^2 \nonumber\\
& = & -g_{\mu\nu}A^\mu A^{*\nu}.
\eea

We immediately arrive at our final result:
\bea
\label{emrad}
E_\mathrm{EM} & = & \int\frac{d^3k}{(2\pi)^3}\frac{e^2}{2}(-g_{\mu\nu}) \left( \frac{p'^\mu}{k\cdot p'} - \frac{p^\mu}{k\cdot p} \right) \left( \frac{p'^\nu}{k\cdot p'} - \frac{p^\nu}{k\cdot p} \right) \nonumber\\
& = & \int\frac{d^3k}{(2\pi)^3}\frac{e^2}{2} \left( \frac{2p\cdot p'}{(k\cdot p')(k\cdot p)}-\frac{m^2}{(k\cdot p')^2}-\frac{m^2}{(k\cdot p)^2} \right).
\eea

\subsubsection{Proca}
When we include an external current in the Proca Lagrangian we have
\be
\mathcal{L}_\mathrm{Proca} = -\frac{1}{4}F^{\mu\nu}F_{\mu\nu} + \frac{\mu^2}{2}A^\mu A_\mu - j_\mu A^\mu.
\ee
The EOM are then
\be
\label{preoma}
\partial_\alpha F^{\alpha\beta}+\mu^2A^\beta = j^\beta.
\ee
For a conserved current, $\partial_\beta j^\beta = 0$, we have a consistency equation
\bea
\partial_\beta \left( \partial_\alpha\partial^\alpha A^\beta - \partial_\alpha\partial^\beta A^\alpha + \mu^2 A^\beta \right) & = & \partial_\beta j^\beta \nonumber\\
\Rightarrow \partial_\alpha\partial^\alpha \partial_\beta A^\beta - \partial_\beta\partial^\beta \partial_\alpha A^\alpha + \mu^2\partial_\beta A^\beta & = & 0 \nonumber\\
\Rightarrow \partial_\beta A^\beta & = & 0.
\eea
Our EOM, \eq{preoma}, become
\be
\partial^2 A^\beta + \mu^2 A^\beta = j^\beta.
\ee
Fourier analysis immediately leads one to
\be
A^\mu(x) = \int \frac{d^4k}{(2\pi)^4} e^{-ik\cdot x}\frac{-ie}{k^2-\mu^2}\left( \frac{p'^\mu}{k\cdot p'+i\epsilon} - \frac{p^\mu}{k\cdot p-i\epsilon} \right).
\ee
The poles in this occur at $k\cdot p' = 0$, $k\cdot p = 0$, and $k^2 - \mu^2 = 0$; the pole structure is exactly the same as for the massless case, \fig{fig1}, but with the radiation poles giving a dispersion relation of
\be
k^0_\pm = \pm\sqrt{\vec{k}^2+\mu^2}.
\ee

Let's examine the $t<0$ case and make sure we recover the correct Yukawa potential in the rest frame.  For $t<0$ we pick up only the $k\cdot p = 0$ pole.  Then
\bea
A^\mu(x) & = & ie\int \frac{d^3k}{(2\pi)^3} e^{-i\left( \frac{\vec{k}\cdot\vec{p}}{p^0} \right) t} e^{i\vec{k}\cdot\vec{x}} \left( \frac{2\pi i}{2\pi} \right) \frac{1}{-\vec{k}^2-\mu^2} \frac{ (p^0,\vec{0})^\mu}{p^0} \nonumber\\
& = & e(1,\vec{0})^\mu \frac{2\pi}{(2\pi)^3} \underbrace{\int_0^\infinity k^2dk \int_0^\pi \sin\theta d\theta \frac{1}{k^2+\mu^2} e^{ikr\cos\theta}}_{\displaystyle{\pi e^{\mu r}/r}} \nonumber\\
& = & \frac{e}{4\pi r}e^{-\mu r}(1,\vec{0})^\mu.
\eea

Let's now examine the $t>0$ solution.  Again we'll ignore the uninteresting $k \cdot p' = 0$ Yukawa pole contribution.  We find that
\bea
A^\mu(x) & = & -ie\frac{-2\pi i}{2\pi} \int \frac{d^3k}{(2\pi)^3} \left\{\phantom{\frac{p'^\mu}{k_+^0p'^0 - \vec{k}\cdot\vec{p}'}}\right. \nonumber\\
& & \frac{1}{2k^0_+} \left( \frac{p'^\mu}{k_+^0p'^0 - \vec{k}\cdot\vec{p}'} - \frac{p^\mu}{k_+^0p^0 - \vec{k}\cdot\vec{p}} \right)e^{-ik_+^0t}e^{i\vec{k}\cdot\vec{x}} \nonumber\\
& & \left. \frac{1}{2k^0_-} \left( \frac{p'^\mu}{k_-^0p'^0 - \vec{k}\cdot\vec{p}'} - \frac{p^\mu}{k_-^0p^0 - \vec{k}\cdot\vec{p}} \right)e^{-ik_-^0t}e^{i\vec{k}\cdot\vec{x}} \right\}.
\eea

We can therefore write $\pr{A}^\mu$ in the form of \eq{complexa} with
\be
\label{prcomplexa}
\mcpr{A} = \frac{-e}{k^0} \left( \frac{p'^\mu}{k\cdot p'} - \frac{p^\mu}{k\cdot p} \right),
\ee
where, from now on, we will implicitly take $k^0 = k^0_+$.  However because $k^\mu k_\mu = \mu^2$ instead of 0 as in the E\&M case we cannot immediately write $|\vec{E}|^2+|\vec{B}|^2$ as $\vec{\mc{E}}\cdot\vec{\mc{E}}^*$ as we did before.  Additionally, since we do not have gauge invariance like in the E\&M case, we do not have transversality of the fields; there are three polarization vectors (whose sum in fact obeys $\sum_{\lambda=1}^{3}\epsilon^\mu_\lambda\epsilon^{*\nu}_\lambda = -g_{\mu\nu}+k^\mu k^\nu/\mu^2$).  As a result we won't use any of the tricks exploited in the E\&M calculation; rather we will plug our solution of $\mcpr{A}$ into the formula for $\pr{E}$, \eq{prt}, and brute force manipulate it.

We are interested in finding quantities of the form $\int d^3x Q^2(x)$ where $Q(x) = \Re \int d^3k \mc{Q}(k)\exp(-ik\cdot x)/(2\pi)^3$.  Then
\bea
\int d^3 Q^2(x) & = & \frac{1}{4} \int d^3x \frac{d^3k}{(2\pi)^3} \frac{d^3k'}{(2\pi)^3} \bigg( \mc{Q}(k)e^{-ik\cdot x} + \mc{Q}^*(k)e^{ik\cdot x} \bigg) \nonumber\\
& & \phantom{\frac{1}{4} \int d^3x \frac{d^3k}{(2\pi)^3} \frac{d^3k'}{(2\pi)^3}} \times \bigg( \mc{Q}(k')e^{-ik'\cdot x} + \mc{Q}^*(k')e^{ik'\cdot x} \bigg) \nonumber\\
& = & \frac{1}{4} \int \frac{d^3k}{(2\pi)^3} \mc{Q}(k)\mc{Q}(-k)e^{-2ik^0t} + \mc{Q}^*(k)\mc{Q}^*(-k)e^{2ik^0t} \nonumber\\
& & \phantom{\frac{1}{4} \int \frac{d^3k}{(2\pi)^3}} + 2\mc{Q}(k)\mc{Q}^*(k) \nonumber\\
& = & \int\frac{1}{2} \frac{d^3k}{(2\pi)^3} \mc{Q}(k)\mc{Q}(-k) \cos(2k^0t) + \mc{Q}(k)\mc{Q}^*(k),
\eea
where we are taking $k=(k^0,\vec{k})$ and $-k=(k^0,-\vec{k})$.  We get the last line because, as we will see below, $\mc{A}^{*\mu} = \mc{A}^\mu \; \Rightarrow \; \mc{Q}(k)\mc{Q}(-k) = \mc{Q}^*(k)\mc{Q}^*(-k)$ for all the quantities of interest.

From \eq{complexe} we see that
\bea
\vec{\mc{E}}_k \cdot \vec{\mc{E}}_{-k} & = & \big[i(\zk\vmca{k} - \vk\zmca{k})\big]\big[i(\zk\vmca{-k}+\vk\zmca{-k})\big] \nonumber\\
& = & \zk\zmca{k}\vk\cdot\vmca{-k}+\vk^2\zmca{k}\zmca{-k}-\left.\zk\right.^2\vmca{k}\cdot\vmca{-k}-\zk\zmca{-k}\vk\cdot\vmca{k} \nonumber\\
& = & \vk^2\zmca{k}\zmca{-k}-\left.\zk\right.^2\vmca{k}\cdot\vmca{-k} \\
& = & \vec{\mc{E}}^*_k \cdot \vec{\mc{E}}^*_{-k},
\eea
and
\be
\vec{\mc{E}}_k \cdot \vec{\mc{E}}^*_{k} = \left.\zk\right.^2\vmca{k}^2 + \vk^2\left.\zmca{k}\right.^2 - 2\zk\zmca{k}\vk\cdot\vmca{k}.
\ee
From \eq{complexb} we find
\bea
\vec{\mc{B}}_k \cdot \vec{\mc{B}}_{-k} & = & \big[i(\vk\times\vmca{k})\big]\big[-i(\vk\times\vmca{-k})\big] \nonumber\\
& = & \vk^2\vmca{k}\cdot\vmca{-k} - \vk\cdot\vmca{k}\vk\cdot\vmca{-k} \\
& = & \vec{\mc{B}}^*_k \cdot \vec{\mc{B}}^*_{-k},
\eea
where we have used the vector identity $(\vec{a}\times\vec{b})\cdot(\vec{c}\times\vec{d}) = (\vec{a}\cdot\vec{c})(\vec{b}\cdot\vec{d})-(\vec{a}\cdot\vec{d})(\vec{b}\cdot\vec{c})$, and
\be
\vec{\mc{B}}_k \cdot \vec{\mc{B}}^*_{k} = \vk^2\vmca{k}^2-(\vk\cdot\vmca{k})^2.
\ee

Plugging these pieces into \eq{prt} we find that
\bea
\pr{E} & = & \frac{1}{4}\int\frac{d^3k}{(2\pi)^3} \left[ \vk^2\zmca{k}\zmca{-k}-\left.\zk\right.^2\vmca{k}\cdot\vmca{-k}+\vk^2\vmca{k}\cdot\vmca{-k}\right. \nonumber\\
& & + \left.(\vk\cdot\vmca{k})(\vk\cdot\vmca{-k})+\mu^2\zmca{k}\zmca{-k}+\mu^2\vmca{k}\cdot\vmca{-k}\right]\cos2\zk t \nonumber\\
& & + \left.\zk\right.^2\vmca{k}^2+\vk^2\left.\zmca{k}\right.^2-2\zk\zmca{k}\vk\cdot\vmca{k}+\vk^2\vmca{k}^2-(\vk\cdot\vmca{k})^2 \nonumber\\
& & + \mu^2\left.\zmca{k}\right.^2+\mu^2\vmca{k}^2.
\eea

First let's prove that the coefficient of the $\cos$ term is 0.  It's clear that the $\vmca{k}\cdot\vmca{-k}$ terms all cancel as $\vec{k}^2+\mu^2 = \zks$.  The remaining part is
\bea
\label{cancel}
\zks\zmca{k}\zmca{-k}-(\vk\cdot\vmca{k})(\vk\cdot\vmca{-k}) & = & (\zk\zmca{k}-\vk\cdot\vmca{k})(\zk\zmca{-k}+\vk\cdot\vmca{-k}) \nonumber\\
& & + \, \zk\zmca{-k}\vk\cdot\vmca{k}-\zk\zmca{k}\vk\cdot\vmca{-k} \nonumber\\
& = & (k^\mu\mc{A}_\mu)(\zk\zmca{-k}+\vk\cdot\vmca{-k}) \nonumber\\
& = & 0,
\eea
where we note that the pieces added to the end of the first line sum to zero.  
Using the definition of $k^0$ to combine terms we have the $\pr{E}$ integrand as
\bea
2\zks \vmca{k}^2 + \zks\left.\zmca{k}\right.^2&&\nonumber\\
-2\zk\zmca{k}\vk\cdot\vmca{k}-(\vk\cdot\vmca{k})^2 & = & 2\zks\vmca{k}^2-2(\vk\cdot\vmca{k})^2 \nonumber\\
& = & 2 \zks \left( \vmca{k}^2-\left.\zmca{k}\right.^2 \right) \nonumber\\
& = & -2 \zks \mathcal{A}^\mu\mathcal{A}_\mu,
\eea
where we have repeatedly exploited $k^\mu\mathcal{A}_\mu = 0$, which implies that
\be
\left(k^\mu \mathcal{A}_\mu \right)^2 = \zks\left.\zmca{k}\right.^2 - 2k^0\zmca{k}\vec{k}\cdot\vmca{k}+\left(\vec{k}\cdot\vmca{k}\right)^2 = 0.
\ee
With $\sum_{\lambda=1,2,3} \epsilon_\lambda^\mu\epsilon_\lambda^{\nu*}= -g^{\mu\nu}+k^\mu k^\nu/\mu^2$, $k^\mu \mathcal{A}_\mu = 0$, and \eq{prcomplexa} we finally arrive at
\bea
\pr{E} & = & 2\zks\sum_{\lambda=1,2,3}\left|\epsilon_\lambda\cdot\mathcal{A}\right|^2 \nonumber\\
\label{procarad}
& = & \int\frac{d^3k}{(2\pi)^3}\frac{e^2}{2}\sum_{\lambda=1,2,3}\left|\epsilon_\lambda\cdot\left( \frac{p'}{k\cdot p'} - \frac{p}{k\cdot p} \right)\right|^2 \\
& = & \int\frac{d^3k}{(2\pi)^3}\frac{e^2}{2} \left( \frac{2p\cdot p'}{(k\cdot p')(k\cdot p)}-\frac{m^2}{(k\cdot p')^2}-\frac{m^2}{(k\cdot p)^2} \right).
\eea

\section{Field Theory Calculation}
\subsection{Bremsstrahlung Radiation}
We wish to calculate the field theory analog of the previous classical calculations.  Following Peskin pgs.\ 182-183 \cite{Peskin:1995}, consider the diagrams in \fig{qftbrem} contributing to bremsstrahlung radiation.

\bfig[!ht]
\centering
\includegraphics[width=\columnwidth]{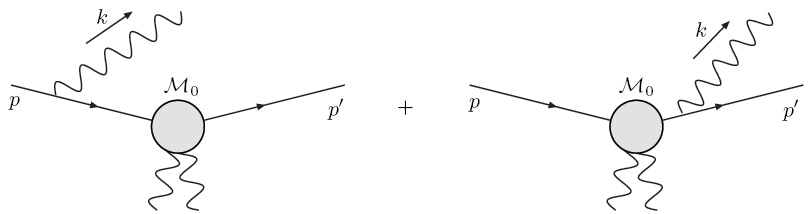}
\caption{\label{qftbrem}Diagrams contributing to the probability of emitting a bremsstrahlung photon.}
\efig

Then, taking $\mathcal{M}_0$ to denote the part of the amplitude that comes from the external field ``tickling'' the charge, we find that
\bea
i\mathcal{M} & = & -ie\bar{u}(p') \left(\mathcal{M}_0(p',p-k) \frac{i(\FMslash{p}-\FMslash{k}+m)}{(p-k)^2-m^2}\gamma\cdot\epsilon^*(k)\right. \nonumber\\
\label{qfteqn}
& & \phantom{ -ie\bar{u}(p') } + \left. \gamma\cdot\epsilon^*(k) \frac{i(\FMslash{p}'+\FMslash{k}+m)}{(p'+k)^2-m^2} \mathcal{M}_0(p'+k,p)\right)u(p).\nonumber\\
\eea
In order to make contact with the classical limit we assume a soft photon: $|\vec{k}|\ll |\vec{p}'-\vec{p}|$.  Then we can take
\be
\mathcal{M}_0(p'+k,p)\approx\mathcal{M}_0(p',p-k)\approx\mathcal{M}_0(p',p),
\ee
and we can ignore the $\FMslash{k}$ in the numerators.  

Using \eq{peskin} we have that \eq{qfteqn} becomes
\be
\label{softdijet}
i\mathcal{M} = \bar{u}(p')\big[\mathcal{M}_0(p',p)\big]u(p)\cdot\left[ e \left( \frac{p'\cdot\epsilon^*}{p'\cdot k} - \frac{p\cdot\epsilon^*}{p\cdot k} \right) \right].
\ee
Summing over polarization states, integrating over phase space, and taking the conditional probability yields
\be
\langle N_\gamma \rangle = \int\frac{d^3k}{(2\pi)^3}\frac{1}{2\omega}e^2\sum_{\lambda}\left|\epsilon_\lambda \cdot \left( \frac{p'}{k\cdot p'} - \frac{p}{k\cdot p} \right) \right|^2.
\ee

For massless QED we have that in a diagram that satisfies the Ward identity $\sum_{\lambda} \epsilon^\lambda_\mu \epsilon^{*\lambda}_\nu = -g_{\mu\nu}$.  Then we have that the energy radiated, to lowest order, is
\be
\label{qftradEM}
E^\mathrm{QFT}_\mathrm{EM} = \int\frac{d^3k}{(2\pi)^3}\frac{e^2}{2} \left( \frac{2p\cdot p'}{(k\cdot p')(k\cdot p)}-\frac{m^2}{(k\cdot p')^2}-\frac{m^2}{(k\cdot p)^2} \right),
\ee
in exact agreement with the classical E\&M result, \eq{emrad}.

For massive Proca we have the additional longitudinal polarization that the photon can radiate into; the polarization sum is modified to $\sum_{\lambda} \epsilon^\lambda_\mu \epsilon^{*\lambda}_\nu = -g_{\mu\nu}+k_\mu k_\nu/\mu^2$.  Additionally $k^2=m_\gamma^2\ne0$ in the propagators.  The soft photon limit corresponds to the limit of small $k^0,\vec{k}$, hence $m_\gamma^2\ll 2p\cdot k$.  To the same level of approximation, then, we ignore both the photon momentum in the numerators and the photon masses in the denominators of the propagators.  Then we have that
\bea
\label{qftrad}
E^\mathrm{QFT}_\mathrm{Proca} & = & \int\frac{d^3k}{(2\pi)^3}\frac{e^2}{2} \left( \frac{2p\cdot p'}{(k\cdot p')(k\cdot p)}-\frac{m^2}{(k\cdot p')^2}-\frac{m^2}{(k\cdot p)^2} \right. \nonumber\\
& & \phantom{\int\frac{d^3k}{(2\pi)^3}\frac{e^2}{2}} + \left. \frac{(k\cdot p')^2}{(k\cdot p')^2\mu^2} + \frac{(k\cdot p)^2}{(k\cdot p)^2\mu^2} - \frac{2 (k\cdot p')(k\cdot p)}{(k\cdot p')(k\cdot p)\mu^2} \right) \nonumber\\
& = & \int\frac{d^3k}{(2\pi)^3}\frac{e^2}{2} \left( \frac{2p\cdot p'}{(k\cdot p')(k\cdot p)}-\frac{m^2}{(k\cdot p')^2}-\frac{m^2}{(k\cdot p)^2} \right);
\eea
since the extra terms all cancel, we may use the equation interchangeably for either the massive or massless case.
\section{Comparison}
\subsection{Full Result: Massive Field, Charge, with Recoil}
We wish to make a connection to the zeroth order radiation emitted from the production of a hard parton at midrapidity.  To that end, let's consider the massive quark and field configuration first.  Using Ivan's notation: a four-vector is not bold-faced (e.g. $p$, $k$), a three-vector is bold-faced with a vector symbol (e.g. $\rv{p}$, $\rv{k}$), and a transverse two-vector is bold-faced without a vector symbol (e.g. $\wv{p}$, $\wv{k}$).  Minkowski four-vectors are written with parentheses, (); light-cone four-vectors with brackets, [].  In this notation our momenta of interest are
\bea
p & = & [\Mq,\Mq,\wv{0}] \\
p' & = & [(1-x)\Ep,\frac{\wv{k}^2+\Mq^2}{(1-x)\Ep},-\wv{k}] \\
k & = & [x\Ep,\frac{\wv{k}^2+\mg^2}{x\Ep},\wv{k}].
\eea

In order to compute the Proca result we must convert these into three vectors.  Doing so faithfully and using these in \eq{procarad} yields to lowest order, when expanding in powers of $1/\Ep$,
\be
\label{procabrem}
\frac{dE}{d^3k} = C \frac{4 (1-x)^2 \big( \wv{k}^2 + (1-x)^2\mg^2 \big)}{\big( \wv{k}^2 + (1-x)^2\mg^2 + x^2\Mq^2 \big)^2},
\ee
where $C = \big( 2 (2\pi)^3 \big)^{-1}$ is a constant that will be have the same value for the rest of this section.  There are two interesting features in this result that are different from Magda's zeroth order massive QCD result (ignoring the additional difference from the error in her published result that she corrected in her thesis).  The first is the overall factor of $(1-x)^2$ that does not reproduce Ivan's massless result, which is $C/\wv{k}^2$.  While this may be swept under the rug as $1-x\approx 1$ in the small-$x$ limit, we will actually recover this overall factor below.  The other difference is the $(1-x)^2\mg^2$ term in the numerator: this cannot be ignored.  Nevertheless this filling in of the ``dead cone'' in the massive field case is not so far-fetched; the additional longitudinal polarization could easily create a mode to emit radiation collinearly.

Plugging the above momenta into \eq{qftrad} and expanding in powers of $1/\Ep$, the lowest order QFT calculation gives
\be
\frac{dE}{d^3k} = C \frac{4 (1-x)^2 \big( \wv{k}^2 + (1-x)^2\mg^2 \big)}{\big( \wv{k}^2 + (1-x)^2\mg^2 + x^2\Mq^2 \big)^2},
\ee
exactly reproducing the classical limit.  This gives us quite a bit of confidence in our results.

\subsection{Massive Field and Charge, No Recoil}
Interestingly the result when there is no recoil,
\be
p' = [\Ep,\frac{\wv{k}^2+\Mq^2}{\Ep},-\wv{k}],
\ee
with the other momenta the same, is
\be
\frac{dE}{d^3k} = C \frac{4 \big( (1+x)^2 \wv{k}^2 + \mg^2 \big)}{\big( (1+x)^2\wv{k}^2 + \mg^2 + x^2\Mq^2 \big)^2}.
\ee
\subsection{Massive Charge and Massless Field with Recoil}
In the case of a massless field the QFT and classical results are necessarily the same; we have
\be
\frac{dE}{d^3k} = C \frac{4 (1-x)^2 \wv{k}^2}{\big( \wv{k}^2 + x^2\Mq^2 \big)^2}.
\ee
Everything is consistent with taking $\mg\rightarrow 0$ from our previous result.  Again we see the overall factor of $(1-x)^2$ out front.  We're tempted to simply take $\Mq\rightarrow 0$ to recover the completely massless limit.  However in the massless limit we must actually solve a completely different problem: for a massless charge we cannot take it to be at rest for $t<0$.  Perhaps of anecdotal interest the totally massless with recoil bremsstrahlung result, using
\bea
p & = & [\Ep,0,\wv{0}] \\
p' & = & [(1-x)\Ep,\frac{\wv{k}^2+\Mq^2}{(1-x)\Ep},-\wv{k}] \\
k & = & [x\Ep,\frac{\wv{k}^2+\mg^2}{x\Ep},\wv{k}],
\eea
is
\be
\frac{dE}{d^3k} = C \frac{x^2}{\wv{k}^2};
\ee
this is clearly not what we are interested in.

What we will do is to solve a problem more relevant to finding the zeroth order contribution than bremsstrahlung radiation: beta decay radiation (pair production).

\section{Pair Production}
\subsection{Classical}
We now take as our current a (charge) pair production
\be
j^\mu(x) = e\int d\tau \left[ \frac{dy_1^\mu(\tau)}{d\tau}\delta^{(4)}\big((x-y_1(\tau)\big) - \frac{dy_2^\mu(\tau)}{d\tau}\delta^{(4)}\big((x-y_2(\tau)\big) \right],
\ee
where we have
\be
y_{1,2}^\mu(\tau) = \left\{ \begin{array}{ll} 0, \qquad & \tau<0 \\ \frac{p_{1,2}^\mu}{m_{1,2}}\tau, \qquad & \tau>0 \end{array}\right.;
\ee
we have allowed ourselves the freedom to have both difference masses and different momenta for the two (opposite) charges.

The Fourier transform of our current is
\bea
\tilde{j}^\mu(k) 
& = & \int d^4x e^{ik\cdot x} j^\mu(x) \nonumber\\
& = & e \lim_{\epsilon\rightarrow0^+} \int_{0}^{\infinity}d\tau \left[ \frac{p_1^\mu}{m_1}e^{\tau \left( -\epsilon+i\frac{k\cdot p_1}{m_1} \right)} - \frac{p_2^\mu}{m_2}e^{\tau \left( -\epsilon+i\frac{k\cdot p_2}{m_2} \right)} \right] \nonumber\\
& = & ie \left[ \frac{p_1}{k\cdot p_1 + i\epsilon} - \frac{p_2}{k\cdot p_2 + i\epsilon} \right].
\eea

The only difference classically from the bremsstrahlung result is that the ``$k\cdot p = 0$'' pole is shoved below the real axis.  Since we throw away these poles anyway we know that we may safely use our previously derived equations for the radiated energy lost, \eq{emrad} and \eq{procarad}.  All we need to do is alter the $p$ momentum appropriately before plugging in.

\subsection{QFT}
We'd like to know whether consideration of pair production, \fig{qftbeta}, results in an amplitude different from the usual bremsstrahlung diagrams, \fig{qftbrem}.  The amplitude from \fig{qftbeta} is
\bea
i\mathcal{M} & \approx & -ie \bar{u}(p')\bar{u}(p)\gamma\cdot\epsilon^*(k)\left[ \frac{i(\FMslash{p}+\FMslash{k}+m)}{(p+k)^2-m^2}\right. \nonumber\\
& & \qquad - \left.\frac{i(\FMslash{p'}+\FMslash{k}+m)}{(p'+k)^2-m^2} \right] \mathcal{M}_0 u\cdots u,
\eea
where we see that the sign difference between the terms no longer comes from the switch in sign of the $(p-k)^2$ vs.\ the $(p'+k)^2$ term in the denominator but rather from the opposite sign of charge.  We may safely employ our previous derivation, \eqtwo{qftradEM}{qftrad}, using momenta associated with pair production.

\bfig[!ht]
\centering
\includegraphics[width=4.25 in]{QFTBetaDecayIIb2.png}
\caption{\label{qftbeta}Diagrams contributing to the probability of emitting a photon in a pair production event.}
\efig

\subsection{Comparison}
\subsubsection{Massive Field, Charge, with Recoil}
It turns out with
\bea
p & = & [\Mq^2/\Ep,\Ep,\wv{0}] \\
p' & = & [(1-x)\Ep,\frac{\wv{k}^2+\Mq^2}{(1-x)\Ep},-\wv{k}] \\
k & = & [x\Ep,\frac{\wv{k}^2+\mg^2}{x\Ep},\wv{k}],
\eea
the Proca result, \eq{procabrem}, is unchanged.
\subsubsection{Massless Field and Charge with Recoil}
With
\bea
p & = & [0,\Ep,\wv{0}] \\
p' & = & [(1-x)\Ep,\frac{\wv{k}^2}{(1-x)\Ep},-\wv{k}] \\
k & = & [x\Ep,\frac{\wv{k}^2}{x\Ep},\wv{k}],
\eea
the E\&M result is
\be
\frac{dE}{d^3k} = C \frac{4 (1-x)^2}{\wv{k}^2},
\ee
where we now find that the limiting form of the massive result is now recovered.

\clearpage

\end{document}